\documentclass[a4paper, 11pt, twoside,numbers		]{ThesisStyle}

\def\bra#1{\mathinner{\langle{#1}|}}
\def\ket#1{\mathinner{|{#1}\rangle}}

\usepackage{mathrsfs} 

\newcommand{\varone}{\mathcal{E}_1}
\newcommand{\vartwo}{\mathcal{E}_2}
\newcommand{\varthree}{\mathcal{B}}

\makeatletter

\newcommand\IntroMatname{Introductory Material}

\newcounter{IntroMat}
\renewcommand\theIntroMat{\roman{IntroMat}}
\@addtoreset{equation}{IntroMat}
\@addtoreset{table}{IntroMat}
\@addtoreset{figure}{IntroMat}

\newenvironment{IntroMat}[2][\IM@temp]
   {\gdef\IM@temp{#2}%
    \global\let\@chapapp@old=\@chapapp
    \global\let\thechapter@old=\thechapter
    \gdef\@chapapp{\IntroMatname}
    \gdef\thechapter{\theIntroMat}%
    \IM@chapt{#1}{#2}%
    \addtocontents{toc}{\protect\contentsline {chapter}{\vskip -2.8em \@plus-\p@}{}{}}
    \adjustmtc\minitoc%
   }%
   {%
    \global\let\@chapapp=\@chapapp@old
    \global\let\thechapter=\thechapter@old}

\newcommand\IM@chapt[2]{%
    \if@openright\cleardoublepage\else\clearpage\fi
    \thispagestyle{plain}%
    \global\@topnum\z@
    \@afterindentfalse
    \ifnum \c@secnumdepth >\m@ne
        \if@mainmatter
            \refstepcounter{IntroMat}%
            \addcontentsline{toc}{IntroMat}%
                {\protect\numberline{\IntroMatname~\theIntroMat:}#1}%
        \else
            \addcontentsline{toc}{IntroMat}{#1}%
        \fi
    \else
        \addcontentsline{toc}{IntroMat}{#1}%
    \fi
    \chaptermark{#1}%
    \addtocontents{lof}{\protect\addvspace{10\p@}}%
    \addtocontents{lot}{\protect\addvspace{10\p@}}%
    \if@twocolumn
        \@topnewpage[\@makechapterhead{#2}]%
    \else
        \@makechapterhead{#2}%
        \@afterheading
    \fi}

\newcommand*\l@IntroMat[2]{%
  \ifnum \c@tocdepth >\m@ne
    \addpenalty{-\@highpenalty}%
    \vskip 1.0em \@plus\p@
    \settowidth\@tempdima{\bfseries \IntroMatname~XX:~}%
    \begingroup
      \parindent \z@ \rightskip \@pnumwidth
      \parfillskip -\@pnumwidth
      \leavevmode \bfseries
      \advance\leftskip\@tempdima
      \hskip -\leftskip
      #1\nobreak\hfil \nobreak\hb@xt@\@pnumwidth{\hss #2}\par
      \penalty\@highpenalty
    \endgroup
  \fi}

\def\toclevel@IntroMat{0}

\makeatother
\usepackage{natbib}                       % Put References at the end of each chapter
\usepackage[english]{babel}
\usepackage[english]{layout}
\usepackage[utf8]{inputenc}
\usepackage[left=3.1cm,right=2.6cm,top=2.6cm,bottom=2.6cm,includehead,headheight=13.6pt]{geometry}
\usepackage{lscape}
\usepackage[normalem]{ulem}
\usepackage{aurical}% for Lukas Svatba
\usepackage{mathtools,amsfonts,amssymb,empheq,mathrsfs,amsmath}
\usepackage{icomma}                              % Pour les virgules en séparateur de décimale
\usepackage{relsize}                             % Pour la taille des symboles mathématiques
\usepackage{euscript}
\usepackage{cancel,nicefrac}                     % Barré et nicefrac
\DeclareMathAlphabet {\mathcal}{OMS}{cmsy}{m}{n} % Restaure mathcal par défaut (modifié par fourier => identique à mathscr)

\usepackage[locale = FR,detect-all,%
            range-units=single,%
            range-phrase=~\textendash~,%
            output-decimal-marker={,},%
            exponent-product={.},%
            retain-zero-exponent=true,%
            list-pair-separator ={ et },%
            list-final-separator={ et }]{siunitx}
\DeclareSIUnit\erg {erg}
\DeclareSIUnit\dyn {dyn}
\DeclareSIUnit\G   {G}
\DeclareSIUnit\Msol{\ensuremath{\mathrm{M_\odot}}}
\DeclareSIUnit\Rsol{\ensuremath{\mathrm{R_\odot}}}
\DeclareSIUnit\Lsol{\ensuremath{\mathrm{L_\odot}}}
\DeclareSIUnit\sr  {sr}
\DeclareSIUnit\uma {uma}
\DeclareSIUnit\j   {jour}
\DeclareSIUnit\js  {jours}
\DeclareSIUnit\an  {an}
\DeclareSIUnit\ph  {photon}
\DeclareSIUnit\phs {photons}

\usepackage{graphicx,booktabs,colortbl}
\usepackage{multirow}
\usepackage{rotating}                            % Sideways of figures & tables
\graphicspath{{Figures/}}
\usepackage[usenames,dvipsnames]{xcolor}
\usepackage{tikz,pifont}                % Pour les figures et les \ding
\usetikzlibrary{fadings,calc,decorations.pathmorphing}
\tikzset{snakearrow/.style={->,>=latex,decorate,decoration={snake,amplitude=.4mm,segment length=1mm,post length=1.4mm}}}
\tikzset{waveline/.style={decorate,decoration={snake,amplitude=0.5mm,segment length=8mm,pre length=2mm}}}
\usepackage[textfont={small}]{caption}        % Pour personnaliser les légendes
\definecolor{gray}    {gray}{0.3}
\definecolor{grayc}   {gray}{0.9}
\definecolor{darkblue}{rgb} {0,0,0.7} 
\definecolor{darkred} {rgb} {0.7,0,0} 

\usepackage{fancybox}                            % Cadre de texte
\usepackage{enumerate,enumitem}
\usepackage[final]{pdfpages}
\usepackage{framed,multicol}
\usepackage[multiple]{footmisc}
\newlength\leftbarsep
\usepackage{moreverb}
\usepackage{minitoc}
\mtcsettitle{minitoc}{Contents}
\usepackage[title,header,titletoc]{appendix}

\newcommand\starchapter[2]{\chapter*{#1\label{#2}}
                           \markboth{#1}{}
                           \addstarredchapter{#1}
                           \vskip -20pt
                           \noindent\hrulefill
                           \vskip 30pt}

\addto\captionsfrench{%
    \let\toctitre\contentsname
    \renewcommand{\contentsname}  {\textsc{\toctitre}}
    \let\loftitre\listfigurename
    \renewcommand{\listfigurename}{\textsc{\loftitre}}
    \let\lottitre\listtablename
    \renewcommand{\listtablename} {\textsc{\lottitre}}}

\usepackage{tabto}
\usepackage{setspace}

\let\minitocORIG\minitoc
\renewcommand{\minitoc}{\minitocORIG\vspace{1.5em}}
\usepackage{fancyhdr}                            % Fancy Header and Footer
\let\headruleORIG\headrule
\renewcommand{\headrule}{\color{black} \headruleORIG}

\usepackage{colortbl}
\arrayrulecolor{black}
\fancypagestyle{plain}{\fancyhead{}
                       \fancyfoot{}
                       }
\def\<{\ensuremath{\mskip-.5\thinmuskip}}
\newlength\subtl
\newlength\subbl
\makeatletter                                    % Clear Header Style on the Last Empty Odd pages
\def\cleardoublepage{\clearpage\if@twoside \ifodd\c@page\else%
  \hbox{}%
  \thispagestyle{empty}%                         % Empty header styles
  \newpage%
  \if@twocolumn\hbox{}\newpage\fi\fi\fi}
\makeatother
\newlength\lengindent
\PassOptionsToPackage{hyphens}{url}\usepackage[hyperindex=,hypertexnames=false]{hyperref}
\hypersetup{bookmarksopen=true,                  % Seehypertexnames=false hyperref documentation for information on those parameters
            pdftitle=" o",
            pdfauthor=" o", 
            pdfsubject=" o",                     % subject of the document
            pdfmenubar=true,                     % menubar shown
            pdfhighlight=/O,                     % effect of clicking on a link
            colorlinks=true,                     % couleurs sur les liens hypertextes
            pdfpagemode=None,                    % aucun mode de page
            pdfpagelayout=SinglePage,            %ouverture en simple page
            pdffitwindow=true,                   % pages ouvertes entierement dans toute la fenetre
            linkcolor=darkblue,                  % couleur des liens hypertextes internes
            citecolor=darkred,                   % couleur des liens pour les citations
            urlcolor=darkblue}                    % couleur des liens pour les url

    {\newpage\vspace*{\fill}\thispagestyle{empty}}
    {\vspace{\fill}}
\newcommand{\bibliosection}%{\null}
    {\null\vspace*{\fill}\bibliographystyle{apsrev4-1}
    {\footnotesize\bibliography{bibli.bib}}}
\renewenvironment{minipage}[1][]{\noindent\begin{minip}[#1]}{\end{minip}}

\newcommand*\cleartoleftpage{\clearpage\ifodd\value{page}\hbox{}\newpage\fi}

\usepackage[hyphenbreaks]{breakurl}
\usepackage{empheq}

\def\pablo#1{{\textcolor{blue}{\bf [#1]}}}        % addition by MB

\title{Discrete-time quantum walks and gauge theories\textendash}
\author{Pablo Arnault}

\newcommand{\motivations}[1]{%
\begin{tikzpicture}
\node[draw,thick,text width=14.05cm,inner sep=6mm] (titlebox)%
{\normalsize#1};
\node[fill=white] (MOTIVATIONS) at (titlebox.north) { \large MOTIVATIONS};
\end{tikzpicture}
}

\newcommand{\titlebox}[1]{%
\begin{tikzpicture}
\node[draw,thick,text width=14.05cm,inner sep=6mm] (titlebox)%
{\normalsize#1};
\node[fill=white] (ABSTRACT) at (titlebox.north) { \large ABSTRACT};
\end{tikzpicture}
}

\chardef\_=`_

\begin{document}

\abovedisplayskip6pt
\belowdisplayskip7pt
\begin{titlepage}

\setlength\parindent{0pt}
\null\vspace{-2cm}
\includegraphics[height=1.2cm]{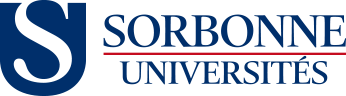} \hfill
\includegraphics[height=1.2cm]{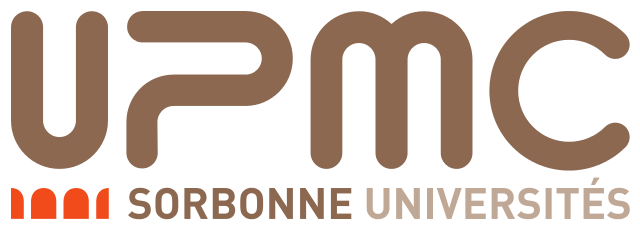} \hfill
\includegraphics[height=1.5cm]{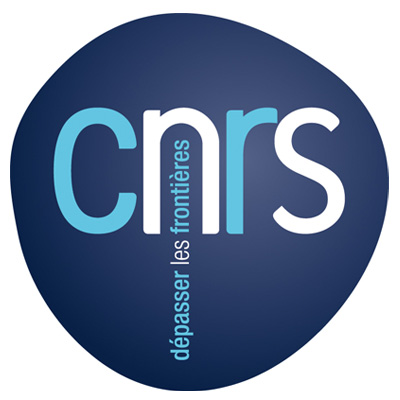} \hfill
\includegraphics[height=1.5cm]{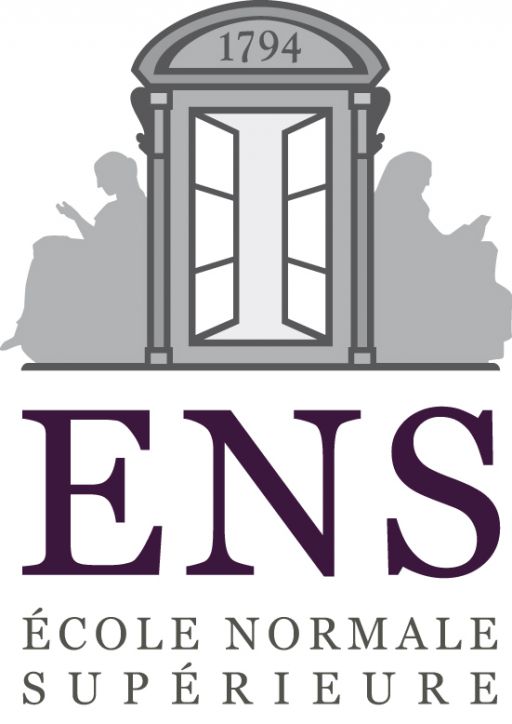} \hfill
\includegraphics[height=1.5cm]{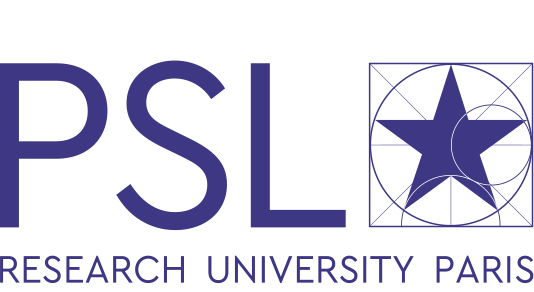} \hfill \\
\begin{center}
\vspace{-0.6cm}
\includegraphics[height=1.4cm]{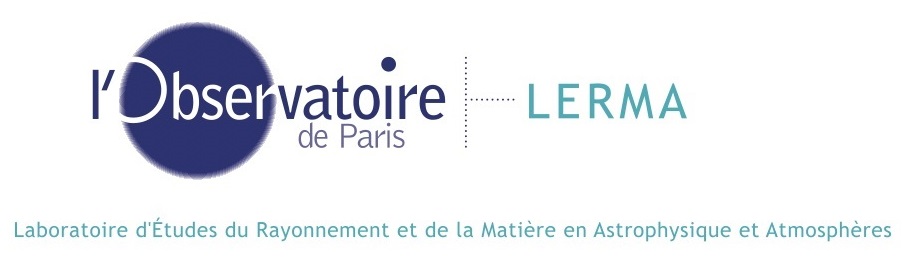} \ \
\includegraphics[height=1.4cm]{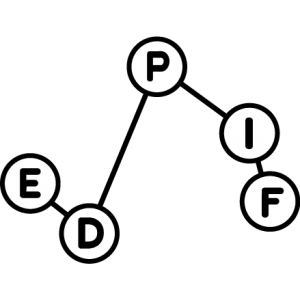} \ \
\includegraphics[height=1.4cm]{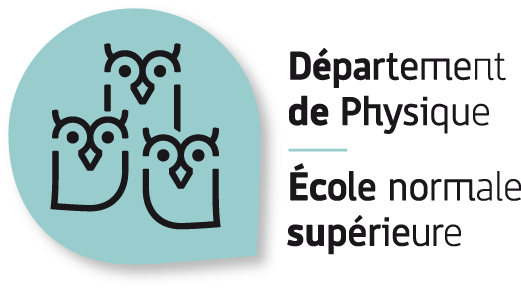}
\end{center}

\begin{center}
\vspace{9mm}
{\huge Université Pierre et Marie Curie} \\
{\large (UPMC)} \\
\vspace{3mm}
\begin{large}
École Doctorale de Physique en \^Ile-de-France \\
(EDPIF) \\
\vspace{3mm}  
{\itshape \normalsize Laboratoire d'Études du Rayonnement et de la Matière en Astrophysique et Atmosphères} (LERMA UMR8112)
\end{large}
\vspace{14mm}
        
\textbf{{ \Huge Discrete-time quantum walks and \\ gauge theories}}
\vspace{5mm}
\begin{large}
            ~\\by Pablo Arnault\\ 
            \vspace{17mm}
            PhD thesis in Theoretical Physics \\
            based on four scientific publications \cite{AD15,AD16,ADMDB16,AD17}
            \vspace{3mm}
            
            Supervisor: Fabrice Debbasch (LERMA, UPMC)\\
            Co-supervisor: Marc Brachet (LPS ENS, CNRS)\\
        \end{large}
        \vspace{3mm}  
        \normalsize
        Defended publicly on 18 September 2017\\
    at the Université Pierre et Marie Curie, in front of the following jury:\\~\\
        \begin{tabular}{@{}l@{~~}l@{$\qquad$}c@{$\qquad$}r@{}}\arrayrulecolor{blue!40}
            \addlinespace
            \toprule[.5pt]
            \addlinespace
Ms. & Martine Ben Amar   &     Professor          & President \\
            \addlinespace
            Mr. &  Armando Pérez   & Professor  & Rapporteur \\
            Mr. &  Pablo Arrighi   &     Professor       & Rapporteur \\
            \addlinespace
            Mr. &  Andrea Alberti  &    Principal Investigator  & Examiner \\
            \addlinespace
            Mr. & Fabrice Debbasch   &    ``Maître de Conférence''       & Thesis supervisor \\
            Mr. & Marc Brachet  & ``Directeur de Recherche'' & (Invited) Thesis co-supervisor \\
            %\addlinespace
            %Mr. & Yutaka Shikano & Associate Professor & \emph{Invited} \\
            \bottomrule[.5pt]            
            \addlinespace
        \end{tabular}
        \vspace{0.5cm}
        \vfill
        \begin{minipage}[b]{2.5cm}
            \includegraphics[width=\linewidth]{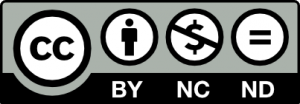}    
        \end{minipage}$\quad$\begin{minipage}[b]{.68\linewidth}
            Except where otherwise noted, this work is licensed under\\
            \url{http://creativecommons.org/licenses/by-nc-nd/3.0/}\vspace{0pt}\null
        \end{minipage}
    \end{center}

\end{titlepage}
\cleardoublepage

\cleardoublepage

\thispagestyle{empty}

\vspace*{\fill}

\vspace{-5cm}

\null\hfill{ \Large{``~Trachte~ich~denn~nach~\emph{Gl\"ucke}?~Ich~trachte~nach~meinem~\emph{Werke}!~'' }}

\null\hfill{In \emph{Also sprach Zarathustra}, by F. Nietzsche, last chapter, ``Das Zeichen". }

\vspace*{\fill}

\cleardoublepage

\thispagestyle{empty}

\vspace*{\fill}

\vspace{-5cm}

\null\hfill{ \Large{ \emph{A mis  padres}}}

\vspace*{\fill}

\cleardoublepage

\newgeometry{left=3.1cm,right=2.6cm,top=0.9cm, bottom=1.2cm}

\thispagestyle{empty}
\null\hfill{\Large\bf Aknowledgments} \\

\noindent
I first thank Fabrice, for being himself, for his energy. He has been a good master. With his help, I improved my mastery of this delicate and precise tool that rationality is. I learned, day after day, how to make it a first-rank partner to develop ideas, but also deeply nourish them, i.e. how to, thanks to reasoning, overcome present intuition, how to trivialize knowledge and make it join this bath of fluctuations that one's intuition is. This dynamics, that is, one's daily growth by these juggles between induction and deduction, the associated captures, and eventual metamorphoses of oneself, is to me a source of the highest pleasures. I also thank Marc for his great kindness and wisdom. I joyfully join with Fabrice's and Marc's curiosity and extrovert enthusiasm for research.

I thank the members of my jury, Martine Ben Amar, Armando Pérez, Pablo Arrighi, and Andrea Alberti, for the following reasons. Martine, for her enthusiastic and rich-in-physics course on non-linear physics, for her human qualities, and for her enthusiasm for  presiding my jury. Armando, for accepting with great simplicity and enthusiasm to join the jury as a rapporteur. Pablo, for his inspired research, that I know too little and in which I will be happy to dig in the near future, for his most serene dynamism, and for accepting to be a rapporteur for my thesis. Eventually, I express my most profound gratitude and admiration to Andrea, obviously -- regarding the first item -- for inviting me to join his group for two months, but beyond that, for his dazzlingly clever and dynamic leadership; he is, both as a scientist and as a human being, one of my greatest (living) sources of inspiration. I am also indebted to Dieter Meschede for the above-mentioned invitation. I warmfully thank Yutaka Shikano for his invitation to join him in Okazaki, and for organizing the Workshops on Quantum Simulation and Quantum Walks. I also thank the Japan Society for the Promotion of Science for the great opportunities it offers to foreigners.

I express my most sincere gratitude to Giuseppe Di Molfetta, former student of Fabrice, for his energetic support and wise advice during these three years. I am also indebted to Lionel De Sá for his extremely efficient and repeated help regarding practical computing issues. I eventually thank Claire Chevallier for helping me to prepare several presentations. I also wish to thank other members of my lab, that I see -- or used to see -- daily, for the following reasons. Chantal Stehle, for her benevolence and advice. Andrea Sciardi, for his serenity, humor, empathy and kindness. Dorian Zarhoski for his great sympathy and kindness. Xavier Fresquet, for his dynamism, curiosity, sympathy and professionalism. Of course, I thank my lab mates, Benjamin Khiar, Loïc Nicolas, Mathieu Drouin and Julien Guyot, for all these moments of fun we've had together and for all our discussions, more or less  senseful, but always sincere. I eventually thank Franck Delahaye for his zest for life and very fun company, and Nora Roger for her dynamism and sympathy.

I would also like to thank several friends. Justin Gabriel and Florencia Di Rocco, for their genius; they are a most inspiring source in how to live my researcher's life, and their company fills me with joy. Guillaume Ferron, for his never-failing friendship, strength and tenderness. Pierre Hollier-Larousse, for the many moments we shared in our early studies, for the completeness of his friendship and his support. Jérémy Saâda, extraordinary neighbor, for his unsettling cleverness and kindness, and for his repeated help with (too) {many} everyday life issues; he also inspires me as a human being. Baptiste Pecorari for his imagination and musical creativity, Amaury Dumoulin for his literary virtuosity, and both for their deep sense of friendship.

I cannot thank my parents better than by loving them from my deepest inner self, as an adult. Besides the infinite love and tenderness that each of them has filled me with, I particularly thank my father for the model of creativity he has been during my childhood and, oh, so much, he still is, as well as my mother for the admirable synthesis of love and rigor that she has achieved in my education, the sense of justice she has taught me, the wisdom of her advice, her constant and energetic support, and the viscerality of her worries for my well-being. 

My mind eventually lies on Sophie Elzière.

\restoregeometry

\cleardoublepage

\newgeometry{left=2.0cm,right=1.6cm,top=0.7cm, bottom=0.7cm}

\setcounter{footnote}{0}

\thispagestyle{empty}
\hfill{\Large \bf Remerciements} \\

\noindent
Je remercie tout d'abord Fabrice, pour celui qu'il est, pour son énergie. Il a été un bon maître. Avec son aide, j'ai perfectionné ma maîtrise de cet outil délicat et précis qu'est la rationalité. Jour après jour, cette-dernière et moi-même avons renforcé notre partenariat, ma raison m'a appris à l'apprivoiser, et a encouragé ce coéquipier de premier rang, indispensable au scientifique ; j'ai compris que cette rationalité n'{\oe}uvrait pas seulement au développement d'une idée donnée, mais contribuait aussi profondément à l'émergence de nouvelles idées  ; j'ai en effet appris, par sa pratique, à dépasser l'intuition présente, à trivialiser la connaissance pour lui faire rejoindre ce grand bain de fluctuations, multi-échelles à l'occasion, qu'est l'intuition de chacun. Cette dynamique, à savoir, la croissance quotidienne du soi par ces jongles entre induction et déduction, les captures associées, et nos éventuelles métamorphoses, est pour moi une source des plus hauts plaisirs. Je remercie également Marc pour sa grande gentillesse et sa sagesse, et me joins enfin allègrement à l'enthousiasme extraverti que lui et Fabrice partagent pour la recherche.

Je remercie les membres de mon jury, Martine Ben Amar, Armando Pérez, Pablo Arrighi, et Andrea Alberti, pour les raisons suivantes. Martine, pour son cours de physique non linéaire, enthousiaste et riche de phénomènes physiques, pour ses qualités humaines, et enfin pour son enthousiasme à l'idée de présider mon jury. Armando, pour avoir accepté avec simplicité et enthousiasme d'être rapporteur de mon manuscrit. Pablo, pour sa recherche inspirée, que je connais trop peu et dans laquelle je serais heureux de me plonger mieux dans un avenir proche, et pour son dynamisme des plus sereins. Enfin, j'exprime ma plus profonde gratitude et admiration envers Andrea, évidemment -- en ce qui concerne le premier point -- pour m'avoir invité à rejoindre son groupe pendant deux mois, mais surtout, plus largement, pour sa direction de groupe à l'intelligence et au dynamisme éblouissants ; il est, aussi bien en tant que scientifique qu'être humain, l'une de mes plus grande sources d'inspiration (vivantes). J'exprime également ma gratitude à Dieter Meschede pour l'invitation sus-mentionnée. Je remercie chaleureusement Yutaka Shikano pour son invitation à le rejoindre à Okazaki, et pour l'organisation des Séminaires sur la Simulation Quantique et les Marches Quantiques. Je remercie également la Société Japonaise pour la Promotion de la Science, pour les précieuses opportunitées qu'elle offre aux étrangers.

J'adresse un grand merci à Giuseppe Di Molfetta, ancien doctorant de Fabrice, pour son soutien énergique et les sages conseils qu'il m'a donnés pendant ces trois années. J'exprime toute ma gratitude à Lionel De Sá pour son aide répétée et des plus efficaces dans la résolution de problèmes informatiques pratiques. Je remercie aussi Claire Chevallier pour son aide à la préparation de diverses présentations. Je tiens également à remercier d'autres membres de mon laboratoire que je cotoie -- ou cotoyais -- quotidiennement, pour les raisons suivantes. Chantal Stehle, pour sa bienveillance et ses conseils. Andrea Sciardi, pour sa sérénité, son humour, son empathie et sa gentillesse. Dorian Zahorski, parti maintenant depuis plus d'un an pour d'autres horizons, pour sa grande sympathie et sa gentillesse. Xavier Fresquet, pour son dynamisme, sa curiosité, sa sympathie et son professionalisme. Bien entendu, je remercie mes copains de labo, Benjamin Khiar, Loïc Nicolas, Mathieu Drouin et Julien Guyot, pour tous les moments de rigolade que nous avons passés ensemble et toutes nos discussions, plus ou moins sensées, mais toujours sincères. Je remercie enfin Franck Delahaye pour sa joie de vivre et sa très sympathique compagnie, ainsi que Nora Roger pour son dynamisme et sa sympathie.

J'aimerais également remercier plusieurs amis. Justin Gabriel et Florencia Di Rocco, pour leur génie~; il sont une source des plus inspirantes pour ma vie de chercheur, et leur compagnie m'emplit de joie. Guillaume Ferron, pour son amitié inaltérable, sa force et sa tendresse. Pierre Hollier-Larousse, pour tout ce que nous avons partagé pendant nos jeunes années d'étude, pour la complétude de son amitié et son soutien. Jérémy Saâda, voisin extraordinaire, pour son intelligence et sa gentillesse déconcertantes, et pour son aide  répétée pour de (trop) nombreux soucis du quotidien ; il m'inspire également en tant qu'être humain. Baptiste Pecorari pour son imagination et sa créativité musicale, Amaury Dumoulin pour sa virtuosité littéraire, et tous deux pour leur sens profond de l'amitié.

Je ne peux mieux remercier mes parents qu'en les aimant du plus profond de mon être, en adulte. Outre l'amour et la tendresse infinis dont chacun d'eux m'a comblé, je remercie tout particulièrement mon père pour le modèle de créativité qu'il a été, pendant toute mon enfance, et qu'il est, ô combien, encore aujourd'hui pour moi, ainsi que ma mère pour l'admirable synthèse d'amour et de rigueur qu'a été son éducation, le sens de la justice qu'elle m'a inculqué, la sagesse de ses conseils, son soutien constant et énergique, et la viscéralité de son souci pour mon bien-être. 

Enfin, mon esprit se pose sur Sophie Elzière.

\restoregeometry
\cleardoublepage

\thispagestyle{empty}
\null\hfill{\Large\bf Abstract} \\

\noindent
A quantum computer, i.e. utilizing the resources of quantum physics, superposition of states and entanglement, could furnish an exponential gain in computing time. A simulation using such resources is called a quantum simulation \cite{Trabesinger2012, Schaetz2013, Georgescu2014, Johnson2014}. The advantage of quantum simulations over classical ones is well established at the theoretical, i.e. software level \cite{Childs2009, CGW13, Arrighi2011,Arrighi2012,Wiese2013,Zohar2013, Banerjee2013, Rico2014,dimolfetta:tel-01230891,Succi2015, Dalibar2015, Hamilton2016,AD17}. Their practical benefit requires their implementation on a quantum hardware. The quantum computer, i.e. the universal one (see below), has not seen the light of day yet, but the efforts in this direction are both growing and diverse. Also, quantum simulation has already been illustrated by numerous experimental proofs of principle, thanks too small-size and specific-task quantum computers or simulators \cite{Bloch2012, scholl:tel-01165961, Aidelsburger2016, Martinez16, OMalley2016, Tame2014}. Quantum walks (QWs) are particularly-studied quantum-simulation schemes, being elementary bricks to conceive any quantum algorithm, i.e. to achieve so-called universal quantum computation \cite{Arrighi2011,Arrighi2012, Childs2009,CGW13}. Eventually, at a purely theoretical level, quantum simulation already provides a broad understanding of many quantum dynamics in terms of (quantum) information-theoretic principles.

The present thesis is a step more towards a simulation of quantum field theories based on discrete-time QWs (DTQWs). First, it is indeed shown, in certain particular cases, how DTQWs can simulate, in the spacetime-continuum limit, the action of a Yang-Mills gauge field on fermionic matter, and the retroaction of the latter on the gauge-field dynamics \cite{AD15,AD16,ADMDB16}. The suggested schemes preserve gauge invariance on the spacetime lattice, i.e. not only in the continuum. In the (1+2)-dimensional Abelian case, lattice gauge-invariant equivalents to Maxwell’s equations are suggested, which are consistent with the current conservation on the lattice; also, outside the continuum limit, phenomena typical of the presence of a lattice appear, such as Bloch oscillations. In the (1+1)-dimensional non-Abelian case, a lattice gauge-covariant version of the non-Abelian field strength is suggested; also, short-time agreement with classical (i.e. non-quantum) trajectories is shown. Second, it is shown how this DTQWs-based fermionic matter can be coupled to relativistic gravitational fields of the continuum, i.e. to curved spacetimes, in 1+2 dimensions \cite{AD17}; an application is presented, where the 2-dimensional space is the polarization plane of a linear plane gravitational wave.

\cleardoublepage

\thispagestyle{empty}
\null\hfill{\Large\bf Résumé} \\
\setcounter{footnote}{0}

\noindent
Un ordinateur quantique, i.e. utilisant les ressources de la physique quantique, superposition d'états et intrication, pourrait fournir un gain exponentiel de temps de calcul. Une simulation utilisant ces ressources est appelée simulation quantique \cite{Trabesinger2012, Schaetz2013, Georgescu2014, Johnson2014}. L’avantage des simulations quantiques sur les simulations classiques est bien établi au niveau théorique, i.e. software \cite{Childs2009, CGW13, Arrighi2011,Arrighi2012, Wiese2013, Zohar2013, Banerjee2013, Rico2014,dimolfetta:tel-01230891, Succi2015,Dalibar2015,Hamilton2016, AD17}. Leur avantage pratique requiert leur implémentation sur un hardware quantique. L’ordinateur quantique, sous-entendu universel (cf. plus bas), n’a pas encore vu le jour, mais les efforts en ce sens sont croissants et variés. Aussi la simulation quantique a-t-elle déjà été illustrée par de nombreuses expériences de principe, grâce à des calculateurs ou simulateurs quantiques de taille réduite et à tâche spécifique \cite{Bloch2012, scholl:tel-01165961, Aidelsburger2016, Martinez16, OMalley2016, Tame2014}. Les marches  quantiques (MQs) sont des schémas de simulation quantique particulièrement étudiés, étant des briques élémentaires pour concevoir n’importe quel algorithme quantique, i.e. pour le dénommé calcul quantique universel \cite{Arrighi2011, Arrighi2012, Childs2009, CGW13}. À un niveau purement théorique, la simulation quantique fournit déjà une large compréhension de nombreuses dynamiques quantiques à partir de principes de théorie de l'information (quantique).

La présente thèse est un pas de plus vers une simulation des théories quantiques des champs basée sur les MQs à temps discret (MQTD). Dans un premier temps, il est en effet montré, dans certains cas particuliers, comment les MQTD peuvent simuler, à la limite au continuum d'espace-temps, l'action d'un champ de jauge Yang-Mills sur de la matière fermionique, et la rétroaction de cette-dernière sur la dynamique du champ de jauge \cite{AD15,AD16,ADMDB16}. Les schémas proposés préservent l’invariance de jauge au niveau du réseau d’espace-temps, i.e. pas seulement au continuum. En dimension 1+2 d'espace-temps et dans le cas abélien, des équations de Maxwell invariantes de jauge sur réseau sont proposées, compatibles avec la conservation du courant sur le réseau ; apparaissent aussi, en dehors de la limite au continuum, des phénomènes typiques de la présence d'un réseau, tels que des oscillations de Bloch. En dimension 1+1, une courbure non-abélienne, définie et covariante de jauge sur le réseau, est proposée. Dans un deuxième temps, il est montré comment cette matière fermionique à base de MQTD peut être couplée à des champs gravitationnels relativistes du continuum, i.e. à des espaces-temps courbes, en dimension 1+2 \cite{AD17} ; une application est présentée, où l'espace de dimension 2 est le plan de polarisation d'une onde gravitationnelle linéaire plane.

\cleardoublepage

\thispagestyle{empty}
\null\hfill{\Large\bf Notes on the typography used in the manuscript}\\

\noindent
Books, scientific memoirs or reports, Master or PhD theses, are noted in italic font.

The doubled quotation marks, ``...'', are used either to quote somebody else's words, or the name of a scientific paper, and are always associated to a reference -- unless sufficiently well-known --, while the non-doubled ones, `...', are used for two different -- although obviously non-orthogonal -- purposes: (i) to stress that the quoted word or group of words is being explained, or (ii) to stress that the quoted entity is not used in its most-used meaning, and/or to warn the reader not to understand this linguistical entity too rigorously or formally, because it is merely used to give some intuition of what is being explained.
The reader should be able to determine, from the context, in which situation we are.

The word `$N$-dimensional' will be abbreviated `$N$D' throughout the manuscript.
\cleardoublepage

\thispagestyle{empty}
\null\hfill{\Large\bf Notes on arXiv's v2 of this manuscript}\\

\noindent
Modification of (i) the first paragraph of Section \ref{sec:Checkerboard_intro}, and of (ii) Section \ref{subsec:Feynman's_quantum_walks}, which are both about Feynman's checkerboard. 
These modifications have been triggered by a conversation I had with Ted Jacobson in November 2017.
In Section \ref{subsec:Feynman's_quantum_walks}, I have in particular modified several of my statements regarding Jacobson's work, which were (sometimes very) misleading, and given important precisions on his work.
The take-away message of the modifications taken altogether is that Bialynicki-Birula's 1994 (1+3)D discretization \cite{BB94a}, is a seminal DTQW (although this name is not used in the paper) which introduces, in particular, the now well-known coin operations with angular parameters, and has thus in-built unitarity, contrary to both Feynman's original scheme and quantum lattice Boltzmann methods, in which one must normalize the scheme to ensure unitarity (regarding this matter, see the 2015 paper by Succi, ``Quantum Boltzmann is a quantum walk'' \cite{Succi2015}).
Bialynicki-Birula's work -- and followings, including the work by G. M. D'Ariano's team, the work by C. M. Chandrasekhar, and of course the 3D generalization (not published) of the results of the present thesis --, is currently one of the most interesting (1+3)D generalizations of Feynman's original checkerboard (see Section \ref{subsec:Feynman's_quantum_walks} for more details).
The differences between D'Ariano's schemes and those of the present thesis are studied in Perez' 2016 work, see \url{https://doi.org/10.1103/PhysRevA.93.012328}.
From 2017 to today, 11 November 2020, a lot of work has been done regarding the multiparticle upgrade of such schemes: see the works by Arrighi et al., \url{https://doi.org/10.1007/s11128-019-2555-4}, Bisio et al., \url{https://arxiv.org/abs/1912.09768}, Farrelly and Streich, \url{https://arxiv.org/abs/2002.02643}.
Let us also not forget the very extensive and more computational-efficiency oriented work by Jordan et al., \url{https://arxiv.org/abs/1112.4833}, which is however not based on QCA.

\vspace{0.01cm}

\null\hfill{\emph{Paris, 11 November 2020}}

\cleardoublepage

\thispagestyle{empty}
\null\hfill{\Large\bf Notes on arXiv's v3 of this manuscript}\\

\noindent
Modification of (i) the last sentence of the first paragraph of Appendix \ref{app:on_quantum_non_locality}, starting by ``The current overwhelmingly-dominant vision, [...]'', and ending now by ``[...] as classically understood.'', as well as of (ii) the very beginning of Footnote 1 of the corresponding page.
These modifications aim at correcting a huge epistemic error which is quite typical and that I had made: the uncertainty principle of quantum mechanics must not be understood as a limitation of our ability to know the objects of Nature (here, position and momentum), it changes the ontology of these objects (that is, position and momentum as classically understood \emph{do not exist} prior to measurement).
Of course, if we try to formulate the results of quantum mechanics with classical notions, we will end up with epistemic nonsenses such as that into which I ran, as well as many others do.
I thank Étienne Klein for reminding me these facts very clearly in an interesting talk of his (in French), on quantum mechanics and the nature of objects, see \url{https://youtu.be/EtswlNoCogU}.
The part which deals with the issue I have mentioned runs from min 15:29 to min 20:14.
I have used Klein's terminology ``ontologic revolution'' to speak about quantum mechanics in my two new sentences.
Klein also reads, in his talk, a beautiful and interesting quote by Nietzsche, from min, let's say 0:00 to have an introduction of the quote, to min 4:34.

\vspace{0.01cm}

\null\hfill{\emph{Paris, 31 October 2021}}

\cleardoublepage

\dominitoc
\pagenumbering{roman}

\tableofcontents
\cleardoublepage

\iffalse
\listoffigures
\cleardoublepage

\listoftables
\cleardoublepage
\fi

\mainmatter

%%%%%%%%%%%%%%%%%%%%%%%%%%%%%%
%%  Alias bibliographiques  %%
%%%%%%%%%%%%%%%%%%%%%%%%%%%%%%
\defcitealias{NRFortran}{Numerical Recipes}
\defcitealias{NRF77}{Numerical Recipes}
\defcitealias{NRF90}{Numerical Recipes}
\defcitealias{IUPAC}{IUPAC}%International Union of Pure and Applied Chemistry

\starchapter{\textsc{Introduction}}{Chap:Intro}
%
\iffalse
This thesis is essentially written in a physicist's style, but contains some mathematical and computational touches of paint well suited to the interdisciplinarity of quantum simulation and quantum information. The directing criterion in the organization of the thesis is modularity. Each part is associated to a given research axis, and contains introductory material. The introductory material contains 
\fi
%
\setcounter{footnote}{0}

\noindent
The guideline of the research work on which this thesis is based is the discretization of relativistic quantum\footnote{The research work presented in this thesis deals only with \emph{first-quantized} field theories, which are actually often referred to as \emph{classical}. The second quantization of our schemes is certainly one of the next steps to take.} field theories (QFTs) with discrete-time quantum walks (DTQWs). 
The research field known as lattice gauge theories (LGTs),  now established for several decades since the seminal work of Wilson on the confinement of quarks in 1974 \cite{Wilson74}, already provides an extensive framework to discretize QFTs\footnote{LGTs exist in a full-fledged second-quantized framework.}, which have found numerous applications \cite{book_Rothe}. 
One may then wonder about the purpose of DTQWs-based discretizations of QFTs. A way to answer this question is to first introduce DTQWs and, more generally, quantum walks (QWs), as well as quantum computing and quantum simulation. %\\
%

%\vspace{0.3cm}
\noindent 
\section{Feynman's checkerboard and very brief history of quantum computing} 
\label{sec:Checkerboard_intro}
 
\noindent
The history of DTQWs is usually traced back to Feynman's attempts to build a fermionic path integral. 
We refer the reader to Schweber's historical review on the subject, where notes on the discretization of the (1+1)D Dirac equation, taken by Feynman in 1946, are reproduced \cite{Schweber86a}; these discretizations are a particular family of nowaday's standard one-dimensional DTQWs\footnote{The DTQWs that appear in Feynman's work have weakly coupled spin components, since he was focused on their continuum limit. This is explained in Subsection \ref{subsec:continuum}. We refer the reader to Subsection  \ref{subsec:Feynman's_quantum_walks} for more details on Feynman's work, and that which followed; in that subsection, we also explain that Feynman did not normalize his discrete scheme -- which is hence not unitary (again, this can be understood by the fact that unitarity is anyways recovered in the continuum limit, which is what interested Feynman) --, but that this can be done, and ends up exactly on a certain DTQW.}. 
These attemps, however, did not satisfy Feynman above one spatial dimension. 
In the 60's, Berezin \cite{book_Berezin} was able to formulate path integration for fermions, as functional integration over Grassmann variables, and this method became standard \cite{Weinberg_QFT1}. 
Since then, several answers to Feynman's original attempts have been given.
The most conclusive one may be Bialynicki-Birula's (1+3)D unitary one-particle scheme of 1994 \cite{BB94a}, a seminal (1+3)D DTQW, built with the now well-known chirality-rotation matrices, i.e., coin operations, see Subsection \ref{subsec:Feynman's_quantum_walks} for more details. 
However, Bialynicki-Birula's work does not deal in detail with the problem of building a path integral from this one-particle scheme.
Bialynicki-Birula's work does mention the possibility of including Yang-Mills gauge couplings, but does not give much details about it.
In the present thesis, we treat this topic in more depth\footnote{That being said, let us mention that Bialynicki-Birula provides a certain discretization of Maxwell's equations based on his DTQW-based discretization of the Dirac equation, which we do not explore in the present thesis (we give a different one, see Ref.\ \cite{AD16}), but which is pursued in more depth in the work of D'Ariano and collaborators.}.
Designing multiparticle schemes with gauge interactions, grounded on such DTQW-type discretizations of the Dirac equation, is the matter of current research.

Generally speaking, QWs model quantum transport on graphs. 
Seminal formulations, in discrete \cite{ADZ93a,Meyer96a} and continuous \cite{FG98a} time, appeared in the 90's\footnote{The scientific papers on which this thesis is based deal only with DTQWs.}. 
The first connection between both formulations was made in 2006 \cite{Strauch06b}. 
QWs are a topic at the crossing between quantum computing and quantum simulation\footnote{The intuitive difference between computing and simulation is elaborated on and usefully phrased in terms of a few key concepts by Johnson, Clark and Jaksch \cite{Johnson2014}; we shall review such concepts further down.}, as their history reveal. 
Quantum computing is the conception of algorithms with, instead of the logical bits, i.e. two-state variables, used by classical algorithms, two-state variables that behave quantumly, i.e. that can actually be in any (normalized) linear superposition of the two states; such two-state quantum variables are called \emph{logical} qubits. A quantum algorithm uses the coherence and entanglement between logical qubits as a mathematical resource to speedup a given computational task. We shall discuss the meaning of quantum simulation further down. 
QWs were identified as suitable elementary schemes to build quantum algorithms in the late 90's \cite{FG98a} and early 2000's\footnote{The first milestone quantum algorithms such as Shor's \cite{Shor94,Shor97} or Grover's \cite{Grover96} appeared, if not before, at least independently of the rise of quantum-walk schemes, in the mid 90's.} \cite{Ambainis2001}. 
They enabled, then, to build faster versions of key classical algorithms, such as database search \cite{AKR2005} or element distinctness \cite{Amb07a}, and were eventually understood, in the 2010's, as being a possible elementary basis to conceive \emph{any} quantum algorithm \cite{Childs2009,CGW13}, that is, a basis for so-called universal quantum computation. 
For them to be useful, quantum algorithms must run on a universal quantum computer, i.e. hardware, in which quantum coherence and entanglement between \emph{physical} qubits, i.e. two-state physical quantum systems, is ensured. 
The idea of quantum computers was seminaly developed by Feynman, who pointed out in 1982 \cite{Feynman1982} the potential exponential computing speedup provided by such a hypothetical device. 
Feynman's original paper is written in the perspective of using such computations to study physical phenomena, more precisely local phenomena\footnote{An important answer to Feynman's conjecture was given by Lloyd's (digital) ``universal quantum simulators" in 1996 \cite{Lloyd1996}. In that work, it is shown that a quantum computer would indeed fournish an exponential gain in the computing time needed to simulate any \emph{local Hamiltonian}. It is said that a quantum computer would simulate \emph{efficiently} this local Hamiltonian, while it is strongly believed that a classical computer cannot. Note however that, as mentioned in Ref. \cite{Farrelly15}, Lloyd's result is only valid for a non-vanishing spatial-lattice step, i.e. the result does not imply that the continuous-space theory can be simulated efficiently. In Ref. \cite{Farrelly15}, DTQWs converging faster towards their continuum limit are suggested. A local Hamiltonian is a Hamiltonian $H = \sum_{k=1}^l H_k$, where $H_k$ acts on a Hilbert space of finite dimension $m_k$ (spatial lattice) which does not grow with the size of the input (locality). The typical example is tight-binding Hamiltonians.}. A convenient way to define local phenomena in a very broad sense, is to demand that they can be described by partial differential equations\footnote{Differential calculus, developed in the 17th century by Newton and Leibniz, provided the appropriate tool to describe local phenomena,  described further down in the main text, and is a key historical step in the development of physics. Physics as a scientific discipline, in the terms science is understood today, is indeed usually traced back to Galileo, that is, not much before the advent of differential calculus. Galileo already pointed out the primordiality of mathematics in the study of natural sciences, and his famous (rephrased) quote, namely, that ``the book of Nature is written in the language of mathematics", is always to medidate. 
\iffalse
In Galileo's official statement, the author of this book was God, but this leaves unanswered the question of the nature of such a god; is it a human-like architect working rather top to down, as in (monotheist) religions, or is it completely identifiable to nature and is rather a bottom-up mechanics, which could however work top to down, i.e. with top-down equations, thanks to quantum non-locality, given the current cosmological theory in which spacetime points causaly disconnected currently could be connected in the past, and carry mutual information if entanglement is ensured in the time evolution? Note that the top-down and bottom-up pictures are, if not independent from the nature of the god of interest, at least conceivable in both god pictures given above. We can give an atheist and more contemporary version of the question about the author of the aforementioned book, written with mathematics: is the author Nature itself, i.e. are mathematics indeed spoken by Nature, or is the author nobody but the human being, i.e. are mathematics merely our best tool to interpret Nature; 
\fi
The ambiguity of the preposition `of' in `of Nature' underlines the metaphorical question of whether the author of this book is (i) Nature itself, in which case the human being would rather be a rune unraveler with a small interpretative part in the translation, or (ii) in fact nobody but the human being, who would then rather be a writer whose work results from an anthropomorphic and rational understanding of Nature. In other words, are mathematics a language actually spoken by Nature, i.e. underlying natural phenomena independently of humans, or are they merely our best tool to understand Nature? This is a rough introductory scheme for the epistemological debate between realism and idealism. It is amusing to note that Galileo's year of death coincides with Newton's birth.}\textsuperscript{,}\footnote{Non-localities may be used to model some phenomena, which translate mathematically as \emph{integro}-differential terms in the PDE.} (PDEs), or discrete versions of such equations. That being said, such mathematics can account for different physical types of locality. %\\
%

%\vspace{0.3cm}
\noindent
\section{Two notions of locality: mesoscopic near-neighbors locality, and relativistic locality}

\noindent
Let us introduce two physical notions of locality. Before the birth of relativity, many physical laws were constructed as local, in the following meaning: they modeled, thanks to PDEs, little-by-little transport resulting from near-neighbors interactions between tiny samples of matter, which have a typical so-called mesoscopic\footnote{A mesoscopic scale is a scale much smaller than the macroscopic scale, i.e. that of everyday-life objects, but still much bigger than the microscopic scale, i.e. that of atoms and molecules. At mesoscopic scales, one can often consider that a given physical medium made of atoms and molecules, is actually a continuous medium, whose physical properties can thus often be described thanks to PDEs. See \url{http://s2.e-monsite.com/2010/03/16/02/meca_flu.pdf}, Chapter 1, for a short explanation on the possibility to distinguish clearly these three scales.} length scale, and are considered point particles on macroscopic scales. We will refer to such a locality notion as the \emph{mesoscopic near-neighbors locality}\footnote{This notion of locality accounts for contact interactions in everyday life, often with solid objects, which are a particular type of continuous media.}. 
Example of such laws are the Navier-Stokes equation \cite{N1822,S1845} in fluid mechanics, or the heat equation in thermodynamics\footnote{The original \emph{Analytic theory of heat}, written by Fourier in French, and published in 1822, is available at \url{https://archive.org/details/thorieanalytiq00four}.}.
With the advent of special relativity around 1905, and then general relativity around 1915, it emerged what we will call \emph{relativistic locality}, which has the following meaning: 
in these theories, the speed of light is an upper bound for the speed of any physical phenomenon or interaction; in the standard continuous spacetime, this translates into the need to consider, instead of the pre-relativistic Galilean spacetime transformations, Lorentz transformations to relate, and thus define, inertial frames. We will clarify this simple picture further down.

Both locality notions have the following common feature, which is often implied by the word `locality': both define a causal neighborhood around each point in space. In Appendix \ref{app:causal}, we elaborate on these two types of causal neighborhood, which are of a very different nature. Now, it is no surprise that the above-mentioned pre-relativistic mesoscopic-transport equations do not verify relativistic locality, i.e. neither these equations are Lorentz invariant, nor the speed of the mesoscopic particles is bounded. That being said, this is not a problem for all applications of these  equations in which the maximum reachable speed is much smaller than the speed of light, which is the case for most macroscopic phenomena on Earth\footnote{Note that these equations we are mentioning do not exhaust all possible types of mesoscopic transport, i.e. they do not always apply. For example, there are some phenomena which actually exhibit relativistic effective features, such as, typically, an upper bound speed, which is usually much smaller than the speed of light and depends on the properties of the medium. The conception of such equations can demand and/or lead to an understanding of the mathematics of relativity. Examples of such equations are relativistic diffusion equations \cite{DR98}, which can govern, e.g., heat transport in some media, or the sine-Gordon equation \cite{II2013}, which can model, e.g.,  wave transport in long biological molecular chains or membranes.}.

Let us now consider relativistic locality and non-relativistic limits. Maxwell's equations, which were precisely the starting point of relativity, verify relativistic locality. However, the pre-Maxwellian versions of Coulomb's force between electric charges and Ampère's force between electric currents, do not verify relativistic locality. These two laws were thought as resulting from some instantaneous so-called ``action at a distance'', already evoked by Newton to refer to his gravitational law, which also does not verify relativistic locality. Such an action at a distance was in astonishing contrast with the intuitive mesoscopic near-neighbors locality of everyday's life. First with Maxwell's equations and then parallely to the advent of special relativity, the electromagnetic interactions were eventually understood to be mediated at the speed of light. The pre-Maxwellian Coulomb's and Ampère's laws were derived as non-relativistic limits of relativistic counterparts derived from Maxwell's equations, namely the forces derived from the so-called ``retarded potentials''. Similarly, the gravitational interaction was understood, parallely to the advent of general relativity, to be mediated at the speed of light\footnote{When such a result was obtained, the Newtonian gravitational law had, obviously, already been derived as a non-relativistic limit of the general-relativistic gravitational law. (Otherwise the gravitational interaction is simply instantaneous.)}. The relativistic electromagnetic and gravitational laws, and the associated non-relativistic limits, are said `fundamental' in the sense that -- as linked to the nature of spacetime -- they are considered scale-independent, at least above the smallest length scale experimentally probed, which is much below the microscopic scale\footnote{The smallest length scale that can be probed is often considered to be $10^{-18} \, \mathrm{m}$. This length scale can be probed at the LHC, with beam collisions involving an initial kinetic energy of the order of $hc/10^{-18} \simeq 4 \, \mathrm{TeV}$.}. 

Now, several of the  pre-relativistic mesoscopic-transport equations were also, since the advent of relativity, gradually made compatible, through certain non-relativistic limits, with relativistic locality.  Such connections are, however, difficult ones to make, and they are still the subject of current research. Indeed, building relativistic counterparts to the pre-relativistic mesoscopic-transport equations often requires, not only implementing fundamental non-relativistic limiting procedures, but also conceiving approximations, for the fundamentally-relativistic microscopic collisional dynamics, which preserve, at the mesoscopic scale, relativistic locality. Two relevant examples of such relativistic laws are (i) relativistic diffusion equations \cite{DMR97}, and (ii) relativistic versions of the Navier-Stokes equation \cite{Freisthler2017}, which may eventually be derived from relativistic versions of the Boltzmann equation \cite{CK2002}. Ultimately, one must perform ab initio computations of quantum molecular dynamics to derive the microscopic collision rules. The possibility of viewing pre-relativistic mesoscopic-transport equations as non-relativistic limits of relativistic counterparts is thus strongly underlaid by all these attempts to preserve relativistic locality, which is valid at the microscopic scale, along the several approximation steps made to capture only the effects which are relevant at macroscopic scales.

As a sum up, we may say that many pre-relativistic physical laws, either fundamental, i.e. considered scale-independent above the smallest length scales experimentally probed, or effective, i.e. emergent only at macroscopic scales, and depending on the smoothing of the microscopic dynamics, were, with the advent of relativity, derived as non-relativistic limits of relativistic laws, and were thus made subject to standard relativistic locality. Also, most of the conclusive theories that developed afterwards, up to nowadays, can be made compatible with standard relativistic locality\footnote{Lorentz covariance is still taken as a principle in many current theoretical works, since experimental tests on the violation of Lorentz covariance are overwhelmingly seen as disproving such a violation, and any new physical theory must anyways reproduce the -- possibly merely effective -- relativistic locality that rules the many phenomena which have been tested, which is a strong constraint. However, nothing prevents, theoretically, from actually forgetting about relativistic locality, or at least its standard version that we have been referring to and which is associated to Lorentz invariance, at small spacetime scales still unprobed, provided we recover this standard relativistic locality at higher, probed scales. See further down for more details.}\textsuperscript{,}\footnote{See Appendix \ref{app:on_quantum_non_locality} for a discussion on so-called \emph{quantum non-locality}.}, as well as information theories\footnote{The fact that the transmission of information must involve physical processes, and should thus, among other properties, not be superluminal, has become a paradigm in information theory; see e.g. \cite{Landauer1991} and \url{http://adamilab.msu.edu/wp-content/uploads/Reprints/2011/Adami2011.pdf}. Quantum mechanics shows that information can be \emph{encoded} non-locally, but cannot be \emph{transmitted} faster than light, a result known as the no-communication, or no-signaling theorem \cite{Peres2004}.}. %\\

%\vspace{0.3cm}
\noindent
\section{Extending the ideas of relativity: locality principle on networks}

\noindent
Now, the previous simple picture used to define relativistic locality actually results from the interplay between three ingredients: two distinct principles, namely (i.a) the relativity principle, i.e. that all physical laws must be the same in all inertial frames, which are postulated to exist, and (i.b) the requirement that the speed of light be the same in all inertial frames, and (ii) asumptions on spacetime, namely isotropy and homogeneity\footnote{For a derivation of the upper-bound property of the speed of light from these assumptions, see, e.g., \cite{Rouge}.}. In special relativity, inertial frames are indeed related, and in a way thus defined by Lorentz transformations, but in general relativity, Lorentz transformations are replaced by generic diffeomorphisms which must be Lorentzian only locally, i.e. at each spacetime point. Now, although the upper-bound property of the speed of light results from the interplay between these three ingedients, it can be tempting to infer a \emph{locality principle}, i.e. to \emph{postulate} the upper-bound property, since it is an easy way to (i) ensure causality in physical phenomena, and (ii) avoid the need to make asumptions on the properties of spacetime. In a broad sense, such a locality principle requires the existence of \emph{any type} of bounded causal neighborhood.  The upper bound for the speed of phenomena and interactions is typically the speed of light in the context of fundamental physics. This upper bound can also take other values if we wish to model other types of finite-speed interactions on graphs.

Several research works, aiming at grounding physics on quantum-information principles, are based on such a locality principle. These works have suggested derivations of standard relativistic locality, i.e. Lorentz transformations in continuous spacetime, from \emph{local evolution rules on an abstract network}. The network is then typically mapped to a discrete spacetime. In addition to verifying the locality principle, these evolution rules also have a flavor of the aforementioned other notion of locality: they are formulated as near-neighbors interactions between the nodes of the network, which is in conceptual and technical analogy with, (i) indeed, the near-neighbors interactions between the mesoscopic samples of matter which form fluids, or, even more, between the cells of their lattice-gas versions \cite{Hardy1976,FHP86,FHHLPR87,Meyer96a,Yepez2016}, but also with (ii) tight-binding condensed-matter models. These discrete schemes also satisfy a kind of relativity principle, i.e. they are invariant under changes of observer, at the network level, in a meaning which is necessarily different from the standard relativistic one, and also specific to each work \cite{DAriano2016,Arrighi2014}. We will briefly comment on  such works further down.

Eventually, let us make a remark before coming back to our initial topic: relativistic locality is, as mesoscopic near-neighbor locality, also reflected by the use of PDEs, but in a field-theory framework. As an example for this, consider any fundamental-fields dynamics, such as Maxwell's equations, already mentioned, the Klein-Gordon equation, or the Dirac equation, among others\footnote{The locality, in terms of PDE description, of the Schrödinger equation, is to bee seen as as residue of the fundamental relativistic locality of the Dirac or the Klein-Gordon equation.}\textsuperscript{,}\footnote{By the way, since we are speaking of fields, another remark is in order: the non-localizability of fields does, in a sense, a priori not necessarily affect locality, since, if localizability is a property related to a quantity $A$, locality is a property related to the variations $\Delta A$ of $A$.}.  %\\

%\vspace{0.3cm}
\noindent
\section{On digital and analog quantum simulation}

\noindent
The idea of discretizing physical laws described by PDEs naturally appeared with the development of computers since the Turing machine.
Cellular automata (CA), initially developed by Ulam and Neumann in the 40's, were designed for this purpose\footnote{In addition to be local in their standard definition, CA are also often homogeneous, i.e. the local rules that are applied at each time step are the same everywhere. One can however relax this constraint. The general question of (i) whether a given CA might be associated a continuum counterpart, i.e. a PDE (the reverse is usually feasible), and of (ii) the generality of such a PDE, has been extensively studied, and is not straightforward; for a review, see \url{https://arxiv.org/pdf/1003.1983.pdf}.}, and were then quantized\footnote{Feynman explicitly mentions cellular automata in his seminal paper on quantum simulation, but ``[doesn't] want to force" \cite{Feynman1982} such a vision, i.e. his work should not be merely regarded as a way to quantize CA, but also algorithms that are not phrased as CA.} \cite{Feynman1982,Grossing_Zeilinger_88}. 
Meyer's seminal paper on DTQWs \cite{Meyer96a}, in which he introduces them as a particular family of single-particle unitary quantum cellular automata\footnote{There is a substantial litterature on two-particle DTQWs \cite{Huberman2017,Berry2011,Carson2015}, including experiments \cite{Sansoni11a}. Now, we know that the framework of second quantization is, on the one hand, useful to describe quantum systems with large numbers of particles \cite{Mahan2000}, and, on the other hand, fundamentally necessary in a relativistic framework \cite{Weinberg_QFT1}. In the sense that such second-quantized formulations of DTQWs are still lacking, as well as works on many-particle DTQWs, current works on DTQWs should still currently rather be seen as \emph{single-particle} unitary QCA. The work led by Boettcher and Portugal on the non-perturbative renormalization of DTQWs \cite{Boettcher2013,Boettcher2016}, is certainly to be used in future second-quantization developments.} (QCA), namely partitioned (or staggered) QCA, and which is, rather than quantum-computing, quantum-simulation oriented, became a reference in the aforementioned quantum-computing arena. 
That being said, since the 70's and 80's, i.e. already before the raise of quantum computing, and then still rather independently from this new subject in the 90's and 2000's, many CA and QCA schemes were developed for simulation purposes. 
Around the 2010's, Arrighi and Grattage demonstrated the intrinsic universality of partioned QCA, i.e. QWs, in the design of quantum-simulation schemes  \cite{AG09,Arrighi2011,Arrighi2012}, a topic that will be discussed further down.
In 2015, the well-known quantum lattice Boltzmann equation was eventually explicitly identified as a QW by Succi \cite{Succi2015}. 
Now, as built out of QWs, these numerical schemes are naturally implementable by quantum algorithms that a quantum computer would naturally run. 
But these QCA schemes are still implemented, nowadays, with classical algorithms on classical computers, which eventually brings us back to the topic of quantum simulation, on which we shall now comment more lengthily. 

As defined by many, quantum simulation is the simulation of physical phenomena with a system utilizing these features of quantum mechanics which are absent from the so-called classical world, such as quantum coherence and entanglement \cite{Trabesinger2012,Schaetz2013,Johnson2014}. I shall use this definition. 
The other obvious definition is that of simulating a system which displays such quantum features \cite{Georgescu2014}. 
This second definition does thus not focus on the computing power of quantumness to simulate \emph{any} physical phenomenon, although both definitions often `interact', simply because simulating systems where all elementary entities are described by functions, as quantum mechanics requires, instead of a few coordinates, as in classical mechanics, is, generically, computationally way more costly, as Feynman seminaly discussed \cite{Feynman1982}. 
Before digging into the notion of quantum simulation, let us first make it clear that of simulation.

Linguistically, the word simulation is close to that of mimic. Does the simulation mimic something? A simulation is aimed at obtaining information about a physical system of interest, that we shall sometimes merely call `system' for brevity. It does so through a mathematical \emph{model}, namely, a mathematical object which is thought to describe the system. The model is meant for being tested, i.e. its `behavior' must be compared to that of the system. This is done by comparing data provided by a so-called \emph{simulation of the model} -- i.e. a computation associated to a particular realization of the model, which is litteral -- to data provided by an experimental measure on the system; this comparison enables to state up to which accuracy the model describes the system. The data out of the simulation are, ultimately, numbers, but they are not `produced' the same way whether this simulation is digital or analog, terms that will be defined below.

It is useful to introduce the concept of hypothetical physical system, or \emph{ideal system}, which is induced by the mathematical model.
In the case where this mathematical model is a dynamical PDE, the simulation will typically explore, starting from an input initial state, the state of the ideal system at successive instants.
Epistemologically, the simulation, as an isolated task, gives results only on the ideal system described by the model, and not on the actual system. To obtain results on the latter, on must, as mentioned, compare the results of the simulation to that of experiments on the actual system.
For it to be part of the scientific method, quantum simulation must, as classical simulation, be subject to this rule.
%

\iffalse
Note that we haven't yet introduced the notion of computer, which is different from that of simulation. In practice, however, current classical simulations often rely on the computing power of some machine, namely the computer.
%
\fi
A simulation performed thanks to a computer is called a digital simulation, see further down for more details. There exist another type of simulation, called analog simulation. We shall discuss the differences between both.
Note that, although classical analog simulation is currently maybe less topical (at least it is not the topic of the present work), since the computing power of current classical computers often makes them better tools for simulating physical phenomena, the digitality or analogity of a simulator are features which are independent from its quantumness.

Before clarifying the concept of digital simulation, one must make it clearer that of computation and computer. 
A computation is a series of operations with \emph{numbers}, aimed at delivering a numerical result. This means any entity carrying out the computation must be able to represent, manipulate, and output \emph{integers}, whatever its functioning may be. This entity can be a human being, but one can envisage a device that carries out these computations. Such a device is called a \emph{computer}, or \emph{digital device} (the `digits' are the symbols with which we write integers, and hence any number, at least approximatively).
There relies the difference between a digital and an analog device: the latter is not designed to perform operations between integers, and its output is a certain \emph{physical quantity} that often belongs to a certain physical continuum.
\iffalse
\textsuperscript{,}\footnote{Let us try to elaborate on the continuous nature of the output of an analog device: this output can indeed not be decomposed into elementary outputs \emph{which are known to result from an internal functioning constructed by humans as a discrete network with internal interactions}, but we obviously know that part of the output is made of an integer number of atoms of more elementary particles. That being said, spacetime is still viewed as continuous.}. 
\fi
The `numerical output of an analog simulation', mentioned above\footnote{We must test the mathematical model and thus extract numbers from the analog simulation.}, is thus indirect, via a \emph{physical measurement}\footnote{The result of a physical measurement is typically interpreted as a real number which cannot be known exactly but only up to some \emph{accuracy}, which depends, among other things, on the analog device.}\textsuperscript{,}\footnote{Note that an analog device may be used only for its physical output, i.e. the measurement of the physical output is needed for certain purposes only.}.
Now, what we often refer to as `computers' in current everyday-life, are \emph{universal} computers, i.e. able to carry out so-called \emph{universal} (classical) computation. Vaguely speaking, this means they can run any program. More fundamentally -- but at this stage not less vaguely -- this means a universal computer is able to realize any type of `automatic task' if we give it the instructions in an `appropriate way'\footnote{This statement obviously calls for mathematical definitions of the notions of `automatic task' and `appropriate way'. The first answer historically given to such questions is that of Turing, with his now-called \emph{Turing machine}, conceived in the late 30's \cite{Turing1937,Turing1938}, and which is still a reference, at `least' pedagogical, and at `most' fundamental through the Church-Turing thesis. In Appendix \ref{app:on_computers}, we briefly elaborate on the notions of computation, computers, and computational universality, both the standard classical notions and their quantum counterparts.}. Universal computation is achieved by (i) implementing well-chosen series of \emph{elementary operations} (called \emph{primitives}), which (ii) belong to a certain well-chosen finite set, and which can often be viewed as equivalent to logical operations, also called \emph{logical gates}. A digital device, or computer, always implements, at least implicitly, certain logical gates, but is not always universal\footnote{Vaguely speaking, this means that either the set of elementary operations is not well chosen, or so is the way of chaining them, or both.}, in which case its computational limitations are often manifest\footnote{All current quantum computers are non-universal, essentially because of the too poor scalability of current quantum technologies. Achieving the construction of a universal quantum computer thus demands to meet scalability challenges. This is briefly elaborated on at the end of Appendix \ref{app:on_computers}.}. 

Now, on the one hand, a digital simulation is simply a computation that `explores' a mathematical model, i.e., typically, that provides a particular numerical realization of this litteral model.
The physical functioning of a digital simulator has, a priori\footnote{
\iffalse
A priori, this is even less the case of a universal digital simulator, i.e. a universal computer used for simulation purposes, in the following sense: (i) the larger the spectrum of phenomena a hardware can simulate, (ii) the smaller is the subset of physical operations that it performs to simulate a given phenomenon, among all the physical operations it \emph{can} perform, and thus (iii) the smaller is the probability that the mathematical logic underlying the physical operations that \emph{can} be performed, be that -- if there is one -- underlying the evolution of the system of interest. This does not preclude, however, since the invoked probability gets smaller with the increase of universality, but never vanishes, 
\fi
If one is able to reproduce all physical phenoma with a universal computer, one is actually naively tempted to say that nature might be ruled, not only mathematically, but even physically, by processes similar to that which happen in a hardware, up to some accuracy.  Such statements can be given a scientifical meaning. They inspire, among others, Arrighi \cite{AG09} and D'Ariano \cite{DAriano2016}, works which fall within the broader subject of natural computing, in which the computing rules are inspired by physical phenomena. We briefly pursue such questions further down in the present introduction.}, nothing to do with that of the ideal system, and thus of the actual system.
On the other hand, if an analog simulator, also called emulator, cannot explore the mathematical model by performing computations, as the digital simulator does, what does it do? The analog simulator is a machine that directly \emph{mimics} the ideal system. 
To sum up, before digging into the concept of analog simulation, while both analog and digital simulation, on the one hand, are used to learn about a physical system, and, on the other hand, epistemologically only provide, as isolated tasks, results on an ideal system induced by a mathematical model, they work differently: analog simulation directly mimics the ideal system while digital simulation explores the mathematical model thanks to an algorithm. 
However, what does mimicking an ideal system actually mean? 

The analog simulator is built out of elementary physical devices and implements interactions between those devices. Now, by interacting between themselves, the elementary devices usually mimic, this time, \emph{actual} elementary systems, i.e. systems, such as atoms, that have been scientifically proved to behave \emph{effectively} -- i.e. in the way they interact with each other -- as the devices. These elementary devices are then assembled according to the mathematical model of interest, in order to mimic the related \emph{ideal} system\footnote{A widely used (digitally-simulated) analog simulator is the electronic-circuit simulator SPICE, i.e. the Simulation Program with Integrated Circuit Emphasis. SPICE can be used on the online platform PartSim: \url{https://www.google.fr/search?client=safari&rls=en&q=spice+simulator&ie=UTF-8&oe=UTF-8&gfe_rd=cr&ei=8OFHWfPkJ4jAaJj_qsAL}. It is used by companies to check the performances of microelectronic circuits to be built. This (digitally-simulated) analog simulator is classical, i.e. it does not use quantum superposition and entanglement as a resource to simulate an ideal system. A class of developing classical analog simulators, this time actual (i.e. not digitally-simulated), are the so-called synthetic molecular motors, which are synthesized (macro)molecules that mimic natural (macro)molecules which convert chemical energy of the living being they belong to, into mechanical energy, in order (i) to perform certain operations, such as DNA transcription and RNA translation, or DNA replication, or (ii) to transport other molecules (these are the so-called ``cargo proteins''). Natural molecular motors are usually chemically driven. The synthetic ones can also be light driven. Synthetic molecular motors can be used for nano-technologies, and not necessarily to simulate and study natural molecular motors. A widespread type of analog quantum simulators are cold-atom optical-lattice simulators. The lattice is formed by laser-induced standing waves. Atoms usually jump from one site to a neighboring one by quantum tunneling, and this dynamics can mimic electron transport in condensed matter. The following review by Jaksch and Zoller may be consulted: \url{https://arxiv.org/abs/cond-mat/0410614}.}.
We have stated above that, before comparing the data out of the analog simulator to that out of the physical system of interest, the simulation only provides information about the ideal system. If the experimental test is passed by the simulator, one can eventually say that it mimics, not only the ideal, but the \emph{actual} system of interest. 

While this protocol must indeed be followed, the way it has just been formulated might give a naive picture of, not only analog simulation, but also, more broadly, of empirical sciences. Indeed, there is sometimes, rather than a final series of tests after the simulator is built, a feedback loop between the ongoing construction of the simulator and the physical system of interest. In other words, the reliability of the analog simulator as a tool to mimic the \emph{actual} system of interest, i.e. the validity of the mathematical model to describe this system, is sometimes tested, at increasing levels of detail, \emph{along the way} in the construction of the analog simulator. By this feedback process we build confidence on the analog simulator, so as to use it as a starting point for future experiments, and not to have to start from scratch, i.e. from the elementary devices. 
Because many of the elementary devices have often been very well tested with respect to previous physical systems of interest, one may want to set up the feedback loop at a (conceptual) scale which is intermediate between that of the elementary devices and that of the final machine, and eventually large with respect to that of the elementary device; to give a naive pedagogical example, one may want to test (i) the whole unfinished simulator every, let's say, to give a number, five devices assembled, if we have to assemble one hundred, or (ii) only the devices on which one has less confidence from the above-mentioned past tests.
\iffalse
\footnote{If I were an experimentalist, I may have not feel the need to theorize all this. Does the fact that I do so proves that I am at the right place as a theorist, or that I should rather be an experimentalist. The answer is probably in between. These lines are a way for me to get closer to experiments. The best way of doing so is probably, however, if one is not an experimentalist, to read experimental papers, and, since this is rather difficult and frustrating, to collaborate with experimentalists. I thank D. Meschede and A. Alberti for their invitation to visit their group, which has inspired me much.}. 
%\footnote{The word `physical' must here be opposed to `mathematical', although `physical' is often redundant with `system'. The property of being focused on a specific system is less relevant to the definition of analog quantum s%imulation, since one can envisage a digital quantum simulator simulating a single type of phenomenon, but again, it does so by explicitly implementing logical gates.}. 
%
\fi

The characteristics of an analog simulator may be summed up in the following sentence: an analog simulator is a simulator that works by analogy, that is, a mapping which is not, at least explicitly, underlaid by logic, but is instead a phenomenal copy of the ideal system.
This mapping a priori depends on the ideal system, so that, in this sense, one has to build, roughly speaking, a machine for each ideal system\footnote{This is `roughly speaking' because there are many intermediate scales. One can indeed think of an analog simulator that would be able to simulate a given physical theory, but which would still not be universal. Can we push this further and think of a universal analog quantum simulator, e.g., an analog equivalent to Lloyd's universal digital quantum simulator? On the one hand, digital simulation focuses on the mathematical model, so that it is scale independent, i.e., typically, it can simulate any type of PDE or Hamiltonian, regardless of the associated physical scales. On the other hand, analog simulation focuses on the ideal system induced by the mathematical model, so that it is a priori more likely to depend more strongly on the scale, which might be a brake on its universality potential.}. 

We can now come back to our first question: does the simulation, in its most general meaning, mimic something? Since the analog simulator is said to mimic (i) the ideal system, (ii) the actual systems that have inspired the latter and, eventually, if experiments are conclusive, (iii) the actual system of interest, can we similarly say that the digital simulator mimics, first the ideal system, and then, if the experimental test is passed, the actual system? This reveals that the notion of mimic still demands to be clarified. 

The digital simulator does the following: it runs an algorithm which reproduces the behavior of an ideal system, but \emph{only} through the associated mathematical model. Now, one can envisage a machine ruled by this digital simulator, that \emph{effectively} mimics the ideal system. Such a digitally-ruled machine is called a \emph{robot}\footnote{Although the purpose of such a machine is to be a phenomenal copy of some existing phenomenon, so that the machine is, \emph{in such an external point of view}, an analog machine, it is obviously \emph{not} and analog \emph{simulator}, because it is ruled by a \emph{digital} simulator, i.e. its \emph{internal functionning} is digital, while the internal functioning of the analog simulator is \emph{also} analog. Generally speaking, an analog device can either rule or be ruled by digital device.}. Now, if, regarding the scientific method, some digital simulator is, effectively, i.e. via a hypothetical associated robot, as good a simulator as some analog simulator, should we less speak of `mimic' for the digital simulator, i.e. for the \emph{internal functioning} of the robot, than for the mimic of the ideal system's internal functioning which is realized by the analog simulator? Shoudn't we rather conclude that the digital simulator is as good a mime as the analog simulator, regarding, if not the `metric', i.e. the position of each physical component of the internal `machinery' of the actual system of interest, at least the `topology' of this internal machinery, i.e. the way its physical components interact with each other, e.g. which ones are connected, and which ones are not? This series of  linguistic questions underlines both (i) the scale dependence of the notion of mimic, and (ii) the question of whether the mimic refers to objects, the `metric', or to relations between these objects, the  `topology'. The difficulty is that these two questions intertwine. To be made scientific, the eventual question of whether logic might rule any extra-human phenomenon should be formulated in terms of measurement of physical quantities, and, because the notion of extra-humanity and that of measurement are difficultly compatible, the question is, stated as such, rather metaphysical\footnote{The current scientific approach to make these notions compatible is that of decoherence in quantum mechanics \cite{var96a,book_Schlosshauer}.}. Note that natural computing, i.e. the design of algorithms inspired by natural phenomena, is a growing field of research. The previous lines aim at illustrating that linguistic and metaphysical questions can (i) be used as a source of inspiration for science, and (ii) eventually be formulated in scientific terms\footnote{While the -- at least intuitive -- knowledge of Popperian epistemology is certainly needed to practice science, it is, by construction, explicitly not sufficient, by excluding from the start, from the analysis, human inductive cerebral processes.}.

In their review on quantum simulation, Johnson, Clark and Jaksch \cite{Johnson2014} note that it would be unreasonable to demand to a quantum simulator more accuracy in its accordance with the mathematical model it simulates, than with the system of interest, which requires comparing data out of the simulator with experimental data out of the system of interest.
While this remark obviously also holds in the case of simulations with classical computers, it is particularly topical in the context of quantum simulation, in a sense given by the following two reasons: first, current quantum simulators can handle at most a few tens, or hundred of physical qubits\footnote{One can speak of physical qubit even if the simulator is analog and not digital, since a physical qubit is merely a two-level quantum system, and analog simulators are often built out of such elementary systems.}, so that the typical \emph{accuracy} that can be achieved if one wants to perform the simulation in a reasonable time can be a limiting factor, while accuracy is not a limiting factor for many standard tasks demanded to classical computers\footnote{The `good question' may be, regarding this matter, merely the naive one: can current quantum computers fulfill the job of classical computers faster? A still up-to-date answer to this question is that the few quantum computers which are currently used are specific-task quantum computers.}, and second, quantum \emph{error} correction is not standard in current quantum simulation, but rather an intensively-developing research field\footnote{Because keeping coherence and entanglement between an increasing number of physical qubits is nowadays, essentially, the major technological problem faced in the design of a quantum computer, quantum error correction is a topic of major importance.}, while errors on current classical computers are rather extremely rare\footnote{Note that we have here, very naturally, and without, I think, bothering the reader, opposed quantum simulation, which can be analog or digital, to classical computation, which is digital, and this for two reasons: on the one hand, as we mentioned above, classical analog simulators are a \emph{much}-less-used tool to study physics than classical computers, while on the other hand, analog quantum simulators are sometimes -- i.e. essentially when it comes to focusing on a specific problem for which the analog simulator has been designed --, better tools than classical computers, thanks to their quantumness.}. %\\ 
%

%\vspace{0.3cm}
\noindent
\section{On the quantum simulation of quantum field theories}

\noindent
Since the possibility of universal quantum computation, although recently established theoretically, is currently limited by the size of the quantum hardware, and is rather seen as a long-term aim, analog quantum simulation attracts much attention for short- and mid-term research, and is the subject of an increasing number of works. In such a context, several theoretical works on the quantum simulation of QFTs, have been carried out. Let us first mention two research groups which are very active in this research field, and whose approaches are essentially the same: on the one hand, the collaboration between Cirac, Zohar and Reznik, and on the other hand, Zoller's team\footnote{These two groups involve other researchers than those mentioned.}. These two groups have carried out intensive theoretical and experimental work on the quantum simulation of standard LGTs, in a strong short- and mid-term experimental perspective. These works eventually yielded seminal experimental evidence several months ago; the title of the paper reporting the results, namely ``Real-time dynamics of lattice gauge theories with a few-qubit  quantum computer" \cite{Martinez16}, already gives some additional precisions on the setup, which simulates the Schwinger mechanism. Let us also mention the theoretical work led by D'Ariano on the quantum simulation of standard continuum QFTs \cite{DAriano2016}. The general aim of this work is (i) to recover QFT from quantum-information principles, in the perspective that (ii) such rules might further account for sub-Planckian physics, i.e. phenomena below the Planck length -- and, of course, deliver known super-Planckian physics\footnote{The smallest length scale ever probed is often considered to be $10^{-18} \, \text{m}$, thanks to particle-physics experiments led at the Large Hadron Collider (LHC). The scale of the Planck length being $10^{-35} \, \text{m}$, there are actually, above the Planck length, still $35-18=17$ orders of magnitude to be probed before reaching the Planck length. Known super-Planckian physics thus lies above $10^{-18} \, \mathrm{m}$.} -- \hspace{-0.1cm}, i.e. might be a setup to overcome the ``logical clash" \cite{DAriano2016} between general relativity and QFT. A notable step has been made in this direction: \emph{free} QFT is recovered as a continuum limit of a certain family of QCA, essentially DTQWs for fermions, and entangled pairs of such DTQW-fermions for bosons. Second quantization is evoked. Einstein's relativity principle, ensured in the continuum, is replaced, at the discrete level, by requiring the ``invariance of the eigenvalue equation of the automaton/walk under the change of representations" \cite{DAriano2016}, without needing to refer to spacetime. Recall that the aim of this work is to get free of as many physical principles as possible in the formulation of the theoretical setup, i.e. to axiomatise QFT with  mathematical quantum-information constraints. The work led by Arrighi is in the same vein, and it moreover suggests discretizations for curved spaces. By nature, this work goes much further in the mathematical analysis of the suggested PDE-discretization schemes\footnote{In Ref. \cite{Arrighi_higher_dim_2014}, for example, the observational covergence of the homogeneous (i.e. free) Dirac DTQW in arbitrary spacetime dimension, is rigorously proven.}. As well as the work by D'Ariano, this work includes recovering continuous-spacetime relativistic transformations in the continuum limit of underlying unitary transformations acting on logical qubits \cite{Arrighi2014}. Eventually, note that non-Abelian gauge theories have recently been phrased in the quantum-information tensor-network language\footnote{Tensor networks were  introduced by Penrose, see \url{http://homepages.math.uic.edu/~kauffman/Penrose.pdf}.} \cite{Tagliacozzo2014,Rico2014}. The reader may want to have a look at the following two presentations, respectively by D'Ariano and Chandrashekar, which review current results on the DTQW-based simulation of (free) QFT: \url{http://www.qubit.it/people/dariano/} and \url{http://www.iopb.res.in/~iscqi2016/slides/conference/d4/chandrasekhar.pdf}; the presentation by Chandrashekar includes recent results on the topology of such schemes. Several original works coauthored, among others, by Di Molfetta and Pérez, and dealing with quantum simulations, with DTQWs, of phenomena related to high-energy physics, have been published within the last months \cite{Bru2016b,Molfetta2016,MrquezMartn2017}; these include the DTQW-based quantum simulation of neutrino oscillations \cite{Molfetta2016} and DTQW-based Kaluza-Klein-like frameworks and phenomenology, by Bru et al. \cite{Bru2016}.

The topic of quantum simulation, however, goes well beyond the quantum simulation of quantum field theories in the narrow sense. Quantum simulation is, in particular, now widely used to study condensed-matter phenomena \cite{dalibard10a,Goldman2014}. %\\

%\vspace{0.3cm}
\noindent
\section{Outline of the thesis}

\noindent
This thesis is organized as follows. It is divided into two parts. Part \ref{Part:Yang-Mills} is devoted to the in-progress construction of DTQWs-based lattice Yang-Mills gauge theories, while Part \ref{Part:Gravitation} presents DTQWs able to simulate, in the continuum limit, Dirac dynamics in curved spacetime. Each part is preceded by some introductory material.

Part \ref{Part:Yang-Mills} contains two chapters. Chapter \ref{Chap:Electromagnetic_2D_DTQWs} is based on two publications, Refs. \cite{AD15} and \cite{AD16}, dealing with the coupling of 2D DTQWs to 2D electromagnetic fields, while  Chapter \ref{Chap:Non-Abelian_1D_DTQWs} is based on a single publication, Ref. \cite{ADMDB16}, dealing with the coupling of 1D DTQWs to a generic non-Abelian U($N$) Yang-Mills gauge field. All together, these three papers should enable the reader to visualize the in-progress construction of DTQWs-based lattice Yang-Mills gauge theories. 

The aim of Introductory Material \ref{IntroMat1}  is twofold. First, several well-established results on 1D homogeneous and inhomogeneous DTQWs, are presented, in Sections \ref{sec:homogeneous} and \ref{sec:inhomogeneous}, respectively; this allows to set up the notations that will be used in the rest of the manuscript. Second, several results established by seminal works related to DTQWs, are reviewed and discussed: Section \ref{sec:seminal} deals with Feynman's checkerboard and Aharonov's quantum random walks, while Section \ref{sec:Wick} aims at situating quantum walks in the interrelations between several relativistic and non-relativistic physical systems related by analytical continuation.

Before presenting, in Section \ref{sec:two_publications}, the two publications \cite{AD15,AD16}, Chapter \ref{Chap:Electromagnetic_2D_DTQWs} begins with Section \ref{sec:higher}, which is a relatively detailed chronological review of historical milestone results on quantum walks, quantum computing and quantum simulation. The quantum-information era that we enter every day a little more, relies on the power of such results.

The reader is strongly invited, before taking a close look, in Section \ref{sec:two_publications}, at Publications \cite{AD15} and \cite{AD16},  to read the first introductory subsection, namely Subsection \ref{subsec:recap_1D}, which is a state-of-the-art `recap' on 1D DTQWs coupled to electric fields, and their interpretation as such, which is threefold.

Publication \cite{AD15} is presented in Subsection \ref{subsec:B}. A new family of DTQWs is presented, which coincides, in the continuum limit, with the Dirac dynamics of a relativistic spin-1/2 fermion coupled to a constant and uniform magnetic field. Standard relativistic Landau levels are extended from the continuum Dirac dynamics to the lattice situation, where the walker is indeed `sensitive' to the non-vanishing step of the spacetime lattice. Such Landau levels for DTQWs are built perturbatively in the step, around the continuum situation. We eventually demonstrate, through numerical simulations, that the parameter interpreted as the magnetic field in the continuum limit, has, beyond both (i) the continuum limit and (ii) perturbative effects of the step, qualitatively the same confining properties as a standard magnetic field, if small compared to $2\pi$, the size of the Brillouin zone. The possibility of quantum simulating  condensed-matter systems by DTQWs is also discussed.

Publication \cite{AD16} is presented in Subsection \ref{subsec:EB}. A new family of DTQWs is presented, which coincides, in the continuum limit, with the Dirac dynamics of a relativistic spin-1/2 fermion coupled to a generic electromagnetic field in dimension 2. We extend the electromagnetic interpretation beyond the continuum limit, by showing the existence of (i) a lattice gauge invariance, (ii) a lattice invariant quantity, and (iii) conserved currents on the lattice. This allows us to suggest lattice equivalents to Maxwell's equations. Summing up, we build a DTQW-based Abelian U(1) lattice gauge theory, in 1+2 dimensions. We recover, through numerical simulations, known phenomena, both standard ones, associated to standard electromagnetic fields \cite{book_Jackson,Kolovsky10,KolovskyMantica14}, and a less standard one, associated to periodically-driven systems \cite{Goldman2014,Hamilton2014,Khemani2016}, which, in particular, matches qualitatively with recent results on DTQWs, both in one \cite{mesch13a, ced13} and two \cite{Yalcinkaya2015,Bru2016} dimensions. 

Before presenting, in Section \ref{sec:paper3}, Publication \cite{ADMDB16}, Chapter \ref{Chap:Non-Abelian_1D_DTQWs} begins with Section \ref{sec:historical_applications}, which is a compact but in-detail referenced chronological review of the historical steps which led to non-Abelian Yang-Mills gauge theories, from the discovery of electricity and magnetism, to chromodynamics and standard lattice gauge theories.

The reader might be, on the one hand, interested in Subsection \ref{subsec:schematic}, and is strongly suggested, on the other hand, to read Subsection \ref{subsec:walk_op_non_ab} before taking a close look at Publication \cite{ADMDB16}, which is commented and then reproduced in Subsection \ref{subsec:comment_paper3}. Indeed, Subsection \ref{subsec:schematic} tries to give a schematic picture of non-Abelian gauge theories, without equations, while Subsection \ref{subsec:walk_op_non_ab} shows how the DTQW on which Publication \cite{ADMDB16} is grounded, is a very natural extension of its Abelian counterpart \cite{DMD14}.

Publication \cite{ADMDB16} is introduced in Subsection \ref{sec:paper3}, but commented and reproduced in Subsection \ref{subsec:comment_paper3}. A new family of 1D DTQWs is presented, which coincides, in the continuum limit, with the unidimensional Dirac dynamics of a relativistic spin-1/2 fermion coupled to a generic non-Abelian U($N$) Yang-Mills gauge field. The gauge-theory interpretation is extended beyond the continuum limit, by showing the existence of, (i) not only a U($N$) gauge invariance on the lattice, but also (ii) an associated lattice gauge-covariant quantity, which delivers, in the continuum, the field strength associated to a standard U($N$) Yang-Mills gauge field. We show, through numerical simulations, that the classical, i.e. non-quantum regime, is recovered at short times.

Part \ref{Part:Gravitation} contains solely Chapter \ref{Chap:Gravitational_2D_DTQWs}, which is based on Publication \cite{AD17}. It is preceded by Introductory Material \ref{IntroMat2}, which aims at giving a pedagogical, general, and essential view of the results of Refs. \cite{DMD13b,DMD14}, that is, of how to make a DTQW propagate in a (1+1)D curved spacetime, to speak in vague but pointful terms, that are clarified in this Introductory Material \ref{IntroMat2}. The publication on which Chapter  \ref{Chap:Gravitational_2D_DTQWs} is based, is an extension of the results presented in this introductory material.

Chapter \ref{Chap:Gravitational_2D_DTQWs} directly presents and comments Publication \cite{AD17}. A new family of 2D DTQWs is presented, which coincides, in the continuum limit, with the Dirac dynamics of a relativistic spin-1/2 fermion propagating in a generic (1+2)D curved spacetime. The particular case where this spacetime corresponds to the polarization plane of a gravitational wave (GW), is focused on. We show how the probability of presence of the DTQW in this plane is affected by a pure shear GW. In the continuum limit, the GW modifies the eigen-energies of the walker by an anisotropic factor which is computed exactly. For wavelengths close to the spacetime-lattice step, the probability of presence is modified non trivially; the minimal comment on such an influence is that the net effect of the GW is maximal, by far, for short wavelengths, comparable to two or three  spacetime-lattice step.

Eventually, the conclusion opens the present work to various related works and possible further studies. 

\part{DTQWs interacting with Yang-Mills fields \label{Part:Yang-Mills}}
   %%%%%%%%%%%%%%%%%%%%%%%%%%%%%%
%%  Alias bibliographiques  %%
%%%%%%%%%%%%%%%%%%%%%%%%%%%%%%
\defcitealias{NRFortran}{Numerical Recipes}
\defcitealias{NRF77}{Numerical Recipes}
\defcitealias{NRF90}{Numerical Recipes}
\defcitealias{IUPAC}{IUPAC}%International Union of Pure and Applied Chemistry

\setcounter{footnote}{0}

\setcounter{section}{0}

\begin{IntroMat}[\hspace{-0.2cm}one-dimensional DTQWs]{One-dimensional DTQWs} \label{IntroMat1}

{\noindent
\motivations{
The aim of this introductory material to the first part of the manuscript, is twofold. First, we discuss some well-established results on (1D) DTQWs, which allows us to set up the notations that will be used in the rest of the manuscript. Second, we review some of the material contained in seminal works related to DTQWs. 
}    \\ }
                     
\section{Homogeneous DTQWs} \label{sec:homogeneous}
    
\subsection{Notations}    \label{subsec:notations}
     In this first paragraph, we set up the notations for the standard DTQW on a 1D graph. Consider a graph whose vertices (also called nodes) are countable, so that one can label each vertex by an integer $p$, and which is two-regular, i.e. for which each vertex $p$ has two neighbors, $p-1$ and $p+1$. Such a graph is called `1D graph' because the $p$'s can be viewed as coordinates on a curve, either (i) a line when the number of vertices is infinite, which in practice corresponds to the situation where the walker never reaches the boundaries, or (ii) a circle with a periodic boundary condition in the opposite case; one might also think of a segment but then the graph is not two-regular at the extremities, and one must set different evolution rules at these boundaries, such as, for example, absorbing or reflecting boundaries. A vertex $p$ will thus also be referred to as a position on the curve, or a site. Unless otherwise mentioned, we consider the case of a line, that we orient in the direction of growing $p$'s to the right. The evolution of the walker on the graph is labelled by a discrete time $j \in \mathbb{N}$. Since the walker is quantum, its state belongs to a Hilbert space $\mathcal{H}$ on the field of complex numbers; we denote this state $\ket{\Psi_j}$, in a Schrödinger picture and using Dirac notations. The Hilbert space is the tensorial product $\mathcal{H} = \mathcal{H}_{\text{spin}} \otimes \mathcal{H}_{\text{position}}$ of the Hilbert space $\mathcal{H}_{\text{position}}$ spanned by the position basis $(\ket p)_{p \in \mathbb{Z} }$, and of a finite-dimensional Hilbert space $\mathcal{H}_{\text{spin}}$ accounting for an internal degree of freedom that we call spin\footnote{The word `spin' is used, a priori, \emph{solely} because we deal with a \emph{two-state} quantum mechanical degree of freedom, i.e. this degree of freedom is, a priori, and a fortiori in a 1D context, \emph{not} the standard spin in 3D space, which is associated to the behavior of the quantum system under 3D rotations, i.e. elements of SO(3). However, this does not preclude from using the mapping SU(2) $\rightarrow$ SO(3) to \emph{visualize} the behavior of the two-state degree of freedom we are dealing with under SU(2) operations. See below for information on this mapping.  One often speaks of \emph{pseudo-spin}.}; the spin (Hilbert) space is also called `coin space', the word `coin' being used in analogy with the coin tossing of classical random walks, and the elements of $\mathcal{H}_{\text{spin}}$ are sometimes called `coins'. The minimal dimension for the coin space of a DTQW on a 1D graph is two, and we consider this case from now on, unless otherwise mentioned.  Each element of the coin space can be identified to its components on a certain basis $(\ket \uparrow, \ket \downarrow)$ that we call up-down basis; these components form a couple of complex numbers, that we will note as a column vector so as to perform matrix operations properly. The state of the walker can be decomposed on what we may call the up-down position basis of $\mathcal{H}$, $( \ket s \otimes \ket p )_{s\in \{\uparrow,\downarrow\}, p\in \mathbb{Z}} $:
\begin{equation}
\ket {\Psi_j}
= \sum_{\substack{s \in \{\uparrow,\downarrow\} \\ p \in \mathbb{Z}}} \psi_{j,p}^{s} \ket s \otimes \ket{p} 
= \sum_{p \in \mathbb{Z}}  \left(\psi_{j,p}^{\uparrow} \ket \uparrow + \psi_{j,p}^{\downarrow} \ket \downarrow \right) \otimes \ket{p} \, . 
\end{equation}     
The probability for the walker to be located at time $j$ on site $p$ with spin $s$ is $|\psi_{j,p}^{s}|^2$. 

The evolution of the walker on the graph is governed by a one-step evolution operator $\hat{W}$, also called walk operator,
\begin{equation} \label{eq:protocol}
\ket {\Psi_{j+1}} = \hat{W} \ket{\Psi_j},     
\end{equation}
which is the composition  of two operators $\hat{S}$ and $\hat{U}$,
\begin{equation}
\hat{W} = (\hat{U} \otimes \hat{I}_{\mathcal{H}_\text{position}}) \hat{S} \, ,
\end{equation}
where $\hat{I}_{\mathcal{H}_\text{position}}$ is the identity operator acting on $\mathcal{H}_{\text{position}}$. 

The operator $\hat{S}$ is a spin-dependent shift of the walker to the neighboring sites. We note by $S_{\text{position}}$ its representation on the up-down position basis of $\mathcal{H}$. $S_{\text{position}}$ acts on the two-component `wavefunction'\footnote{We put quotation marks because $\Psi_j$ is only defined on the spatial (1D) lattice, i.e. it is rather a sequence than a function of $p$. It will indeed become a standard wavefunction in the continuum limit, see Subsection \ref{subsec:continuum}.} $\Psi_j:p \mapsto \left( \psi_{j,p}^{\uparrow},\psi_{j,p}^{\downarrow}  \right)^{\top}$, where $\top$ denotes the transposition, as
\begin{equation} \label{eq:shift}
(S_{\text{position}} \Psi_j)_p  
%= \langle \begin{matrix} \langle \uparrow, p+1 | \Psi_j \rangle \\ \langle \downarrow, p-1 | \Psi_j \rangle \end{matrix} \rangle
= \left( \begin{matrix} \psi_{j,p+1}^{\uparrow} \\ \psi_{j,p-1}^{\downarrow} \end{matrix} \right)  ,
\end{equation}
so that the upper spin component moves left in time, and the lower moves right. A key feature of this operator is that it is entangling; in formal terms, this means it cannot be factorized, i.e. written as the tensorial product of an operator acting on the position space and another acting on the spin space.

The operator $\hat{U}$, which acts on $\mathcal{H}_{\text{spin}}$, is a rotation of the walker's spin up to a global phase, and is called `coin operator' (or operation).  Written in a particular basis of the coin space, the coin operation is represented by a $2\times 2$ unitary matrix,
%\footnote{One can also say that `the components of $\hat{U}$ on this particular basis form a $2\times2$ unitary matrix'.}
the unitarity being (in particular) needed for the total occupation probability of the walker, $\Pi_j \equiv \sum_{p \in \mathbb{Z}} \left(|\psi_{j,p}^{\uparrow}|^2 + |\psi_{j,p}^{\downarrow}|^2\right)$, to be preserved in time, i.e. $\Pi_j$ independent of $j$. The set U(2) of $2\times2$ unitary matrices is a Lie group, and admits several parametrizations. In particular, any matrix $U$ of U(2) can be written as the product of a global phase\footnote{The word `phase' will be used either for the angle or the associated complex exponential, but the proper meaning should be obvious from the writing.} $e^{i\alpha} \in \text{U(1)}$ by a matrix $\bar{U}$ of SU(2), the subgroup of U(2) whose matrices have unit determinant\footnote{For a proof of this result in the general case of U($N$), see Appendix \ref{app:unitary_group}, Section \ref{sec:dimN}.}. To each matrix of SU(2), one can associate a rotation matrix belonging  to SO(3)\footnote{The mapping $\text{SU(2)} \rightarrow \text{SO(3)}$ is not one to one but two to one, $\pm U \in \text{SU(2)} \mapsto R(U) \in \text{SO(3)}$, which reflects different topologies: SU(2) is simply connected, while SO(3) is doubly connected, which can be seen by the fact that SO(3) is topologically (i.e. homeomorphic to) a 3D ball with antipodal points identified -- this is the 3D equivalent of identifying the ends of a real interval, which yields a circle, which is doubly connected. SU(2) is the universal covering group of SO(3), and it covers SO(3) twice. This accounts for the fact that the spin does not behave exactly as a vector under rotations.} (this is why I said above that $\hat{U}$ is a rotation up to a phase); this mapping accounts for the classical representation of the spin, that is, an orientation in the 3D (Euclidean) space, i.e. a point on the so-called Bloch sphere. SU(2) can thus be parametrized with the Euler angles of SO(3): we write the matrix representation of $\hat{U}$ on basis $(\ket \uparrow, \ket \downarrow)$ as
\begin{equation} \label{eq:coin_operation}
U^{\text{Euler}}(\alpha,\theta,\xi,\zeta) = e^{i\alpha}
\underbrace{\begin{bmatrix}
e^{i\xi} \cos \theta & e^{i\zeta} \sin\theta \\ -e^{-i\zeta} \sin\theta  & e^{-i\xi} \cos \theta 
\end{bmatrix}}_{\bar{U}^{\text{Euler}}(\theta,\xi,\zeta) \ \in \ \text{SU(2)}} \ 
\end{equation}
to match with the notations of Refs. \cite{DMD12a,DDMEF12a,DMD14}, where 
\begin{equation} \label{eq:mapping_set}
\{ (\alpha,\theta,\xi,\zeta) \in \Sigma = [0, \pi[ \times [0, \pi/2] \times [0, 2\pi[^2 \} \,  
\end{equation}
is a possible set\footnote{This set is natural given the parametrization used.} to map the whole Lie group U(2) one to one\footnote{Except when $\theta=0$ or $\pi/2$, for which this parametrization is not unique.}\textsuperscript{,}\footnote{Note that this set is not what we call a `chart' (or, rather, `chart codomain', the `chart domain' being the space subset which is being mapped), because it is not open. In particular, this set gives no information about the topology of U(2). Only an atlas, i.e. an open cover of U(2) (collection of open sets whose union contains U(2)) together with its corresponding mapping functions to the codomains, namely the charts, can describe the topology of U(2). Because this topology is not trivial, i.e. not that of a cartesian product, but yet simple enough, any atlas of U(2) is made of at least two charts (not more).}, a result proved in Appendix \ref{app:unitary_group}, Section \ref{sec:dim2}, Subsection \ref{subsec:param}, for another possible parametrization set\footnote{One can easily adapt the proof to the above set.}, namely 
\begin{equation} \label{eq:mapping_set_2}
\{ (\alpha,\theta,\xi,\zeta) \in \Sigma' = [0, \pi[ \times [-\pi/2, 0] \times ]-2\pi, 0] \times [0, 2\pi[ \, \} \, ,
\end{equation}
which is more appropriate to link $\theta$, $\xi$ and $\zeta$ to standard Euler angles of SO(3) for an active rotation, as shown in Subsection \ref{subsec:link_Euler}. In the following, we will either use the first set, $\Sigma$, or forget to choose the angles within this set when such details are not needed. Note that $U^{\text{Euler}}(\alpha,\theta,\xi,\zeta)$ is in SU(2) if and only if $\alpha = \text{k} \pi, \text{k} \in \mathbb{Z}$. The well-known Hadamard coin operation, or Hadamard transform, which transforms a spin up or down into an equal superposition of spin up and down\footnote{By `equal superposition', we mean as usual a superposition whose coefficients have equal modulus. The Hadamard transform is not the only equally-weighting operation, since the phases in front of the equal-modulus coefficients can be modified.}, is obtained for $\theta=\pi/4$ and $\alpha=\xi-\pi=\zeta-\pi=\pi/2$, and is thus not in SU(2).
It is useful to have in mind the following expression:
\begin{equation} \label{eq:M_matrix}
\bar{U}^{\text{Euler}}(\theta,\xi,\zeta) = e^{i\frac{\xi+\zeta}{2} \sigma_3} e^{i\theta \sigma_2} e^{i\frac{\xi-\zeta}{2} \sigma_3} \ ,
\end{equation}
having introduced two of the three Pauli matrices,
\begin{equation}
\sigma_1 = \begin{bmatrix}
0 & 1 \\ 1 & 0
\end{bmatrix}  , \ \ \ \
\sigma_2 = \begin{bmatrix}
0 & -i \\ i & 0
\end{bmatrix}  , \ \ \ \
\sigma_3 = \begin{bmatrix}
1 & 0 \\ 0 & -1
\end{bmatrix} .
\end{equation}

The DTQW is said homogeneous in space (resp. time) when the coin operation does not depend on space (resp. time), i.e. when the parameters of the coin operation, namely $\alpha$, $\theta$, $\xi$ and $\zeta$ in the parametrization we use, do not depend on $p$ (resp. $j$). The DTQW is said homogeneous when both homogeneities are satisfied (hence the title of the present subsection).

After $n$ time steps, the walker catches a global phase $n\alpha$ which does not affect the different occupation probabilities, typically the spin-traced one $||\psi_{j,p}||^2 \equiv |\psi_{j,p}^{\uparrow}|^2 + |\psi_{j,p}^{\downarrow}|^2$, or the two spin-dependent ones involved in the latter, $|\psi_{j,p}^{\uparrow}|^2$ and $|\psi_{j,p}^{\downarrow}|^2$.   

An explicit practical way to write the one-step evolution equation, (\ref{eq:protocol}), is
\begin{equation} \label{eq:protocol_explicit}
\left(
\begin{matrix}
\psi^{\uparrow}_{j,p} \\ \psi^{\downarrow}_{j,p}
\end{matrix}
\right)
=
e^{i\alpha}
\begin{bmatrix}
e^{i\xi} \cos \theta & e^{i\zeta} \sin\theta \\ -e^{-i\zeta} \sin\theta  & e^{-i\xi} \cos \theta 
\end{bmatrix}
\left( \begin{matrix} \psi_{j,p+1}^{\uparrow} \\ \psi_{j,p-1}^{\downarrow} \end{matrix} \right) .
\end{equation}
A widespread convention for this one-step evolution equation is to apply the coin operation before the shift, and to make the upper (lower) spin component move right (left) in time. This convention is equivalent to ours up to a time-reversal operation, as detailed in Appendix \ref{app:time_reversal}. We used our convention rather than this widespread one because it is simpler in the following sense: when the coin operation depends on position, applying the shift after the coin operation shifts the parameters of the coin operation and makes the equations more cumbersome.

\subsection{Solution of the homogeneous DTQW on the line by Fourier transform} 

From Eq. (\ref{eq:protocol}), we obtain
\begin{equation} \label{eq:n-step_state}
\ket {\Psi_n} = \hat{W}^n \ket {\Psi_0} \, .
\end{equation} 
Since the positions $p$ are discrete, the representation $W_{\text{position}}$ of $\hat{W}$ on the up-down position basis of $\mathcal{H}$ is a matrix. This matrix has a finite dimension which is easy to determine, as explained in the next paragraph. Hence, solving the previous equation reduces to computing powers of the finite-dimensional matrix $W_{\text{position}}$, which can formally always be done. 

The kind of spatial transport we consider, namely, the spin-dependent shift operation given in Eq. (\ref{eq:shift}), implies that the maximal position $p_{\text{max}}$ that a walker localised at $j=0$ on $p=0$ can reach, is constrained by $|p_{\text{max}}| \leq n$. Hence, to evolve an arbitrary initial wavefunction $\Psi_{j=0}$ up to time $n$, one needs in the end a Hilbert space $\mathcal{H}$ with dimension $D(n)=2(2n+\Delta)$, where the overall factor $2$ is the dimension of the coin space, $\Delta \in \mathbb{N}$ is the support of $\Psi_{j=0}$, and the factor 2 in front of $n$ is because the walker moves both left and right at each time step. One should thus compute, to evolve the walker from time $j-1$ to $j$, the  $D(j)$-dimensional matrix product $W_{\text{position}} \times W^{j-1}_{\text{position}}$. It is computationally faster to first diagonalize the $D(n)$-dimensional matrix $W_{\text{position}}$ and then to compute the $D(n)$-dimensional matrix power, which reduce to scalar powers; this diagonalization is always possible since this matrix is unitary by construction.

The operation  $U^{\text{Euler}}(\alpha,\theta,\xi,\zeta)$ is a $2\times2$ matrix with non-vanishing determinant, whose diagonalization is straightforward. The shift operation $\hat{S}$ is already diagonal in the up-down basis of the coin space: indeed, its `components' on this basis, that we denote by $S=[\hat{S}^{ab}], (a,b) \in \{\uparrow,\downarrow\}^2$, are\footnote{We also define $U = [\hat{U}^{ab}]$, independently of any parametrization of U(2), and then $U^{\text{Euler}}$ is the function that maps $(\alpha,\theta,\xi,\zeta)$ to $U=U^{\text{Euler}}(\alpha,\theta,\xi,\zeta)$. In the case of $\hat{S}$, the `components' on the spin basis are operators acting on the position Hilbert space.} 
\begin{equation} \label{eq:shift2}
S = 
\begin{bmatrix}
\sum_p |p-1\rangle\langle p| & 0 \\
0 & \sum_p |p+1\rangle\langle p|
\end{bmatrix} .
\end{equation} 
$S$ is manifestly non-diagonal in the position basis, for example $\langle p-1 | S | p \rangle = \text{diag}(1,0) \neq 0$. When $D(n)$ is big, and in the case where the coin operation involves no parameters\footnote{By parameters, I mean numerically-undetermined literal quantities, i.e. variables belonging to $\mathbb{R}$.} but only numbers, one can perform the diagonalization of $W_{\text{position}}$ numerically, within some finite numerical precision, in a reasonable time, provided that $D(n)$ is not too big. If we ask for an infinite numerical precision or if the coin operation contains parameters, the diagonalization can be done with a formal-computation software, but this task rapidly becomes very demanding when $D(n)$ increases, and the resulting expressions for the eigenvalues and eigenvectors are often unpractical, i.e. extracting valuable information from them can be difficult\footnote{This is even more the case if the coin operation depends on time or position.}.

Now, for homogeneous walks, the coin operation does not depend on the position $p$, so that we can can give a simple analytical solution to our problem by Fourier transform. The previous discussion is still useful because it holds in the case of spatially inhomogeneous walks, where going to Fourier space does not simplify the problem analytically. Taking the Fourier transform of Eq. (\ref{eq:protocol_explicit}) yields a simple eigenvalue problem for a $2\times2$ matrix, whose solution we give in the next paragraph. The Fourier transform of a sequence $s$ on $\mathbb Z$ takes as an argument a wavevector $k\in\mathbb{R}$ belonging to an interval of length $2\pi$.  We choose the interval to be $[-\pi,\pi[$, in order to have a simple mapping between this wavevector and the notion of momentum of standard quantum mechanics\footnote{In standard quantum mechanics, one typically integrates over \emph{components of the momentum on some axis}, which can be positive or negative, hence the use of this symmetric interval, although any interval of length $2\pi$ can be chosen.}. The wavevector $k$ will be referred to as the quasimomentum. See Appendix \ref{app:quasimomentum} for a discussion on the use of such a terminology. In the continuum limits considered further down in this work, the quasimomentum is replaced by the standard notion of momentum. The wavelength of the modes is by definition $\lambda = 2\pi/|k|$\footnote{This wavelength is bounded by $2$ from below and goes up to infinity. One should not be surprised that it is possible to generate any function on a lattice with a superposition of modes having wavelengths which cannot go below $2$, because the smallest length scale on which a sequence can vary on a lattice or, equivalently, the smallest period of a periodic sequence on this lattice, is $2$.}. We can write this Fourier transform as
\begin{equation}
\tilde{s}_k = \sum_{p\in\mathbb Z} s_p \, e^{-ikp} \, ,
\end{equation}
so that the inverse transform reads
\begin{equation}
s_p = \frac{1}{2 \pi}\int_{-\pi}^{\pi} dk \, \tilde{s}_k \, e^{ikp}  \, .
\end{equation}
More formally, this corresponds to introducing the quasimomentum basis $(\ket k)_{k\in[-\pi,\pi)}$ and the quasimomentum operator $\hat{k}$ which acts on the quasimomentum basis as $\hat{k}\ket k = k \ket k$ and satisfies $e^{i\hat{k}}=\sum_{p \in \mathbb{Z}} |p-1\rangle\langle p|$, so that the shift operation, which reads
\begin{equation} \label{eq:shift_compact}
S = 
\begin{bmatrix}
e^{i\hat{k}} & 0 \\ 0 & e^{-i\hat{k}}
\end{bmatrix}  ,
\end{equation}
is diagonal in the quasimomentum basis. We note 
\begin{equation} \label{eq:shift_fourier}
S_k \equiv \bra k S \ket k = \text{diag}(e^{ik},e^{-ik}) \, .
\end{equation}

One is thus led to the diagonalization of the $2\times2$ unitary matrix $W_k(\alpha,\theta,\xi,\zeta) \equiv U^{\text{Euler}}(\alpha,\theta,\xi,\zeta) S_k$, which is straightforward.
%\footnote{This angle performs rotations in the plane of the Bloch sphere which is orthogonal to the 3D direction defined by the up-down spin basis, called quantization axis; in other words $\zeta$ specifies a basis in the aforementioned plane. Since the DTQW dynamics couples the spatial degrees of freedom to the spin components on the quantization axis, a change in $\zeta$ has no effect on the dynamics. One can see this formally by looking at Eqs. (\ref{eq:protocol_explicit}) and (\ref{eq:M_matrix}): the passage matrix $P(\zeta_1,\zeta_2)$ used to go from a coin operation with some $\zeta_1$ to another one with some $\zeta_2$, chains out thanks to the relation $P(\zeta_1,\zeta_2)P^{-1}(\zeta_1,\zeta_2)=1$ when applying the walk operator several times, so that, in the end, going from $\zeta_1$ to $\zeta_2$ can be viewed as a change of spin basis with no change in the coin operation, which has no effect on the dynamics.}
% \footnote{Since the shift operation is diagonal in the coin space, see Eq. (\ref{eq:shift2}), $e^{-i\frac{\zeta}{2}\sigma_3}$ and $S$ commute. Hence, the one-step evolution equation can be written $\Psi'_{j+1}=U(\alpha,\theta,\xi,0)S \Psi_j$, with $\Psi'_j=e^{-i\frac{\zeta}{2}\sigma_3}}\Psi_j$. Thus, choosing some $\zeta \neq 0$ for the coin simply corresponds to a change of spin basis with no effect on the spatial dynamics;.
A priori, one can deduce from Eqs. (\ref{eq:M_matrix}) and (\ref{eq:shift_fourier}) that the eigenvalues of $W_k(\alpha,\theta,\xi,\zeta)$ do not depend on $\zeta$.
Since $\alpha$ is just a global phase with no effect on the spatial dynamics, we set it to zero. Computing the eigenvalues of $W_k(\alpha=0,\theta,\xi,\zeta)$ yields
\begin{equation} \label{eq:eigenvalues}
\lambda_k^{\pm}(\theta,\xi) = \cos \theta \cos(k+\xi) \mp i \sqrt{1 - \cos^2 \theta \cos^2(k+\xi)} \, .
\end{equation}
One can easily compute the eigenvectors \cite{Venegas_review}.

Let us give more details about the role played by the angles $\zeta$ and $\xi$. Using Eq. (\ref{eq:M_matrix}), the one-step evolution equation can be written (with $\alpha=0$) as
\begin{equation} \label{eq:protocol_xi_zeta}
\Psi_{j+1} = e^{i\frac{\xi+\zeta}{2} \sigma_3} e^{i\theta \sigma_2} e^{i\frac{\xi-\zeta}{2} \sigma_3} S \, \Psi_j \, ,
\end{equation}
where $S$ is given  by Eq. (\ref{eq:shift2}). One can rewrite the former equation as
\begin{equation}
\Psi'_{j+1} = e^{i\frac{\xi}{2} \sigma_3} e^{i\theta \sigma_2} e^{i\frac{\xi}{2} \sigma_3} S \, \Psi'_j \, ,
\end{equation}
where 
\begin{equation} \label{eq:chnage_zeta}
\Psi'=e^{-i\frac{\zeta}{2}\sigma_3}  \Psi \, .
\end{equation}
This means that choosing an angle $\zeta \neq 0$ is equivalent to the change of spin basis given by the previous equation\footnote{To be fully explicit, the SU(2) matrix $e^{-i\frac{\zeta}{2}\sigma_3}$ corresponds to the SO(3) active rotation (we rotate the spin) of axis $\boldsymbol{e}_z$ (vertical unit vector, oriented from bottom to top) and angle $\zeta$, if the counterclockwise, i.e. trigonometric rotation direction, is counted positively, which is the standard convention. Equivalently, it is a rotation of the spin \emph{basis} by the opposite angle $-\zeta$.}, with manifestly no effect on the spatial dynamics. This can be understood on the Bloch sphere: $\zeta$ performs rotations in the plane of the Bloch sphere which is orthogonal to the axis corresponding to the up-down spin basis, called quantization axis\footnote{This was pointed out to me by A. Alberti.}; since the DTQW dynamics couples the spatial degree of freedom to the spin components on the up-down spin basis, the choice of basis in the plane orthogonal to the quantization axis is irrelevant to the dynamics. For $\xi$, the situation is a bit different: Eq. (\ref{eq:protocol_xi_zeta}) can be written
\begin{equation}
\Psi''_{j+1} = e^{i\frac{\zeta}{2} \sigma_3} e^{i\theta \sigma_2} e^{-i\frac{\zeta}{2} \sigma_3} 
\begin{bmatrix}
e^{i(\hat{k}+\xi)} & 0 \\ 0 & e^{-i(\hat{k}+\xi)}
\end{bmatrix} \Psi''_j \, ,
\end{equation} 
where 
\begin{equation}
\Psi''=e^{-i\frac{\xi}{2}\sigma_3} \Psi \, .
\end{equation}
Choosing an angle $\xi \neq 0$ is thus equivalent to both (i) the change of spin basis given by the previous equation\footnote{Corresponding to the SO(3) active rotation of axis $\boldsymbol{e}_z$ and angle $\xi$.}, which has no effect on the spatial dynamics, and (ii) a shift of the wavevector by $\xi$, which appears in the expression of the eigenvalues, Eq. (\ref{eq:eigenvalues}).

Let us now focus on the angle $\theta$, which is the only one that induces superpositions of spin up and down\footnote{$\theta \neq 0$ produces non-diagonal elements in $U^{\text{Euler}}(\alpha,\theta,\xi,\zeta)$.}, and hence entangles the spatial and spin degrees of freedom through the alternance of repeated spin-dependent shifts and coin operations. We will call it \emph{mixing angle}, \emph{spin coupling} or simply \emph{coupling}.

\subsection{Energy spectrum and Hamiltonian of the DTQW} 

Because $W_k(\alpha,\theta,\xi,\zeta)$ is unitary, its eigenvalues have unit modulus, so that we can write
\begin{equation}
\lambda_k^{\pm}(\theta,\xi=0)=e^{-iE^{\pm}_k(\theta)} \, ,
\end{equation}
where we have introduced the \emph{eigenenergies}\footnote{This quantities are `mathematically' eigenfrequencies. In this quantum-mechanical context, one would associate  to these frequencies an energy through the Planck-Einstein relation, i.e. energy = $\hbar$ $\times$ frequency. Since we work with the natural units, energy and frequency have the same value. One could also envisage a notion of energy `simply' related to signal processing; essentially, this means that the proportionality constant between energy and frequency would have a different value than $\hbar$. I don't know to which extent the DTQW scheme could model classical propagation with complex notations. The following 2000 paper by Spreeuw, \url{https://arxiv.org/pdf/quant-ph/0009066.pdf}, constructs a classical analog to quantum-information processing with classical wave optics; one must note, however, that the scaling of the suggested scheme is very poor, i.e. it is basically a classical scaling (exponential growth), which is precisely what one wishes to avoid with quantum simulation. In the present thesis, DTQWs are rather used for their quantumness than for their classical-simulation potential, but the quantumness of our schemes should certainly be investigated in depth.}
\begin{equation} \label{eq:dispersion_relation}
E^{\pm}_k(\theta) = \pm f_k(\theta) = \pm 2 \arctan \frac{ \sqrt{1 - \cos^2 \theta \cos^2 k}}{1 + \cos \theta \cos k} \, .
\end{equation}
Because of symmetry properties of $f_k(\theta)$, it is enough to study this function for $k \in [0,\pi/2]$ (recall that $\theta \in [0,\pi/2]$).
One can easily show using trigonometric formulae that
\begin{equation}
f_k(\theta) = \arccos (\cos \theta \cos k) \ \ \  \text{for} \ \ (k,\theta) \in [0,\pi/2]^2 \ ,
\end{equation}
so that, for example, $f_k(\theta=0)=k$ and $f_{k=0}(\theta)=\theta$. As functions of $k$, the eigenenergies are called \emph{dispersion relations}, \emph{energy spectrums} or \emph{energy bands}, the last denomination being widely used in condensed-matter physics. Here, we have two bands, $E^+(\theta)$ and $E^-(\theta)$. The \emph{energy gap} is the minimum value of the difference between the two bands, $\text{min}_k \{ E^+_k(\theta)-E^-_k(\theta) = 2f_k(\theta) \}$. This gap is then a function of $\theta$. It occurs at $k=0$ and is given by  $2f_{k=0}(\theta)=2\theta$. The gap then only vanishes when $\theta=0$, i.e. when the coin operation is the identity: this is the trivial situation where the upper and lower spin components of the wavefunction do not interfere; the upper is transported left and the lower right, at opposite constant speeds (one space step every time step) and without deformation. When $\theta=\pi/2$ (resp. $k=\pi/2$), the eingenenergies do not depend on $k$ (resp. $\theta$) and equal $\pm \pi/2$.

\iffalse
Figure \ref{}
\fi

One can formally define the Hamiltonian $\hat{H}$ associated to the walk operator by
\begin{equation}
\hat{W} = e^{-i\hat{H}} \ .
\end{equation}

\subsection{Continuum limit for homogeneous DTQWs} \label{subsec:continuum}

Consider the effect of a weak coupling, $\theta \ll \pi/2$, on the dynamics of the large-wavelength Fourier modes ($\lambda \gg 1$, i.e. $|k| \ll \pi$) of an initial walker $\Psi_{j=0}$. Taylor expanding Eq. (\ref{eq:dispersion_relation}) at leading order in $\theta$ and $k$ yields
\begin{equation} \label{eq:linearised_band}
 E_k^{\pm}(\theta) = \pm \sqrt{k^2+\theta^2} \, \, + \, \, \text{higher order terms} \ ,
\end{equation}
which is a relativistic dispersion relation with mass $\theta$. We say that the dispersion relation (\ref{eq:dispersion_relation}) has a \emph{Dirac cone} at $k=0$ when $\theta=0$. In the following, we are going to show that, in the limit of infinitesimally-small coupling, wavevectors and frequencies, which corresponds to a continuous limit in both space and time, the dynamics of the walker is that of spin-1/2 relativistic particle in free space. We call this limit the continuum limit.

To be coherent with the papers I coauthored and with the litterature of Di Molfetta et al., I will use $\zeta=\pi/2$, keeping $\alpha=\xi=0$ for aforementioned reasons ($\alpha$ and $\xi$ are irrelevant to the dynamics). This coin operation reads
\begin{align} \label{eq:standard_coin}
C(\theta) &= U^{\text{Euler}}(\alpha=0,\theta,\xi=0,\zeta=\pi/2) = \begin{bmatrix}
\cos \theta & i \sin \theta \\ i \sin \theta & \cos \theta
\end{bmatrix} \, .
\end{align}
It is \emph{balanced} (or \emph{unbiased}) for $\theta = \pi/4$, i.e. it transforms each spin component, up or down, in an equally-weighted superposition of both, as the Hadamard transform does, but differs from the latter in the sign of the determinant ($C(\theta)$ do belong to SU(2)).

Let us come back to the dynamics of the large-wavelength Fourier modes of the walker: one can push this to the continuous-space limit. To handle this limit formally, we introduce a (dimensionless) space step $\epsilon_x \in \mathbb{R}^+$, assume that $\Psi_{j,p}$ coincides with the value taken by a function $\Psi_j: p \mapsto  \Psi_j(x_p)$ at space coordinate $x_p = p \, \epsilon_x$, and let $\epsilon_x$ go to zero. Now, \emph{provided that $\theta \ll \pi/2$}, then a small wavenumber implies, from Eq. (\ref{eq:linearised_band}), small eigenenergies i.e. large eigenperiods, which can be pushed to the continuous-time limit, so that we can write derivatives with respect to time and, in the end, describe the evolution of those large-wavelength and large-period modes by partial differential equations (PDEs). We then also introduce a dimensionless time step $\epsilon_t\in \mathbb{R}^+$, assume that $\Psi_j(x_p)$ coincides with the value taken by a function $\Psi(\cdot,x_p): j \mapsto  \Psi(t_j,x_p)$ at time $t_j = j \, \epsilon_t$, and let $\epsilon_t$ go to zero with $\epsilon_x$. 

The condition $\theta \ll \pi/2$ seems necessary to go to the continuous-time limit, as underline above; let us evaluate this formally. We allow $\theta$ to read
\begin{equation}
\theta = \theta_0 + \Delta \theta \, ,
\end{equation}
where $\theta_0$ is the limit of $\theta$ when $\epsilon_t$ and $\epsilon_x$ go to zero, and $\Delta \theta$ goes to zero with them. To handle the continuum limit formally, we write
\begin{equation}
\Delta \theta =  \epsilon_m \bar \theta \, ,
\end{equation}
i.e. we \emph{scale} $\Delta \theta$ by some $\epsilon_m \in \mathbb{R}^+$ (the index $m$ will make sense further down) that goes to zero with $\epsilon_t$ and $\epsilon_x$,  $\bar \theta$ being an arbitrary real parameter. Using the coin operation (\ref{eq:standard_coin}), Eq. (\ref{eq:protocol_explicit}) reads
\begin{align} \label{eq:walk_expanded}
\left( \begin{matrix}
\psi^{\uparrow}(t_j+\epsilon_t,x_p) \\
\psi^{\downarrow}(t_j+\epsilon_t,x_p)
\end{matrix}
\right) =  \begin{bmatrix}
\cos (\theta_0 + \epsilon_m \bar \theta) & i \sin (\theta_0 + \epsilon_m \bar \theta) \\ i \sin (\theta_0 + \epsilon_m \bar \theta) & \cos (\theta_0 + \epsilon_m \bar \theta)
\end{bmatrix} 
\left(
\begin{matrix}
\psi^{\uparrow}(t_j,x_p+\epsilon_x) \\
\psi^{\downarrow}(t_j,x_p-\epsilon_x) 
\end{matrix}
\right) .
\end{align}
Assume that $\Psi$ is differentiable at least once in both $t$ and $x$, and that both derivatives are continuous, so that we can Taylor expand the left- and right-hand sides of the previous equation at first order in $\epsilon_t$, $\epsilon_x$, and $\epsilon_m$. The expansion must be valid for any $(\epsilon_t,\epsilon_x, \epsilon_m)$, in particular for $(0,0,0)$, so that the zeroth-order terms of the expansion must balance:
\begin{equation} \label{eq:zeroth-order_constraint}
\left(
\begin{matrix}
\psi^{\uparrow}(t,x) \\
\psi^{\downarrow}(t,x)
\end{matrix}
\right) =
\begin{bmatrix}
\cos \theta_0 & i \sin \theta_0 \\ i \sin \theta_0 & \cos \theta_0
\end{bmatrix} 
\left(
\begin{matrix}
\psi^{\uparrow}(t,x) \\
\psi^{\downarrow}(t,x)
\end{matrix}
\right) .
\end{equation}
This is possible only if $C(\theta_0)$ is the identity matrix, i.e. if 
\begin{equation}
\theta_0 = 0 \, ,
\end{equation}
actually $\theta_0 = n\pi$ with $n \in \mathbb{Z}$, but considering $\theta \in [0, \pi/2] $ is enough to parametrize any coin operation (see the mapping set for SU(2), \ref{eq:mapping_set}). This zeroth-order constraint, namely
\begin{equation}
C(\theta) \xrightarrow[\epsilon_x,\epsilon_t \rightarrow 0]{} \mathbf{1}_2 \, ,
\end{equation}
is the proper formal way to express the informal small-coupling condition, $\theta \ll \pi/2$.

\iffalse
, this weak coupling is a necessary condition for either (i) the large-wavelength ($\rightarrow$ continuous-space) modes of an initial walker to have, at any time $t$, large periods with respect to the time step $\epsilon_t$ (their frequency spectrum, Fourier transform in time on some time window $[0,t]$, is dominated by frequencies much smaller than $2 \pi / \epsilon_t$) or, in a complementary point of view, (ii) for the low-frequency  ($\rightarrow$ continuous-time)  modes of a walker to have, at any time $t$, large wavelengths w.r.t the step of the lattice (their wavenumber spectrum, Fourier transform in space on some space window $x_{\text{max}}$, is dominated by wave numbers much smaller than $2 \pi / \epsilon_x$). 
\fi

%Let us now see what PDE one obtains for $\Psi$ once the existence of the continuum limit is ensured by the balance of the zeroth-order terms. 
If the zeroth-order constraint is satisfied by the coin operation, then the limit to the continuum is ensured, i.e. the zeroth-order terms of the above-mentioned Taylor expansion cancel each other. The PDE statisfied by $\Psi$ is obtained by retaining, on both sides of this Taylor expansion, the first-order terms corresponding to the small parameter that goes to zero the slower, either $\epsilon_t$, $\epsilon_x$ or $\epsilon_m$. We write $\epsilon_t = \epsilon_m^\delta$ and $\epsilon_x = \epsilon_m^{\rho}$, where the exponents $\delta > 0$ and $\rho > 0$ account for the fact that $\epsilon_t$, $\epsilon_x$ and $\epsilon_m$ may tend to zero differently. The richest PDE is actually obtained when $\delta = \rho = 1$, i.e. when all three small parameters go to zero at the same `speed', and reads  
\begin{subequations} \label{eq:Dirac_PDE}
\begin{align} 
(\partial_0 - \partial_1) \psi^{\uparrow} & = i \bar \theta \psi^{\downarrow}   \label{eq:Dirac_PDE_1} \\
(\partial_0 + \partial_1) \psi^{\downarrow} & = i \bar \theta \psi^{\uparrow} \ ,  \label{eq:Dirac_PDE_2}
\end{align}
\end{subequations}
where $\partial_0 = \partial_t$ and $\partial_1 = \partial_x$. This equation can be recast as
\begin{equation} \label{eq:Dirac_free}
(i \gamma^{\mu} \partial_{\mu} - m)\Psi = 0 \ ,
\end{equation}
where (hence the index $m$ in $\epsilon_m$)
\begin{equation}
m \equiv -\bar \theta \, ,
\end{equation}
the index $\mu$ is summed over from $0$ to $1$, and the gamma matrices have been defined as $\gamma^0 \equiv \sigma_1$,  $\gamma^1 \equiv i\sigma_2$. This equation is the flat-spacetime Dirac equation, with Lorentzian-metric convention $[\eta^{\mu\nu}]= \text{diag}(+,-)$, for a relativistic spin-1/2\footnote{In the definition of the DTQW, the word `spin' has been used solely because of the two-state (or, more generally, finite-dimension) property of the coin degree of freedom. Now that (i) the 1D space is not a lattice anymore, but \emph{continuous}, and that the walker dynamics is governed by the Dirac equation, one can speak of \emph{(Dirac) spinor in (1+1)D}, and in this sense use the word `spin 1/2'. However, this `spin 1/2' is \emph{not} the usual spin 1/2 that particles `living' in 3D space can have, but actually corresponds to so-called \emph{chirality}, and one often speaks of \emph{chiral fermions}. Indeed, in one spatial dimension rotations do not exist (only reflections), and the (1+1)D spin 1/2 can thus not be associated to the behavior of the system under 3D rotations. For information on spinors and the Dirac equation in 1+1 dimensions, see, e.g., the following notes, by N. Wheeler, \url{http://www.reed.edu/physics/faculty/wheeler/documents/Classical\%20Field\%20Theory/Miscellaneous\%20Essays/A.\%202D\%20Dirac\%20Equation.pdf}, and the following, by R. Davies, for the construction of spinors in various dimensions, \url{http://www.rhysdavies.info/physics_page/resources/notes/spinors.pdf}. In 2D space, there is no notion of chirality, and the two-state internal degree of freedom of the (1+2)D Dirac wavefunction has another physical status. In 3D space, there is both chirality and standard spin 1/2 associated to 3D rotations, and the  (1+3)D Dirac wavefunction has a four-state internal degree of freedom.} fermion of mass $m$. The gamma matrices satisfy the so-called Clifford algebra, $\gamma^{\mu} \gamma^{\nu} + \gamma^{\nu} \gamma^{\mu} = 2 \eta^{\mu \nu}$. The weak-coupling limit of large-wavelength and low-frequency modes is thus described by an effective fermionic classical-field equation.

We already mentioned that a choice $\zeta \neq \pi/2$  can be viewed as a spin-basis change with no change in the coin operation $C(\theta)$, namely  $\Psi \rightarrow e^{i\frac{-(\zeta-\pi/2)}{2}\sigma_3}\Psi$ (active-rotation presentation), which does not affect the dynamics. If we do not absorb $\zeta$ in the spin basis, this results, in the continuum limit, in a complex-valued mass $m$ \cite{DMD14}.
A choice $\xi \neq 0$ can be viewed as both a spin-basis change, namely $\Psi \rightarrow e^{-i\frac{\xi}{2}\sigma_3}\Psi$ (active-rotation presentation), and a substitution $k \rightarrow k'=k+\xi$, so that the small-wavenumber condition needed to go to the continuous-time limit becomes $|k+\xi| \ll \pi$ if we keep the variable $k$, i.e. the Dirac cone obtained for $\theta=0$ is located at $k=-\xi$.

As a final remark regarding homogeneous QWs (not only DTQWs, but also CTQWs), I would like to mention a publication I like a lot, namely Ref. \cite{Patel2005}, which I recommend as a concise and big-picture introduction to QWs, that should obviously be completed by more detailed reviews.
    
\section{Inhomogeneous DTQWs} \label{sec:inhomogeneous}
    
\subsection{Introduction}    
    
We are now interested in making the parameters of the coin operation, $\alpha$, $\theta$, $\xi$ and $\zeta$, time and position dependent: $\alpha_{j,p}$, $\theta_{j,p}$, $\xi_{j,p}$ and $\zeta_{j,p}$.  The walker's dynamics on the spacetime lattice can now be determined by Fourier transform only if none of those parameters is space dependent.   
     
\subsection{Continuum limit for inhomogeneous DTQWs and electric coupling} \label{subsec:continuum_limit_inhomogeneous}

We have seen previously that the condition for the continuum limit to exist is that the coin operation tends to the identity matrix when $\epsilon_t$ and $\epsilon_x$ go to zero. 
This constrains, in addition to the previous condition on $\theta_0$, i.e. $\theta_0=0$, (i) the angles $\alpha$ and $\xi$ to tend to space- and time-independent values in the continuum limit, and (ii) such values to belong to a certain discrete set \cite{DMD14}; the angle $\zeta$, which may depend on $j$ and $p$, is not constrained \cite{DMD14} (no difference from the homogeneous case, i.e. $\zeta \neq \pi/2$ yields a complex-valued mass, but this time $\zeta$ can depend on $j$ and $p$.). Let us then, as we did for $\theta$, assume that
\begin{equation}
\alpha_{j,p} = \alpha_0 + {\Delta \alpha}_{j,p} \, , \ \ \ \ \ \ \ \ \ \ \ \xi_{j,p} = \xi_0 + {\Delta\xi}_{j,p} \, ,
\end{equation} 
where $\alpha_0$ and $\xi_0$ are possible values for the angles in the continuum limit, and
\begin{equation}
\Delta{\alpha}_{j,p} = \epsilon_A {\bar \alpha}_{j,p} \, , \ \ \ \ \ \ \ \ \ \ \ \Delta{\xi}_{j,p}  = \epsilon_A \bar \xi_{j,p} \, ,
\end{equation} 
where we have introduced some $\epsilon_A \in \mathbb{R}^+$ (the index $A$ will make sense further down) that goes to zero with $\epsilon_t$, $\epsilon_x$ and $\epsilon_m$, and  $\bar \alpha$ and $\bar \xi$ may have any dependence on time $j$ and position $p$.  For simplicity, let us make the possible  \cite{DMD14} choice
\begin{equation}
\alpha_0 = \xi_0=0 \, .
\end{equation}
\noindent
We assume that ${\bar \alpha}_{j,p}$ and ${\bar \xi}_{j,p}$ coincide with the values ${\bar \alpha}(t_j,x_p)$ and ${\bar \xi}(t_j,x_p)$ taken by continuous functions ${\bar \alpha}$ and ${\bar \xi}$ of time and space. We choose $\zeta$ of the form $\zeta = \zeta_0 + \Delta\zeta$,  with $\zeta_0=\pi/2$ to be coherent with the previous example, and $\Delta \zeta = - \Delta \xi$, for  at least two reasons, explained below, and which are independent from the continuum limit.
The coin operation obtained can be written
\begin{subequations} \label{eq:electric_walk_intro}
\begin{align}
U^{\text{Euler}}(\alpha,\theta,\xi,\zeta) &= U^{\text{Euler}}(\Delta \alpha, \Delta \theta, \Delta\xi, \pi/2-\Delta \xi)  \\
 &= e^{i\Delta \alpha} C(\Delta\theta) F(\Delta\xi) \, ,
\end{align}
\end{subequations}
where we have introduced the spin-dependent phase-shift operation,
\begin{equation} \label{eq:phase_shift}
F(\omega)= \begin{bmatrix}
e^{i\omega} & 0 \\ 0 & e^{-i\omega}
\end{bmatrix} .
\end{equation}
At each time step, not only the walker's spin is rotated by an amount $\Delta \theta$, but the walker also catches a local spin-independent phase $\Delta \alpha_{j,p}$ and a local spin-dependent one, $\pm \Delta \xi_{j,p}$. The PDE obtained in the continuum limit is, setting all four small parameters equal,  i.e. $\epsilon_t=\epsilon_x=\epsilon_m=\epsilon_A$,
\begin{equation} \label{eq:Dirac_coupled}
(i \gamma^{\mu} D_{\mu} - m)\Psi = 0 \ ,
\end{equation}
where $D_{\mu} = \partial_{\mu} - i A_{\mu}$ is the covariant derivative, with
\begin{equation}
A_0 = \bar \alpha \, , \ \ \ \ \ \ \ \ \ \ \  A_1 = - \bar \xi \, ,
\end{equation}
(hence the index $A$ in $\epsilon_A$). This is the same Dirac equation as Eq. (\ref{eq:Dirac_free}) but, this time, for a fermion coupled,  through a charge $q=-1$, to an electromagnetic potential with covariant components $A_{\mu}$. The mass $m$ can also depend on time and space, i.e. the angle $\theta$ may also depend on $j$ and $p$ through $\bar \theta$.

As long as $\Delta\zeta$ goes to zero with the other small parameters, any choice of $\Delta\zeta$ delivers the same continuum limit, because, in the Taylor expansion in the small parameters, the term containing $\Delta \zeta$ is multiplied by $\epsilon_m$, and is then of order more than one in $\epsilon_m$. The reason why we have chosen  $\Delta\zeta = - \Delta \xi$ is theoretical, and explained in the next subsection\footnote{In the case one is only interested in simulating the continuum limit, I see a posteriori another reason which leads to choose $\Delta\zeta = - \Delta \xi$, without invoking the theoretical reason explained in the next subsection, and which is both experimental and computational: as manifest in Eq. (\ref{eq:M_matrix}), shifting $\xi$ by $\Delta \xi$ without simultaneously shifting $\zeta$ by $-\Delta \xi$ at each time step involves performing \emph{two} spin-dependent phase shifts of the form (\ref{eq:phase_shift}), instead of a single one with the retained choice $\Delta \zeta = - \Delta \xi$.}.

\subsection{Lattice gauge invariance of DTQWs} \label{subsec:gauge_inv}

We know that the Dirac equation (\ref{eq:Dirac_coupled}) is invariant under local phase rotations: when we perform the substitution $\Psi(t,x) \rightarrow \Psi'(t,x)=e^{-i\phi(t,x)}\Psi(t,x)$ in (\ref{eq:Dirac_coupled}), where $\phi$ is some local phase, the equation is kept invariant provided that we also shift the 2-vector potential by the 2-gradient of the local phase,
\begin{equation} \label{eq:Abelian_gauge_transfo}
A_{\mu} \rightarrow A'_{\mu} = A_{\mu} - \partial_{\mu} \phi \, .
\end{equation}
The electric field derived from the electromagnetic potential is the same for any function $\phi$: this property, which already exists in classical electrodynamics, is called \emph{gauge freedom}, and the potential $A_{\mu}$ is then called \emph{gauge potential}. This so-called \emph{gauge invariance} is that associated to charge conservation through Noether's theorem.  We can view this invariance property with the gauge principle\footnote{A review, by Jackson and Okun, on the historical ``roots'' of the gauge principle formulated by Weyl in 1929 \cite{Weyl1929}, can be found at \url{https://arxiv.org/vc/hep-ph/papers/0012/0012061v1.pdf}. The post-Weyl developments are only mentioned. To Weyl, this gauge principle was a first step in the developement of a unified theory of electromagnetism and gravity, on which he worked. See the following paper on the spin connection in Weyl space, \url{http://www.weylmann.com/connection.pdf}.}: the free Dirac equation, (\ref{eq:Dirac_free}), is invariant under \emph{global} phase rotations; for it to be invariant under \emph{local} phase rotations, one must introduce a gauge field which transforms as written above under local phase rotations.

One is thus naturally led to evaluating the gauge invariance of 1D DTQWs under local phase rotations on the spacetime lattice. It turns out that the one-step evolution equation, (\ref{eq:protocol_explicit}), is invariant \cite{DMD14} under the substitution $\Psi_{j,p} \rightarrow \Psi'_{j,p}=e^{-i\phi_{j,p}}\Psi_{j,p}$, provided that we replace the spacetime-dependent angles by
\begin{subequations} \label{eq:transfo_grid}
\begin{align}
\alpha'_{j,p} &= \alpha_{j,p} + \Sigma_{j,p} - \phi_{j+1,p} \\
\xi'_{j,p} &= \xi_{j,p} + \Delta_{j,p}\\
\zeta'_{j,p} &= \zeta_{j,p} - \Delta_{j,p} \\
\theta'_{j,p} &= \theta_{j,p} \, ,
\end{align}
\end{subequations}
where
\begin{equation}
\Sigma_{j,p} =  (\phi_{j,p+1} + \phi_{j,p-1}) / 2 \, , \ \ \ \ \ \ \ \Delta_{j,p} =  (\phi_{j,p+1} - \phi_{j,p-1}) / 2 \, .
\end{equation}
The angle $\theta$ is manifestly gauge invariant. The sum $\xi + \zeta$ is also gauge invariant. This means that we don't need the variable $\xi + \zeta$ to have a gauge invariant DTQW; we thus set it to zero, i.e. we take $\zeta= - \xi$, which explains the choice made in Subsection \ref{subsec:continuum_limit_inhomogeneous}. 
Transformations (\ref{eq:transfo_grid}) with the choice $\zeta=-\xi$ thus provide a discrete transformation for the phases, which delivers, in the continuum, the standard gauge transformation of the electric 2-potential \cite{DMD14}, which is formally made explicit by Eq. (\ref{eq:discrete_gauge_inv}) in Subsection \ref{subsec:recap_1D}.

\section{Reviewing seminal papers} \label{sec:seminal}

\subsection{Feynman's checkerboard} \label{subsec:Feynman's_quantum_walks}

As far as I know, the first appearance of DTQWs in the litterature is due to Feynman. In 1965, the book \emph{Quantum Mechanics and Path Integrals} \cite{FeynHibbs65a}, written by Feynman and Hibbs, was published; it is based on notes taken by the latter while attending lectures of the former. In Problem 2-6 of this book, one considers a particle that, at each time step, can only move one step to left or to the right, at the velocity of light (which is $1$ in natural units). At each time step, the particle has a probability amplitude $i \epsilon_t m$ of changing its direction, where $\epsilon_t$ is the (dimensionless) time step and $m$ the (dimensionless) mass of the particle. In Feynman's work, the time step is $\epsilon=\epsilon_t m =\epsilon_t/(1/m)$, i.e. it is given in units of $1/m$ \cite{FeynHibbs65a}. This model is known as the Feynman checkerboard (or chessboard), and corresponds to the following particular finite differentiation of the Dirac equation in dimension 1, written by Feynman himself in his notes of 1946 on the ``Geometry of Dirac's Equation in 1 dimension'', reproduced in Fig. 8 of Ref. \cite{Schweber86a}:
\begin{subequations}
\begin{align}
\psi_{L}(t,x) &= \psi_L(t-\epsilon,x-\epsilon) + i \epsilon \, \psi_R(t-\epsilon,x-\epsilon) \\
\psi_{R}(t,x) &= \psi_R(t-\epsilon,x+\epsilon) + i \epsilon \, \psi_L(t-\epsilon,x+\epsilon) \, .
\end{align}
\end{subequations}
The subscript `$L$' (resp                                                                                                                    . `$R$') is for `\emph{coming} from the left (resp. right)', i.e. `\emph{going} to the right (resp. left)'. These equations are those of a DTQW \emph{except for the non-unitarity of the coin operation}: indeed, they can be recast as
\begin{equation}
\Psi(t,x) = \left(S^{\text{s}}_{\text{position}} K^{\text{s}} \, \Psi(t-\epsilon,\cdot)\right)_x \, ,
\end{equation}
where
\begin{equation}
\Psi(t,x) = \left( \begin{matrix}
\psi_L(t,x) \\ \psi_R(t,x)
\end{matrix} \right) ,
\end{equation} 
the spin-dependent shift operation acts as
\begin{equation}
\left(S^{\text{s}}_{\text{position}} \Psi(t,\cdot)\right)_x = \left( \begin{matrix}
\psi_L(t,x-\epsilon) \\ \psi_R(t,x+\epsilon)
\end{matrix} \right)  ,
\end{equation} 
and the \emph{non-unitary}\footnote{This non-unitarity can be fixed by dividing $K^{\text{s}}$ by $\sqrt{1+\epsilon^2}$. A normalization procedure of this type, but more general, is also used in the more general quantum lattice Boltzmann methods  (QLB) \cite{Succi1993}, to ensure the unitarity of the scheme. The coin operations resulting from (i) such a way of discretizing the Dirac equation (i.e., the way of Feynman and of QLB methods) and (ii) then ensuring unitarity (which is only done in QLB methods, not in Feynman's notes), can be written in the standard angular form with cosines and sines, see Eq.\ \eqref{eq:coin_operation}, through a certain mapping \cite{Succi2015}.} coin operation is given by
\begin{equation}
K^{\text{s}} = \begin{bmatrix}
1 & i\epsilon \\ i \epsilon & 1
\end{bmatrix} .
\end{equation}
The superscript `s' is for standard, because I use in this thesis a different convention for the one-step evolution equation, in which the coin operation is applied \emph{after} the shift, but both conventions are equivalent up to a time-reversal operation, as detailed in Appendix \ref{app:time_reversal}. See also the last paragraph of Subsection \ref{subsec:notations} for comments on these two conventions.

This dynamics yields a family of possible zig-zag paths (in the $(x,t)$ plane) taking the particle from a given starting position to a given final one. To each path corresponds a probability amplitude of taking it, and the problem suggests to show that the sum of all these probability amplitudes yields, in the continuous-spacetime (or continuum) limit $\epsilon \rightarrow 0$, the standard (retarded) propagator for the Dirac equation in dimension 1. This propagator is a $2 \times 2$ matrix. The discrete-spacetime version suggested in the problem \cite{FeynHibbs65a} is noted $[K_{\beta\alpha}(\epsilon)], (\alpha,\beta) \in \{ L,R\}^2$, and is given by
\begin{equation}
K_{\beta\alpha}(\epsilon) = \sum_{R} N_{\beta\alpha}(R) (i\epsilon)^R \, ,
\end{equation}
where $N_{\beta\alpha}(R)$ is the number of paths with $R$ corners whose first (resp. last) step is in the $\beta \ (\text{resp.} \ \alpha)$ direction, and $(i\epsilon)^R$ is the probability amplitude of each such path. The problem suggests to find the expressions of the $N_{\beta\alpha}$'s and to take the continuum limit, but no solution is provided in the book.  

Narlikar \cite{Narlikar71} gives expressions for the four $N_{\beta\alpha}$'s, which involve combinatorial factors.
Gersch \cite{Gersch1981} solves the problem without computing the $N_{\beta\alpha}$'s, thanks to a mapping between $K_{\beta\alpha}$ and the partition function of an Ising model. Jacobson and Schulman \cite{JS84} also provide expressions for the $N_{\beta\alpha}$'s, without citing Narlikar, whose work may have been unknown to them\footnote{Jacobson wrote in 1985 a review on the subject, citing the previous related papers he had coauthored until then \cite{Jacobson1985review}. More than a review, this work actually contains a new formulation of Jacobson's previous work on the subject, see the next paragraph for more details on this.}. Kauffman and Noyes \cite{KN96} provide explicit wavefunction solutions for the DTQW-fashion finite-differentiated Dirac equation, which involve the same kind of combinatorial factors as those aforementioned. Kull and Treumann \cite{KT99} also provide wavefunction expressions in discrete spacetime, similar to those written in \cite{KN96}, and with combinatorial arguments similar but slightly different from those used to derive the $N_{\beta\alpha}$'s in the previous works; they show that the continuum limit of these wavefunctions concides with solutions of the Dirac equation, but they do not check this at the discrete level, i.e. on a finite-differentiated Dirac equation, as \cite{KN96} does. While reviewing works \cite{JS84} and \cite{KN96}, Earle \cite{E08} points out some errors they contain, and gives in particular correct expressions for the few-steps paths. All works recover the appropriate continuum limit.

 In Ref.\ \cite{J84}, Jacobson provides a certain (1+3)D generalization\footnote{This generalization can, in addition, incorporate non-Abelian Yang-Mills couplings.}, which is however (ii) \emph{not} on a lattice but where the steps can be taken in any direction on a sphere\footnote{This (1+3)D generalization was actually also formulated on a lattice, but in Jacobson's PhD thesis, which is unpublished.}, and (i) not unitary\footnote{Although this can be fixed, as mentioned in the paper and as developed in Jacobson's PhD thesis, which is unpublished.}. This scheme requires superluminal (although bounded) motion,  as extensively discussed in the paper\footnote{This fact is also mentioned by Plavchan in his informal paper ``Feynman's checkerboard, the Dirac equation and spin'', at \url{http://www.brannenworks.com/plavchan_feynmancheckerboard.pdf}. I thank A. Alberti for sending me this reference.}. In a private communication I had with him, Jacobson wrote me that this superluminality is ``necessary for the convergence to the continuum result''. He also wrote me: ``I hoped for steps only at the speed of light, until I understood that was impossible, without forward and backward in time steps. This is one thing I did in the paper [...] Feynman's Checkerboard and Other Games. It is not just a review, as it has this totally new formulation in it.". A work of 2016 by Foster and Jacobson \cite{Foster2016} also presents a (1+3)D generalization; this scheme is on a lattice, is non-unitary, and also requires superluminal (although bounded) motion.

A recent (2014) rigorous mathematical formulation of a 3D Dirac-equation checkerboard-like path integral, by W. Ichinose \cite{Ichinose2014}, also requires superluminal motion (the speed can even be infinite in this case). In this paper, it is pointed out that Feynman \cite{Feynman1985book} noted that the same superluminal motion could happen in the path integral of a photon, which is a (massless) \emph{bosonic} particle (with spin 1). An earlier (2005) work  by T. Ichinose \cite{Ichinose2005} first treated the same problem in radial coordinates, with distance $r$ to the origin, where there is a possibly-superluminal local speed $c(r)=r$ (in natural units). The following works should also be reviewed for further research: \cite{Rosen1983,Ord1993,SmithJr95}.

Let us also mention the work of Riazanov \cite{Riazanov58}, prior to the Feynman checkerboard, which already ``exploits the tie between spin and translation" \cite{J84} to build, thanks to a ``fith parameter" \cite{J84} (in addition to the four coordinates of spacetime) which is ``integrated out in the end" \cite{J84}, a propagator for the (1+3)D Dirac equation. Gaveau et al. \cite{Gaveau1984} and then Karmanov \cite{Karmanov1993} investigate the link between the aforementioned continuum-limit propagators and random processes, through analytic continuation, see Subsection \ref{sec:Wick}.

Let me put an end to this section and the questions of (i) connecting Feynman's checkerboard to DTQWs and (ii) generalizing it to 1+3 dimensions, by citing an important reference: Bialynicki-Birula's 1994 paper \cite{BB94a}, is probably the first (1+3)D generalization of Feynman's checkerboard whose unitarity does not have to be fixed by any normalization but is in-built, thanks to the use of chirality-rotation matrices, i.e., coin operations, such as those we use for DTQWs. Bialynicki-Birula's scheme can hence, in particular, be considered as a seminal DTQW. I mention this work in relation to other ones below in this thesis: the work was extensively reconsidered separately by C.\ M.\ Chandrashekar and by G.\ M.\ D'Ariano and his team. According to Jacobson (private communication), Bialynicki-Birula's scheme may also, as his scheme, be superluminal, for the same reason, namely, it is necessary for the convergence to the continuum limit.

\subsection{Aharonov's quantum random walks}

In their paper of 1990 entitled ``Quantum random walks'' \cite{ADZ93a}, Aharonov, Davidovich and Zagury introduce what can be viewed as a standard DTQW in which we measure and reinitialize the spin at each time step. Kempe's review \cite{Kempe_review} provides details on this work, in particular on the part in which, thanks to the large-wavepacket approximation, the effects of the quantum interferences between the left- and right-going walkers are pedagogically exemplified (Paragraphs 4 and 5 of the original paper). In Appendix \ref{app:Aharonovs_scheme}, I give extensive formal material to treat the theoretical part of the original paper (from the beginning up to Paragraph 6, included), adding some physical insights. The conclusions are the following. 

The authors show that in the absence of coin operation, their protocol essentially corresponds, effectively, to a classical random walk on the line. Indeed, from a given time step to the next one, the wavefunction is simply shifted one step to the right or to the left with some probabilities $|c^+|^2$ and  $1-|c^+|^2$, respectively, without deformation (but with a possible phase change). 

Introducing an appropriate coin operation enables the following effect to appear: if, after successive spin measurements, the classical counterpart walker, which has followed a classical random path, is at some distance from its original position, it is possible for the quantum-walker \emph{mean position} to be much further away. This applies to any classical path, including single-direction-going ones (example taken by the authors), but this does not mean that the quantum walker can overcome the well-known light-cone position bound of DTQWs (the particularities of Aharonov's scheme with respect to standard non-measured DTQWs change nothing to this essential feature of DTQWs, determined by the spin-dependent shift operation). Indeed, imagine a wavepacket whose initial support is bounded and large with respect to the lattice step; its edges\footnote{If the support is not bounded, and this merely a technical point, one tends to the effective speed of light in the limits $x\rightarrow \pm \infty$, where $x$ is the position of the walker.} will move at the effective speed of light (with opposite directions), but this does not prevent the center of the wavepacket to go faster than the speed of light \emph{for some finite time (not forever)}. A subtlety regarding this matter is that, to show this effect regarding the center of the wavepacket, one does an approximation of non-deformation of the wavepacket, while this deformation is necessary to account for the respect of the light-cone position bound.

\section{Wick rotations in relativistic and non-relativistic frameworks} \label{sec:Wick}

It is known that the Schrödinger equation\footnote{For English translations of the original four-part paper, go to \url{http://www.physics.drexel.edu/~bob/Quantum_Papers/Schr_1.pdf}, \url{http://www.physics.drexel.edu/~bob/Quantum_Papers/Schr_2.pdf}, \url{http://www.physics.drexel.edu/~bob/Quantum_Papers/QAEP3.pdf}, and \url{http://www.physics.drexel.edu/~bob/Quantum_Papers/QAEP4.pdf}.} \cite{Schrodinger1926} can be obtained from the diffusion equation\footnote{The diffusion equation is satisfied, in particular, by the expectation value of the Wiener process, a result known as the Feynman-Kac formula \cite{kac1951}.} \cite{Fick1855Allemand,Fick1855,Einstein1905} by analytic continuation\footnote{Also called Wick rotation, which maps the (1+3)D Minkowski spacetime to the 4D Euclidean space \cite{Wick1954}.}, as well as the (\emph{fermionic}) relativistic version of the first, the Dirac equation, is obtained, \emph{in the case of one spatial dimension}, by analytic continuation of the telegrapher's equations\footnote{The telegrapher's equation is satisfied, in particular, by the expectation value of a certain kind of Poisson process, studied by Kac et al. \cite{Kac74}.} \cite{Gaveau1984}. The discrete-spacetime version of the relativistic mapping is the persistent classical random walk yielding the DTQW\footnote{This mapping is recalled in a work of 2013 by Boettcher, Falkner and Portugal \cite{Boettcher2013}, on the non-perturbative renormalization of quantum walks. A work of 2016, by Boettcher, Li and Portugal, keeps exploring the direction of non-perturbative renormalization \cite{Boettcher2016}. Portugal has published, in 2013, a book entitled \emph{Quantum walks and search algorithms}.}. There is also a discrete-spacetime version of the non-relativistic mapping: the \emph{non-biased} classical random walk yields, by analytic continuation, the CTQW \cite{FG98a}. At the discrete level, the CTQW as a non-relativistic limit of the DTQW was derived by Strauch in 2006 \cite{Strauch06a}.

At the path-integral level, analytical continuation is used:
\begin{enumerate}

\item In the non-relativistic framework, to compute probability amplitudes in quantum mechanics, by mapping the Feynman measure to the Wiener measure\footnote{It is actually the standard way to give the Feynman measure a definite meaning.}\textsuperscript{,}\footnote{Note that the path-integral formulation of quantum mechanics was developed by Feynman during his PhD. A scan of the original PhD dissertation is available at \url{http://cds.cern.ch/record/101498/files/Thesis-1942-Feynman.pdf?version=1}.} \cite{Wiener1923}.

\item In the relativistic framework:
\begin{enumerate}
\item For bosons, to compute either scattering amplitudes in bosonic quantum field theory, in which the on-shell free fields obey the Klein-Gordon equation, or, in the case we already have quantum results, their Wick-rotated equivalents, namely the correlators in bosonic statistical field theory, in which the equilibrium non-interacting field obeys the massive (or screened) Laplace (or sourceless Poisson) equation. Recall that these dynamical or equilibrium equations are obtained by functional optimization of the corresponding path integral.

\item For fermions, to achieve the same computational aim as in the bosonic case. 

\ \ \ \ Regarding the on-shell fields, the fermionic case is that which is described in the first paragraph of this subsection\footnote{In the case of bosons, the passage from the Klein-Gordon equation to its non-relativistic limit, which is also the Schrödinger equation, is derived in \cite{Huepe2000}, in the more general case of an additional non-linear term. Is there a stochastic process associated to a solution of the Euclidean version of the Klein-Gordon equation, as provided by the Feynman-Kac formula in a non-relativistic context? I haven't seen such a mapping in the litterature. One can define relativistic stochastic processes (RSP) \cite{DMR97}, which satisfy an equation that yields, in the  non-relativistic limit, a kind of diffusion equation \cite{DR98}, but this equation satisfied by the RSP cannot, in its present form, be mapped mapped to the Klein-Gordon equation by analytic continuation.}. In a Feynman-checkerboard approach, difficulties arise in spatial dimensions greater than one (the one-dimensional case is solved by Feynamn's original checkerboard, as mentioned above). Let us stress that the starting-point problem does not deal with the functional integration, but with the on-shell (or equilibirum) fields: indeed, the association, through analytical continuation, between the Dirac equation and the telegrapher's equations, \emph{is only done in one spatial dimension}; Gaveau and Jacobson \cite{Gaveau1984} underline that, in the higher-dimensional case, ``it is not clear how to obtain an underlying stochastic process (related by analytic continuation)''. It seems we deal here with a matter of structure of the Dirac equation in more than one dimension, or, in other words, with the spacetime view of \emph{relativistic} quantum mechanics\footnote{The problem is certainly deeper than a mass-coupling issue in the (1+3)D Dirac equation (the mass couples two Weyl equations in the Weyl representation). The problem might also, at least in the massless case, not be a fermion/boson poblem, since in one of his books \cite{Feynman1985book}, Feynman notices the presence of superluminal paths in the photonic path integral, as pointed out by W. Ichinose \cite{Ichinose2014}, while photons are bosons, not fermions. Massless fermions are, as photons (massless spin-1 bosons), helical particles (the helicity, projection of the spin along the momentum, is a good quantum number); their wavefunctions transform the same way under Lorentz transformations, through the same little group, which is $\mathrm{SE}(2)=\mathrm{ISO}(2)$ in 1+3 dimensions; actually it is sometimes said that the little group for massless fermions is the double cover of $\mathrm{ISO}(2)$ (if we define the little group at the level of the half-integer-spin representation and not the original Lorentz group), but, again, it is not clear to me that the problem comes, in the massless case, from this double cover, because of this note by Feynman. In the Weyl representation of the Clifford algebra, massless fermions are described by two independent (no mass coupling) Weyl equations.}, that was precisely not clear to Feynman because of this higher-dimensions problem, which is why he did not include his one-dimensional checkerboard model \cite{FeynHibbs65a} in his ``Space-time approach to \emph{non-relativistic} quantum mechanics'' \cite{Feynman1948}.  

\ \ \ \ To overcome those difficulties in the higher-dimensional fermionic path integral, the standard approach is that of Grassman variables\footnote{A German-to-English translation of Grassman's paper on the application of his exterior product (also called wedge product) to mechanics is available at \url{http://neo-classical-physics.info/uploads/3/0/6/5/3065888/grassmann_-_mechanics_and_extensions.pdf}. This product is one of the tools of Cartan's exterior calculus, which is a powerful framework to do differential geometry on fiber bundles, either with flat base manifolds, as in classical electromagnetism, or with curved base manifiolds, as in general relativity, or `simply' on, e.g., curved surfaces of materials. Cartan developed the exterior derivative, that acts on the exterior product pretty much like the standard derivative acts on a standard product, but with an additional dimension-dependent antisymmetrical property.}  \cite{Candlin1956, book_Berezin, DeWitt}. 
It is still an open question whether the Feynman checkerboard can be properly extended to more than one spatial dimension, as an alternative to the Grassman-variables approach  -- it may actually rather be another view of Grassman's variables, but which may be very enlightening. An interesting paper by Gaveau and Schulman \cite{GS89} tighten the link between the checkerboard picture and the Grassman-variables approach, but does not achieve the aim of formulating the ``Dirac-particle path integral as a sum over path -- possibly constrained -- on an \emph{ordinary} manifold'' \cite{GS89}. In Fig. \hspace{-0.2cm} 1 of this paper is given a nice diagram depicting the relations discussed in the present section. I reproduce it in Fig. \ref{fig:interrelations}.
\end{enumerate}
\end{enumerate}

\iffalse
Note that in the non-relativistic framework, the use of the Wick rotation is historically rather `statistical to quantum' (use random walks and statistical mechanics to prove quantum results), while both ways have been used to find results in the relativistic framework. 
\fi

Note also that, while the Schrödinger equation\footnote{Plus an additional spin term in the presence of an external magnetic field.} limits the Dirac equation in the non-relativistic approximation, the diffusion equation limits the telegrapher's equation at long times\footnote{The propagating signal, having ballistic spread, is `attenuated more and more', i.e. with a diffusion coefficient, and its spread becomes diffusive.},  i.e. in the limit where the propagation speed of the signal goes to infinity\footnote{A good short lecture chapter by G. Randall on (i) the continuum limit of the persistent classical random walk, namely the telegrapher's equations, and (ii) the propagation (or ballistic) and diffusion limits (or, by Wick rotation, ultrarelativistic and non-relativistic limits) of the latter, can be found at \url{https://ocw.mit.edu/courses/mathematics/18-366-random-walks-and-diffusion-fall-2006/lecture-notes/lecture10.pdf}.}.

\begin{figure}[t!]
\centering
\includegraphics{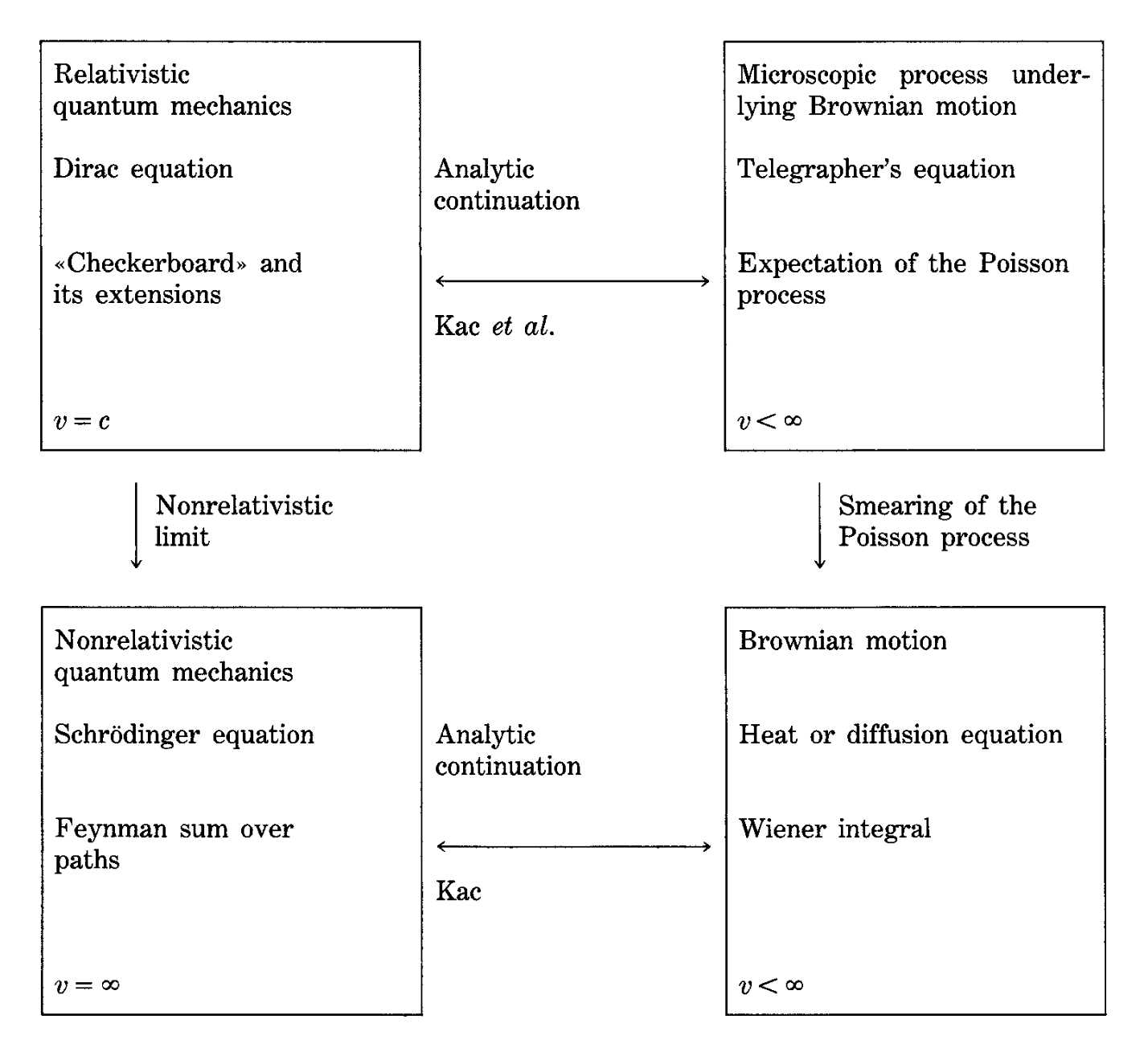}
\caption[Mappings between quantum and statistical mechanics]{``Interrelations between physical systems", from \cite{GS89}. The extensions of the checkerboard to more than one spatial dimension are still unsatisfying currently, see the discussion above.}
\label{fig:interrelations}
\end{figure}

\end{IntroMat}
%\cite{DMD14}

%\bibliographystyle{unsrt}
%\bibliography{bibli.bib}
   %%%%%%%%%%%%%%%%%%%%%%%%%%%%%%
%%  Alias bibliographiques  %%
%%%%%%%%%%%%%%%%%%%%%%%%%%%%%%
\defcitealias{NRFortran}{Numerical Recipes}
\defcitealias{NRF77}{Numerical Recipes}
\defcitealias{NRF90}{Numerical Recipes}
\defcitealias{IUPAC}{IUPAC}%International Union of Pure and Applied Chemistry

\chapter{\textsc{Two-dimensional DTQWs in electromagnetic fields \\ (Publications \cite{AD15} and \cite{AD16})} \label{Chap:Electromagnetic_2D_DTQWs}}

\vspace{-0.5cm}

\minitoc

\section{Higher-dimensional (discrete-time) quantum walks}
\label{sec:higher}

\noindent
\motivations{ 
The aim of this long introduction to Publications \cite{AD15} and \cite{AD16}  is to give a relatively detailed historical review of important results on (higher-dimensional) quantum walks, quantum information and quantum simulation.
}  \\

\noindent
There are several ways to extend DTQWs to higher-dimensional coin and/or node spaces. Historically, the first extensions to higher dimensions that we encounter in the litterature under the name of `quantum walks' are those developed by quantum-algorithmics authors, in the early 2000's, but particular cases of such extensions had actually already appeared before in the community of physics simulation, and more precisely, of cellular automata, in the 90's, as we will detail below. Let us first come back a bit in time.

\subsection{From the Turing machine to Meyer's and then Ambainis's one-dimensional quantum walks}

\noindent
The principle of modern computers is the Turing machine, proposed by Turing, mathematician and one of the first computer scientists, in his seminal paper of 1936 \cite{Turing1937}.  
Cellular automata (CA) were then introduced by both the mathematician Ulam and the mathematician and physicist Von Neumann \cite{VonNeumann66}, to simulate, with computing machines, (classical) physics governed by local rules.
Feynamn, physicist, added his contribution to the computing era by  proposing in 1982 a model to quantize these CA \cite{Feynman1982}, which was formally developed by Deutsch \cite{Deutsch1985}. The term `quantum cellular automaton' (QCA) appeared in a paper by Grössing and Zeilinger \cite{Grossing_Zeilinger_88} in 1988, although their model differs from that of Feynman and Deutsch. Meyer, mathematical physicist, entered quantum computing and quantum simulation with the aforementioned background. The main part of his work is nowadays still at this crossing between quantum simulation and quantum computing.
In his seminal paper ``From quantum cellular automata to quantum lattice gases'' \cite{Meyer96a}, Meyer shows how to obtain what are now called 1D DTQWs as a particular family of 1D reversible (i.e. unitary\footnote{Unitarity implies reversibility, and the converse also holds in standard quantum mechanics, see the following discussion, \url{https://physics.stackexchange.com/questions/150733/unitarity-of-a-transformation-and-reversibility-imply-one-another}. The relaxation of the unitarity constraint is usually explicitly mentioned.}) QCA\footnote{Again (see above the definition I give for classical cellular automata), and as commonly done, the locality of the evolution rules is implied in what I call QCA.}. In particular, he explains that imposing unitarity to homogeneous 1D QCA, results in a trivial dynamics. That is why \cite{Meyer96a} (i) Grössing and Zeilinger relax the unitarity constraint in their paper \cite{Grossing_Zeilinger_88}, and (ii) Meyer weakens the homogeneity constraint, as follows. First, (ii.a) Meyer weakens the strict homogeneity\footnote{This means (I use Meyer's notations) that at time $t+1$ and discrete position $x$ on the line, the state $\phi(t+1,x)$ of the QCA is obtained from the states which are at $t$ on positions $x-1$, $x$ and $x+1$, namely $\phi(t,x-1)$, $\phi(t,x)$ and $\phi(t,x+1)$, in a manner which is the same for each $\phi(t+1,x)$ on the space line; to be more precise, $\phi(t,x) = w_{-1}\phi(t-1,x-1) + w_{0}\phi(t-1,x) + w_{+1}\phi(t-1,x+1) $, with $w$'s that do not depend on $x$ at $t+1$ (for any $t$).} to a one-every-two-cells homogeneity. To preserve unitarity, the only case that can be retained is that of `non-interacting' pairs of cells, i.e. each pair evolves independently from the other pairs, with no quantum superpositions of pairs. This can be viewed as creating the spin degree of freedom, but each spin is still independent from the others. Second, (ii.b) Meyer adds to the model the alternate evolution (sometimes called partitioning or staggered rule), characteristic of DTQWs, and performed by what we refer to nowadays as the spin-dependent shift, which makes the pairs of cells (i.e. the spins) interact (i.e. superpose) and allow propagation to occur. He thus connects his work to  Feynman's checkerboard, recalling the link with the Dirac equation.
  Computational work on 1D DTQWs is then continued by computer scientists, the denomination `quantum walk' appearing for the first time in a paper title in the paper ``One-dimensional quantum walk'' by Ambainis et al. \cite{Ambainis2001}, in 2001. 
  
\subsection{Higher dimensions in, essentially, a quantum-computing perspective: exponentional speedup of this research field in the 2000's?} 

\subsubsection{Higher-dimensional coin space in a combinatorial abstract node space} \label{subsubsec:higher_dim_node_space}

\noindent
This 1D-DTQW model \cite{Ambainis2001} is then quickly extended to arbitrary $d$-regular graphs, by Dorit Aharonov, Ambainis, Kempe and Vazirani \cite{Aharonov2001}. In this model, the coin operation acts, at each of the $n$ nodes of the graph, on qudits (generalization of qubits i.e. spins-1/2), belonging to a $d$-dimensional coin space, and the qudit-dependent shift transports each component of the qudit to one of the $d$ different neighbors. In this work, the vertex (i.e. position) space is abstract, and is not connected to physical dimensions. The authors (i) define the basic tools to study the spreading properties of DTQWs on such graphs, among which the mixing time, and (ii) prove  the existence of a disappointing lower bound in this mixing time: DTQWs can \emph{mix} at most `only' (almost) quadratically faster than their classical counterparts, classical random walks (CRWs) (mixing time of $O(n \log n)$ for DTQWs, vs. $O(n^2)$ for CRWs). This `bad news' is then reexamined  on the particular case of an hypercube\footnote{The hypercube is widely used in algorithmics, to handle computations on bit strings. It is one of the regular graphs for which the number of nodes satisfies $n=2^d$. For any $d$-regular graph with $n$ nodes, the number of edges is $e=nd/2$, but this is still not sufficient to characterise the hypercube.} by Moore and Russell \cite{Moore2002}, who show that the same bound exists for continuous-time quantum walks (CTQWs). These results are however quickly followed by better news in a paper by Kempe \cite{Kempe2003}: DTQWs \emph{hit} exponentially faster than CRWs (first evidence of exponential speedup enabled by DTQWs). Previous  and similar but somehow weaker results were given for CTQWs by Fahri and Guttman in their seminal paper \cite{FG98a}, and then reviewed by Childs and the two previous authors \cite{Childs2002}: these authors found an exponentially-smaller `local' (\emph{I} use this word) hitting time on some particular graphs. 

\subsubsection{Higher-dimensional physical node space}

\noindent
Shortly after the publication of this generalization of DTQWs by computer scientists, a different extension to higher dimensions was proposed by the physicists Mackay, Bartley, Stephenson and Sanders \cite{Mackay2002}. 
%Note that Bartley and Sanders have written an introductory book to quantum information \cite{}.

 This extension is explicitly viewed as a way to move a discrete-time quantum walker in higher-dimensional physical spaces. For a $D$-dimensional physical space, the coin space is built as the tensorial product of $D$ two-dimensional coin spaces, and has thus dimension $2^D$. The spin-dependent shift then moves a walker located at some node, not to its $2D$ nearest neighbors in the `natural' $D$-dimensional hypercubic lattice induced by the $D$ elementary qubit-dependent shifts, but to its $2^D$ $D$-th nearest neighbors, and the walker does thus not explore this whole `natural' hypercubic lattice, but only one of the $D$ independent sublattices generated by such a transport (which are generically not hypercubic). For example in two dimensions, the $2$- (i.e. next-) nearest neighbors of the node $(0,0)$, are $(1,1)$, $(-1,-1)$, $(1,-1)$ and $(-1,1)$. In $D=2$, the number of nearest beighbors and of $D$-th nearest neighbors happen to be equal (to 4) so that, apart from the fact that there is a (unique) useless sublattice, the coin space has a dimension which \emph{is} the minimal one to move a walker on a square lattice, which is the natural hypercubic lattice in $2D$ (square lattice); but this is just a coincidence: the lattice explored, although square, is \emph{not} the original one induced by the $D=2$ elementary qubit-dependent shifts. This coincidence dissapears for $D>2$. In $D=3$, for example, the number of nearest neighbors is $2D=6 < 2^D=8$, the number of next-next-nearest neighbors. To fully explore a given cubic lattice, one could use the generalization described in the previous subsubsection, with a coin space having dimension $2D=6$. The sublattice explored by the present generalization is not cubic, but \emph{body-centered cubic}, and this scheme is natural to explore such a lattice. Note that this body-centered cubic sublattice can only be $d=2^D$ regular either if it is infinite, i.e., in practice, in the bulk, or if we consider periodic boundary conditions, which makes the $D$-dimensional lattice a $D$-dimensional torus.

After recalling the definition of two equally-weighting transformations, namely the discrete Fourier transform (DFT) and the Grover operator (GO), the authors show that the entanglement between the spatial dimensions, produced by these transformations (thanks to their alternate action with the spin-dependent shift),  can serve to reduce the rate of spread of the DTQW, with respect to an evolution by the $D$-tensorized Hadamard gate, a separable transformation which thus produces no entanglement between the spatial dimensions. Since the extensions to higher dimensions are done in a systematic way, the rate of spread always increases with $D$ for a given transformation.

This seminal paper about the spreading properties of DTQWs in higher \emph{physical} dimensions, was followed by another study by Tregenna, Flanagan, Mail and Kendon \cite{Tregenna2003}. In this paper are presented short but conceptually comprehensive and systematic studies of some phenomenal spreading properties of quantum walks on low-degree graphs, from two to four, as well as some comparisons with CTQWs; it is in particular shown how, thanks to the coin degree of freedom, one can better monitor the propagation in DTQWs than in CTQWs. 
\iffalse
A notable work of this team formed by Tregenna and Kendon is that ``Decoherence can be useful in quantum walks'' on the line, the cycle and the hypercube \cite{Kendon2003}, which is underlined in Kendon's review of 2006 on quantum walks \cite{Kendon2006}.
\fi

\subsubsection{Quantum walks for spatial search and universal computation}

\noindent
The results described in the two previous sections regarding the spreading properties of quantum walks, are bricks with which to elaborate more complex algorithms. Spatial search is one of those. 

\paragraph{$\bullet$ Early quantum algorithms, Grover's search and followings} \hspace{3cm}

\vspace{0.2cm}

\noindent
The recent review of 2016 by Giri and Korepin \cite{GK16}, gives a nice straight-to-the-point introduction to quantum search algorithms derived from Grover's. 

Before getting into the precise subject, the authors present, through some of the early quantum algorithms, some of the essential features of quantum algorithms that may be used to speedup computing power (provided these algorithms run on a quantum computer). 
First, they describe the very basic Deutsch algorithm, which enables to find, in single query to the quantum oracle (quantum query), whether some function of a single qubit and taking at most two different values (Boolean function) is constant or balanced, while one needs two queries to a classical oracle (classical query) for such a task to be fulfilled; this illustrates quantum parallelism. Second, they show that this number of quantum queries does not grow with the number $k$ of input qubits\footnote{More precisely, the input is a series of $k$ bits, namely a  $k$-bit, and this $k$-bit is treated as a $k$-qubit by the algorithm, i.e. it is viewed as a basis vector of a $2^k$-dimensional Hilbert space.} if one uses the Deutsch-Jozsa algorithm\footnote{The funtion is said balanced if it takes one of the two possible values for exactly half of all the possible inputs, i.e. half of the possible $k$-bits.}, while one needs $2^{k-1} +1$ classical queries in the worst case. Rather than the classical equivalent to the Deustch algorithm, it is more efficient to use, to fulfill the proposed task with large inputs on a classical computer, an algorithm which is similar but fulfills the job in only $d$ classical queries; the authors show that the quantum version of this algorithm, variation of the Deutsch-Jozsa algorithm called the Bernstein-Vazirani algorithm, ensures the same performance of a single quantum query. The authors do not discuss Shor's famous quantum algorithm of 1994 \cite{Shor94, Shor97}, which is (i) not a search algorithm, and (ii)  widely studied in the litterature. Shor's algorithm enables prime-number factorization of an integer $n$ in a time (i.e. number of steps, there is no oracle here) $O((\log n)^3)$; we say that the algorithm is polynomial, with exponent 3, in the physical input size, which is $\log n$. To be precise, the algorithm actually runs in a time $O((\log n)^2(\log \log n) (\log \log \log n))$ (which is better than $O((\log n)^3)$). This  is a big improvement with respect to the best known classical algorithm performing such a task, ``Gordon's adaptation'' \cite{Shor97} of the so-called number field sieve \cite{Gordon1993}, which does the job in a sub-exponential time $O(\exp(c(\log n)^{1/3}(\log \log n )^{2/3}))$. 

%, and the time can even be reduced from this $O((\log n)^3)$ to $O((\log n)^2 (\log \log n) (\log \log \log n))$ using fast multiplication \cite{Beckman1996}.

Giri and Korepin then  present in depth many variants of Grover's famous spatial search algorithm of 1996 \cite{Grover96}; there is in particular a distinction between full and partial database search, the latter being achieved by the Grover-Radhakrishnan-Korepin algorithm. The basic important result presented in Grover's seminal paper is the following: instead of the $O(n)$ classical queries that are needed to find a given element in a database of $n$ elements, Grover's quantum-mechanical algorithm only needs $O(\sqrt{n})$ quantum queries, which is a quadratic improvement. It can be shown, and this is an important restrictive result to have in mind, that $O(\sqrt{n})$ is a lower bound for \emph{any} quantum search algorithm  \cite{Bennett1997,Boyer1998,Zalka1999}.

\paragraph{$\bullet$ Spatial search with DTQWs: AA, SKW, AKR, AKR-Tulsi, MNRS} \hspace{3cm}

\vspace{0.2cm}

\noindent
We have seen that Grover's algorithm finds a given item among $n$ in a time $O(\sqrt{n})$, which is a lower bound for any quantum search algorithm, and any DTQWs-based search algorithm will \emph{not} go below this bound. One of the aims of DTQWs-based search algorithms is to search only with local operations, Grover's algorithm being ``highly non-local'' \cite{Shenvi2003}. The first DTQW-based search algorithm to reach the aforementioned lower bound is the SKW algorithm, suggested by Shenvi, Kempe and Whaley \cite{Shenvi2003} for a search on the hypercube (abstract space). This paper was then followed by the well-known AKR algorithm, by Ambainis, Kempe and Rivosh \cite{AKR2005}, that manages to adapt the previous algorithm to hypercubic lattices of dimension $D$ (physical space, explored by standard lattices, which \emph{have much lower spreading capacities than the hypercube of dimension $d= \log n$}), containing a number of nodes $n=( \! \sqrt[D]{n} )^D$, the optimal performance of $O(\sqrt{n})$ being reached for $D\geq 3$.  In $D=2$, the running time is  $O(\sqrt{n} \log n)$, and the authors show that it is sufficient to have a two-dimensional (instead of a four-dimensional) coin space to reach this lower bound. The AKR algorithm combines (i) the monitoring capacities  of coins in DTQWs \cite{Tregenna2003, Shenvi2003}, and (ii) the scheme developed by Aaronson and Ambainis \cite{Aaronson2005}, in order to surmount the difficulty of implementing Grover's algorithm on a 2D grid\footnote{As he explicitly stated in the abstract of his paper, Benioff \cite{Benioff2002} considered null the possibility that the implementation of Grover's algorithm on a 2D grid could outperform classical search on such a grid. As explicitly stated in the abstract of their paper, Aaronson and Ambainis \cite{Aaronson2005} disproved Benioff's claim.}  (they also add to this scheme `amplitude amplification' to increase its probability of success). These results \cite{Aaronson2005, Shenvi2003, AKR2005} were compiled and phrased in a highly simplified framework by Szegedy \cite{Szegedy}, as quantum counterparts to classical Markov chains. 

Further developments have focused on increasing the probability of success of the search (without increasing the time search). All these developments were used by Magniez, Nayak, Roland and Santha, to construct their MNRS algorithm \cite{Magniez2007}, a simplified version of which is given (i) in Santha's review of 2008 \cite{Santha08quantumwalk}, which deals with many computing problems other than search, and (ii) in an updated version of \cite{Magniez2007}, namely \cite{Magniez2011}. Some of the latest developments in increasing the probability of success are the 100\% chances of success achieved in the SKW algorithm \cite{Potoek2009}, and the time $O(\sqrt{n \log n})$ reached in the AKR algorithm in 2011 \cite{Ambainis2013} (thus disproving Szegedy's ``probably optimal" \cite{Szegedy} referring to the $O(\sqrt{n} \log n)$); this last result is achieved by managing to get rid of the amplitude amplification included in the original AKR algorithm, which demands an $O(\sqrt{\log n})$ running time, but with no need of any other modification of the original algorithm, as previously done by Tulsi \cite{Tulsi2008} and later by Krovi et al. \cite{Krovi2010} to reach the same $O(\sqrt{n \log n})$. In the first three pragraphs of his paper \cite{Tulsi2008}, Tulsi gives a clear and fast sum-up of the `previous episodes'. 
 
\paragraph{$\bullet$ Parallel and/or joint research with CTQWs: search problem and universal computation, by Childs and others} \hspace{3cm}

\vspace{0.2cm}

\noindent
Here we report some of the key steps reached by CTQWs in the run `between' continuous- and discrete-time quantum walks in both the quest for the optimization of the search problem and universal quantum computation. Quickly after Kempe's result on the exponential hitting-time speedup provided by DTQWs, Childs et al. \cite{Childs2003} published a paper showing that CTQWs can be exponentially fast in oracle-type problems, in particular, search (Kempe's result dealt with a non-oracular problem). Two years later, in 2004, Childs and Goldstone \cite{Childs2004} show via CTQWs that the lower bound of $O(\sqrt{n})$ can be reached on the hypercube and in the \emph{`physical'} search problem, for lattices of dimension $D \geq 5$, while in lower dimensions only $O(\sqrt{n} \, \mathrm{polylog} \,   n)$ is reached, a worse performance than the solution provided one year later by Ambainis et al. with DTQWs \cite{Aaronson2005} and then by the AKR algorithm \cite{AKR2005}, already discussed above.  

While many quantum-computing teams were busy improving the AKR algorithm, from 2005 to 2008, Childs kept working with CTQWs, ``perhaps easier to define'' \cite{Childs2009}, and published in 2009 his ``Universal computation by quantum walk'' \cite{Childs2009}, showing CTQWs can be used as computational primitives for \emph{any} quantum algorithm, and not only search problems. This paper was followed in 2012 by an extensive generalization to multi-particle quantum walks \cite{CGW13}.

In 2006, Strauch precisely connected the continuous- and the discrete-time quantum walks \cite{Strauch06b}, underlying that, ``at least in [the] simple case''\cite{Strauch06b} he treats, the coin degree of freedom is irrelevant to the speedup provided by DTQWs-based algorithms, and that the speedup is only due to a ``simple interference process'' \cite{Strauch06b}. This work was  followed by an in-depth generalization to arbitrary graphs by Childs \cite{ChildsCD2009}. In 2010, the results of the first paper by Childs on universal quantum computation via CTQW, were recast in the framework of DTQW, by Lovett et al. \cite{Lovett2010}.

\paragraph{$\bullet$ Recent developments in quantum computing} \hspace{3cm}

\vspace{0.2cm}

\noindent
In 2014, a 3-nearest-neighbors CTQW-algorithm has been designed for spatial search on graphene lattices \cite{FG2014, Foulger_thesis_2014}.
A recent work by Chakraborty, Novo, Ambainis and Omar \cite{Chakraborty2016}, show, using CTQWs, that ``spatial search by quantum walk is optimal for almost all graphs'', meaning that ``the fraction of graphs of $n$ vertices for which this optimality holds tends to one in the asymptotic limit'' \cite{Chakraborty2016}.

A recent paper by Wong \cite{Wong2016} proves with CTQWs that the Johnson graph $J(n,3)$ supports fast search (i.e. in a time $O(\sqrt{n})$), thus generalizing the same results obtained previously for (i) $J(n,1)$ with the CTQW version of Grover's algorithm by Childs and Goldstone \cite{Childs2004}, and 
(ii) $J(n,2)$. Note that Johnson graphs were already used as support by Ambainis to develp his ``Quantum walk algorithm for element distinctness'' \cite{Ambainis04quantumwalk, AmbainisElementDistinctness14,Amb07a}.

\subsection{Higher dimensions in the perspective of simulating physical phenomena}

\subsubsection{The community of non-linear physics and kinetic theory of gases: FHP, LBE, QBE}

\paragraph{$\bullet$ On the history of fluid dynamics, up to the Boltzmann equation}\hspace{3cm}

\vspace{0.2cm}

\noindent
To learn about the history of fluid mechanics ``From Newton's principles to Euler's equations'', see the paper by Darrigol and Frisch at \url{https://www-n.oca.eu/etc7/EE250/texts/darrigol-frisch.pdf}.  In his \emph{General Principles of the Motion of Fluids}\footnote{A scan of the original manuscript, written in French and published by the Prussian Royal Academy of Sciences and Belles-Lettres, in Berlin, is available on The Euler Achive at \url{http://eulerarchive.maa.org//docs/originals/E226.pdf}. Ref. \cite{U08} is an English translation adapted from Burton's by Frisch.}, published in 1757, Euler derives two out of the three equations which are nowadays known as the Euler equations for inviscid (i.e. non-viscous) fluids. 

These two equations are (i) the momentum conservation with local pressure and generic external forces, and (ii) the mass conservation. These original equations are the current most general ones for inviscid fluids\footnote{This is true in the hytohesis of local thermodynamical equilibrium; otherwise, one can use, in some situations, the more general framework of so-called extended thermodynamics, a branch of non-equilibrium thermodynamics.}. In particular, they \emph{are} valid for compressible fluids. The question of the solutions of these equations is a different one. We have (i) five scalar unknowns, namely the three components of the Eulerian speed field, the pressure, and the density, and (ii) four scalar equations: three given by the momentum conservation, and one by the mass conservation (also called continuity equation). This means that even if we manage to make Euler's seminal system linear (with, for example, some ansatz for the unknowns), it will be under-determined\footnote{For a general definition of over- and under-determination of a system of PDEs, see \url{https://www.ljll.math.upmc.fr/frey/cours/UdC/ma691/ma691_ch3.pdf}. The author of the following notes on the subject, \url{https://www.researchgate.net/file.PostFileLoader.html?id=56498d8b5cd9e3774f8b4569&assetKey=AS:296325611573251@1447660939976}, feels ``we should not attach too much importance to this terminology''. One can put the Euler equations and, more generally, the Navier-Stokes equations, into the form of a quasilinear system.}, except for incompressible fluids, whose density is a constant. Euler already mentions, after summing up his 3+1 equations, that a fifth equation should exist\footnote{I wonder whether such a statement is made by thinking rather `physically' or `mathematically', thinking of the determination of linear systems; probably both. I should investigate on the thermodynamics of Euler's time.} between pressure, density, and, in Euler's words ``the heat of the fluid particle'', which is nowadays called the internal specific (i.e. per-unit-volume) energy, essentially related to the temperature of the fluid (proportional, for ideal gases). Thermodynamics developed later, in the whole 19th century, and at Euler's time the distinction between the concepts of `heat' and `temperature' was at least not properly formalized. 

The equation between these three quantities (which also involves the Eulerian speed field), is the energy-balance equation. A first relation of this kind enabled Laplace to make the first correct calculation of the sound speed \cite{L1816}. This condition was later recognized as (i) an adiabatic condition \cite{CM13}, in the context of a developing thermodynamics, and, more precisely, as (ii) an isentropic\footnote{Note that Kelvin's circulation theorem holds for isentropic or barotropic fluids.} condition (which is an idealized adiabatic condition), after Clausius introduced the concept of entropy in 1865 \cite{C1865}: indeed, the equation can simply be stated as the conservation of the entropy of any fluid particle. 

 To focus on the convective term of the momentum-conservation equation, which is the non-linear term, the pressure term is sometimes dropped, which leads to the (inviscid) Burgers equation. In addition to be physically motivated in some cases, this approximation often simplifies the search for solutions for the speed field. 

After deriving the two aforementioned equations, Euler shows that, when the fluid is incompressible, the integral form of the momentum-conservation equation yields Bernouilli's original theorem, published in 1738, and which was only valid for incompressible fluids, although nowadays one can write down a compressible version. For a pre-Eulerian history of partial differentiation, see the paper by Cajori at \url{http://www.math.harvard.edu/archive/21a_fall_14/exhibits/cajori/cajori.pdf}. 

In his \emph{Memoir on the Laws of the Motion of Fluids} of 1822, Navier \cite{N1822} introduced the notion of viscosity, but ``he did not developed the concept of shear''\footnote{See Huilier's note on Navier at \url{http://www.daniel-huilier.fr/Enseignement/Histoire_Sciences/Histoire.pdf}.}. The present form of the Navier-Stokes (momentum-conservation) equation was written by Stokes in 1845\footnote{A scan of the original paper \cite{S1845} can be found at \url{http://www.chem.mtu.edu/~fmorriso/cm310/StokesLaw1845.pdf}.}. The Navier-Stokes equations describe the fluid motion at the so-called mesoscopic scale, which is much greater than the microscopic scale, but still much smaller than the macroscopic scale\footnote{See \url{http://s2.e-monsite.com/2010/03/16/02/meca_flu.pdf}, Chap. 1, for a quick discussion on these three scales.}.

The Boltzmann equation, derived in 1872 \cite{Boltzmann1872, Boltzmann2012}, describes the evolution of an out-of-equilibrium fluid (gas or liquid), which emerges at mesoscopic and macroscopic scales from the collisional dynamics of the classical microscopic particles it is made of. It is an equation on the (one-point) probability distribution function of the fluid particles. From this Boltzmann equation, one can derive\footnote{Maxwell first realized this, see the following online scan of his \emph{Scientific papers}, \url{http://strangebeautiful.com/other-texts/maxwell-scientificpapers-vol-ii-dover.pdf}.} \cite{M1890, E17, CC52a} mesoscopic-scale equations (such as the standard Navier-Stokes equation), by taking the so-called hydrodynamical limit\footnote{See the following notes by, respectively, Golse, \url{http://www.cmls.polytechnique.fr/perso/golse/Surveys/FGEcm04.pdf}, and Villani, \url{http://archive.numdam.org/article/SB_2000-2001__43__365_0.pdf}, which presents results obtained by Bardos, Golse, Levermore, Lions, Masmoudi and Saint-Raymond.} \cite{L09}, which essentially consists in a ``patching together of equilibria which are varying slowly in space and time'' \cite{FHP86}. A recent paper (2016) by Chow et al.   \cite{Chow2016} gives an analytical solution for the $N$-dimensional compressible Euler equations with damping, for a barotropic (pressure only function of the density) fluid, with the pressure as function of the density given by a power law.

\paragraph{$\bullet$ The lattice Boltzmann equation (LBE)}\hspace{3cm}

\vspace{0.2cm}

\noindent
The following few lines on the history of the lattice Boltzmann equation are strongly inspired from Scholarpedia's sum-up on the subject, see \url{http://www.scholarpedia.org/article/Lattice_Boltzmann_Method}. Experts on the discretization of the Boltzmann equation noted that the first \emph{simplified} discretizations of the Boltzmann equation, from the 60's to the mid 80's, were rather used to find analytically-tractable solutions of the Boltzmann equation, and not as alternatives to the discretizations of the Navier-Stokes equation in a numerical-computation perspective. The idea of simulating fluid dynamics with lattice-gas models was introduced in 1986 by Frisch, Hasslacher and Pomeau, with their well-known FHP lattice-gas cellular automaton \cite{FHP86}. This model runs, in dimension 2, on a hexagonal lattice, to fulfill an isotropy condition, unsatisfied by the preceding HPP automaton developed by Hardy, de Pazzis and Pomeau \cite{Hardy1973,Hardy1976} in the 70's; in 3 dimensions, a suitable lattice is given by the projection of a face-centered-hypercube of dimension 4 on one of the coordinate axis \cite{FHHLPR87}. The collision rules at each vertex are inspired by a discrete Boltzmann model \cite{Harris1966}. Here is a 1989 review on ``Cellular automaton fluids"  \cite{RajLakshmi1989}. Several pitflaws of the FHP automaton lead to the development, in the late 80's and 90's, of the so-called lattice Boltzmann methods\footnote{A short review of the book by Succi \cite{succi2001lattice}, written by Yeomans, is available at \url{http://physicstoday.scitation.org/doi/pdf/10.1063/1.1537916}.} \cite{succi2001lattice}.

\paragraph{$\bullet$ The quantum (lattice) Boltzmann equation (QBE): a DTQW!}\hspace{3cm}

\vspace{0.2cm}

\noindent
In 1992, Succi and Benzi published their ``Lattice Boltzmann equation for quantum mechanics'' \cite{Succi1993}, inspired (in part) by the FHP automaton \cite{FHP86}. The central idea of this important paper is that ``the non-relativistic Schrödinger equation ensues from the relativistic Dirac equation under the same conditions which govern the passage from the LBE to the Navier Stokes equation'' \cite{Succi1993}. To accept this statement without further clarifications, recall that the LBE is a \emph{simplified} discrete Boltzmann equation, and this simplification procedure gets rid of some microscopical degrees of freedom of the original Boltzmann equation. Of course, one still has to find a suitable discretization of the Dirac equation, and this is done via an operator-splitting technique. It is straightforward to implement in this scheme a minimal coupling to an external Abelian potential. In 2015, the (1+1)D QBE is explicitly identified with a (1+1)D quantum walk by Succi \cite{Succi2015}, the a straightforward generalization to (1+3)D, with steps taken alternatively in the 3 spatial directions of the lattice, and not at the same time as in \cite{MBSS2002}, which enables to write these equations sticking to the `physical' low-dimensional coin spaces associated to spinors, as opposed to the `algorithmical' ones, built by tensorial products of qubits \cite{MBSS2002}.

\subsubsection{From particle physics to cellular automata: Wolfram, 't~Hooft, Bialynicki-Birula, Yepez}

\noindent
After a PhD in particle physics, obtained in 1979 at the age of 20, Wolfram turned into complex systems and cellular automata as a tool to model them. He developed the formal-computation software Mathematica and the search engine Wolfram Alpha. His ideas and findings on cellular automata are summarized in his book \emph{A New Kind of Science} \cite{Wolfram2002}.
In the mid 80's, Wolfram had already developed many ideas on cellular automata, which inspired other theoretical physicists. His seminal paper on ``Cellular automata as models of complexity'' \cite{Wolfram1984} is cited in the FHP-automaton paper \cite{FHP86}.

In 1988, 't Hooft introduced a cellular automaton to describe quantum sytems \cite{Hooft1988}. In 2014, he wrote a book on the \emph{Cellular Automaton interpretation of Quantum Mechanics} \cite{Hooft2014}; for an online free version, go to \url{https://arxiv.org/abs/1405.1548}. In 1994, Bialynicki-Birula wrote the ``Weyl, Dirac, and Maxwell equations on a lattice as unitary cellular automata'' \cite{BB94a}.

\subsubsection{Note}

\noindent
More authors and works are evoked in the introduction, and discussed more lengthily in the conclusion. The authors include the teams Cirac-Zohar-Reznik, Zoller's, D'Ariano-Perinotti-Bisio, Arrighi-Facchini-Di Molfetta, Chandrasekhar-Busch, to cite only some of the members of each team. All these works deal with relativistic aspects. A huge amount of works deal with non-relativistic quantum simulation, including, among those working with cold atoms, Zoller's team, the collaboration Bloch-Cirac, the team Meschede-Alberti, and that of Dalibard. Among the teams working with photons, let us mention the team Silberhorn-Sansoni and the team Mataloni-Sciarrino, to cite only a few.
\newpage
\iffalse
\subsection{Hamiltonian formulations of DTQWs}

\subsection{Experimental implementation of quantum walks}

\section{Driven quantum systems: theory and experiments}

\subsection{Driven quantum walks}
\fi

\section{Publications \cite{AD15} and \cite{AD16}: DTQWs in electromagnetic fields} \label{sec:two_publications}

\titlebox{
Both these publications deal with 2D DTQWs interacting with electromagnetic fields. Publication \cite{AD15} is devoted to relativistic Landau levels in a constant and homogeneous artificial (or synthetic) magnetic field. The study is perturbative in the step of the lattice, around the continuum limit, i.e. the wavepackets are large with respect to the step. We also present, through numerical simulations, some phenomenology demonstrating the confining properties of the magnetic field, beyond the continuum limit  and the previous perturbative effects. Publication \cite{AD16} exhibits an electromagnetic gauge theory on the (1+2)D spacetime lattice for DTQWs, so that these DTQWs interact with \emph{general} 2D electromagnetic fields. Lattice counterparts of Maxwell's equations are suggested, which preserve the charge current conservation on the lattice. For weak electric and magnetic fields, i.e. tiny fractions of $2\pi$, the Brillouin zone, known phenomena are recovered, such as the $\boldsymbol{E} \times \boldsymbol{B}$ drift or Bloch oscillations. For high fields, i.e. sizeable fractions of the Brillouin zone, the spreading properties strongly depend on whether or not the values of the electric and magnetic fields are rational fractions of the Brillouin zone.
} \\

\noindent
I give below a compact presentation of these two publications \cite{AD15,AD16}.

Let us first give some background: first, an extended `recap' of the 1D case \cite{DDMEF12a,DMD14} (Subsection \ref{subsec:recap_1D}), and  second, the expression of the 2D DTQW-operator with electromagnetic coupling (Subsection \ref{subsec:walk_2D}). This 2D walk operator is the central ingredient of the two publications, whose initial purpose and first achievement is to extend the 1D results to dimension 2.

Recall that the the one-step evolution is given by Eq. (\ref{eq:protocol}), that is (I make explicit the time-dependence of the walk operator, which is in order in the cases we are going to deal with):
\begin{equation} \label{eq:protocol_2}
\ket {\Psi_{j+1}} = \hat{W_j} \ket{\Psi_j}.    
\end{equation}

\subsection{State-of-the-art `recap' on the 1D electric DTQW} \label{subsec:recap_1D}

\noindent
In dimension 1, there can be no magnetic field\footnote{Unless one embeds this single dimension into a higher-dimensional space, which is sometimes called `effective 1D', i.e. the motion is constrained to be one-dimensional but the actual physical space has a higher dimension. This is not the case here, the scheme is strictly 1D.}, and the 1D DTQW-operator with electric coupling reads (see Eqs. (\ref{eq:electric_walk_intro})):
\begin{equation} \label{eq:1D_walk_operator}
W_j^{\text{1D}} = e^{i\Delta \alpha_j} U(\Delta \theta, \Delta \xi_j) \, S \, ,
\end{equation} 
where $\Delta \alpha_{j,p}$ is an overall (i.e. spin-independent) local (i.e. time- and position-dependent) phase shift,  the spin-dependent shift $S$ is given by Eq. (\ref{eq:shift_compact}), and where I have introduced a 1D\footnote{In contrast with coin operations acting on coin spaces greater than 2 and are associated to multidimensional spin-dependent shifts.} coin operation with mixing angle $\Delta\theta$ and spin-dependent local phase shift $\pm\Delta \xi_{j,p}$,
\begin{equation} 
U(\Delta\theta,\Delta\xi_{j,p}) = C(\Delta\theta) \, F(\Delta\xi_{j,p}) \, ,
\end{equation}
with the standard coin operation $C(\theta)$ and the spin-dependent phase-shift operator $F(\xi)$, given respectively by Eqs. (\ref{eq:standard_coin}) and (\ref{eq:phase_shift}). We write the mixing angle and the phases as (see below why)
\begin{subequations}
\begin{align}
\Delta\theta &= - \epsilon_m \, m \\
\label{eq:deltaalpha} 
\Delta\alpha_{j,p} &= \epsilon_A \, (A_0)_{j,p}  \\
\label{eq:deltaxi} 
\Delta\xi_{j,p} &= \epsilon_A \, (A_1)_{j,p} \, . 
\end{align}
\end{subequations}
Note that $\Delta\theta$ could be chosen $(j,p)$-dependent.

The electric\footnote{There is no magnetic field in this 1D scheme.} interpretation of this walk is threefold. The first interpretation is explained in depth in Subsection \ref{subsec:continuum_limit_inhomogeneous}\footnote{This first interpretation is only valid in the limit of both (i) wavefunctions whose width is large with respect to the spatial-lattice step, and which vary over time periods which are large with respect to the time step, and (ii) a weak gauge field, i.e. $\epsilon_A|A^{\mu}_{j,p}| \ll 2\pi$ for $\mu = 0,1$ and for all $j$'s and $p$'s.}, the second is introduced in Subsection \ref{subsec:gauge_inv}, and the third is a generalization, to an arbitrary electric field, of the interpretation adopted in Refs. \cite{Bauls2006,mesch13a,ced13} in the case of a non-generic constant and uniform electric field\footnote{The paper of 2006 by Ba\~nuls et al. \cite{Bauls2006} shows that this constant and uniform electric field can be obtained through either a spatial or a temporal discrete derivative, but the discrete gauge invariance is not exhibited (which is linked to the fact that the case of an arbitrary electric field is not treated). In the `Electric-quantum-walk' experiment with neutral atoms in a 1D optical lattice, by Genske et al. \cite{mesch13a}, the electric field is obtained with the (discrete) \emph{spatial} derivative; the paper by Cedzich et al. \cite{ced13} provides theoretical interpretations of the experimental results, which are far beyond the continuum situation, and deal with revivals of the initial state, due to the periodicity of the system (this periodicity dissapears in the continuum limit).}:
 
\begin{itemize}
\item[$\bullet$]  First, considering $\Psi_{j,p}=\Psi(\epsilon_l \, \! j, \epsilon_l \, \! p)$, where $\epsilon_l=\epsilon_x=\epsilon_t$ is the spacetime-lattice step (the subscript `$l$' is for `lattice'), choosing $\epsilon_m=\epsilon_A=\epsilon_l=\epsilon$\footnote{If these small parameters differ from each other by proportionality factors only, it changes nothing to the equation obtained in the continuum limit except from such proportionality factors in front of either $A$, $m$, or the lattice coordinates. However, one of these small parameters can be, for instance, a power of another small parameter, possibly non-integer, which makes the continuum limit change; see Subsection \ref{subsec:continuum}.}, and taking the continuum limit $\epsilon \rightarrow 0$, the one-step evolution equation (\ref{eq:protocol_2}) with walk operator (\ref{eq:1D_walk_operator}) yields the (1+1)D Dirac equation for a relativistic spin-1/2 fermion of mass $m$ and charge $-1$, coupled to an electric potential with covariant time and spatial components $A_0$ and $A_1$, see Eq. (\ref{eq:Dirac_coupled}).

\item[$\bullet$]  Second, the  walk satisfies the following gauge invariance \cite{DMD14} on the spacetime lattice with nodes labelled by $(j,p)$ (we have recast the results of Subsection \ref{subsec:gauge_inv}): Eq. (\ref{eq:protocol_2}) with the walk operator (\ref{eq:1D_walk_operator}) is invariant under the following substitutions:
\begin{subequations}\label{eq:discrete_gauge_inv}
\begin{align} 
\Psi &\rightarrow \Psi' = \Psi e^{-i\phi} \\
\label{eq:discrete_gauge_inv_gauge_field}
A_{\mu} & \rightarrow A'_{\mu} = A_{\mu} - d_{\mu} \phi \ ,
\end{align}
\end{subequations}
where $\phi_{j,p}$ is an arbitrary local phase change, and where we have introduced the following discrete derivatives (finite difference operators):
\begin{equation} \label{eq:finite_diff}
d_0 = (L - \Sigma_1)/\epsilon_A \ , \ \ \ \ \ \ \
d_1  = \Delta_1/ \epsilon_A ,
\end{equation}
where, for any $(j,p)$-dependent quantity $Q$,
\begin{subequations}
\begin{align}
(L Q)_{j,p} &= Q_{j+1,p} \\
(\Sigma_1 Q)_{j,p} &= \frac{Q_{j,p+1}+Q_{j,p-1}}{2} \\
(\Delta_1 Q)_{j,p} &= \frac{Q_{j,p+1}-Q_{j,p-1}}{2} \ .
\end{align}
\end{subequations}
\noindent
Note that $d_1$ is a standard finite difference in the spatial direction (defined over two sites), while $d_0$ can be viewed as the mean between two standard finite-differentiated convective derivatives (with Eulerian speed equal to $1$), one going forward, and the other backwards\footnote{This was pointed out to me by D. Meschede.}:
\begin{equation}
(d_0Q)_{j,p} = \frac{1}{2 \epsilon_A} \Big[ \big\{(Q_{j+1,p}-Q_{j,p}) + (Q_{j,p}-Q_{j,p-1}) \big\} + \big\{ (Q_{j+1,p}-Q_{j,p}) + (Q_{j,p}-Q_{j,p+1}) \big\} \Big] \, .
\end{equation}
\noindent
In the continuum limit, the finite-difference operators $d_0$ and $d_1$ yield the standard partial derivatives $\partial_0$ and $\partial_1$, and the lattice gauge invariance (\ref{eq:discrete_gauge_inv}) yields the standard gauge invariance of the Dirac equation in (1+1)D continuous spacetime.

Eventually, one can define the following lattice equivalent of the electric\footnote{It contains no magnetic field.} tensor in 1+1 dimensions,
\begin{equation} \label{eq:lattice_1D_strength_tensor}
(F_{\mu\nu})_{j,p} = (d_{\mu} A_{\nu})_{j,p} - (d_{\nu} A_{\mu})_{j,p} \ ,
\end{equation}
where $(\mu,\nu) \in \{0,1\}$. This quantity is (i) antisymmetric by construction, (ii) invariant under (\ref{eq:discrete_gauge_inv}), and (iii) its continuum limit yields the usual electric tensor. Moreover, we can check the following conservation equation on the lattice:
\begin{equation} \label{eq:current_cons}
d_0 J^0 + d_1 J^1 = 0 \ ,
\end{equation}
where $J^{\mu}$ is invariant under (\ref{eq:discrete_gauge_inv}) and has the same expression as the Dirac current in (1+1)D continuous spacetime (but is defined on the lattice).

\item Third, the walk operator (\ref{eq:1D_walk_operator}) can be obtained from the standard walk having only an angle $\Delta\theta\neq0$, i.e. with $\Delta\alpha_{j,p}=\Delta \xi_{j,p}=0$, by implementing a `Bloch phase' $e^{i(\Delta\alpha_{j,p}\pm\Delta\xi_{j,p})}$ \footnote{I use the same denomination as in \cite{mesch13a}: the so-called Bloch phase is that which implements the electric potential, namely, with contravariant components, $\sim (\Delta\alpha_{j,p}, \Delta\xi_{j,p})$.}, which reminds of that acquired by tight-binding electrons driven, in dimension 1, by a superimposed arbitrary electric field, where the $+$ is for the electron tunneling in one direction, and the $-$ for the other direction. To be precise on the tight-binding procedure in the presence of electromagnetic fields: one mathematically  \emph{adds} the \emph{scalar} potential $\sim \Delta\alpha_{j,p}$ to the tight-binding Hamiltonian, and performs, to account for the presence of the \emph{vector} potential, a Peierls substitution on the hopping (off-diagonal) matrix elements, namely $t \rightarrow t \, e^{\pm\Delta\xi_{j,p}}$, see Eq. (12) in \cite{Graf1995}. In the case of a constant and uniform electric field, this condensed-matter system yields the well-known Bloch oscillations of the electrons, with period inversely proportional to the electric field. Despite the phenomenal similarities between such a DTQW-scheme and the 1D tight-binding Hamiltonian with superimposed electric field, there are two \emph{fundamental} differences. First: in the DTQW, the particle, which has a spin (or pseudo-spin), undergoes a spin-dependent transport, while in the tight-binding Hamiltonian, the left-right jump has \emph{nothing} to do with spin, it is due to \emph{tunneling}\footnote{The DTQW-scheme is intrinsically chiral, and this has nothing to do with the presence or absence of a magnetic field. In the case of tight-binding Hamiltonians, one can include spin-orbit corrections \cite{Jaffe1987,Barreteau2016} when a magnetic field is applied, but again the left/right jump has  \emph{nothing} to do with that, it is due to tunneling. (Note that the probabilities of going left or right by tunneling can be modified by a superimposed magnetic field \emph{only} if the latter breaks the translational symmetry of the system; this might be a way of simulating a biased coin operation with such a condensed-matter system, or, rather, a simulation of this condensed-matter system.)}. Second, here (i) the Peierls flux exponential is implemented, not on the Hamiltonian, as standardly done with tight-binding Hamiltonians, but on the (one-step) evolution operator, and (ii) the implementation of the scalar potential is not equivalent either, although there is a closer matching\footnote{To the exponential implemented on the DTQW one-step evolution operator, $e^{i\Delta\alpha_{j,p}}$, corresponds a scalar potential $\sim\Delta\alpha_{j,p}$, which is precisely that \emph{added} (no multiplication like for the Peierls substitution) to the tight-binding Hamiltonian.}, because of the generic non-commutativity of the one-step evolution operators. Eventually, there is a third difference whose `importance' (to define) I haven't completely evaluated yet: in the DTQW-scheme, time is discrete, while it is continuous in tight-binding Hamiltonians.

\end{itemize}

\subsection{Walk operator for the 2D DTQW with electromagnetic (i.e. Abelian) coupling} \label{subsec:walk_2D}

Now the walker $\ket{\Psi_j}$ lives, in real space, on a 2D lattice with nodes $(p_1,p_2) \in \mathbb{Z}^2$.

We build the 2D DTQW operator by doing a 1D walk in the first spatial direction, with spin-dependent phase shift $\Delta\xi^{(1)}_{j,p_1,p_2}$, followed by another 1D walk but in the other spatial direction, with spin-dependent phase shift $\Delta\xi^{(2)}_{j,p_1,p_2}$\footnote{Similar constructions can be found, in a quantum-computing perspective (if we are to categorize), in the AKR paper (2005) \cite{AKR2005} and in the paper by Di Franco et al. (2011) \cite{DiFranco2011}, and, in the perspective of simulating quantum physics with cellular automata, in the papers by Succi et al. (1993) \cite{Succi1993}, Bialynicki-Birula (1994) \cite{BB94a}, Yepez (2002) \cite{Yepez2002Dirac}, Strauch (2006) \cite{Strauch06a}, and, more recently, Arrighi et al. (2015, 2016) \cite{Arrighi_higher_dim_2014,AF16} and Succi et al. (2015) \cite{Succi2015}. Note the following papers by Yepez: these two rather old ones (2002) \cite{Yepez2002Schro, Yepez2002Burgers}, and this recent proceeding of 2016 \cite{Yepez2016}.}:
\begin{equation}\label{eq:walk_operator_2D}
W_j^{\text{2D}} = e^{i\Delta \alpha_j} \, \, \left[ U \! \left(f^-(\Delta\theta), \Delta \xi^{(2)}_j\right)  \, S^{(2)} \right] \, \,\left[ U \! \left(f^+(\Delta \theta), \Delta \xi^{(1)}_j\right) \, S^{(1)} \right] \, ,
\end{equation}
where the mixing angle $\Delta \theta$ is encoded through
\begin{equation}
f^{\pm}(\Delta\theta) = \pm \frac{\pi}{4} + \frac{\Delta \theta}{2} \, .
\end{equation}

\subsection{Publication \cite{AD15}: Landau levels for discrete-time quantum walks in artificial magnetic fields} \label{subsec:B}

\titlebox{
In this publication, we present a new family of 2D DTQWs which coincides, in the continuum limit, with the Dirac dynamics of a relativistic spin-1/2 fermion coupled to a constant and uniform magnetic field. This allows us to extend standard relativistic Landau levels from the continuum Dirac dynamics to the lattice situation, where the walker is indeed `sensitive' to the non-vanishing step of the spacetime lattice. Such Landau levels for DTQWs are built perturbatively in the step, around the continuum situation. We eventually demonstrate, through numerical simulations, that the parameter interpreted as the magnetic field in the continuum limit, has, beyond both (i) the continuum limit and (ii) perturbative effects of the step, qualitatively the same confining properties as a standard magnetic field. The possibility of quantum simulation of condensed-matter systems by DTQWs is also discussed.
} \\

%{\begin{center} 
\subsubsection{Short but detailed review and comments} 
%\end{center}}

%\vspace{-0.2cm}
%\hrule 

%\subsubsection{Short but detailed review and comments}
%\vspace{0.2cm}
In Section 2 of the paper, we introduce the above walk, Eq. (\ref{eq:walk_operator_2D}), but \emph{only} in the particular case
\begin{equation}
\Delta\alpha_{j,p_1,p_2}=0 \ , \ \ \ \ \ \ \Delta\xi^{(1)}_{j,p_1,p_2} = 0 \ , \ \ \ \ \ \ \Delta\xi^{(1)}_{j,p_1,p_2} = B p_1 \epsilon \ ,
\end{equation}
where $B \in \mathbb{R}$ is the parameter that is interpreted as a constant and uniform magnetic field in the continuum limit\footnote{Outside the continuum limit, the confinement properties of $B$, exhibited in Section 5 of the paper, also endow $B$ with a phenomenal \emph{qualitative} magnetic interpretation.}, derived in Section 3 of the paper.

Note this technical formal point: the two coin matrices introduced in the paper do \emph{not} correspond to the two $U$'s of the above equation, (\ref{eq:walk_operator_2D}), but after expanding the matrix products in the paper, one can recast the equations exactly as (\ref{eq:walk_operator_2D}). This is so simply because the correspondence between the DTQW-equations and the two matrices chosen to write these equations in a compact form is not one to one, but one to many\footnote{When writing this paper we had not figured out the suitable compact form of Eq. (\ref{eq:walk_operator_2D}) because we had not figured out the minimal possible gauge invariance on the lattice. That compact form is given in our next paper \cite{AD16}; it enables, in particular, a straightforward correspondence between the lattice gauge fields and those in the continuum.}.

Section 4 is devoted to a study of how the non-vanishing step of the spacetime lattice modifies the eigenstates of the continuum-limit Hamiltonian, which are the well-known relativistic Landau levels. The study is perturbative in the step $\epsilon$, the zeroth order being the continuum situation. The zeroth-order eigen-energies are well-known; they vary as the square root of the level\footnote{Have in mind that this square-root dependence is that expected for relativistic Dirac fermions or gapped graphene-like quasiparticles (which behave effectively exactly as the former), in contrast with the linear dependence expected for non-relativistic particles.}.  There are two usual choices for the eigenstate basis of the Landau levels. We choose the so-called Hermite basis, for which the momentum along  $y$ is a good quantum number\footnote{The other eigenvector basis is the Laguerre basis, for which the good quantum number is a certain gauge-invariant angular momentum, see \cite{HRR93} and Appendix \ref{app:gauge_invariant_gen} for an  expression of the gauge-invariant generator of an arbitrary symmetry of the electromagnetic field.} (this is possible in the chosen Landau gauge); this choice is natural because translations are symmetries of the mesh on which the walker lives, and the walker `sees' this mesh at higher orders. In Appendix B of the paper, we compute analytically the first-order corrections to (i) the eigen-energies (which vanish) and to (ii) the Hermite eigenvectors, and we numerically check in the body of the paper that these corrections are correct\footnote{I have checked that, for the Laguerre basis at zeroth order, the DTQW-scheme converges when the lattice step goes to zero, but the convergence is not improved when I add the first-order corrections to the eigenvectors, maybe because these corrections only make sense if the rotational symmetry of space is not broken for a non-vanishing step, while this is \emph{not} the case in our scheme, which uses a square lattice.} (see Figs. 4 and 5 of the paper). We also provide, in  Fig. 3, a typical (relative) change induced by the corrections on the probabilities of presence of the five first Landau levels. These plots may be useful for, e.g., a comparison with future experimental data.

Section 5 aims at giving a flavor of the phenomena produced by our scheme well beyond the continuum limit (the perturbative computation can intrinsically only go, a priori, slightly beyond). We show that the parameter $B$ has the same qualitative confinement properties as a standard magnetic field in continuous spacetime.

In the conclusion (Section 6), we stress that, if we view the DTQW as a way to discretize the Dirac equation, (i) this discretization is a naive  symmetric one, and suffers from fermion doubling, which happens at the boundaries of the Brillouin zone, but that (ii) this does not preclude the use of this discretization around the Dirac cone, located at the center of the Brillouin zone, i.e. for small vavectors (in comparison with the size of the Brillouin zone), which is the only limiting case where the Dirac-fermion interpretation, i.e. the continuum limit, makes sense.

\includepdf[pages={-}]{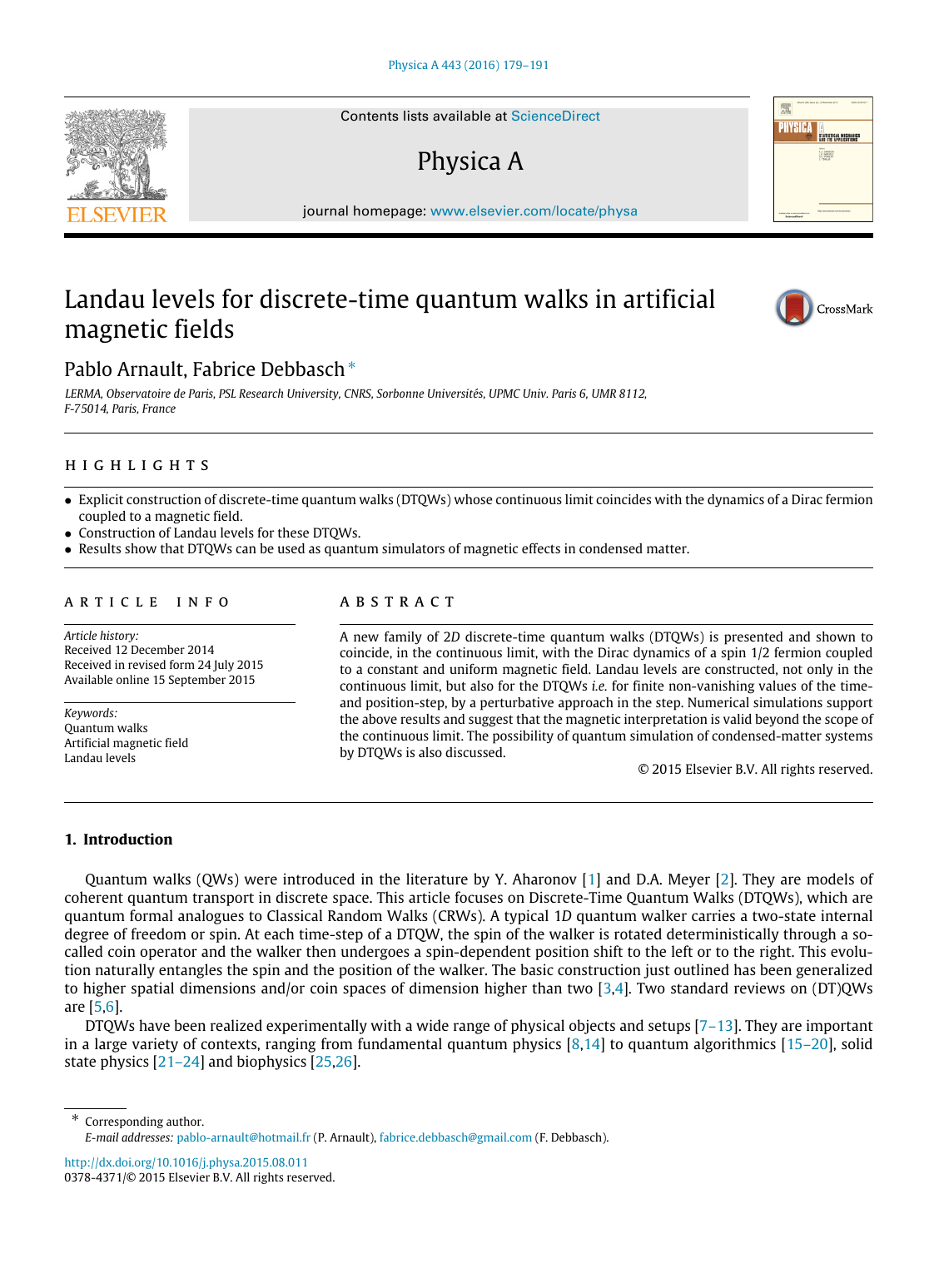}

\cleardoublepage
\subsection{Publication \cite{AD16}: Quantum walks and discrete gauge theories} \label{subsec:EB}

\titlebox{
In this publication, we present a new family of 2D DTQWs which coincides, in the continuum limit, with the Dirac dynamics of a relativistic spin-1/2 fermion coupled to a generic electromagnetic field in  dimension 2. We extend the electromagnetic interpretation beyond the continuum limit, by showing the existence of (i) a lattice gauge invariance, (ii) a lattice invariant quantity, and (iii) conserved currents on the lattice. This allows us to suggest lattice equivalents to Maxwell's equations. Summing up, we build a DTQW-based Abelian U(1) lattice gauge theory, in (1+2)D. We recover, through numerical simulations, known phenomena, both standard ones, associated to standard electromagnetic fields, and a less standard one, associated to periodically-driven system, which, in particular, matches qualitatively with recent results on DTQWs, both in one and two dimensions. 
} \\

\subsubsection{Short but detailed review and comments}

This is the paper where, in Section II, we give the proper compact form of the 2D DTQW-operator with generic electromagnetic coupling, Eq. (\ref{eq:walk_operator_2D}). The continuum electromagnetic interpretation is given in Section II: the continuum limit of the scheme is that of a spin-1/2 Dirac fermion, with mass $m$ and charge $-1$, coupled to an electromagnetic potential with contravariant components $A_0 \sim \Delta\alpha$, $A_1 \sim -\Delta\xi^{(1)}$ and $A_2 \sim - \Delta\xi^{(2)}$, where, as for the 1D case, the symbol `$\sim$' means that between the $A_{\mu}$'s and the phase shifts, there is a factor $\epsilon_A$ (see Eqs. (\ref{eq:deltaalpha}) and (\ref{eq:deltaxi})) which is taken equal to the spacetime-lattice step $\epsilon_l$ and goes to zero in the continuum, where the discrete coordinate $p_1$ ($p$ in the paper) becomes the continuous one $x$ ($X$ in the paper), and $p_2$ ($q$ in the paper) becomes $y$ ($Y$ in the paper). In addition to this formal continuum limit, shown on the evolution equation for the quantum particle, we show numerically that the limit is correct, as follows: we run the MQTD scheme with an initial condition whose time evolution through the Dirac equation (for constant and uniform crossed electric and magnetic fields) can be computed in closed form, and we compare this explicit continuum time evolution to the MQTD time evolution for decreasing spacetime-lattice steps, see Figs. 1 and 2 of the paper.

After a generalization, in Section III, of the 1D results on the lattice gauge invariance (Eq. (8) of the paper generalizes Eq. (\ref{eq:discrete_gauge_inv})), and gauge invariants (Eq. (11) of the paper generalizes Eq. (\ref{eq:lattice_1D_strength_tensor})), we present, in Section 4, a lattice version of (the inhomogeneous\footnote{The homogeneous ones are satisfied \emph{by construction} on the lattice.}) Maxwell's equations (Eq. (20) of the paper), which ensure the current conservation on the lattice  (Eq. (17) of the paper). These lattice Maxwell equations yield the standard ones in the continuum limit. This means we can theoretically quantum simulate, in the continuum limit, a first-quantized Abelian gauge theory: the spin-1 bosons\footnote{The spin-1 boson is, in this (1+2)D case, $(A_0,A_1,A_2)$, and thus has three internal states.} evolve through Maxwell's equations, and the fermions through the Dirac equation.

Section 5 is devoted to the phenomenal properties of the 2D electromagnetic DTQW beyond the continuum limit. A detailed discussion of the two different `small parameters' is now in order. The first small parameter, $\epsilon_l$, is the spacetime-lattice step. The second small parameter, $\epsilon_A$, is that associated to the lattice gauge field. The continuum limit requires that both $\epsilon_l$ and $\epsilon_A $ tend to zero. But one can envisage taking only one of the two parameters as actually small, either (i) large wavefunctions, i.e. $\epsilon_l \simeq 0$, or (ii) a weak gauge field\footnote{The weak-gauge-field condition is that needed to Taylor expand $\exp(i\epsilon_A A^{\mu}_{j,p_1,p_2})$ around zero, i.e. $\epsilon_A|A^{\mu}_{j,p_1,p_2}| \ll 2\pi$ for $\mu = 0,1,2$ and for any $(j,p_1,p_2)$. This condition thus enforces the variations of the $\epsilon_A A^{\mu}_{j,p_1,p_2}$'s between two points of the spacetime lattice to be much smaller than $2\pi$ as well. Conversely, the picture is different: choosing small variations, i.e., in our case, $\epsilon_A E$ and $\epsilon_A B$ $\ll 2\pi$, ensures the weak-gauge-field condition \emph{only for some time}.}\textsuperscript{,}\footnote{One can obviously refine this picture by `dezooming' only on either the temporal or the spatial lattice. Such a distinction is necessary for a mapping to CTQWs.}: 

\begin{enumerate}
\item If we are in situation (ii), we recover known phenomena. First, we recover  the classical $\boldsymbol{E} \times \boldsymbol{B}$ drift \cite{book_Jackson}, even if (i) is not satisfied -- i.e. even if the walker `sees' the edge of the Brillouin zone --, which is a priori non-obvious; see Figs. 4 (bottom propagating front) and 5 of the paper. Second, we recover tight-binding-like phenomena: Bloch oscillations for a vanishing magnetic field in Fig. 3, and top propagating front \cite{Kolovski03, Kolovsky04, KolovskyMantica14} for a non-vanishing magnetic field in Fig. 4. In contrast with the classical $\boldsymbol{E} \times \boldsymbol{B}$ drift, these tight-binding-like phenomena \emph{demand} that (i) be unsatisfied at some point in the dynamics, i.e. that the walker `see' the edge of the Brillouin zone\footnote{The $\boldsymbol{E} \times \boldsymbol{B}$ drift is `blind' to (i), while the tight-binding-like phenomena demand non-(i).}. These tight-binding phenomena are qualitatively not that unexpected (see the third item in Subsection \ref{subsec:recap_1D}).

\item If (ii) is not satisfied, i.e. if the gauge-field components `see' the edge of the Brillouin zone, i.e. reach sizeable fractions of $2\pi$, then the dynamics is strongly dependent on whether the electric and magnetic field are \emph{exact} rational fractions of $2\pi$, see Figs. 6 and 7 of the paper. These features have already been studied in full mathematical depth in dimension 1, for a sole electric field, by Cedzich et al. (2013) \cite{ced13}, and have been studied more recently in dimension 2 for a sole magnetic field by Yal{\c{c}}{\i}nkaya et al. (2015) \cite{Yalcinkaya2015}, and, for a sole electric field, by Bru et al. (2016) \cite{Bru2016}.
\iffalse
The results given by Cedzich et al. \cite{ced13} for the 1D DTQW in a constant and uniform electric field, are summarized in Table.  of the paper, and very briefly summarized in Appendix . 
\fi
\end{enumerate}

%\newpage\null\thispagestyle{empty}\newpage

%\clearpage{\thispagestyle{empty}\cleardoublepage}
\includepdf[pages={-}]{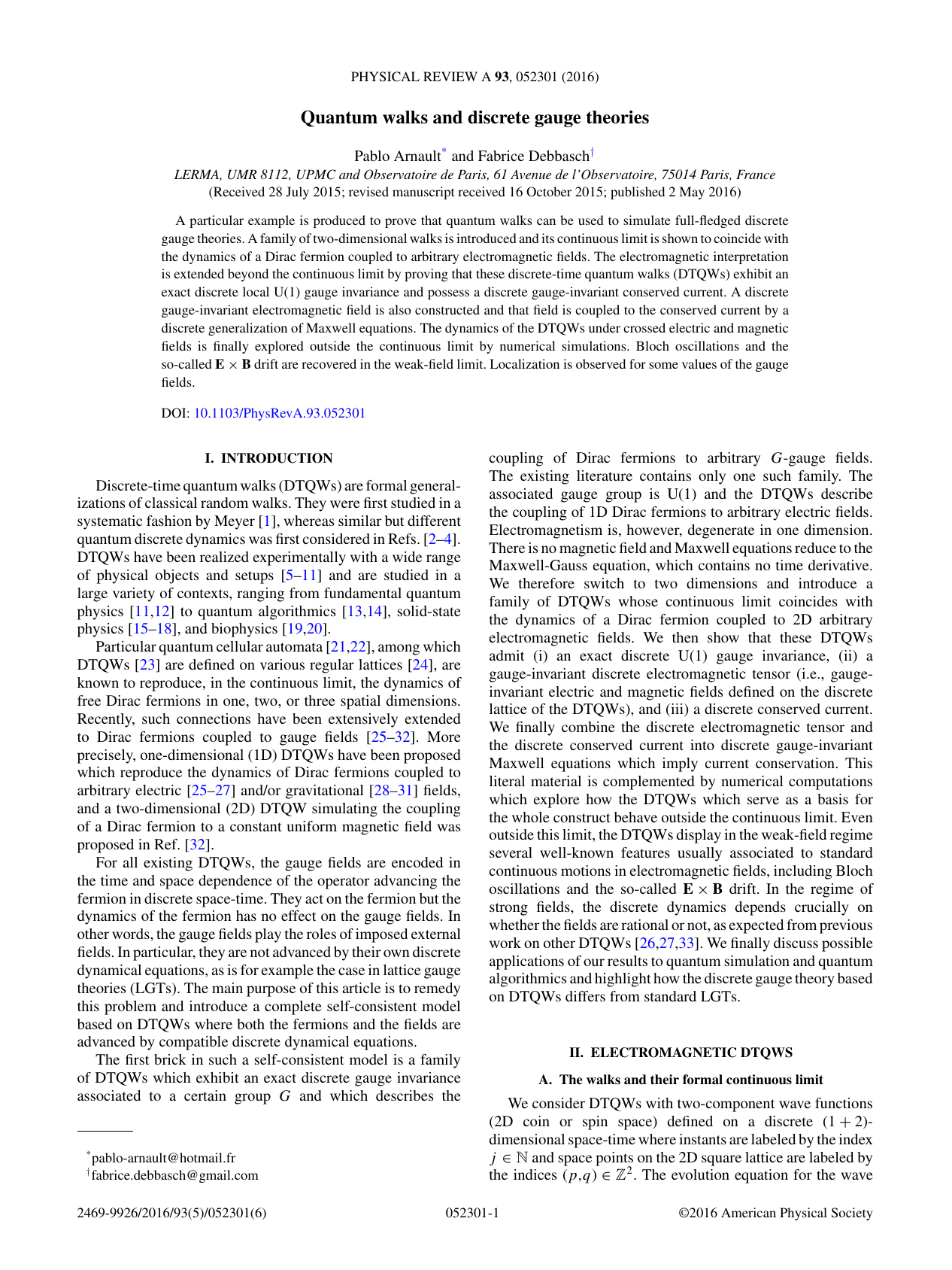}
   %%%%%%%%%%%%%%%%%%%%%%%%%%%%%%
%%  Alias bibliographiques  %%
%%%%%%%%%%%%%%%%%%%%%%%%%%%%%%
\defcitealias{NRFortran}{Numerical Recipes}
\defcitealias{NRF77}{Numerical Recipes}
\defcitealias{NRF90}{Numerical Recipes}
\defcitealias{IUPAC}{IUPAC}%International Union of Pure and Applied Chemistry

\chapter{\textsc{One-dimensional DTQWs in non-Abelian Yang-Mills fields \\ (Publication \cite{ADMDB16})} \label{Chap:Non-Abelian_1D_DTQWs}}
{
\minitoc

\iffalse
Before presenting the paper I coauthored in Section \ref{sec:paper3}, I will first make, in the following section, a recap on `standard' historical applications of non-Abelian gauge theories
\fi

\iffalse
, and then briefly present, in Section\ref{sec:LGT_quantum_simu}, standard lattice gauge theories and the very recent topic of their quantum simulation\footnote{
\fi

\section{On non-Abelian Yang-Mills gauge theories} \label{sec:historical_applications}

\motivations{
The aim of this long introduction to Publication  \cite{ADMDB16} is to give a review on the history of Yang-Mills gauge theories, from the discovery of the electric and magnetic forces to standard lattice gauge theories. 
}  \\

\noindent
Originally introduced in 1954 by Yang and Mills \cite{Yang1954} to describe the attractive force that holds protons and neutrons together in the atomic nucleus\footnote{For some notes on the (nuclear) isospin degree of freedom and its history, see Appendix \ref{app:isospin}.}, non-Abelian Yang-Mills gauge theories soon appeared to be a suitable framework to describe the weak force and the strong force between quarks\footnote{Nucleons are made of quarks, hold together by strong interactions. The electromagnetic force is described by an \emph{Abelian} Yang-Mills gauge theory. A gauge theory is a theory in which the state of the system is invariant under some group of spatiotemporally-dependent (i.e. local) transformations of some of its degrees of freedom. A given such transformation relates two different possible \emph{gauges} to describe the same state of the system. Gravity can also be viewed as a gauge theory, but the attemps to formulate it as a Yang-Mills-type gauge theory remain unconclusive.}\textsuperscript{,}\footnote{Yang-Mills theories historically arose in a quantum framework. They can however be developed non-quantumly, see Ref. \cite{Boozer2011}. In second-quantized physics, they have to be renormalized, i.e. they only make sense if the `physical', i.e. measured coupling constant of the interaction, is considered as a function of the energy scale of the physical phenomenon. This energy scale is simply the input total energy of the system if one can consider this system isolated during the phenomenon; in the case of a particle-physics experiment, this total energy is the total kinetic energy given to the particles by accelerating them, before we make them collide. The proof that non-Abelian Yang-Mills theories of the electroweak type are renormalizable was given in 1971 by 't Hooft, first in the massless \cite{tHooft1971} and then in the massive \cite{Hooft1971b} gauge-field case.}, see Subsection \ref{subsec:particle_physics}. Non-Abelian Yang-Mills gauge theories have also found applications in condensed-matter physics \cite{Berche2012}.

%, see Subsection \ref{subsec:condensed_matter_physics}. 

\subsection{Fundamental forces and particle physics} \label{subsec:particle_physics}

\subsubsection{The electromagnetic force, a long-range force}

\paragraph{{$\bullet$ Pre field-notion works}}

\subparagraph{$\star$ \emph{Electrostatics and Coulomb's force law between electric charges}}\hspace{3cm}

\vspace{0.2cm}

\noindent
In 1785, Coulomb stated, in his \emph{First Memoir on Electricity and Magnetism}\footnote{See the following scan of the original document, written in French, \url{https://books.google.fr/books?id=by5EAAAAcAAJ\&pg=PA569\&redir_esc=y\#v=onepage\&q\&f=false}.}, the law which was then to be named after him: the force exerted by a macroscopic or mesoscopic point-like sample of matter of total charge $q_1$, on another one of total charge $q_2$ located at distance $r$ from the first sample, is given by\footnote{Strictly speaking, a sufficient condition for this law to hold is the spherical symmetry of the charge distribution of each sample. This is what, strictly speaking, `point-like' means.}
\begin{equation}
\boldsymbol{F}_{\text{Coulomb}} = k_{e} \frac{q_1 q_2}{r^2} \boldsymbol{u} \, ,
\end{equation}
where $\boldsymbol{u}$ is the unit vector from the first to the second point-like sample, and $k_e$ is Coulomb's constant, which was later recognized to be
\begin{equation}
k_e=\frac{1}{4\pi\varepsilon_{0}} \, ,
\end{equation}
where $\varepsilon_0$ is the vacuum permittivity. The experiment made to prove such a law used a torsion balance. The $1/r^2$ dependence had already been suggested in 1762 by Cavendish, but his experiment was not fully conclusive.

\subparagraph{$\star$ \emph{Magnetostatics and Ampère's force law between electric currents}}\hspace{3cm}

\vspace{0.2cm}

\noindent
In 1820, {\O}rsted made the famous discovery that a magnetic needle is acted on by an electric current\footnote{Both the original Latin version and the English translation of his 1820 paper ``Experiments on the effect of a current of electricity on the magnetic needle", can be found in the 1981 book by Franksen, \emph{H. C. Ørsted – a man of the two cultures}.}. Soon after, the same year, the experimental and theoretical work of Biot and Savart\footnote{See Biot's speech to the French Academy of science, reproduced in the following scan of the second volume of a collection of memoirs of physics published by the French Society of physics in 1885, page 80 and followings, \url{http://www.ampere.cnrs.fr/bibliographies/pdf/1885-P286.pdf}. See in particular the footnote page 80 for historical details on Biot and Savart's scientific notes. The formula giving the magnetic field (this notion appeared later, in the work of Faraday) created by a current distribution was later (and is currently) named after Biot and Savart. By generating such a magnetic field, this current distribution acts on a charge distribution through the (magnetic part of the) Lorentz force.}, and then Ampère\footnote{See the reproduced speech that follows that of Biot in the aforementioned collection. This work gave its name, in particular, to this Maxwell equation which involves the dynamics (i.e. the time derivative) of the eletric field.}, lead to the expression of the force exerted by a current element on another one, now called Ampère's force law, and which is the magnetic analogue of Coulomb's force law.

\paragraph{{$\bullet$ Faraday's induction law and concept of (magnetic) field} } \hspace{3cm}

\vspace{0.2cm}

\noindent
The following notes were essentially produced after reading a short paper on the history of the concept of field, by Assis, Ribeiro and Vannucci\footnote{See~\url{http://www.ifi.unicamp.br/~assis/The-field-concepts-of-Faraday-and-Maxwell(2009).pdf}.} [ARV], which contains interesting quotes by Faraday and Maxwell.

In 1831, Faraday presented orally, to the English Royal Society of London, his experiments showing magnetic induction\footnote{See a scan of his  \emph{Experimental researches in electricity}, published in 1844, at \url{https://docs.lib.noaa.gov/rescue/Rarebook_treasures/QC503F211839_PDF/QC503F211839v2.pdf}, or at \url{https://archive.org/details/experimentalrese00faraiala}. I haven't seen a single litteral equation in these notes.}, i.e., roughly speaking,  that moving some magnetic material near charges causes their displacement around the material. In this work, he introduces the notion of `magnetic curves' or `lines of magnetic force', but a ``clear definition" only appeared, according to [ARV], in a 1852 paper. In 1846, he officially used the word `(magnetic) field', but again, a ``clear definition" only appeared, according to [ARV], in a 1850 paper.  Faraday ``does not seem to have utilized the expression electric field in his works" [ARV].

According to Faraday, the magnetic field was ``any portion of space traversed by magnetic power" [ARV], and ``he would probably understand" [ARV] the electric field the same way.

\paragraph{{$\bullet$ Maxwell's equations: a full-fledged dynamics for the electromagnetic field} } \hspace{3cm}

\vspace{0.2cm}

\noindent
In 1864, Maxwell gave a presentation to the English Royal Society of London, on ``A dynamical theory of the electromagnetic field" \cite{Maxwell1865}. In this work, he presented what are nowadays called Maxwell's equations, under the form of 20 equations with 20 unknowns, using quaternions\footnote{Maxwell had already published, in 1861,  his ``molecular-vortices" -- in his own words -- to model  magnetism, which can simply be viewed as a fluid model of electromagnetism. This work is both a graphical development and a mathematization of the ideas of Faraday on magnetic lines of forces. For the original paper, see \url{https://upload.wikimedia.org/wikipedia/commons/b/b8/On_Physical_Lines_of_Force.pdf}. One can actually already extract from this work the modern form of Maxwell's equations, which use vectorial analysis operators, namely the divergence and the curl. In this paper, Maxwell also suggested that light is an electromagnetic phenomenon. For reviews on Maxwell's fluid model of magnetism, see, e.g., the paper by Brady and Anderson \url{https://arxiv.org/pdf/1502.05926.pdf}, and the following presentation \url{https://www.pprime.fr/sites/default/files/pictures/pages-individuelles/D2/germain/Cargese2004.pdf}.}, and showed that, in the absence of charges, one can combine these equations to obtain the same wave propagation equation for both the electric and the magnetic field, whose characteristic speed parameter coincides numerically with the speed of light\footnote{The first conclusive measure of the speed of light, which determines its correct order of magnitude, is that of R{\o}mer and Huygens in 1675. The value was later more accurately measured by Bradley in 1729, Fizeau in 1849, and Foucault in 1862, two years before Maxwell's speech.}. In his \emph{Treatise on Electricity and Magnetism} of 1973, Maxwell rewrites the equations under the form of 8 equations. The equations were later given they modern form\footnote{4 vectorial equations with 2 vectorial unknowns, 1 pseudo-vectorial unknown, and 1 scalar unkown.} by Heaviside in 1884, see \emph{The Maxwellians}, by Hunt.

These equations are the result of a synthesis, into local equations, of the work of Ampère, Gauss, Helmholtz, Thomson and Faraday, among others, and was made possible by the introduction of a lacking piece in Ampère's theorem (see below), the \emph{displacement current}, which corresponds to the time derivative of the electric field $\boldsymbol{E}$, up to a constant multiplicative factor\footnote{The displacement current is the time derivative of the \emph{displacement field} $\boldsymbol{D}$, which is indeed proportional to the electric field only in the case of (i) vacuum (down to the smaller experimentally testable scales, i.e. even below atomic scales) or of (ii) linear, homogeneous and isotropic materials (above some mesoscopic scale). The proportionality relation reads $\boldsymbol{D}=\varepsilon \boldsymbol{E}$, where the proportionality factor $\varepsilon$ is the permittivity of the considered material, or simply $\varepsilon_0$ in the case of vacuum. The permittivity of some material needs to be measured experimentally, and can sometimes be theoretically derived, by suitable mesoscopic averages, as a function of the permittivity of vacuum and microscopical data regarding the molecular composition of the material.}.

The 4 Maxwell equations are the local forms of 4 corresponding integral theorems derived by Maxwell's aforementioned predecessors:
\begin{itemize}

\item[$\mathbf{1.}$] Gauss's flux theorem (1813), which states that the flux of the electric field across a closed surface is proportional\footnote{The proportionality factor is $1/\varepsilon$.} to the total charge contained \emph{within} that closed surface. The local equivalent is that the divergence of the electric field is proportional to the charge density.

\item[$\mathbf{2.}$] 
Gauss's flux theorem for magnetism\footnote{The law already appeared in the work of Lagrange, in 1773.}, which states that the flux of the magnetic field around a closed surface always vanishes, which is equivalent to the non-existence of magnetic monopoles, i.e., in analogy with electric phenomena, of magnetic charges\footnote{Here are some lines from Wikipedia's web page on Gauss's flux theorem for magnetism:
``This idea, of the nonexistence of magnetic monopoles, originated in 1269 by Petrus Peregrinus de Maricourt. His work heavily influenced William Gilbert, whose 1600 work \emph{De Magnete} spread the idea further. In the early 1800s Michael Faraday reintroduced this law, and it subsequently made it into James Clerk Maxwell's electromagnetic field equations."}. The local equivalent is that the divergence of the magnetic field is always zero\footnote{In French teaching, this local equation is sometimes called Maxwell-Thomson equation, because of later work carried out by Thomson: in a problem, Thomson had to add a term to the vector potential he had derived for the divergence of the magnetic field to be zero, because he wanted Maxwell equations to hold. Fitzgerald gave physical interpretations for the addition of such a term. See the book \emph{Cambridge and the rise of mathematical physics}, which cites Buchwald's writings, \url{http://www.hss.caltech.edu/content/jed-z-buchwald}, \url{http://www.persee.fr/doc/rhs_0151-4105_1998_num_51_1_1308}.}.

\item[$\mathbf{3.}$] Faraday's induction law, already mentioned above, which states, roughly speaking, that the time variation of the magnetic flux across a surface induces the surrounding electric charges to move around the magnetic material, about the direction of the average magnetic field generated by the material. The local equivalent is that the time derivative of a magnetic field corresponds to the curl of the electric field. The problems pointed out by Einstein\footnote{For the complete work of Einstein, translated in English, one may go to Princeton's dedicated website, \url{http://einsteinpapers.press.princeton.edu}.} in 1905, regarding the change of point of view\footnote{With Galilean frame changes, which consider time is absolute i.e. does not depend on the observer.} in such an induction experiment \cite{Einstein1905b}, lead him to formulate special relativity, i.e., following earlier work by Poincaré, to realise, with \emph{Gedankenexperiments}, that time is not absolute but depends on the observer, so that Galilean transformations must be substituted by Lorentz transformations\footnote{These transformations had already been introduced by Lorentz, as transformations that led Maxwell's equations invariant, and then Poincaré before Einstein already gave some physical interpretations for the different times involved in the transformations. The key idea of Einstein's work, is to realize that what had to be modified were not Maxwell's equations, but  Newton's dynamical law, up to then sacralized, and this funtamentally needed a replacement of the Galilean transformations relating inertial frames by these Lorentz transformations that leave Maxwell's equations unchanged. To give a rough picture, this idea came naturally with a previous first key idea, namely to abolish the notion of ether, so that the problem could not come from Maxwell's equations since we had \emph{nothing} to know about a hypothetical medium in which electromagnetic waves would propagate, and whose misunderstanding would give us wrong equations.}.

\item[$\mathbf{4.}$] Ampère's circulation theorem, which states that the flux of some current density across a surface determines, through the magnetic permeability\footnote{On can define the magnetic induction $\boldsymbol{H}=\boldsymbol{B}/\mu$, where the magnetic permeability $\mu$ is a scalar in the case of (i) vacuum, $\mu=\mu_0$, or (ii) a linear, homogeneous and isotropic material, as for the relation between the displacement field and the electric field.}, the circulation of the magnetic field along the edge of this surface.  With the additional displacement current introduced by Maxwell, the local version of this theorem becomes the Maxwell-Ampère equation, which states that the magnetic field `rotates' about the direction specified by the vectorial sum of currents and displacement currents (i.e. temporal variations of the electric field).

\end{itemize}

\paragraph{{$\bullet$ Quantum electrodynamics} } \hspace{3cm}

\vspace{0.2cm}

\noindent
With the advent of quantum mechanics, (non-relativistic) quantum electrodynamics developed, with early contributions by Pauli. This enabled to account for emission and absorption spectra of several atoms, and was one of the great successes of quantum mechanics. Also, the theory was formulated as a consequence of Weyl's gauge principle applied to the global phase invariance of quantum mechanics.

To overcome the problems of the Klein-Gordon equation, straightforwardly derivable with the help of the correspondence principle, Dirac managed to write down a relativistic quantum-mechanical equation for fermions, in which one could implement minimal coupling, and this equation successfully addressed the deviations from experiments that were obtained by deriving the spectrum of the hydrogen atom from the Schrödinger equation.

On pre-QFT relativistic quantum electrodymanics (QED), one may consult Bjorken and Drell's book.
For `the' history of second quantization and relativistic quantum electrodynamics, one may refer to Schwinger's book, \emph{QED and the men who made it}. Landau and Lifshitz's book may also be consulted. 

\iffalse
\paragraph{{$\bullet$ The Lorentz force}} \hspace{3cm}

\vspace{0.2cm}

The modern form of the Lorentz force is the result of a appeared after Maxwell's equations
\fi

\iffalse
\paragraph{{$\bullet$ Remarks on the electromagnetic force} } \hspace{3cm}

The 
\fi

\vspace{0.2cm}

\subsubsection{The weak force, a short-range force much weaker than the electromagnetic force}

\paragraph{$\bullet$ Radioactivity, beta decay, and Fermi's theory of beta decay}
\hspace{3cm}

\vspace{0.2cm}

\noindent
In 1896, Becquerel, who was working on phosphorescence, noticed that some of his uranium samples emitted radiations, even the non-phosphorescent ones\footnote{He wondered whether the radiations emitted by phosphorescent bodies previously exposed to light were X-rays, which had been discovered (`by chance') one year earlier, by R{\"o}ntgen. To Becquerel, the light emitted by phosphorescent bodies was similar to that emitted by cathodic tubes previously exposed to X-rays, and might thus have been of the same nature. The experiment was the following: Becquerel put his samples on a photographic plate, protected by paper, and wanted to see whether the samples would imprint the plate. Some of the samples had been exposed to light previously, others not. The only samples that imprinted the plaque were the uranium samples, regardless of their previous exposition to light. The plate, called ``étiquette bleue",  was a successful dry photographic plate commercialized by the Lumière borthers. Here are many details on the experiment and the context, in French, by Basdevant, \url{https://www.bibnum.education.fr/sites/default/files/BECQUEREL_SUR_LES_RADIATIONS_EMISES_BASDEVANT.pdf}.}. Becquerel understood these radiations were not the X-rays he was expecting, and he called them `uranic radiations'. The term `radioactivity' was introduced by Marie Curie in her scientific reports\footnote{Here is a very complete website of the American Institute of Physics on Marie Curie, \url{http://history.aip.org/exhibits/curie/}.}. She and Pierre Curie discovered two other radioactive elements, the radium and the polonium. In 1899, Rutherford distinguished three types of radioactive decays, namely alpha, beta and gamma, according to their penetrating power. In 1901 \cite{Rutherford1902I,Rutherford1902II}, Rutherford and his pupil Soddy understood, analyzing thorium decays, that decays go with a transmutation of the radioactive element, and in 1903, the well-known exponential-decay law for radioactive samples was published\footnote{For the original paper, see \url{http://www.bibnum.education.fr/sites/default/files/rutherford-texte-partie1.pdf}. Some short notes on the 2 men by the American Physical Society, \url{https://www.aps.org/programs/outreach/history/historicsites/rutherfordsoddy.cfm}, extended historical notes on the context of the discovery, by Bonolis, \url{http://www.mediatheque.lindau-nobel.org/research-profile/laureate-soddy\#page=1}, and a detailed review of the paper by Radyanyi, \url{https://www.bibnum.education.fr/sites/default/files/Rutherford-analyse-english.pdf}.} \cite{Rutherford1903}. 
Evidence for a loss of energy during the beta decay accumulated between 1911 and 1927, and in 1934 \cite{Fermi1934Ger,Fermi1934Eng}, Fermi published his famous theory of beta decay (now called beta-minus decay), for the change of a neutron into a proton by emission of an electron (the beta-minus radiation), in which a hypothetical particle, which now corresponds to the antineutrino, is emitted\footnote{On the intertwined history of neutrinos and neutrons, with Pauli's 1930 contribution and Chadwick's 1932 discovery of the neutron, see the following historical paper, \url{http://physicstoday.scitation.org/doi/pdf/10.1063/1.2995181}.}; in Fermi's picture, the interaction was a contact interaction with no range, but this interaction is now described, as any weak interaction, by a non-contact interaction with short range. This four-fermion interaction was the first version of a weak interaction.

\paragraph{{$\bullet$ Parity violation, chiral vector-minus-axial ($V-A$) theory and $\mathrm{CP}$ violation} } \hspace{3cm}

\vspace{0.2cm}

\subparagraph{$\star$ \emph{Parity violation and the chiral $V-A$ theory with its charged weak bosons $W^{\pm}$ }} \hspace{3cm}

\vspace{0.2cm}

\noindent
In 1956 Lee and Yang \cite{Lee1956} suggested that Fermi's weak interaction might break parity\footnote{Parity is the transformation that flips all spatial coordinates. It transforms a vector into its opposite, and is in this sense viewed as a `mirror' transformation. The idea of parity-symmetric physical processes was formalized by Wigner in 1927, see \url{http://www.digizeitschriften.de/download/PPN252457811_1927/PPN252457811_1927___log30.pdf} for the original German paper, and \url{http://link.springer.com/chapter/10.1007\%2F978-3-662-02781-3_7} for the collected works of Wigner. Parity is a symmetry of classical gravitational laws, electromagnetic laws, was a symmetry of the laws involving the nuclear and strong forces at the time, and is still.}, and the experimental proof was given one year later, by Wu et al.  \cite{Wu1957}.

Together with the development of perturbative renormalization theory\footnote{For some general ideas on perturbative renormalization theory, see Appendix \ref{app:renormalization}.}, this parity violation lead, the same year, Marshak and Sudarshan\footnote{Some short notes by Sudarshan on the history of the discovery, at \url{http://quest.ph.utexas.edu/Reviews/VA/VA.pdf}.} \cite{Sudarshan1957,Sudarshan1958}, and then Feynman and Gell-Mann \cite{Feynman1958}, to introduce chirality\footnote{For massless particles, chirality and helicity, the projection of the spin on the momentum, coincide, and `replace' the spin, since the latter is not a good quantum number for such free particles (their little group is ISO(2) and not SO(3)). For massive particles, chirality is not fixed, so that one often uses the Dirac representation, which is not irreducible, and combines left and right chiralities, which are interchanged in some weak interactions.}\textsuperscript{,}\footnote{Only left chirality was introduced, the right chirality being linked to the neutral $Z^0$ boson, see below.} and develop the chiral $V-A$ theory, which introduces two massive and charged mediators for the weak force, namely $W^{\pm}$. This is the first unified theory of what is now called the weak interaction.

See the following notes by Lessov on the history of the weak interaction from Fermi's beta decay to the chiral $V-A$ theory, \url{https://arxiv.org/pdf/0911.0058.pdf}.

\subparagraph{$\star$ \emph{ $\mathrm{CP}$ violation}} \hspace{3cm}

\vspace{0.2cm}

\noindent
Landau \cite{Landau1957} pointed out that CP, the composition of parity, P, and charge conjugation (i.e. substituting particles by their antiparticles), C, was preserved by the $V-A$ theory of the weak interaction, and this seemed to solve the problem raised by the parity violation: the `good transformation', i.e. the `symmetry' of nature, was CP and not P\footnote{See this article of the CERN Courier on the ``CP violation's early days'', at \url{http://cerncourier.com/cws/article/cern/57856}. In the paper by Lee and Yang \cite{Lee1956} questioning the conservation of parity in weak interactions, there is a remark about the CP symmetry.}. In 1964, however, Christenson, Cronin, Fitch and Turlay \cite{Christenson1964}, showed experimentally that the CP symmetry was broken in the kaon decay (these kaons rarely decay into two pions).

\paragraph{{$\bullet$ Glashow's electroweak unification and its neutral weak boson $Z^0$}} \hspace{3cm}

\vspace{0.2cm}

\noindent
One needs a consistent framework to describe situations were the weak and the electromagnetic forces are involved in interactions involving comparable energy exchanges.

Schwinger suggested the idea of the electroweak unification through a third weak (massive) \emph{neutral} boson, but he gave the task of developing this idea to his PhD student, Glashow, and a paper presenting such a unification through a $Z^0$ gauge boson was eventually published in 1961\footnote{For his 1979 Nobel prize lecture, see \url{http://www.nobelprize.org/nobel_prizes/physics/laureates/1979/glashow-lecture.pdf}, and here follow interesting historical details compiled by Brooks, \url{http://www.quantum-field-theory.net/electroweak-unification/}.} \cite{Glashow1961}.

\iffalse
The question of the huge magnitude difference between 
\fi

\vspace{0.2cm}

\paragraph{{$\bullet$ Spontaneous symmetry breaking, Higgs mechanism and electroweak Standard Model} } \label{par:electroweak} \hspace{3cm}

\vspace{0.2cm}

\noindent
For historical (and technical) details on the concept of spontaneous symmetry breaking, see the following presentation by Jona-Lasinio,  \url{http://www.gakushuin.ac.jp/univ/sci/phys/notice/JonaLasinioNov09.pdf}. The historical steps by which this phenomenon acquired its maximal recognition and was named that way begin with Heisenberg's 1959 suggestion that the concept might be relevant in particle physics, and then Nambu's \cite{Nambu1960} discovery of such a mechanism in BCS\footnote{After Bardeen, Cooper and Schrieffer.} superconductivity. With Jona-Lasinio \cite{Nambu1961}, Nambu brought the concept to particle physics, after which Goldstone \cite{Goldstone1961} developed it, which ended up in the formulation of Goldstone's theorem, with Salam and Weinberg  \cite{Goldstone1962}. The theorem is the following. Consider a system in an excited state, which has a degenerate energy groundstate. To minimize its energy, the system goes to one of these groundstates. Goldstone's theorem states that this is necessarily accompanied by the apparition of an additional scalar mode in the spectrum of the system, for each generator of the broken symmetry, called a Goldstone boson. The theorem states that such a mode should be \emph{massless}.

Anderson first proposed a way to give a mass to Goldstone modes in a non-relativistic framework, and figured out some consequences in particle physics \cite{Anderson1963}. One year later, the relativistic version, needed for particle physics, was developed by three independent groups, Brout-Englert \cite{Englert1964}, Higgs  \cite{Higgs1964}, and Guralnik-Hagen-Kibble  \cite{Guralnik1964}. In 1967, Weinberg \cite{Weinberg1967} and Salam\footnote{In the following notes, Salam's contribution is discussed: \url{https://arxiv.org/pdf/1109.1972.pdf}.} \cite{Salam1959,Salam1968} included the Higgs mechanism into Glashow's electroweak unification. The following historical and technical notes by Kibble are very interesting, \url{https://arxiv.org/pdf/1502.06276.pdf}. The unified picture can formally be very briefly summed up as follows:
\begin{enumerate}
\item At high energy, one has the so-called electroweak interactions between particles, which are described by a  non-Abelian $\text{U}_{Y}$(1) $\times$ SU(2) Yang-Mills gauge theory\footnote{In this notation, the $\text{U}_Y$(1) and SU(2) groups are, as usually said, the \emph{abstract} versions of the matrix groups U(1) and SU(2), and  $\times$  denotes the standard `cartesian product', or, rather, `group product', since the denomination `cartesian' is associated to vector spaces, while there is a huge branch of group theory which has nothing to do with vector spaces. If, by $\text{U}_{Y}$(1) and SU(2), one wants to refer to the representations of the abstract groups in terms of linear automorphisms, one would denote the electroweak symmetry group as $\text{U}_{Y}$(1) $\otimes$ SU(2).}\textsuperscript{,}\footnote{Note that $\text{U}$(1) $\times$ SU(2) contains, but is not equal to U(2), see Appendix \ref{app:unitary_group}. In mathematical group-theory terminology, we say that U($N$) is, not the direct, but the semidirect product of SU($N$) by U($1$).}. `The' generator of the gauge group $\text{U}_{Y}$(1) is called \emph{weak hypercharge}, denoted $\hat{Y}$, and is associated to the $B$ gauge boson. The gauge group SU(2) has three generators $\hat{I}_i$, $i=1,2,3$, and is associated to the weak-isospin d.o.f., which belongs to a 2D Hilbert space; each generator is associated to a gauge boson  $W^i$. All gauge fields are real-valued 4-vector fields.
\item At low energy, the electroweak internal gauge symmetry $\text{U}_{Y}$(1) $\times$ SU(2) is spontaneously broken: $B$ and $W^3$ couple through the Weinberg, or weak angle $\theta_W$, accounting for the photon (field) $\gamma$ and the neutral weak boson $Z^0$, as
\begin{equation}
\left( \begin{matrix}
\gamma \\ Z^0
\end{matrix} \right)
=
\begin{bmatrix}
\cos \theta_W & \sin \theta_W \\
- \sin \theta_W & \cos \theta_W
\end{bmatrix}  
\left( \begin{matrix}
B \\ W^3
\end{matrix} \right) \, . 
\end{equation}
The photon field $\gamma$ is the spin-1 bosonic field also known as electromagnetic potential, i.e. the gauge field associated to the electromagnetic gauge group $\text{U}_{\text{em}}(1)$, whose generator is the electric charge operator, given by $\hat{Q}=\hat{I}_3+ \hat{Y}/2$. The remaining electroweak bosons $W^1$ and $W^2$ mix to account for the charged weak bosons, $W^{\pm}$:
\begin{equation}
W^{\pm} = \frac{W^1 \mp W^2}{\sqrt{2}} \, .
\end{equation}
\end{enumerate}
The masses of the weak bosons, provided by the Higgs mechanism, are related by
\begin{equation}
M_Z = \frac{M_W}{\cos \theta_W} \, .
\end{equation}

The theory was validated in two steps: (i) in 1973, the neutral currents caused by the exchange of $Z^0$ bosons were discovered in the Gargamelle bubble chamber, so that the Glashow-Weinberg-Salam Standard model of electroweak interaction became widely accepted\footnote{Glashow, Weinberg and Salam won the Nobel prize in 1979.}, and (ii) in 1983, the three weak bosons were detected at CERN, by an experiment led on the LEP.

\iffalse
Some notes on the specificities of the weak interaction, found on Wikipedia:
``It is the only interaction capable of changing the flavour of quarks (i.e., of changing one type of quark into another).
It is the only interaction that violates parity. It is also the only one that violates CP.
It is mediated by force carrier particles that have significant masses, an unusual feature which is explained in the Standard Model by the Higgs mechanism.".
\fi

\subsubsection{The strong force, a short-range force much stronger than the electromagnetic force}

\iffalse
Most of the mass of a common proton or neutron is the result of the strong force field energy; the individual quarks provide only about 1\% of the mass of a proton.
\fi

\paragraph{$\bullet$ Atomic model, atomic nucleus and the strong nuclear force}\hspace{3cm}

\vspace{0.2cm}

\noindent
The concept of atom, as well as the word, is usually traced back to Democritus, philosopher, 5th century before Christ, who suggested that matter is composed of small indivisible particles. A century later, Aristotle, philosopher, stated instead that any piece of matter is a continuum that can be divided ad infinitum. Because of the huge influence of Aristotle in occidental thinking, this continuum vision prevailed up to the 19th century.

In the second half of the 18th century, the chemist Lavoisier stated the principle of matter conservation in chemical reactions, and in 1808, Dalton suggested a chemistry-grounded atomic model of matter.

In 1897, Thomson concluded from its experiments on cathod rays that negatively charged particles, later called `electrons', can be detached from atoms. He suggested in 1904 the plum pudding model of atom \cite{Thomson1904}, in which the paste is positively charged, and the dried grapes inside are the electrons. The model ``purposely"\footnote{This is Bohr's word \cite{Bohr1913I}.} avoids the problem of energy loss through Larmor radiation \cite{Larmor1897}, by suggesting the atom should contain a number of electrons one thousand times greater than the numerical value, in grams per mole, of the mass of the atom, which he realized to be impossible in 1906.

In 1911 \cite{Rutherford1911}, Rutherford deduced from its 1909 experiments on the deviation of alpha particles by a one-atom-thick layer of gold\footnote{See the following short videoclip, \url{https://youtu.be/VLU4dntonhE}.}, that the positive charges in the gold foil are concentrated in volumes much smaller than that of the atom. He thus suggested what is now called Rutherford's atomic model: the atom is made of a positively charged tiny nucleus, around which the electrons move. In his 1911 paper, Rutherford mentions the necessity to determine experimentally whether the center is actually positive or negative\footnote{Rutherford mentions Nagaoka's theoretically-grounded planetary or Saturnian model, suggested in 1904, in part by the formal correspondence between the Coulomb and the gravitational forces. Nagaoka tried to fix, in his model, the well-known problem of radiating electrons. In 1908, however, he eventually abandonned his model after (i) the failure in reconciling the model with radioactivity and atomic emission rays and (ii) Thomson pointed out unstable oscillations of the electrons orthogonally to the plane of the orbit (see \url{http://link.springer.com/chapter/10.1007\%2F978-3-540-70626-7_10}). Hope for non-radiating orbits was brought by Ehrenfest's 1910 paper, but this idea was abonned during the rise of quantum mechanics.}.

In 1913 \cite{Bohr1913I, Bohr1913II},  Bohr proposed a model of the atom grounded on Planck's quanta, first suggested in 1900 to account for the black body radiation\footnote{See the following two historical reviews on Planck's seminal work in quantum physics, \url{https://arxiv.org/pdf/physics/0402064.pdf} and \url{http://citeseerx.ist.psu.edu/viewdoc/download?doi=10.1.1.613.4262&rep=rep1&type=pdf}.}, and then used and conceptually developed by Einstein to explain the photoelectric effect, in 1905\footnote{For an English translation of the original paper, go to \url{http://www.esfm2005.ipn.mx/ESFM_Images/paper1.pdf}.}. Bohr's suggestion is, in a semi-classical picture, that only some specific orbits are allowed for the electron in motion around the nucleus, and thus the electron's energy belongs to a discrete set of possible values, i.e., must be `quantized'. This model avoids the radiation problem, and accounts for several atomic emission spectra. These historical landmarks sign the rise of quantum physics.

In 1932, Chadwick discovered the neutron. Within months after this discovery, Heisenberg and Ivanenko proposed neutron-proton models for the atomic nucleus, which included the idea of the (nuclear) isospin, see Appendix \ref{app:isospin}.

\iffalse
isotopic spin conserved in nucleon nucleon interaction

Within months after the discovery of the neutron, Werner Heisenberg[8][9][10] and Dmitri Ivanenko[11] had proposed proton–neutron models for the nucleus.[12] Heisenberg's theory for protons and neutrons in the nucleus was a "major step toward understanding the nucleus as a quantum mechanical system."[13] Heisenberg introduced the first theory of nuclear exchange forces that bind the nucleons. He considered protons and neutrons to be different quantum states of the same particle, i.e., nucleons distinguished by the value of their nuclear isospin quantum numbers. The liquid drop model treated the nucleus as a drop of incompressible nuclear fluid, with nucleons behaving like molecules in a liquid. The model was first proposed by George Gamow and then developed by Niels Bohr, Werner Heisenberg and Carl Friedrich von Weizsäcker. This crude model did not explain all the properties of the nucleus, but it did explain the spherical shape of most nuclei. The model also gave good predictions for the nuclear binding energy of nuclei.
\fi

In 1935, Yukawa\footnote{The original paper was reprinted in 1955 \cite{Yukawa1935}, see \url{https://academic.oup.com/ptps/article-lookup/doi/10.1143/PTPS.1.1}.} \cite{Yukawa1935} suggested an analytical form for the nuclear two-body potential, now so-called Yukawa potential, which is a Coulomb potential screened by a decreasing exponential, which makes the range of the nuclear force `infinitely' smaller than that of the electromagnetic or gravitational forces\footnote{If the \emph{range} is defined through the characteristic scale of variation of the exponential, then the electromagnetic and gravitational forces have infinite range.}. The nuclear force was conceived as transmitted by particles called mesons, which were discovered in 1947. In essence, the modern picture of the nuclear force is the same, but this nuclear force is viewed as a residual force of the underlying so-called strong force between quarks, elementary particles which manifest themselves only\footnote{We shall come back to this fact further down, in Paragraph \ref{par:chromo}.} through so-called hadronic bound states; the particles sensitive to the nuclear, and, more generally, the underlying strong force, are called hadrons. Quarks are viewed as the elementary hadrons, and these hadrons divide into two families of bound states: baryons, made of one or several triplets of quarks, such as protons and neutrons, and mesons, made of one or several quark-antiquark pairs.

\paragraph{$\bullet$ Strong (or color) force}\hspace{3cm}

\vspace{0.2cm}

\subparagraph{$\star$ \emph{Six quarks to account for the variety of hadrons}}   
\label{par:quarks}
\hspace{3cm}

\vspace{0.2cm}

\noindent
With the development of particle accelerators and colliders after World War II, many particles were discovered. In 1953, Nakano and Nishijima \cite{Nakano1953} were able to extract, out of the experimental data, a formula relating the charge, the nuclear isospin, and the baryonic number of particles, in which one had to add a new quantum number, which was then conceptually developed up to 1955 by Nishijima \cite{Nishijima1955}, and eventually called \emph{strangeness}. In 1956, Gell-Mann\footnote{An extensive selection of Gell-Mann's papers is available at \url{http://longstreet.typepad.com/thesciencebookstore/2013/06/important-papers-of-murray-gell-mann-as-they-appear-in-murray-gell-mann-selected-papers-and-published-by-world-scie.html}.} also suggested the same formula, independently, and the formula is currently known as the Gell-Mann--Nishijima formula. The same year, Sakata \cite{Sakata1956} suggested the so-called triplet model of hadrons, in which all known hadrons were made out of a combination of particles from the triplet $(p,n,\Lambda)$.

The Eightfold Way report, written in 1961 by Gell-Mann \cite{GellMann1961}, is an attempt to describe the strong interaction to which were sensitive the eight spin-1/2 baryons that were known when he wrote the paper. In Gell Mann's words, ``the most attractive feature of the scheme is that it permits the decription of eight vector mesons by a unified theory of the Yang-Mills type (with mass term)". In the abstract of his paper publicising the Eightfold Way and formally presenting its mathematics \cite{GellMann1962}, Gell-Mann cites Ne'eman, which developed essentially the same 8-vector-mesons picture, independently \cite{Neeman1961}. The baryons were organized in an octuplet according to charge and strangeness. The model was an alternative to Sakata's triplet.

\iffalse
To better understand the origin of this symmetry, Gell-Mann proposed the existence of up, down and strange quarks which would belong to the fundamental representation of the SU(3) flavor symmetry.
\fi

One year later, the spin-3/2  baryons were viewed as organized in a decuplet, but this required the existence of a still unobserved baryon, which Gell-Mann called $\Omega^-$. In 1964, the $\Omega^-$ was observed \cite{Barnes1964}, and Gell-Mann showed that assuming the hadrons were made of three \emph{flavors} of \emph{quarks}, having fractional charges, would greatly simplify the picture\footnote{The three flavors of quarks were \emph{up}, \emph{down}, and \emph{strange}. The word `flavor' came later. In his paper \cite{GellMann1964}, Gell-Mann explicitly states that this triplet is formally highly reminiscent from Sakata's triplet, and that the quark-picture flavored Lagrangian can be built by analogy with Sakata's-triplet Lagrangian.} \cite{GellMann1964}; this was also realised independently by Zweig\footnote{See his two relevant CERN reports, at \url{http://cds.cern.ch/record/352337/files/CERN-TH-401.pdf}, and \url{http://library.lanl.gov/cgi-bin/getfile?00323548.pdf}.}\textsuperscript{,}\footnote{Zweig orally presented, in a 1980 conference at Caltech, a highly detailed historical picture of the development of the quark model prior to 1964; his conference notes are available at \url{http://authors.library.caltech.edu/18969/1/Origins_of_the_Quark_Model_Final_Zweig\%5B1\%5D.pdf}.}.  The power of the quark picture can be illustrated by the following two equalities, as Gell-Mann and Zweig found out: 
\begin{enumerate}
\item An equality for lowest-flavor-order (i.e. spin-0) mesons:
\begin{equation} \label{eq:meson_group_theory}
\mathbf{3}_F \otimes \bar{\mathbf{3}}_F = \mathbf{8}_F \oplus \mathbf{1}_F \, ,
\end{equation}
which states that hadronic bound states made of quark-antiquark pairs\footnote{Quarks belong to the three-dimensional flavor representation Hilbert space noted $\mathbf{3}_F$, and the antiquarks to its adjoint, noted $\bar{\mathbf{3}}_F$.}, i.e. spin-0 mesons\footnote{Mesons made of more than one quark-antiquark pair have a higher-dimensional spin.}, belong to a spin-0 meson representation Hilbert space of dimension $8 + 1$, i.e. there are $8+1$ basis states, which are the $8+1$ spin-0 mesons, which had already been detected in 1964. The subscript $F$ is for `flavor'.
\item An equality for lowest-flavor-order (i.e. spin-1/2 and spin-3/2) baryons:
\begin{equation} \label{eq:baryon_group_theory}
\mathbf{3}_F \otimes \mathbf{3}_F \otimes \mathbf{3}_F = \mathbf{10}_F \oplus \mathbf{8}_F \oplus \mathbf{8}_F \oplus \mathbf{1}_F \, ,
\end{equation} 
which states that the lowest-order baryons, made of three quarks (each belonging to a flavor representation, noted $\mathbf{3}_F$), organize into a singlet, two octets and a decuplet. The link with the spin-1/2 baryonic octet and the spin-3/2 baryonic decuplet needs however a more in-depth analysis. We refer the reader to first Wikipedia's article on the quark model and then, e.g., Kosmann-Schwarzbach's 2010 book on \emph{Groups and symmetries, from finite groups to Lie groups} \cite{KosmannSchwarzbach2010}.
\end{enumerate}
The modifications brought to Sakata's model \cite{Maki1964} managed to maintain it as a concurrent of the quark model\footnote{Note that Sakata used the simplifying quark picture in several papers \cite{Sakita1964}.}, although it was largely abandoned after the experimental evidence, in 1974 at the SLAC, for the $J/\Psi$ meson, viewed as a new, so-called \emph{charmed}, quark-antiquark pair.

\iffalse
According to Zweig, the discovery of the phi mseon (strange antistrange) is very important.
\fi

In 1973, Kobayashi and Maskawa \cite{Kobayashi1973} showed that the CP violation, by some weak interactions, made inconsistent Maki and Ohnuki's \cite{Maki1964} quartet baryon scheme, which had been introduced to overcome the difficulty of the ``strangeness changing neutral current" \cite{Kobayashi1973} in Sakata's triplet scheme. This can be translated as a need for a third generation of baryons. Note that the quark picture is neither adopted nor mentioned in their paper, a possible sign that this picture was not fully established yet\footnote{I wonder whether this paper is one of the last milestone papers on hadronic physics which does not adopt the quark picture.}.

Indirect experimental evidence for the bottom and top quark was found in 1977 and 1995, respectively, at Fermilab.

\iffalse
2000: tau neutrino

2012: Higgs boson
\fi

\subparagraph{$\star$ \emph{Eight vector bosons mediating the interactions between quarks: the gluons}}\hspace{3cm}

\vspace{0.2cm}

\noindent
When Gell-Mann published his quark picture to simplify the classification of hadrons, hadronic interactions were still viewed as mediated by mesons, but the quark picture soon gave hints about the pitflaws of such a description: interpreting, through the quark picture, the experimental data on the spin-3/2 baryon $\Delta^{++}$, led, in 1965, Greenberg \cite{Greenberg1964} and, independently, Han and Nambu \cite{Han1965}, to suggest that quarks may posses and additional degree of freedom, later called \emph{color}, that should be associated to their interaction through a \emph{color charge}, as the electromagnetic interaction is associated to the electric charge. 

The color degree of freedom belongs to a three-dimensional Hilbert space, and each vector of an arbitrary reference basis is associated a given color. Now, the force carriers are not directly represented by the generators of naive transformations that would preserve the color norm and which would belong to the symmetry group U(3) of the color representation space, that we note $\mathbf{3}_C$, where $C$ is for `color'. Instead, and this is due to the flavor quantum number, the force carriers are represented by \emph{color-anticolor pairs}, in a quantum-mechanical way, i.e. any (normalized) linear combination of such pairs is allowed, provided we built a basis of an appropriated symmetry group with these states. This is the first basic complexity of the so-called \emph{(quantum) chromodynamics}, i.e. the study of the strong interaction through the color-charge picture, which is currently the standard framework. The force carriers are called \emph{gluons}. As in the flavor quark picture for hadrons, see  Eqs. (\ref{eq:meson_group_theory}) and (\ref{eq:baryon_group_theory}), the color picture for gluons holds in the following equality:
\begin{equation}
\mathbf{3}_C \otimes \bar{\mathbf{3}}_C = \mathbf{8}_C \oplus \mathbf{1}_C \, .
\end{equation}
Now, from experimental data, it was deduced that only the octet, namely $\mathbf{8}_C$, was involved in strong interactions, i.e., in modern words, that the strong interaction should be described by the gauge group SU(3) rather than U(3)\footnote{Note that the weak force cannot either be described by a group of transformations preserving the norm of a single  \emph{naive} internal d.o.f.. The complexity of the weak force is described through the electroweak picture, see Paragraph \ref{par:electroweak}.}. The ``Advantages of the color octet gluon picture", rather than the color singlet gluon picture previously considered, were described by Fritzsch, Gell-Mann, and Leutwyler in 1973 \cite{Fritzsch1973}. These findings are related to the suggestion of considering color as independent from flavor, while Han and Nambu's early model \cite{Han1965} intertwined both quantum numbers.

\iffalse
A similar mysterious situation was with the Δ++ baryon; in the quark model, it is composed of three up quarks with parallel spins. In 1964–65, and Greenberg[12] and Han–Nambu[13] independently resolved the problem by proposing that quarks possess an additional SU(3) gauge degree of freedom, later called color charge. Han and Nambu noted that quarks might interact via an octet of vector gauge bosons: the gluons.
\fi

\iffalse
Technically, QCD is a gauge theory with SU(3) gauge symmetry. Quarks are introduced as spinors in Nf flavors, each in the fundamental representation (triplet, denoted 3) of the color gauge group, SU(3). The gluons are vectors in the adjoint representation (octets, denoted 8) of color SU(3). For a general gauge group, the number of force-carriers (like photons or gluons) is always equal to the dimension of the adjoint representation. For the simple case of SU(N), the dimension of this representation is N2 − 1.
In terms of group theory, the assertion that there are no color singlet gluons is simply the statement that quantum chromodynamics has an SU(3) rather than a U(3) symmetry. There is no known a priori reason for one group to be preferred over the other, but as discussed above, the experimental evidence supports SU(3).[7] The U(1) group for electromagnetic field combines with a slightly more complicated group known as SU(2) – S stands for "special" – which means the corresponding matrices have determinant 1 in addition to being unitary.
\fi

\iffalse
\cite{Han1965}
\fi

\paragraph{$\bullet$ Chromodynamics and its difficulties}
\label{par:chromo}
\hspace{3cm}

\vspace{0.2cm}

\noindent
The aforementioned group-theoretical arguments that managed to account for many structural features of the strong interaction were developed, roughly speaking, while standard quantum-field-theory perturbative computations, i.e. perturbative renormalization methods\footnote{See Appendix \ref{app:renormalization} for a general picture of perturbative renormalization theory.}, were essentially failing, due to the particularly high magnitude of this interaction (which is linked to the `strong force' denomination). In 1973,  Wilczek and Gross \cite{Gross1973} and, independently, David and Politzer \cite{Politzer1973}, theoretically described the so-called asymptotic freedom in quantum chromodynamics (QCD): at sufficiently small length scales, i.e. sufficiently high energies, quarks interact weakly, and eventually not at all in the limit of infinite energy, the so-called asymptotic freedom of quarks, which, in essence, makes perturbative computations work at such high-enough energies. This is in strong contrast with quantum electrodynamics (QED), where one has the opposite effect: the lower the typical energy of the interaction, the better perturbative computations work\footnote{This is physically interpreted as a so-called \emph{screening} of the electric charge by electron-positron pairs of the vacuum fluctuations surrounding the electric particle. While this is also the case for the color charge of quarks, the fact that gluons also carry a color charge, while photons do not, and they way they carry it, results in an opposite effect, called \emph{antiscreening}, which accounts for quark confinement.}, while at high energy, these perturbative computations fail, which was quickly pointed out by Landau in the early days of QED, in the 50's.

Now, while understanding such an asymptotic freedom enables to deal with high energies in QCD, the low-energy behaviour of the theory still requires to go beyond perturbative methods. 
\iffalse
At this point, a very brief technical recap is in order. Generally speaking, the aim of a physical theory is to predict the state of a physical system in the future. In a quantum-mechanical framework, this demands to compute the transition probability amplitude to reach a state $\ket{f}$ at time $t$ starting from an initial state $\ket{i}$ at $t=0$. This probability amplitude is given by $\langle f | \Psi_t\rangle = \langle f | e{-i\hat{H}t} | i \rangle$, where $\hat{H}$ is the Hamiltonian of the system. The Hamiltonian is the sum of a free part and of an interaction term, $\hat{H}=\hat{H}_0 + \hat{V}$. By definition, the notion of interaction is a priori `orthogonal' to that of symmetry, and $\hat{V}$ does a priori not commute commute with $\hat{H}_0$ definition, a physical interaction  results 
\fi
Going beyond perturbative methods can be done with so-called lattice gauge theories (LGTs), which construct QFTs on a spacetime lattice. As we mentioned in the introduction, the seminal paper on LGTs is that of Wilson, in 1974 \cite{Wilson74}. In this paper, Wilson shows that such a lattice construction of the theory, which avoids the problems of perturbation theory\footnote{Which does not mean LGTs are free of other problems. Indeed, LGTs have other problems from the start, such as the loss of unitarity.}, enables to predict the confinement of quarks at low energy, a phenomenon for which the (hadron-)physics community was expecting a theoretical explanation since there was -- and there is still -- no evidence for free quarks in experiments\footnote{As we have seen in Paragraph \ref{par:quarks}, quarks were introduced by Gell-Mann and Zweig in 1964, as mathematical objects that greatly simplified the understanding of the hadronic classification and of the gauge-theory structure of the strong force. Two opposed visions emerged after the advent of the quark picture. Since there was no evidence for free quarks, Gell-Mann considered quarks were not real particles, but Feynman still thought the opposite, in the sense that quarks can be associated a classical trajectory through the path-integral framework; Feynman would refer to such `real quarks' as \emph{partons}, i.e. `parts of hadrons'. Such an opposition is still up to date.}. In a few years, the lattice picture quickly yielded many results, and LGTs are now standard.

\iffalse

\subsubsection{GUT and problems}

Alain cones

GUT
 \footnote{One of the current Millenium problems is to find a so-called Grand Unified Theory (GUT) of fundamental interactions. The hope is to manage to merge the strong force with the electroweak force at high energy. The electroweak and strong forces would result from the symmetry breaking $\text{U}(3) \rightarrow \text{U}(2) \times \text{SU}(3)$ through some still unknown mechanism, as the electromagnetic and weak forces result from the symmetry breaking $\text{U}(2) \rightarrow \text{U}(1) \times \text{SU}(2)$ through the Higgs mechanism.}
TEO

dark matter

neutrino oscillations

\fi

\iffalse

\subsection{Condensed-matter physics} \label{subsec:condensed_matter_physics}

Rashba

quantum technologies with non-abelian couplings 

review by this guy

\section{On lattice gauge theories and their quantum simulation} \label{sec:LGT_quantum_simu}

chromo and wilson

schwinger model

Cirac (max planck), muschik, zoller,  (innsbruck)

Our work vs this greek dude
\fi

\section{Publication  \cite{ADMDB16}: Quantum walks and non-Abelian discrete gauge theory} \label{sec:paper3}

\titlebox{
In this publication, we present a new family of 1D DTQWs which coincides, in the continuum limit, with the unidimensional Dirac dynamics of a relativistic spin-1/2 fermion coupled to a generic non-Abelian U($N$) Yang-Mills gauge field. The gauge-theory interpretation is extended beyond the continuum limit, by showing the existence of, (i) not only a U($N$) gauge invariance on the lattice, but also (ii) an associate lattice gauge-covariant quantity, which delivers, in the continuum, the field strength associated to a standard U($N$) Yang-Mills gauge field. We show, through numerical simulations, that the classical, i.e. non-quantum regime, is recovered at short times.
} \\

\noindent
The aim of this paper \cite{ADMDB16} is to extend the previous results on Abelian gauge fields to non-Abelian gauge fields. By simplicity, we have started this work in 1D, and the paper only deals with this case.

\subsection{Schematic picture of a non-Abelian Yang-Mills gauge theory} \label{subsec:schematic}

Before explaining how we build our DTQW with non-Abelian coupling, let us give a sum-up on the standard non-Abelian Yang-Mills gauge theory, which is the continuum-limit situation of our DTQW scheme, as we show in Section III of the paper.

\subsubsection{Gauge-field-induced matter-field dynamics}

Here is a simple picture of such a dynamics.

An Abelian gauge field (i.e. electromagnetic potential) influences the dynamics of a first-quantized matter field $\Psi(\boldsymbol{x},t)$\footnote{We make this presentation in $n$ spatial dimensions for the sake of generality, although we will then limit ourselves to the 1D case, so that the Abelian field only contains an eletric potential and no magnetic potential.} by, \emph{roughly speaking}, modifying its phase locally (i.e. depending on the spacetime location $(\boldsymbol{x},t)$), and through a coupling between this dynamical phase modification and the external (i.e. position) degree of freedom (d.o.f.) of the field\footnote{\label{footnote:phase_modif_picture}A way to see this coupling is by applying the evolution operator that takes $\Psi(\boldsymbol{x},t_1)$ to $\Psi(\boldsymbol{x},t_2)$, which is a complex exponential of the Hamiltonian with (Abelian) gauge coupling: because the gauge field generically depends on position, it does not commute with the momentum, and this creates a non-trivial coupling between the internal $\text{U}(1)$ degree of freedom and the external one; this coupling becomes trivial if the gauge field does not depend on position but only on time.

Note that, as long as we keep it vague, this simple `physical' picture holds whatever the nature of the field, i.e. whether its  spin is half integer or integer: it can be a spin-1/2 field, in a relativistic or non-relativistic framework, but also an integer-spin field, such as the spinless (i.e. scalar) complex field, which, if it describes matter (as opposed to gauge fields), can only describe composite (i.e. non-elementary) particles. 

See Appendix \ref{app:classical_fields_and_spin} to go beyond this simple vague picture and a discussion on the distinction between half-integer-spin and integer-spin (classical) fields.}. This phase (i) is an internal degree of freedom and (ii) is associated to\footnote{The internal degree of freedom is arbitrary, i.e. fixed up to a transformation by an element of $\text{U}(1)$,  in the absence of gauge field. This internal d.o.f. belongs to a half-integer-spin representation space of the $\text{U}(1)$ group.} the Abelian gauge group $\text{U(}1)$, so that it can be called Abelian or $\text{U}(1)$ internal degree of freedom. The magnitude of the interaction is characterized by a charge, specific to the considered field, called Abelian charge.

A non-Abelian gauge field influences the dynamics of the  aforementioned matter field by modifying another internal d.o.f., called non-Abelian or U($N$) internal d.o.f., and through a coupling between this dynamical modification of the non-Abelian internal d.o.f. and the position d.o.f.. The magnitude of the interaction is given by an associated non-Abelian charge. As for any degree of freedom, the non-Abelian internal d.o.f. is formally implemented by tensorial product, which reduce to a multiplication in the Abelian case.

Such a gauge-field-induced matter-field dynamics is described, in the case of half-integer-spin (i.e. fermionic) matter, by the Dirac equation with Yang-Mills (minimal\footnote{Non-minimal couplings have been considered essentially between the gravitational force and some other physical entity, that we describe by a field, either fermionic or bosonic. Such a non-minimal coupling arises naturally when describing the gravitational force by a curvature of spacetime, see the following discussion, \url{http://physics.stackexchange.com/questions/103892/minimal-vs-non-minimal-coupling-in-general-relativity}. The field that couples to gravity can be, e.g., a matter field \cite{Bertolami2009,Wang2010}, an electromagnetic field \cite{Balakin2010}, or an inflaton field \cite{Mahajan2014,Budhi2017} in primordial cosmology. To give a global vague picture, we may say that such couplings have been introduced to tackle two important classes of theoretical problems in physics: (i) describing phenomena where the magnitude of  the gravitational force is comparable to that of another given force, either the electromagnetic, the weak, or the strong one, and, (ii) understanding dark matter and dark energy, which have been invoked in cosmology to account for unexpected observations, as the unexpected galaxy rotation diagrams for dark matter, and the acceleration of universe's expansion for dark energy. For information on modified gravity theories with non-minimal couplings to other fields, see, e.g., the following review of 2014 \cite{Harko2014}, the following same-year talk, \url{http://www.cpt.univ-mrs.fr/~cosmo/SW_2014/PPT/Balakin.pdf}, or this Master thesis, \url{https://fenix.tecnico.ulisboa.pt/downloadFile/395138327112/ThesisFinalN.pdf}, which is however less recent (2009).}) coupling.

To have a better picture of both the Abelian and non-Abelian interactions, the geometrical framework of fiber bundles is useful, see Appendix \ref{app:geometry}: the gauge field is a connection on the vector U($N$)-bundle. A necessary condition for the gauge field to be `Yang-Mills' and not only `non-Abelian', is that it belongs to the Lie algebra of the gauge group. An interesting epistemological paper about the geometrical foundations of classical Yang-Mills theory, by Catren, can be found at \url{http://www.sciencedirect.com/science/article/pii/S1355219808000166}\footnote{This paper aims at going beyond the standard `physical' argument underlying gauge theories, namely that physics must be local and that any degree of freedom of the system should not be constrained to have a the same value at all spacetime points, but should actually be a field defined on spacetime, i.e. that (i) one should be able to modify the value of the parameter locally without being, at least immediatly, affected by very distant objects, and that, (ii) one should still require the invariance property of the system with respect to changes of the degree of freedom once it is made local, by extended relativity principle, as Einstein did to go from special to general relativity.}.

\subsubsection{Dynamics of the gauge field: matter field acting on a gauge field}

The previous picture is not complete, because it does not take into account that matter influences the gauge-field dynamics.

Such a dynamics is described, in the case of an Abelian gauge field, by Maxwell's equations. In the case of a non-Abelian gauge field, Maxwell's equations generalize into the so-called ``Yang-Mills field equations''. In our paper, we do not provide any DTQW equivalent for these equations. Yang-Mills field equations take a simple form when written in the framework of Cartan's calculus (whose central ingredient is the exterior derivative), used covariantly on (Lie-algebra) principal U($N$)-bundles. 

\subsubsection{First-quantized (generically non-Abelian) Yang-Mills gauge theory of interactions}

A generic first-quantized (non-Abelian) Yang-Mills gauge theory can thus be written compactly in 3 (non-scalar) equations: 
\begin{itemize}
\item[$\bullet$] One dynamical equation for the matter field. For a spin-1/2 field, this is the Dirac equation with non-Abelian coupling, see below, Subsubsection \ref{subsubsec:matter_field_formalism}.
\item[$\bullet$] Two dynamical equations for the gauge field:
\begin{enumerate}
\item The so-called \emph{homogeneous} equations, which are (dynamical) constraints on the gauge field, independent of the presence of matter (the matter field does not enter this equation).
\item The so-called \emph{inhomogeneous} equations, which describe how the presence of matter modifies the dynamics of the gauge field. The gauge-field part of this equation can be obtained from the homogeneous equation by Hodge duality.
\end{enumerate}
\end{itemize}
For information on the solutions to this system, and their relevance to physics, see the following discussion, \url{http://physics.stackexchange.com/questions/27309/which-exact-solutions-of-the-classical-yang-mills-equations-are-known}, which contains many references.

\subsubsection{Dynamics of the matter field and non-Abelian gauge invariance: formalism}
\label{subsubsec:matter_field_formalism}

\paragraph{{$\bullet$ Dynamical equation}} \hspace{3cm}

\vspace{0.2cm}

\noindent
The Dirac equation describing the dynamics of a spin-1/2 field $\Psi$ of mass $m$, interacting with a non-Abelian  U($N$) Yang-Mills gauge field, is:
\begin{equation}
 \left[i \gamma^\mu \otimes D_\mu - m \mathbf{1}_{dN} \right]\Psi= 0 \ ,
\label{eq:yangmillseq}
\end{equation}
where $d$ is the dimension of the spin-representation space, and the index $\mu$ is summed over from $0$ to $n$. The covariant derivative $D_\mu = \mathbf{1}_{N} \partial_\mu  - i B_{\mu}$, where the $B_{\mu}$'s  are the $n+1$ spatiotemporal components of the gauge field. This gauge field is valued in the Lie algebra of U$(N)$, which is an $N$-dimensional vector space, and it can thus be decomposed as
\begin{equation}
B_{\mu} = B_{\mu}^k \tau_k \, ,
\end{equation}
where summation over $k$ is implied, and where the $\tau_k$'s are $N^2$ possible generators of U($N$), i.e. a possible basis of its Lie algebra\footnote{U($N$) is both compact and connected; hence, if $U \in \text{U}(N)$, there exist a Hermitian matrix $H$ such that $U = \exp[iH]$. Let us write $H_{ij}=a_{ij}+ib_{ij}$, where $a_{ij}$ and $b_{ij}$ are real numbers. $H$ Hermitian means $a_{ji} = a_{ij}$ and $b_{ji} = - b_{ij}$. The number of $b_{ij}$'s that can be chosen arbitrarily is $(N^2-N)/2$ $b_{ij}$; indeed, to the $N^2$ arbitrary real numbers  that define an arbitrary matrix, one must omit $N$ since the diagonal elements of $b$ must vanish ($b_{ii}=-b_{ii}$), and divide the whole by $2$ because if all $b_{ij}$'s are chosen for, say, $i>j$, then those for $i'=j<j'=i$ are imposed. For $a_{ij}$, the reasoning is the same, and we must add, to $(N^2-N)/2$, the diagonal elements, which can be given an arbitrary real value, so that the number of  $a_{ij}$'s that can be chosen arbitrarily is $(N^2-N)/2 +N$. The set of all Hermitian matrices of dimension $N$, which is the Lie algebra of U($N$), is thus a real vector space of dimension $(N^2-N)/2 + [(N^2-N)/2 + N]=N^2$.}. Here we have set the coupling constant to $g=-1$.

In one spatial dimension, the irreducible representations of the Clifford algebra have dimension $2$, and we can choose $\gamma^0 = \sigma_1$ and $\gamma^1 = i\sigma_2$.

\paragraph{{$\bullet$ Non-Abelian gauge invariance}} \hspace{3cm}

\vspace{0.2cm}

\noindent
Equation (\ref{eq:yangmillseq}) is invariant under the following gauge transformation, which generalizes the Abelian one, i.e. Eq. (\ref{eq:Abelian_gauge_transfo}):
\begin{subequations} \label{eq:non-Abel_gauge_inv_continuum}
\begin{align}
\Psi &\rightarrow \Psi' = (\mathbf{1}_d \otimes G) \, \Psi \\
\label{eq:gauge_field_transfo}
B_{\mu} &\rightarrow B_{\mu}' =  GB_{\mu}G^{-1} - i (\partial_{\mu}G) G^{-1} \ ,
\end{align}
\end{subequations}
where $G \in \text{U}(N)$ is the gauge transformation, that replaces the Abelian one $e^{-i\phi} \in U(1)$ . 

The field strength, that generalizes the electromagnetic tensor (also called Faraday tensor), is given by
\begin{equation} \label{eq:field_strength}
F_{\mu\nu} = \partial_{\mu}B_{\nu} - \partial_{\nu}B_{\mu} - i [B_{\mu},B_{\nu}] \, ,
\end{equation}
where the commutator vanishes in the Abelian case. This field strength is gauge covariant, i.e. its transformation under (\ref{eq:gauge_field_transfo}) solely reflects the basis change in the non-Abelian internal-d.o.f. Hilbert space, due to the change of gauge:  $F'_{\mu\nu}=GF_{\mu\nu}G^{-1}$. In geometric terms, $F_{\mu\nu}$ is the curvature of the vector U$(N)$-bundle, of which the U$(N)$ part of the $dN$-component field $\Psi$ (which also contains a spin part, in the notation we use) is a so-called section.

\subsection{Walk operator for the 1D DTQW with non-Abelian coupling} \label{subsec:walk_op_non_ab}

The starting-point ingredient of our paper is the definition of a DTQW which (i) not only has the standard Yang-Mills matter dynamics as a continuum limit, Eq. (\ref{eq:yangmillseq}), as already announced, but also (ii) satisfies a gauge invariance on the lattice, that generalizes the relations found in the Abelian case, Eqs. (\ref{eq:discrete_gauge_inv}), and which delivers, in the continuum limit, the standard non-Abelian gauge invariance given by Eq. (\ref{eq:non-Abel_gauge_inv_continuum}).

Actually, our 1D non-Abelian DTQW-operator satisfying the above conditions can be constructed by `straightforward' extension of the Abelian case. Indeed, the 1D Abelian DTQW operator can be written (rewrite Eq. (\ref{eq:1D_walk_operator}) as below, replacing the superscript `1D' by `$N=1$', which specifies the dimension of the gauge group):
\begin{equation}
W^{N=1}_j = C(\Delta\theta)
\begin{bmatrix}
e^{ia^+_{j}} & 0 \\
0 & e^{ia^-_{j}}
\end{bmatrix} S \, ,
\end{equation}
with
\begin{subequations}
\begin{align}
e^{ia^{\pm}_{j,p}} &\equiv e^{i[\Delta\alpha_{j,p} \pm \Delta\xi_{j,p}]} \ \in \ \text{U}(1) \\
&\equiv e^{i[(a_0)_{j,p} \pm (a_1)_{j,p}]} \, ,
\end{align}
\end{subequations}
where the $(a_{\mu})_{j,p}$'s are the Abelian gauge-field lattice-spacetime components, related to the standard Abelian gauge-field covariant continuum-spacetime components $(A_{\mu})_{j,p}$'s by 
\begin{equation}
a_{\mu} = \epsilon_A A_{\mu} \, ,
\end{equation}
where $\epsilon_A$ goes to zero in the continuum limit.

The 1D non-Abelian DTQW-operator is then `simply' constructed as:
\begin{equation} \label{eq:walk_non-Abelian}
W^{N}_j = \left(C(\Delta\theta) \otimes \mathbf{1}_N \right)
\begin{bmatrix}
e^{ib^+_{j}} & 0 \\
0 & e^{ib^-_{j}}
\end{bmatrix} \left(S \otimes \mathbf{1}_N \right) \, ,
\end{equation}
which acts on a $2N$-component walker $\Psi_j$, with
\begin{equation}
e^{ib^{\pm}_{j,p}} \equiv e^{i[(b_0)_{j,p} \pm (b_1)_{j,p}]} \ \in \ \text{U}(N) \, ,
\end{equation}
where the $(b_{\mu})_{j,p}$'s are related to the standard non-Abelian Yang-Mills gauge-field covariant continuum-spacetime components  $(B_{\mu})_{j,p}$'s by (this is shown in Section III of our paper):
\begin{equation}
b_{\mu} = \epsilon_A B_{\mu} \, .
\end{equation}

Eventually, the one-step evolution equation with non-Abelian DTQW operator (\ref{eq:walk_non-Abelian}) is invariant under the substitution $\Psi_{j,p} \rightarrow \Psi'_{j,p} = (\mathbf{1}_{d=2} \otimes G_{j,p}) \, \Psi_{j,p}$, provided that the gauge field transforms as
\begin{equation} \label{eq:non-Abel_gauge_inv_lattice}
e^{i(b^{\pm})'_{j,p}}  = G_{j+1,p} \, e^{ib^{\pm}_{j,p}} \, G^{-1}_{j, p\pm1} \nonumber \\ \ .
\end{equation}
This gauge transformation relates the exponentials of the gauge-field components, which are elements of U$(N)$, and cannot be reduced to a relation obtained by replacing, in the standard continuum transformation of the gauge-field components, Eq. (\ref{eq:gauge_field_transfo}), the partial derivative, $\partial_{\mu}G$, by a discrete derivative, as in the Abelian case, see Eq. (\ref{eq:discrete_gauge_inv_gauge_field}). This is so because of the non-Abelian nature of the coupling, which implies that, if $e^{i\omega_1}$ and $e^{i\omega_2}$ are two elements of U($N$), then
\begin{equation} \label{eq:non_commut}
e^{i\omega_1}e^{i\omega_2} \neq e^{i(\omega_1+\omega_2)} \, .
\end{equation}
A related fact is that the natural quantities involved in the gauge invariance are not the lattice equivalents, $b_{\mu}$'s, of the continuum gauge-field components $B_{\mu}$'s, which both belong to the Lie algebra of U($N$), but the $e^{ib^{\pm}}$'s\footnote{And not even the $e^{ib_{\mu}}$'s, because of (\ref{eq:non_commut}).}, which belong to the U($N$) group\footnote{In the Abelian case, the phase $\phi_{j,p}$ belongs to the Lie algebra of U($1$), while here $G_{j,p}$ belongs to U($N$).}. Note however that we can write the lattice gauge transformation, (\ref{eq:non-Abel_gauge_inv_continuum}), as a relation between the Lie-Algebra-valued quantities, $b_{\mu}$'s, thanks to the Baker-Campbell-Hausdorff formula; this is  \emph{by far} more cumbersome, but can be useful to study the differences between the lattice and the continuum situations, and the passage from one to another.

\subsection{Short but detailed review and comments} \label{subsec:comment_paper3}

In Section III, we show that the continuum limit of our DTQW, i.e. the one-step evolution equation with the U($N$) walk operator (\ref{eq:walk_non-Abelian}), yields the standard non-Abelian Yang-Mills equation for matter, Eq. (\ref{eq:yangmillseq}), and that the lattice gauge invariance, Eq. (\ref{eq:non-Abel_gauge_inv_lattice}), yields the standard non-Abelian gauge invariance, Eq. (\ref{eq:gauge_field_transfo}).

Now, a notable result of our paper is the finding, in Subsection II B, of a lattice gauge-covariant quantity, $\mathcal{F}_{j,p}$, which delivers, in the continuum (Section III), the standard field strength of non-Abelian Yang-Mills theory, see Eq. (14) of the paper. We have already mentioned, in the last paragraph of the previous subsubsection, that the lattice gauge transformation, Eq. (\ref{eq:non-Abel_gauge_inv_lattice}), is not expressed naturally in the Lie algebra of U$(N)$\footnote{Such a Lie-algebra-valued expression involves using several times the Baker-Campbell-Hausdorff formula, which is very cumbersome.}, and that the natural quantities involved in this gauge transformation are manifestly the $e^{ib^{\pm}}$'s, which belong to the U($N$) group. It is thus natural to look for a lattice gauge-covariant quantity $\mathcal{F}_{j,p}$ by manipulating this quantities on the lattice; this is how we proceeded, and the result is Eq. (8). In the Appendix, we show that the U($N$) gauge field can be viewed as the combination of an SU($N$) and a U($1$) gauge field, and we show the precise connection with the Abelian case U(1).

In Section II, we first check, in Subsection II B, the validity of our formal computation of the continuum limit, on an explicit solution, and, in Subsection II C, we show short-time-scale agreement with the classical, i.e. non-quantum situation, derived in \cite{Boozer2011}.

\newpage\null\thispagestyle{empty}\newpage

\includepdf[pages={-}]{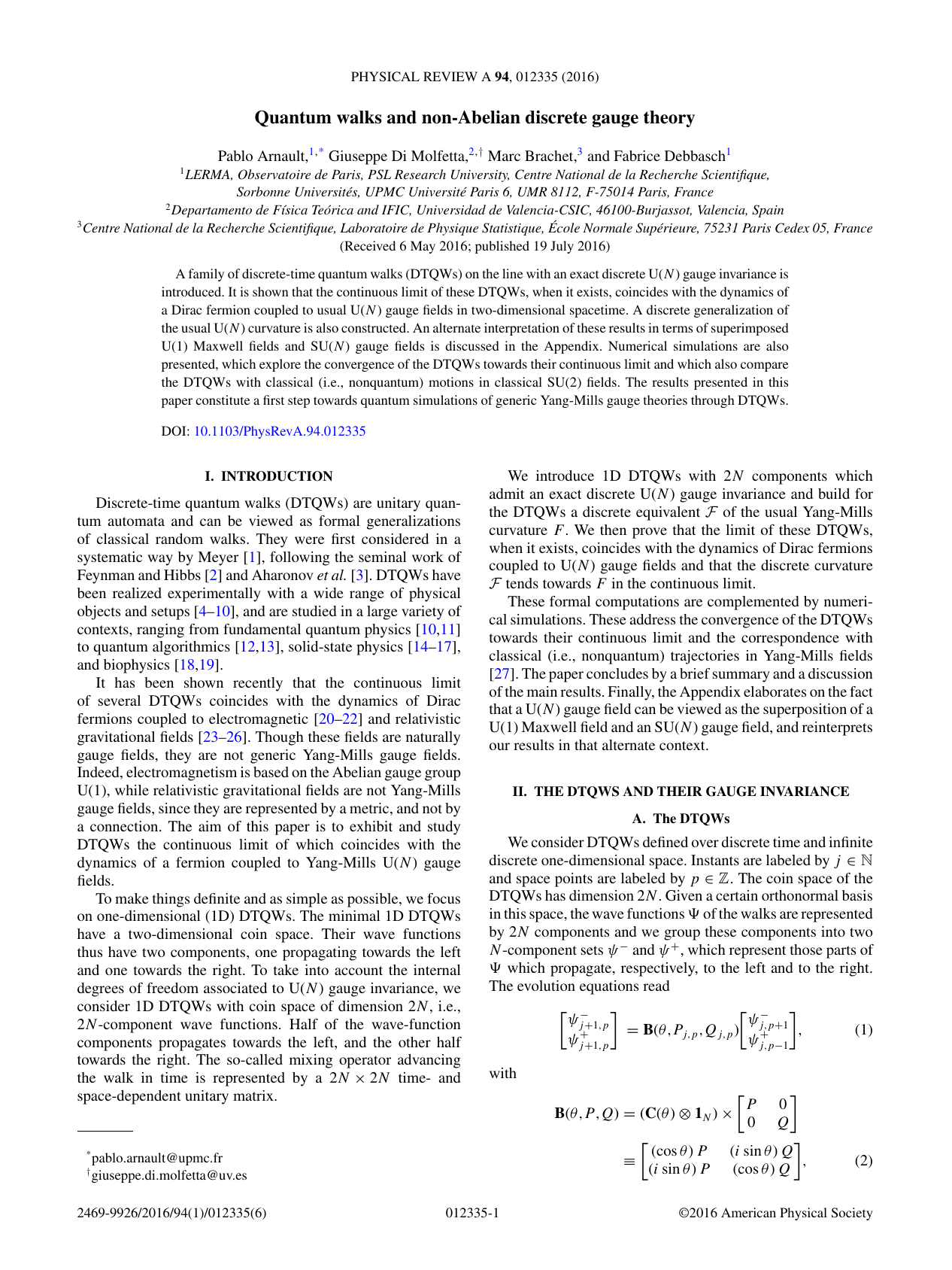}

}

\part{DTQWs interacting with gravitational fields \label{Part:Gravitation}}
   %%%%%%%%%%%%%%%%%%%%%%%%%%%%%%
%%  Alias bibliographiques  %%
%%%%%%%%%%%%%%%%%%%%%%%%%%%%%%
\defcitealias{NRFortran}{Numerical Recipes}
\defcitealias{NRF77}{Numerical Recipes}
\defcitealias{NRF90}{Numerical Recipes}
\defcitealias{IUPAC}{IUPAC}%International Union of Pure and Applied Chemistry

\setcounter{footnote}{0}

\setcounter{section}{0}
{
\begin{IntroMat}[\hspace{-0.2cm}one-dimensional DTQWs in gravitational fields]{One-dimensional DTQWs in gravitational fields} \label{IntroMat2}

       {\noindent
\motivations{
The aim of this introductory material to the second part of the manuscript, is to give a pedagogical, general, and essential view of the results of Refs. \cite{DMD13b,DMD14}, that is, of how to make a DTQW propagate in a (1+1)D curved spacetime, to speak in vague but pointful terms, that must be clarified later. The publication on which the next, third and last chapter is based, is an extension of the results presented in this introductory material.
}       \\ }
                  
\section{Quantum walks as Dirac fermions in (1+1)D curved spacetime} \label{sec:1D_gravity}

\subsection{General relativity and gravitation-induced advecting speed field}

In general relativity, the gravitational field experienced by a particle, due to surrounding mass-energy sources, is encoded in the spacetime deformation\footnote{Deformation with respect to the standard Euclidean spacetime.}, which is mathematically described by a pseudo-Riemmanian space with signature $(+,-,...,-)$, with as many minus signs as the spatial dimensions. In this curved spacetime, the particle follows geodesics, i.e. is a free particle, subject to no force, since the gravitational force is \emph{already} encoded in the curvature of space. Projecting this spacetime deformation, with the tetrad formalism, on some flat, i.e. Minkowskian spacetime, which can be chosen as the local inertial spacetime of the particle (i.e., the spacetime which is tangent to the worldline), gives the particle a \emph{spacetime-varying} speed (in this flat spacetime), i.e. an  \emph{acceleration}, consequence of the gravitational force produced by the projection of the original-spacetime curvature on the flat spacetime. If we take the non-relativistic limit, such a spacetime variation of the speed is that induced by the \emph{Newtonian} gravitational force.

Equation (\ref{eq:Dirac_PDE_2}) tells us that $\psi^{\downarrow}$ is transported at (dimensionless) speed 1 (algebraically on the line), with a deformation induced by the mass-coupling source term $-im\psi^{\uparrow}$ (the mass $m$ couples $\psi^{\uparrow}$ and $\psi^{\downarrow}$).

Indeed, forget about the mass-coupling source term for a moment, and consider the following standard (deformationless) advection equation for an arbitrary quantity $u(t,x)$, at constant and uniform speed $v$:
\begin{equation} \label{eq:advection_eq}
(\partial_t + v\partial_x)u=0 \, .
\end{equation}
Any function $f(x-vt)$ is a solution of this equation, and the equation thus describes the transport of the function $f$ at speed $v$; this transport mechanism is homogeneous, i.e. is the same at every spacetime point because so is $v$.

Imagine now that $v$ is spacetime dependent, i.e. $v(t,x)$, as induced by a gravitational field. Again, any solution of the form $f(x-v(t,x)t)$ is solution of this equation\footnote{This is true provided that $v(t,x)$ itself satisfies the equation, i.e. provided that its convective derivative vanishes, $\frac{dv}{dt}\equiv (\partial_t + v\partial_x)v=0$, which merely states that $v$ is an imposed field, having no dynamics, and that if a point particle is embeded in it at point $(t,x)$, it will move with speed $v(t,x)$.}: the equation thus describes the transport of the function $f$ at speed $v(t,x)$; this transport mechanism is inhomogeneous, i.e. depends on the spacetime point $(t,x)$ through $v(t,x)$.
     
\subsection{Standard DTQWs cannot yield any spacetime-dependent advecting speed field in the continuum limit}

\subsubsection{The DTQW must be inhomogeneous}

Our question is how to obtain the spacetime-dependent transport described by Eq. (\ref{eq:advection_eq}) with local speed $v(t,x)$, for the wavefunction of a quantum particle, in the continuum limit of some DTQW. Technically, the question is how to obtain a spacetime-dependent  speed $v(t,x)$ in front of $\partial_x$ in, e.g., Eq. (\ref{eq:Dirac_PDE_2}), instead of the constant speed 1.

Considering a standard  DTQW, see Eq. (\ref{eq:protocol_explicit}), homogeneous, i.e. whose coin operation does not depend on the (1+1)D spacetime-lattice point $(j,p)$, means that the discrete transport mechanism is the same at each $(j,p)$, and this holds in some possibly-existing continuum limit, i.e. the latter cannot describe any spacetime-dependent transport such as that we wish to obtain.

\subsubsection{The standard `one-step' continuum limit of inhomogeneous DTQWs does not do the job either}

Here we present a quick technical sum-up of why such a continuum limit will not work.

Consider a coin operation (\ref{eq:coin_operation}) with spacetime-dependent entries, $q_{j,p} = ((q_s)_{j,p})_{s=1,4} = (\alpha_{j,p}, \theta_{j,p}, \xi_{j,p}, \zeta_{j,p})$, that we allow to read
\begin{equation}
(q_{s})_{j,p} = (q^{0}_s)_{j,p} + \epsilon_{q} \, (\bar{q}_s)_{j,p} + \mathcal{O}(\epsilon_q^2) \, ,
\end{equation}
which is merely a sufficient condition for the coin operation to have a limit in the continuum, i.e. when the spacetime-lattice step $\epsilon_l=\epsilon_q$, goes to zero. Consider now the Taylor expansion in $\epsilon_l$ of the DTQW one-step evolution equation, Eq. (\ref{eq:protocol_explicit}), with spacetime-dependent coin operation $U^{\text{Euler}}(q(t,x))$, $t= j \, \epsilon_l$, $x=p \, \epsilon_l$. The derivatives $\partial_x \psi^{\uparrow\downarrow}$ can only come from the multiplication of the \emph{first-order} terms $\partial_x \psi^{\uparrow\downarrow}$ by a spacetime-dependent quantity function of the \emph{zeroth-order} angles, $(q^0_s)_{j,p}$. But we know that there is a zeroth-order constraint for the continuum limit to exist, namely $U^{\text{Euler}}(q^0(t,x))=\mathbf{1}_2$, that is to say,
\begin{subequations}
\begin{align}
e^{i(\alpha^0+\xi^0)} \cos \theta^0 &=1 \\
e^{i(\alpha^0-\xi^0)} \cos \theta^0 &=1 \\
e^{i(\alpha^0+\zeta^0)} \sin \theta^0 &=0 \\
e^{i(\alpha^0-\zeta^0)} \sin \theta^0 &=0 \, . \\
\end{align}
\end{subequations}
\noindent
For such constraints to be satisfied, all four zeroth-order angles must be \emph{independent of the spacetime point}. Hence, no standard DTQW can yield any spacetime-dependent transport in the continuum limit we have considered.

\subsection{$k$-step continuum limit and DTQW in curved spacetime}

The previous  `one-step' continuum limit does not allow for any sharp variation of the $(j,p)$-dependent quantities, even on spacetime scales of the order of a few spacetime-lattice sites, because we demand that these $(j,p)$-dependent quantities coincide with smooth functions at \emph{each} spacetime point. Let us take a coarser look and lower this requirement by demanding a coincidence every $k$ spacetime points, $k$ being an integer superior or equal to 2. This defines a larger class of continuum limits, which contains the one-step class. This means that the evolution equation whose Taylor equation we are going to perform is not anymore the one-step equation, Eq. (\ref{eq:protocol_explicit}), but the $k$-step one, i.e.
{\small
\begin{equation} \label{eq:protocol_explicit_2}
\langle p | \! \left(
\begin{matrix}
| \psi^{\uparrow}_{j+k} \rangle \\ | \psi^{\downarrow}_{j+k} \rangle
\end{matrix} \!
\right)
=  \langle p |  {\Bigg[
\underbrace{e^{i\alpha_{j}} \!
\begin{bmatrix}
e^{i\xi_j} \cos \theta_j & e^{i\zeta_j} \sin\theta_j \\ -e^{-i\zeta_j} \sin\theta_j  & e^{-i\xi_j} \cos \theta_j 
\end{bmatrix}}_{U^{\text{Euler}}(q_j)}
\begin{bmatrix}
\sum_p |p-1\rangle\langle p| & 0 \\
0 & \sum_p |p+1\rangle\langle p|
\end{bmatrix} \Bigg]}^{k} \! \!
\left( \begin{matrix} | \psi_{j}^{\uparrow} \rangle \\ | \psi_{j}^{\downarrow} \rangle \end{matrix} \! \right) ,
\end{equation}}
where we have used the suitable notation of a quantum state which is explicit in spin space but implicit in position space.

The zeroth-order constraint for such a $k$-step continuum limit to exist is $(U^{\text{Euler}}(q^0_{j,p}))^k=\mathbf{1}_2$. This stroboscopic view of DTQWs has been introduced by Di Molfetta et al. in \cite{DMD13b}, and the zeroth-order constraints for the continuum limit to exist have been fully examined for $k=2$ in \cite{DMD14}.

We can show that the following coin \cite{DMD13b,DMD14},
\begin{equation}
B(\theta) =
\begin{bmatrix}
-\cos \theta & i \sin \theta \\
-i \sin \theta & \cos \theta
\end{bmatrix} \, , \ \ \ \ \theta(t,x) \, ,
\end{equation}
which satisfies the zeroth-order constraint for $k=2$ (but \emph{not} for $k=1$), yields a DTQW whose 2-step continuum limit describes the dynamics of a (massless) Dirac fermion in a (1+1)D curved spacetime with signature $(+,-)$, with a speed 
\begin{equation}
v(t,x) \propto \cos^2 \theta(t,x) \, ,
\end{equation}
where $t$ is, as in the one-step case, the continuum-limit time corresponding to the discrete one $j$, i.e. $t_j = j \epsilon_t \rightarrow t$ when $\epsilon_t \rightarrow 0$.
The PDE of such a dynamics can be written in  the following compact form,
\begin{equation}
\gamma^{(a)} \left[ e^{\mu}_{(a)} \partial_{\mu} \Phi + \frac{1}{2} \frac{1}{\sqrt{-G}} \partial_{\mu} \left( \sqrt{-G}e^{\mu}_{(a)} \right) \Phi \right] = 0 \, ,
\end{equation}
where $\mu \in \{t,x\}$, $a \in \{0,1\}$, $e^{\mu}_{(a)}$ are the components of the diad, i.e. orthonormal
basis, ($e_0 = e_t$, $e_1 =(\cos \theta) e_x$), on the original coordinate basis $(e_t,e_x)$, $\Phi = \Psi \sqrt{\cos \theta}$ is the spinor which lives in the curved spacetime, $\left[G_{\mu\nu}\right]=\text{diag}(1,-1/\cos^2 \theta)$ is the metric tensor of the curved spacetime, $G=\text{det}\left[G_{\mu\nu}\right]$, and the (flat) gamma matrices are given, as usually chosen in our work, by $\gamma^{(0)} =\sigma_1$, $\gamma^{(1)} = i \sigma_2$. (We know put round brackets around the indices labelling the coordinates on the locally-flat spacetime.)

\section{Application: radial motion of a DTQW in a Schwartzschild blackhole}

In Refs. \cite{DMD13b,DMD14} is shown an application of the previous result to a (1+1)D metric associated to a time and a radial coordinate of a Schwartzschild blackhole. If the walker, which, remember, carries as spin-1/2, starts on the horizon of the blackhole, half of its occupation probability (or density) will collapse on the singularity, while the other half will remain localized on the horizon. If the walker starts below the horizon, i.e. inside the blackhole, it will not escape from it; more precisely, half of its density (i.e. one branch), will end on the singularity at some instant, while the other branch will also end on this singularity, but later. If the walker starts outside the blackhole, one branch will still collapse on the singularity,  while the other branch will escape from the blackhole.

\end{IntroMat}}

   %%%%%%%%%%%%%%%%%%%%%%%%%%%%%%
%%  Alias bibliographiques  %%
%%%%%%%%%%%%%%%%%%%%%%%%%%%%%%
\defcitealias{NRFortran}{Numerical Recipes}
\defcitealias{NRF77}{Numerical Recipes}
\defcitealias{NRF90}{Numerical Recipes}
\defcitealias{IUPAC}{IUPAC}%International Union of Pure and Applied Chemistry

\chapter{\textsc{Two-dimensional DTQWs in gravitational fields \\ (Publication  \cite{AD17})} \label{Chap:Gravitational_2D_DTQWs}}
\minitoc

\iffalse

\section{Gravitation as a gauge theory}

GTG

GGT

\subsection{Gauge unified theory}

\section{Discretizing special and general relativity}

computational GR

Arrighi, Facchini,

D'Ariano, Perinotti, 

\fi

\section{Publication \cite{AD17}: Quantum walks and gravitational waves}

\titlebox{
In this publication, we present a new family of 2D DTQWs which coincides, in the continuum limit, with the Dirac dynamics of a relativistic spin-1/2 fermion in a generic (1+2)D curved spacetime. We then focus on the particular case where this spacetime corresponds to the polarization plane of a gravitational wave (GW). We show how the probability of presence of the DTQW in this plane is affected by a pure shear GW. In the continuum limit, the GW modifies the eigen-energies of the walker by an anisotropic factor which is computed exactly. For wavelengths close to the spacetime-lattice step, the probability of presence is modified non trivially; the minimal comment on such an influence is that the net effect of the GW is maximal for short wavelengths, comparable to the spacetime-lattice step.
} \\

\subsection{Short but detailed review and comments}

This paper, Ref. \cite{AD17}, begins by a 2D generalization of the 1D formal results obtained in Ref. \cite{DMD13b}: indeed, we introduce a 2D DTQW whose continuum limit yields the dynamics of a spin-1/2 Dirac fermion propagating in an arbitrary (1+2)D curved spacetime, see Section III of the paper. The construction is based on the 1D 2-step construction, see Section \ref{sec:1D_gravity} for a sum-up, and is discussed in depth in Section II of the paper. For a general technical presentation of the Dirac equation in curved spacetime, see Appendix \ref{app:spinors}.

We then consider the particular case where the 2D curved space is that corresponding to a polarization plane of a (linear and plane) gravitational wave (GW), see Section IV, and we focus on pure shear GWs in Section V, where we present and discuss the dispersion relation of the DTQW, in Fig. 1. It is important to have in mind that the (1+2)D curved spacetime can be interpreted as such only in the continuum situation, because we have not made any link with a proper notion of curvature on the lattice. The curved metric corresponding to the aforementioned GW is usually treated, in the standard continuum situation, as a small metric deformation of the flat-spacetime Minkowskian metric. We proceed in the same way for the DTQW evolving on the lattice, and all computations are made at first order in the metric perturbation, whose perturbative nature is traced by the small parameter $\xi$.

In Section VI, we analyse two case-studies in this pure-shear-GW background:

\begin{itemize}

\item[$\bullet$] The first case-study, Subsection VI A, presents the continuum situation, i.e. how the pure shear GW modifies the flat-spacetime dynamics of a 2D spin-1/2 Dirac  fermion. Such a continuum situation is reached, from the DTQW lattice situation, by making the spacetime-lattice step, $\epsilon_l$, go to zero, or equivalently, by taking to zero the quasimomentum of the quantum walker, namely $\boldsymbol{q} = \epsilon_l \boldsymbol{Q}$, where $\boldsymbol{Q}$ is the momentum of the associated Dirac fermion of the continuum situation\footnote{Note that, even if $\epsilon_l$ (and thus $\boldsymbol{q}$), goes to zero, $\boldsymbol{Q}$ can have an arbitrarily big modulus.}. We give the corrections to the free (i.e. flat-spacetime) dynamics, due to the metric curvature induced by the GWs. We give the corrections (i) to the one-step Dirac evolution operator, see Eq. (32), and (ii) to the eigen-elements of this Dirac operator, see Eqs  (33) and (34). In particular, the effect of the pure shear GWs on the usual relativistic dispersion relation of Dirac fermions, of the form energy = $\pm$ momentum, is the addition of an anisotropic correction, i.e. whose magnitude depends on the direction in the polarization plane, see Eq. (35).

\item[$\bullet$] The second case-study, Subsection VI B, goes beyond the continuum-limit situation, i.e., to speak in a lattice terminology, explores the whole Brillouin zone, and not only small quantum-walker wavevectors.  The problem we are interested in is the modification, by the GW, of an interferential DTQW matter distribution located in some polarization plane of the wave. In Appendix \ref{app:GWs}  are given technical details and motivations for such a study. We limit ourselves to the modification after a single time step, which is best seen on the relative matter-distribution change, between the initial time and that after on time step, defined by Eq. (37) and given, after an analytical computation, by Eq. (38). Since such a quantity is position dependent in the polarization plane, and we are interested in the maximum effect of the wave, we look for the maximum value, in the polarization plane, of the relative matter-distribution change, see Eq. (40). Such a quantity is maximum for matter modes whose wavelength is comparable to a few lattice steps, see Fig. 4, i.e. the effect of the wave is stronger at small spatial scales comparable to the lattice step, than in the continuum large-scale situation.

\end{itemize}

\includepdf[pages={-}]{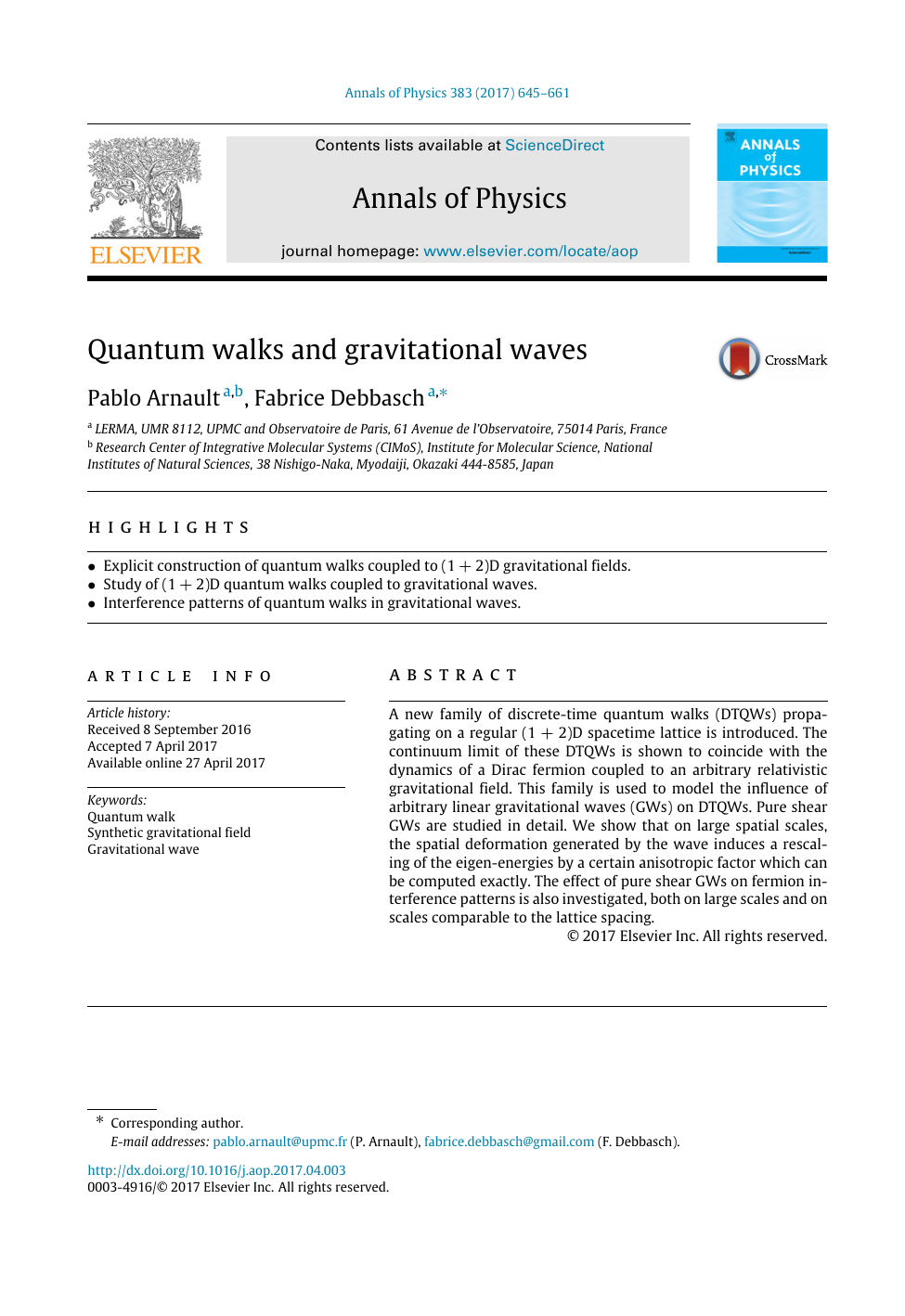}

   \starchapter{\textsc{Conclusion \& perspectives}}{Chap:Concl}

\noindent
The conclusion of the research work on which this thesis is based is essentially twofold. \\

In Part \ref{Part:Yang-Mills}, it has been shown that DTQWs can be used to discretize certain first-quantized Yang-Mills gauge theories \cite{AD15,AD16,ADMDB16}, which extends previous results in lower spacetime and coin-space dimensions \cite{DMD12a,DDMEF12a,DMD14}. The starting point of these discretizations is, indeed, the discretization of the fermionic matter-field dynamics -- i.e., typically, the Dirac equation for spin-1/2 fermions -- under the form of a DTQW, which was seminaly introduced by Feynman in 1+1 dimensions \cite{Schweber86a,FeynHibbs65a}. In addition to deliver the aforementioned field theories in the continuum limit -- which is the minimal requirement for a discretization of such field theories~-- the DTQWs-based discretizations suggested in the present work, not only (i) ensure, by construction, a unitary dynamics on the lattice\footnote{Regarding unitarity in standard LGTs, see Appendix \ref{app:unitarity}.}, but also (ii) preserve, at the lattice level, both (ii.a) Yang-Mills gauge invariance and (ii.b) `continuity', i.e. the conservation of the Yang-Mills charge current, and (iii) can be associated a lattice gauge-invariant quantity\footnote{In the non-Abelian case, this quantity is gauge \emph{covariant}.}~-- which of course delivers the standard Yang-Mills field strength in the continuum. By combining (iii) with (ii), one can eventually construct a lattice equivalent to the gauge-field dynamics\footnote{The \emph{matter-induced}, i.e. \emph{inhomogeneous}, lattice gauge-field dynamics, indeed combines (iii) with (ii),  while the \emph{homogeneous} gauge-field dynamics, i.e. the dynamical constraints, simply combine (iii) with (ii.a). To be more precise, the lattice equivalent suggested for the matter-induced gauge-field dynamics ensures (ii.b).}. 

Let us be more precise on the achievements of this work. The DTQWs-based discretization program described in the previous paragraph, has indeed been achieved for an Abelian Yang-Mills gauge theory in 1+2 dimensions \cite{AD15,AD16}, see Chapter \ref{Chap:Electromagnetic_2D_DTQWs}, and there is work in progress on the Abelian (1+3)D case. In the non-Abelian case, a (1+1)D scheme has been suggested, which couples the matter dynamics to a generic non-Abelian U($N$) Yang-Mills gauge field \cite{ADMDB16}, see Chapter \ref{Chap:Non-Abelian_1D_DTQWs}. While this scheme verifies by construction property (ii.a), which was a necessary starting point to eventually achieve (iii)\footnote{This was much more complicated to achieve than in the Abelian case.}, point (ii.b) is still an open question, as well as the possibility to design a lattice equivalent for the gauge-field dynamics. \\

In Part \ref{Part:Gravitation}, it has been shown that it was possible to extend to 1+2 dimensions \cite{AD17} previous (1+1)D results \cite{DMD13b,DMD14} demonstrating that DTQWs can be used to discretize Dirac dynamics in curved spacetime, while preserving unitarity by construction. This (1+2)D scheme has been used to study how linear plane gravitational waves influence a transverse fermionic matter distribution, while the (1+1)D scheme had been used to simulate fermionic radial motion in a Schwarzchild blackhole. Higher-dimensional extensions of DTQWs-based fermionic dynamics in curved spacetime have also been achieved by Arrighi et al. \cite{AF16}, by pairing DTQWs. The differences between both schemes are discussed in both publications, \cite{AD17} and  \cite{AF16}. \\

Let us now comment on the specifities of the DTQWs-based discretizations of QFTs presented in Part \ref{Part:Yang-Mills} of this manuscript, with respect to other works on the quantum simulation of QFTs. 

While D'Ariano et al. \cite{DAriano2016} and Arrighi et al. \cite{Arrighi2014} have suggested lattice substitutes for Lorentz invariance of standard continuum QFTs, an issue which is not addressed in the present research work, the latter instead provides, not only (i) a lattice equivalent for the Yang-Mills gauge invariance of standard continuum QFTs, but also (ii) a lattice gauge-invariant associated quantity\footnote{In the non-Abelian case, this quantity is gauge \emph{covariant}.}, and (iii) lattice Maxwell's equations ensuring the conservation of the charge current on the lattice, as detailed above. This whole issue of Yang-Mills gauge invariance is not adressed by either  Arrighi et al. or D'Ariano et al.\footnote{However, D'Ariano et al. integrate, to their axiomatic presentation, the seminal work of Bialynicki-Birula which provides a unitary discretization of Maxwell's equations that ensures energy conservation on the lattice \cite{BB94a}.} 

Recall that, while Yang-Mills gauge invariance is standardly associated, through Noether's theorem, to the (Yang-Mills) current conservation, the spacetime-translation subgroup of Poincaré invariance is associated to the conservation of the total energy and momentum, and the Lorentz subgroup is associated to the conservation of the total angular momentum. Generally speaking, the interest of designing a discrete dynamical model satisfying conservation laws, is twofold. If the discrete model is used in order to reproduce a continuum theory, preserving the continuum conservation laws on the lattice makes it unnecessary to compensate for the discrepancies arising otherwise, and which can lead to physical aberrations. Now, independently of the continuum limit of such discrete models, designing them in the perspective of making them satisfy discrete conservation laws -- which are a pilar in all physical theories -- a priori endow these discrete models with, if not a great one, at least some versatility, i.e. makes them, a priori, suitable for various applications, not necessarily connected, at least explicitly, to a continuum situation, but do connected to the conservation of physical quantities, on a discrete background.

In standard LGTs, Yang-Mills gauge invariance was seminaly implemented on the lattice by Wilson \cite{Wilson74}. Indeed, he constructed a gauge-invariant action on the lattice for both the matter field and the gauge field. In particular, the lattice gauge-invariant gauge-field kinetic term yields, in the continuum, essentially the standard `square of the field strength', as needed. Let us now comment on the on-shell dynamics on the lattice. To obtain this lattice on-shell dynamics of either the matter field or the gauge field, one ultimately needs, starting from Wilson's original LGT, to define a variational principle on the lattice, to be applied directly to Wilson's lattice gauge-invariant action. Now, one month after Wilson submitted his seminal paper, Kogut submitted a Hamiltonian formulation of LGT  \cite{KogSuss75a}, which discretizes spatially the Hamiltonian of the system, while keeping time continuous. To obtain the on-shell dynamics of either the matter field or the gauge field with such a formulation, one simply needs to omit from the full Hamiltonian, which describes both the matter-field and the gauge-field dynamics, the irrelevant term for the desired on-shell dynamics. For more information on the differencies between Wilson an Kogut's LGT, see Appendix \ref{app:unitarity}. While many suggestions for the either analog \cite{Wiese2013} or digital \cite{Byrnes2006,Jordan2012,Wiese2013} quantum simulation of LGTs, also ensure Yang-Mills gauge invariance on the lattice and implement lattice gauge-invariant Hamiltonians, a difficulty related to the on-shell gauge-field dynamics was first adressed by Jordan et al. \cite{Jordan2012} and Wiese \cite{Wiese2013}, and then in \emph{extensive} detail by Zohar et al. in 2015 \cite{Zohar2015}. By nature, this difficulty contains both energy-conservation and current-conservation aspects.

This difficulty, that we shall briefly clarify soon, has been one of the experimental challenges that had to be met to achieve the recent tracking of ``Real-time dynamics of lattice gauge theories with a few-qubit quantum computer", led last year at the University of Innsbruck by Martinez et al. \cite{Martinez16}. This experiment digitally quantum simulates the quantum-electrodynamical Schwinger model, i.e. the creation of electron-positron pairs by vacuum fluctuations. Because such a seminal experimental work deals with a Yang-Mills gauge theory which is Abelian, the gauge-field on-shell dynamics corresponds to Maxwell's equations, and, because the model is considered in the minimal spacetime dimension, namely 1+1, these Maxwell's equations reduce to Ampère's law without magnetic field\footnote{We may rather call this a degenerate (general) Ampère's law, since speaking of this law in such terms, i.e. ``Ampère's law'', usually implies the presence of a magnetic field (and sometimes even forgets about the electric field time variation). With no magnetic field, this law simply states that the charge current is given by the time variation of the electric field.} and Gauss's law; while the former is indeed a dynamical equation, the latter is a non-dynamical constraint. The aforementioned theoretical and experimental challenge is the implementation of Gauss's law in this minimal quantum simulation of LGT. Martinez et al.'s experimental work is the concretisation of a large number of theoretical works on the quantum simulation of standard LGTs \cite{Bchler2005,Byrnes2006, Jordan2012,Wiese2013,Zohar2015}. Note the tensor-network reformulation \cite{Pichler2016} of the theory contained in Martinez et al.'s experiment \cite{Martinez16}. See the following talk given in 2013 by Reznik, on his joint work with Zohar and Cirac, 
\sloppy
\burl{http://www.lattice2013.uni-mainz.de/presentations/Plenaries\%20Tuesday/Reznik.pdf}. One of the next steps to take is certainly  to experimentally quantum simulate non-Abelian U($N$) and SU($N$) Yang-Mills gauge theories, based on recent theoretical work, still in the framework of standard LGTs \cite{Zohar2013,Banerjee2013}.

Let us highlight some differences between (standard\footnote{We shall ommit the word `standard' in the following discussion, for the sake of brevity.}) LGTs and the DTQWs-based discretizations of (first-quantized) QFTs that have been discussed in the present manuscript \cite{AD16,ADMDB16}, that we will call DTQW-GTs in the following discussion. As in LGTs, the lattice gauge fields essentially appear, in DTQW-GTs, as exponentials of their continuum counterparts, which reflects the periodicity induced by the lattice, and makes it sufficient to limit both the \emph{lattice} quasimomenta and gauge fields\footnote{In argument of the exponentials.} -- which deliver the standard momenta and gauge fields in the continuum --, to the spacetime Brillouin zone\footnote{The denomination `Brillouin zone' usually refers to the spatial zone, while the temporal zone is often referred to as the `Floquet zone'.}. While on the one hand, as Wilson seminaly pointed out \cite{Wilson74}, such an exponentiation ensures, in LGTs, the local gauge invariance on the lattice, starting from a \emph{standard} finite differentiation of the continuum free theory, the lattice gauge invariance in DTQW-GTs is on the other hand, as the exponentials, in-built\footnote{Provided we consider a sufficiently broad family of inhomogneous DTQWs.}, essentially because of an \emph{in-built (time-evolution) unitarity} which is directly constructed at the level of the evolution operator with no need to define a Hamiltonian. Regarding the question of unitarity in LGTs, see Appendix \ref{app:unitarity}.
This remark leads us to the following important difference: while, in LGTs, the gauge-field exponential is implemented in the Lagrangian \cite{Wilson74} or the Hamiltonian \cite{KogSuss75a}, it is implemented at the level of the evolution operator in DTQW-GTs. This difference is reminiscent of that between tight-biding models and DTQWs schemes regarding the implementation of gauge fields, which has been underlined when discussing the `third' electromagnetic interpretation of electromagnetic DTQW schemes, in Subsection \ref{subsec:recap_1D}.

Another difference must be evoked, between the gauge-field exponentials appearing in LGTs and those appearing in DTQW-GTs. In DTQW-GTs, the gauge fields are, as matter fields, defined on the sites of the lattice, while in LGTs the gauge fields are defined on the links of the lattice. As the aforementioned exponentiation in LGTs, this is necessary to ensure the gauge invariance of the basic finite-differentiated free kinetic term of the continuum theory, which involves products of fields at different lattice sites. More technically, one needs, in order to build the lattice gauge field, to perform an integral of the continuum gauge field over the link before exponentiating it; this yields a so-called parallel transporter, that replaces the continuum gauge field in the lattice theory \cite{Wilson74,KogSuss75a, Kogut83a,Munster2011,book_Rothe}. This integration over the links is not introduced, at least explicitly, in DTQW-GTs. Note eventually that the `discrete differential operators' involved in DTQW-GTs are \emph{not} standard finite differences. 

Eventually, discretizing fermionic fields is notoriously difficult in LGTs \cite{Susskind77a,Munster2011,book_Rothe}. We have mentioned in the discussion of Ref. \cite{AD15} that the scheme presented in this publication exhibits \emph{fermion doubling} since, altough the discretization is not done by standard finite-differentiation, it is still a \emph{symmetric discretization}. This is the case of all DTQW-GTs presented in this manuscript\footnote{In Ref. \cite{Farrelly15} is suggested a way of removing fermion doubling in DTQWs.}. Regarding this matter, among others, I would like to mention the paper of 2016 by Brower et al. \cite{Brower2016}, which constructs finite elements suited for lattice formulations of QFT on curved backgrounds, based on standard LGTs; the construction mixes standard finite-element methods to Regge calculus.

Let us finally sum up possible directions for further studies.

\begin{enumerate} 
\item  {\bf Second quantization} %1

LGTs are fully second quantized, while
no second quantization of DTQW-GTs has been suggested yet. Indeed, although several works have suggested canonical second quantizations for DTQWs, none of them implements, to the best of my knowledge, gauge invariance on the lattice. These works use the framework of QCA to describe multiparticle states, and the standard DTQW is derived has the one-particle case, see, e.g., Refs. \cite{SW04,Bisio2015,DAriano2016}, as well as the PhD manuscript of T. C. Farrelly \cite{Farrelly15}\footnote{Today, 22 August 2017, I am reading this document for the first time -- it is not open-access yet, but should be in a few days. In this work, various paths actually converge, all together, towards the QCA-based second quantization of DTQWs (although there is still work to be done), with a gauge-invariance implementation -- which seems to be the same as ours, see Section 3.4.2 of \cite{Farrelly15} -- but in the Abelian case only, while we have suggested a non-Abelian gauge-invariant scheme \cite{ADMDB16}. From what I have seen, no gauge-invariant quantity is suggested, while we have suggested gauge-invariant (or gauge-covariant) quantities in both the Abelian and the non-Abelian case \cite{AD16,ADMDB16}, as well as lattice equivalents to Maxwell's equations \cite{AD16}. Farrelly's PhD manuscript is based on four publications: two of them indeed deal with DTQW and QCA schemes for fermions \cite{Farrelly2014a, Farrelly2014b}, while the two others deal with quantum thermodynamics. I strongly recommend the reading of Ref. \cite{Farrelly15}.}. 

\item {\bf Mapping between LGTs and DTQW-GTs}
\label{item:LGTs} %2

It would be interesting to investigate to which extent it is possible to map these two a priori different spacetime discretizations of Yang-Mills gauge theories, if there is a mapping at all. Whatever the answer may be, one should discuss the pros and cons of each discretization, with respect to the other. Let us list several items to be investigated regarding those questions.

\begin{enumerate}
\item {\bf \emph{Lagrangian vs. Hamiltonian LGTs}} %2.1

There are two standard formulations of LGTs: (i) the Lagrangian formulation, seminaly introduced by Wilson (W) in 1974 \cite{Wilson74}, which is naturally associated to the path-integral formalism\footnote{The scattering-matrix elements are viewed as statistical-physics correlators through a Wick rotation.}, and  (ii) the Hamiltonian formulation, seminaly introduced by Kogut and Susskind (KS) in 1975\footnote{Both papers were submitted to Phys. Rev. D one month apart.} \cite{KogSuss75a}, which is naturally associated to the canonical formalism\footnote{The scattering-matrix elements are computed as in canonical quantum physics, with Hermitian operators acting on multiparticle Hilbert spaces (so-called Fock spaces).}. In KS-LGTs, the time-evolution unitarity is manifest, as in DTQW-GTs, whereas this unitarity is far from manifest in W-LGTs. We refer the reader to Appendix \ref{app:unitarity} for a few more comments regarding this matter\footnote{Recall that the lattice is a possible regulator for field theories, but not the only one.}. Eventually, it is straightforward to derive, from KS-LGTs, the lattice on-shell dynamics of either the matter field or the gauge field: one indeed simply needs to omit from the full Hamiltonian the irrelevant terms. If one works instead with W-LGTs, one ultimately needs to define a variational principle on the lattice to achieve such a purpose.

\item {\bf \emph{Link variables}} %2.2

In LGTs, the gauge fields are naturally associated to the links of the lattice. In DTQW-GTs, they have been defined on the nodes \cite{AD16,ADMDB16}. In LGTs, each spin component can be transported in both directions on a given lattice axis\footnote{See, e.g., Eqs. (12) and (13) in \cite{ZO17} (1D case): for each spin component, the Hermitian-conjugate term of (12) accounts for the propagation in the other direction.}, while in DTQW-GTs, the upper component is always transported in the same direction along a given axis of the spatial lattice, and the lower component is transported along the other direction along this axis; this is obvious given that the transport is determined by the spin-dependent shift. Now, the gauge field is implemented, in DTQW-GTs, by the spin-dependent phase. Let us write Eq. (1) of \cite{AD16} in the case of vanishing mixing angles $\theta^+$ and $\theta^-$ \footnote{For the scheme to deliver Dirac propagation, these angles must have value $\pm \pi / 4$ at zeroth order in the step of the lattice, i.e. they cannot vanish as we consider here. The present consideration has \emph{only} a pedagogical purpose, since it enables to see how the spin-dependent phase is related to the spin-dependent shift, in \emph{both} directions, putting aside the spin coupling.}:
\begin{align}
\psi^-_{j+1,p,q} &= e^{(\alpha_2)_{j,p,q}} e^{(\alpha_1)_{j,p,q+1}} \psi^-_{p+1,q+1} \\
\psi^+_{j+1,p,q} &= e^{-(\alpha_2)_{j,p,q}} e^{-(\alpha_1)_{j,p,q-1}} \psi^+_{p-1,q-1} \, ,
\end{align}
with
\begin{equation}
(\alpha_{\mu})_{j,p,q} = i \epsilon q (A_{\mu})_{j,p,q} \, ,
\end{equation}
where $\epsilon$ is the lattice spacing, the electric charge equals $-1$ in \cite{AD16}, and $(A_{\mu})_{j,p,q}$ is the covariant $\mu$ component of the continuum gauge field, but defined only on the nodes of the lattice.
In our scheme, $\psi^-$ goes left in time and $\psi^+$ goes right. Now, notice that the gauge field can be associated to the links, exactly as in LGTs. Indeed, let us \emph{define}
\begin{align}
U_{(j,p,q)\rightarrow(j,p+1,q)} &\equiv (U_{\mu=1})_{j,p,q} \equiv e^{i \epsilon q (A_{\mu=1})_{j,p,q}} \\
U_{(j,p,q)\rightarrow(j,p,q+1)} &\equiv (U_{\mu=2})_{j,p,q} \equiv e^{i \epsilon q (A_{\mu=2})_{j,p,q}} \, ,
\end{align}
where the arrow has been defined according to the opposite of the time evolution of the matter field, which is the same as in LGTs\footnote{Indeed, in LGTs, a non-daggered transporter $p \rightarrow p+1$ is conventionally associated, in the gauge-matter coupling term of the Hamiltonian, to a particle annihilation at $p+1$ and a particle creation at $p$. See, e.g., Eqs. (12) and (13) in \cite{ZO17}.}. These two expressions match exactly with Eq. (5.18) of \cite{book_Rothe}. We now see that link variables can be considered in DTQW-GTs, which are the same as in LGTs\footnote{Up to (i) the fact that time is continuous in KS-LGTs, and (ii) the difference mentioned in the next item.}\textsuperscript{,}\footnote{Similar gauge-field link variables are also described in Section 3.4.2 of \cite{Farrelly15}.}. With the help of such a formulation, it may be possible to view the differences between LGTs and DTQW-GTs as a choice of which transporters we choose to apply to build the time evolution, and in which order we choose to apply them, that is, a choice in the way we connect the nodes by the links, from a given instant to the next one. This may as well be another way of viewing the fact that the differences between LGTs and DTQW-GTs substantially lie in the way one discretizes the Dirac equation.

\item {\bf \emph{Evolution-operator vs. Hamiltonian gauge invariance}} %2.3

In KS- (resp. W-) LGTs, the gauge-field exponentials are introduced in the Hamiltonian (resp. Lagrangian). This enables to preserve gauge invariance at the level of the spacetime-discretized Hamiltonian (resp. Lagrangian). In DTQW-GTs, these exponentials are introduced in the evolution operator, so that what appears in the effective Hamiltonian\footnote{This effective Hamiltonian is generically obtained from the evolution operator, defined as a product of unitaries, by using the Baker-Campbell-Hausdorff formula.} is the argument of the exponential, i.e. what will give the gauge field in the continuum. This makes DTQW-GTs a priori more natural in the sense that the fact that one has gauge-field exponentials is only due to the time-evolution unitarity; in other words, one does not need to exponentiate the continuum gauge field to ensure gauge invariance; this gauge invariance is instead ensured by an appropriate definition of the DTQW operator in terms of a product of well-chosen unitaries.

\end{enumerate}

\item {\bf Mappings between tight-binding condensed-matter systems and DTQW schemes} \label{eq:Condensed}

\begin{enumerate}
\item {\bf \emph{An application of Item}} \ref{item:LGTs}

If a mapping is found between LGTs and DTQW-GTs, the present item, \ref{eq:Condensed}, can be seen as an application of such a hypothetical result. Indeed, there are already well-established correspondences between LGT prescriptions to implement gauge fields and the so-called Peierls substitution to implement a magnetic field in (condensed-matter) tight-binding Hamiltonians (TBHs). Now, some of these TBHs can be engineered with several quantum setups, such as, e.g., cold atoms in optical lattices \cite{dalibard10a,Goldman2014}.  A seminal experimental proposal in the field of the quantum simulation of TBHs with cold atoms coupled to synthetic gauge fields, by Osterloh et al. \cite{Osterloh2005}, makes it fully explicit this correspondence between LGTs and cold-atom engineered systems -- in the case of a 2D lattice and for a particular non-Abelian SU(2) gauge field. These correspondences have now been extended to very generic TBHs \cite{Zohar2015}, and underlie all current proposals to quantum simulate high-energy lattice QFTs with low-energy engineered systems.

\item {\bf \emph{Weaker mappings} }

One can also directly try to simulate TBHs with DTQWs, without trying to map, at least explicitly, DTQW-GTs to LGTs. To achieve such a purpose, one may want to build intuition from the quantum simulation of TBHs with LGTs. Notice that TBHs are non-relativistic, while DTQWs are intrinsically relativistic. This is a priori a notable difference, since, from a given instant to the next one, there is, in the relativistic case, a bounded causal neighborhood, while this is not the case in TBHs, i.e. there is a non-vanishing probability that a particle moves of an arbitrary big distance from a given instant to the next one\footnote{This fact was reminded me by A. Alberti. The result is also recalled, and proved, in the introduction (Chapter 1) of Farrelly's PhD manuscript \cite{Farrelly15}. In the sense that they preserve the relativistic causal neighborhood, DTQWs seem to be models a priori more suited to the discretization of continuum-spacetime relativistic dynamics than KS-LGTs, which, since essentially implementing TBHs in continuous time, do not preserve the above-mentioned causal neighborhood. In his introduction, Farrelly reviews several arguments in favor of -- but also some against -- time discretization in quantum simulations of QFTs \cite{Farrelly15}.}\textsuperscript{,}\footnote{Although this probability must, on average, decrease with distance, above some distance threshold.}. One may need to take a non-relativistic limit as in Strauch's paper (infinitely-strong mass coupling, i.e. $\theta \rightarrow \pi/2$ in Eq. (\ref{eq:coin_operation}).) \cite{Strauch06a}, so that the time coordinate of DTQWs and that of TBHs would not be the same. The question would then be how to obtain a TBH from the non-relativistic limit of a DTQW, intead of obtaining the space-discretized Schrödinger equation as in \cite{Strauch06a}. Indeed, while mappings between DTQWs and TBHs have been suggested, these connections are only phenomenal, i.e. there is no rigorous connection between both; see, e.g., \cite{AADSWW11} and \url{http://science.gu.se/digitalAssets/1581/1581243_tosini.pdf}. Let us recall several applications of the coupling of TBHs to gauge fields:
\begin{enumerate}
\item \emph{Abelian case}

These are TBHs with a superimposed electromagnetic field. This system has a myriad of applications, e.g., all the quantum Hall effects. Regarding quantum Hall effects, the two following very complete references may be consulted: \cite{Goerbig_review,Tong2016}.

\item \emph{Non-Abelian case}

A well-known condensed-matter non-Abelian gauge theory is that of the Rashba coupling, which describes how the Pauli spin-orbit term affects TBHs \cite{Berche2012}.
Notice that the spin-orbit coupling contributes to the the energy gaps in the bandstructure of solids; in the case of graphene, e.g., this spin-orbit coupling accounts for a (very) small gap \cite{Konschuh2010}, that can be neglected in a first approximation, which delivers the well-known graphene Dirac `strict' i.e. `massless' cones.
Eventually, an idea regarding the interpretation of the overlap integrals of TBHs in terms of the DTQW parameters is mentioned at the end of Appendix \ref{app:quasimomentum}, which could be investigated further.
\end{enumerate}

\end{enumerate}

\end{enumerate}

\setcounter{section}{0}

\starchapter{\textsc{Publications, internships \& communications}}{Chap:Publications}
\def\gmarge{\advance\leftskip \lengindent}
\setlength\parindent{-\lengindent}

\section*{Publications}

\begin{enumerate}

\item Pablo Arnault and Fabrice Debbasch.  \\
Landau levels for discrete-time quantum walks in artificial magnetic fields. \\
{\bf Physica A}, 443:179--191, 2015. \\
Received 12 December 2014, Revised 24 July 2015, Available online 15 September 2015, Published 1 February 2016. \\
\url{https://doi.org/10.1016/j.physa.2015.08.011}.

\item Pablo Arnault and Fabrice Debbasch.  \\
Quantum walks and discrete gauge theories. \\
 {\bf Phys. Rev. A}, 93:052301, 2016. \\
Received 28 July 2015, Revised manuscript received 16 October 2015, Published 2 May 2016. \\ 
\url{https://doi.org/10.1103/PhysRevA.94.012335}.

\item Pablo Arnault, Giuseppe Di Molfetta, Marc Brachet and Fabrice Debbasch. \\
Quantum walks and non-Abelian discrete gauge theory. \\
{\bf Phys. Rev. A}, 94:012335, 2016. \\
Received 6 May 2016, Published 19 July 2016. \\
\url{https://doi.org/10.1103/PhysRevA.94.012335}.

\item Pablo Arnault and Fabrice Debbasch. \\
Quantum walks and gravitational waves. \\
{\bf Ann. Physics}, 383:645-661, 2017. \\ 
Received 8 September 2016, Accepted 7 April 2017, Available online 27 April 2017. \\
\url{https://doi.org/10.1016/j.aop.2017.04.003}.

\end{enumerate}

\setlength\parindent\lengindent

%\section*{Visitings and communications}

\section*{Internships}

\begin{enumerate}
\item {\bf April and May 2016: 2-month internship} in the team of Dieter Meschede and Andrea Alberti, at the Institute for Applied Physics (IAP), Bonn, Germany, with a grant from the group; I am indebted to Dieter Meschede and Andrea Alberti for both their invitation and financial support. Ongoing collaboration on 2D DTQWs coupled to magnetic fields, with a focus on topological magnetic phases and the bulk-edge correspondance. In-progress experiment, with an analog quantum simulator consisting of pseudo-spin-1/2 cold atoms trapped in a 2D periodic potential generated with laser beams; one can drive the atoms on the 2D lattice, left or right and up or down, according to their pseudo-spin. For details on both (i) the study of topological effects with DTQWs (in the absence of magnetic field), and (ii) on the experimental setup, see the paper by Thorsten Groh et al.  \cite{Groh2016}. More details on the experimental setup are available on the website of the group, \url{http://quantum-technologies.iap.uni-bonn.de/en/quantum-walks/2d-quantum-simulator.html}. 

\item {\bf July and August 2016: 2-month internship} in the team of Yutaka Shikano, at the Center for Integrative Molecular Systems (Cimos), Okazaki, Japan, thanks to a grant of the JSPS Summer Program; I warmfully thank both the JSPS for their support, and Yutaka Shikano for his invitation and additional support. In-progress theoretical work on both (i) a DTQW self-induced curved metric, started by Giuseppe Di Molfetta, and (ii) a close examination of the interference terms in the feed-forward DTQW, which is used for the quantum simulation of anomalous diffusion \cite{SWH14}; possible future collaboration with Norio Konno and Etsuo Segawa on this second project. 

\end{enumerate}

\section*{Communications}

\begin{enumerate}
\item {\bf Poster} on publications \cite{AD15} and \cite{AD16}, at the First International Conference for Young Quantum-Information Scientists (YQIS), November 2015, Institut d'Optique Graduate School (IOGS), Palaiseau, France.  The book of abstracts is available at \url{https://yqis15.sciencesconf.org/conference/yqis15/pages/BoA_YQIS2015_IOGS_2.pdf}.

\item {\bf Poster} on the research projects I was involved in at that time, during the introduction week of the JSPS Summer Program 2016. Don't hesitate to visit the JSPS website for information on its activities and fellowships, at \url{http://www.jsps.go.jp/english/programs/index.html}. For one-page-long texts, each sharing a personal experience, of the French fellows of the JSPS Summer Program 2016, and older ones, go to \url{http://jsps.unistra.fr/les-bourses-de-la-jsps/summer-program/}.

\item {\bf 25-min talk} covering all the publications I coauthored as a PhD student \cite{AD15,AD16,ADMDB16,AD17}, at the Workshop on Quantum Simulation and Quantum Walks (WQSQWs), November 2016, Faculty of Nuclear Sciences and Physical Engineering, Czech Technical University, Prague, Czech Republic. The book of abstracts is available at \url{http://wqsqw2016.phys.cz/WQSQW-abstracts.pdf}.

\item {\bf Poster} giving a compact presentation of the state-of-the-art ongoing construction of DTQW-based lattice gauge theories, at the conference Atomes Froid et Technologies Quantiques (AFTQ), Decembre 2016, \'Ecole Normale Supérieure, Paris, France. For an abstract of the poster, go to \url{https://aftq16.sciencesconf.org/126422/document}. For an abstract of the poster by Muhammad Sajid, which describes the in-progress work of the Meschede-Alberti group on quantum-Hall physics with the 2D DTQW analog simulator, go to \url{https://aftq16.sciencesconf.org/126328/document}.

\end{enumerate}

\cleardoublepage

\part*{Appendix \label{Part:Appendix}}

\begin{appendices}
  %%%%%%%%%%%%%%%%%%%%%%%%%%%%%%
%%  Alias bibliographiques  %%
%%%%%%%%%%%%%%%%%%%%%%%%%%%%%%
\defcitealias{NRFortran}{Numerical Recipes}
\defcitealias{NRF77}{Numerical Recipes}
\defcitealias{NRF90}{Numerical Recipes}
\defcitealias{IUPAC}{IUPAC}%International Union of Pure and Applied Chemistry

%\chapter{\textsc{Compléments mathématiques}\label{Ann:Maths}}% Intermezzo mathématique

\chapter{Causal neighborhoods} \label{app:causal}

\iffalse
\footnote{In such an interpretation, a quantum \emph{system} only `exists' in our minds, and the ontological statements that one can make about the external world only concern quantum \emph{phenomena}, explained logically by this intellectual abstraction that the quantum \emph{system} is.}
\fi

Given a certain time interval $\Delta t$, let us define the \emph{relativistic causal neighborhood} around a point $P$ in physical space as the $P$-centered sphere of radius $R_{\mathrm{rel}} = c \, \Delta t$, $c$ being the speed of light. $R_{\mathrm{rel}}$  is much bigger than many usual macroscopic scales. Let us define a \emph{mesoscopic-transport (or continuum-medium) causal neighborhood} around a point $P$ in some medium, as the $P$-centered sphere of radius $R_{\mathrm{mes}} = v \, \Delta t$, where $v$ is the typical speed modulus of the matter of which the medium is made of. $v$ depends on the initial conditions and on the diverse phenomena which take place within the medium\footnote{These phenomena can be, e.g., convection, viscosity diffusion or heat diffusion. It is often useful to consider a different $v$ for each phenomenon. The terminology `typical speed' used above is actually obscure. One must distinguish between characteristic speeds and average speeds. Convection and diffusion are not related to any characteristic speeds, but one can associate to these phenomena average speeds by $v=d \, / \, \Delta t$, where $d$ is the mean displacement induced by the considered phenomenon over  $\Delta t$.}. 

The simple qualitative picture accounting for the possibility to consider mesoscopic samples of matter as pointlike, is the following. There are microscopic particles, namely atoms and molecules, which are either (i), in the case of fluids, subject to many collisions between each other, due to particular thermodynamical conditions such that the mean free path between two collisions is precisely below the mesoscopic scale, or (ii) well arranged, as in solids, and oscillate around their equilibrium position, with an amplitude which is also (much) below the mesoscopic scale considered. This makes it possible to consider that, on sufficiently short time scales $\Delta t$, a microscopic particle located at the center of some mesoscopic volume cannot escape it\footnote{In the case of solids, $\Delta t $ is actually infinite if the considered system is isolated.}\textsuperscript{,}\footnote{The following examples are instructive. A point $P$ in vacuum can be reached, within a time interval $\Delta t = 1 \, \mathrm{s}$, by any photon located within the $P$-centered sphere of radius $R_{\mathrm{rel}} \simeq 3 \times 10^{8} \, \mathrm{m}$. Consider now that this point $P$ is in a sufficiently dense medium, such as, e.g., the Sun, which is essentially made of Hydrogen, and Helium at its center. Photons collide with the molecules of the medium. Assume that after each collision the new direction of the photon is random. If $P$ is in the radiative zone, it can only be reached by photons contained within the $P$-centered sphere of radius $R_{\mathrm{mes, rad}} = \sqrt{c \, \Delta t \, d} \simeq 145 \mathrm{m}$, where $d \simeq 7 \times 10^{-5} \,  \mathrm{m}$ is the mean free path of a photon in the Sun (i.e. the distance between two collisions). For details on this phenomenon, see, e.g., \url{https://media4.obspm.fr/public/AAM/pages_proba/exo-soleil.html}. In the convective zone, the energy carried by photons is mainly transported by convection of the hydrogen plasma: a point $P$ in the convective zone can only be reached by photons contained within the $P$-centered sphere of radius $\mathrm{max} \{ R_{\mathrm{mes, rad}}, R_{\mathrm{mes, conv}} \}$, where $R_{\mathrm{mes, conv}} = v_{\mathrm{conv}} \, \Delta t$ and $v_{\mathrm{conv}}$ is the average plasma-convection speed.\iffalse Consider, for example, a gas molecule having a speed modulus $v$ corresponding to ambiant temperature. It's mean free path at ambiant pressure is $d_{\mathrm{mes}}\simeq 93 \, \mathrm{nm}$, see \url{}. Within a time interval $\Delta t = 1 \mathrm{s}$between two collisions is $t_{\mathrm{mes}} = d_{\mathrm{mes}}/v$. Now, in \fi}. If, moreover, $\Delta t$ is much bigger than the time scale $T$ of the studied phenomenon, then mesoscopic particles keep their integrity during the phenomenon. It is not necessary, for the mesoscopic-particle or continuum-medium picture to hold, that these mesoscopic particles strictly keep their integrity all along the mesoscopic and macroscopic dynamics, but only that the average displacement of atoms which is induced by the microscopic dynamics be (much) smaller than the average displacement of the physical quantity described by the mesoscopic-transport equation\footnote{These quantities can indeed be (i) the speed of the mesoscopic samples of matter, but also (ii) the temperature, with no macroscopic or mesoscopic matter displacement, or (iii) the density, which may be transported on distances much bigger than those on which matter is transported; this is typically the case of density waves in continuum media. (It may be more intuitive to think of this as a transport of the \emph{spatial variations}, i.e. the \emph{gradient} of these quantities.)}. 

The atomistic collisional picture for fluids was initially developed by Boltzmann in the late 19th century, while the existence of atoms was far from widely accepted yet. If one takes into account special relativity, it is a natural idea to combine this collisional picture with the relativistic locality to which atoms are subject, in order to derive relativistic mesoscopic-transport equations; that being said, achieving this is far from straightforward. This point is briefly developed in the rest of the introduction, i.e. after the reference to this appendix in the introduction.
   %%%%%%%%%%%%%%%%%%%%%%%%%%%%%%
%%  Alias bibliographiques  %%
%%%%%%%%%%%%%%%%%%%%%%%%%%%%%%
\defcitealias{NRFortran}{Numerical Recipes}
\defcitealias{NRF77}{Numerical Recipes}
\defcitealias{NRF90}{Numerical Recipes}
\defcitealias{IUPAC}{IUPAC}%International Union of Pure and Applied Chemistry

%\chapter{\textsc{Compléments mathématiques}\label{Ann:Maths}}% Intermezzo mathématique

\chapter{On computers} \label{app:on_computers}

This appendix aims at mentioning a (very) few important milestones, both theoretical and experimental, in the history of both computation and computers, first classical, and then quantum. The appendix will end with the difficulties that have to be overcome in order to achieve the construction of a  universal quantum computer.

\section{Classical computer}

\subsection{Abstract classical computer}

In his 1936  seminal  paper \cite{Turing1937,Turing1938}, Turing described both (i) a-machines (for `automatic machines'), now called \emph{Turing machines} (TMs), and (ii) the now-called \emph{universal Turing machines} (UTMs). 

\subsubsection{Specific-task abstract classical computer: the Turing machine} \label{sec:TM}

A TM is an abstract machine, i.e. a mathematical object, let us call it $T$, that models a hypothetical actual machine that would implement an algorithmic task, which we first informally define as a given series of unambiguous operations, triggered by an input belonging to a certain set of possible inputs\footnote{We code these inputs essentially with binary numbers, i.e. with a series of zeros and ones.}. The TM always produces an output -- i.e. it always `halt' -- entirely determined by the input (and is in this sense deterministic). A TM can thus be viewed as implementing -- thanks to several `ingredients' and in a certain \emph{way} -- a mathematical function, and, again, the TM is viewed as a model for an actual machine implementing this mathematical function. The \emph{way} this function is implemented by the TM is encoded in the so-called `action table' of the TM, which indicates which operation should be done at each step\footnote{There are in general several possible action tables to compute the same function, and the question of the simplest one naturally arises.}. The action table lists all the possible configurations that the TM may encounter given the set of all possible inputs, and specifies the action to be performed for each of those configurations, in order to eventually make the TM produce the desired output. To implement the instructions of its action table, a hypothetical actual TM only needs to be able to perform 6 elementary operations, called \emph{primitives}. These concrete primitives are \emph{right}, \emph{left}, \emph{print}, \emph{scan}, \emph{erase}, and \emph{nothing/halt}\footnote{One could say that the notion of TM makes this `black box' which is introduced to teach the notion of function to young students, (\emph{more} than) white.}.

We have above informally defined the notion of algorithm. Actually, the notion of TM historically enabled to give a proper \emph{mathematical definition} of an algorithm: the latter is, in this perspective, any series of operations that can either (i) directly be performed by a TM, or (ii) at least reducible to Turing-machine-performable operations. Indeed, Turing eventually came up with the notion of TM by requiring it to capture and be able to realize any task that we intuitively view as an algorithm. Grounded on this notion of TM which, although still abstract, is more than strongly underlaid by a concrete realization, the field of algorithmics quickly developed. In such an abstract-algorithmics context, the Turing-machine primitives often translate into \emph{logical gates}, i.e. elementary logical operations with which one can build any algorithm. 

To sum up, an actual automatic machine (i.e. automaton) realizing a certain task triggered by an input which can be encoded in a series of zeros and ones, that we may call specific-task (classical) computer, can always be described, theoretically, by a certain TM (i.e. having a certain action table and a certain set of possible inputs). The Turing-machine model is thus essentially the most general mathematical model for a specific-task (classical) computer.

From this theoretical point of view, the first legacy of Turing and his Turing-machine model may be, as several authors suggest, to have given the proper way of thinking algorithmically, i.e. the way to design practical mathematical algorithms, that could all be implemented on machines through the same principles, with, if not none, at least only a few technical modifications. I make this remark because the architecture of actual current computers is way more sophisticated and differs, in several aspects, from the too basic one defined by TMs (or, rather, UTMs), as we briefly underline further down.

\subsubsection{Universal abstract classical computer}

\vspace{0.1cm}
\paragraph{$\bullet$ The universal Turing machine}
\hspace{10cm}

\vspace{0.1cm}
\noindent
The obvious limitation of an automaton based on a certain TM is that it can only perform the task encoded in the action table. To perform other TM-type tasks, one has either to (i) build another automaton which realizes another action table, or, if possible, to (ii) reprogram the automaton with another action table. 

Turing thus proposed, in the same above-mentioned 1936 seminal paper \cite{Turing1937}, the now-called UTMs. A UTM $U$ takes as input a certain TM $T$ and a possible input $i$ for $T$. By encoding the action table of $T$ as a standard Turing-machine input, i.e. as a series of one's and zero's, and by (abstractly) storing it in its memory, $U$ can then, by reading this encoded action table together with the input $i$, produce the same output as $T$ would produce with input $i$, which is noted $U(T,i)=T(i)$.

It is important to note that a UTM is not more than a TM in terms of the ingredients it is made of, i.e. a UTM is a TM, with additional functioning specifications \emph{only}. More relevantly: once we have mathematically defined the notion of TM, it is a theorem that there exist a certain TM, that we call UTM, which does the job of any TM -- that we give the UTM as an encoded input -- \emph{as efficiently as the TM itself}\footnote{See, e.g., Theorem 2 in the following notes, \url{https://homes.cs.washington.edu/~anuprao/pubs/CSE531Winter12/lecture1.pdf}. The notion of \emph{efficiency} has a precise meaning in computer science.}. In the discussions about computational power, we will thus from now on often drop the `U' in `UTMs'. 

\vspace{-0.3cm}
\paragraph{$\bullet$ Other models of computation}
\hspace{1cm}

\vspace{0.1cm}
\noindent
As implicitly suggested at the end of Subsubsection \ref{sec:TM}, one can conceive other models of computation than the TM. Let us try to give both an intuitive and general picture of what a \emph{computational model} is.

\vspace{-0.4cm}
\subparagraph{$\star$ Computational model.}
\noindent
A computational model may be viewed as the given of (i) a set of \emph{elementary operations}, called \emph{primitives}, such as that of the TM, listed above, or such as logical gates, and (ii) a \emph{background}, such as the whole TM setup, described above, or a graph, on which one positions and relates the primitives according to some \emph{rules}. All these elements are defined in order for the  computational model to achieve a certain \emph{class of computational aims}. 

\vspace{-0.4cm}
\subparagraph{$\star$ Universal computational model.}
\noindent
The computational model is considered \emph{universal} when the class of computational aims it can fulfill is sufficiently broad according to some \emph{criteria}, or \emph{requirements of reference}. Designing a universal computational model thus demands to find both a set of primitives and a way of chaining them that enable to meet the universality requirements. A standard universality requirement -- and thus definition -- is that the computational model have (at least) the same computational power as a (U\footnote{In the discussions about computational power, we will from now on definitely drop the `U'.})TM, in which case the model is said Turing-complete\footnote{`Having the same computational power as' means (i) `being able to realize the same tasks as' and (ii) `as efficiently as'.}. Other universality criteria may be useful, depending one's computational aims. Taking the TM as a reference is, in particular, underlaid by the so-called \emph{Church-Turing thesis}, which states that TMs should be able to implement any type of `automatic task'. Because the notion of `automatic task' is not (meant as being) mathematically defined in the Church-Turing thesis and its extensions, these theses are not mathematical claims that call for mathematical proofs or disproofs, but rather philosophical or metamathematical claims that call for mathematical definitions for notions such as `automatic task' or `computation'\footnote{This leads to classifying computational models into equivalence classes of robust computability notions, i.e. computability notions which are frequently found to be equivalent from one computational model to the other.}.

An example of widespread computational model is the \emph{circuit model}. This model is not Turing-complete, but its extension, the \emph{circuit-family model}, is more than Turing-complete, i.e. a circuit family (CF) may compute a broader class of tasks than a TM.  By the denomination `circuit model', one sometimes refers to the circuit-family model, since, being not Turing-complete, the mere circuit model is not of sufficient interest. There is a subclass of CFs which is equivalent to (polynomial-time) TMs, namely, \emph{uniform CFs}.

\subsection{Actual classical computer}

\subsubsection{Specific-task actual classical computer}

An example of such a machine is the Turing (electromechanical) bomb, which was build by Turing himself, together with Welchman, in order to decrypt the encrypted messages created by the Nazis during World War II thanks to Enigma machines, which are another example of specific-task computer.

Specific-task computers were later also designed to treat and store the data generated by scientific experiments, e.g., in radioastronomy.

\subsubsection{Universal actual classical computer}

The theoretical description of such machines, i.e. the so-called `stored-program (abstract) computer', is strongly reminiscent of the UTM, although it differs from it in several aspects. 

The essence of stored-program computers is contained in the so-called `Von Neumann architecture', seminaly introduced in 1945\footnote{This architecture was seminaly described in the \emph{First draft on the EDVAC}, a reproduction of which is available at \url{https://web.archive.org/web/20130314123032/http://qss.stanford.edu/~godfrey/vonNeumann/vnedvac.pdf}. Von Neumann explicitly mentioned UTMs as a source of inspiration.}, and both terminologies are sometimes used as synonyms. Stored-program computers encode the input instructions describing a program\footnote{A `program' can be seen as either a less-abstract version of an `algorithm', or simply as a different terminology to refer to algorithms.} as standard data which is stored in their main (or primary) memory, the RAM i.e. random-access memories\footnote{This memory must be quickly accessible by the processor, or CPU (Central Processing Unit), since it stores encoded programs that should be accessible to be run \emph{at an arbitrary time and from an arbitrary location}, hence the denomination `random access'.}, and then read and execute these data as a standard input, as UTMs do with input TMs. The encoding and the execution of the source code -- which is written in a `human' language (usually, english) -- can be performed successively as a single block by an \emph{interpreter}. In this case, the interpreter stores the encoded source code in the RAM before executing it. The encoding and the execution can also be separated tasks. In this second case, the encoding is done by a \emph{compiler}, and the compiled i.e. encoded source code is stored in the so-called `data memory' (to be `opposed' to the RAM), for `future' execution; the file containing the compiled code is directly executable by the computer, whenever wished.

\section{Quantum computer}

\subsection{Abstract quantum computer}

\subsubsection{Specific-task abstract quantum computer}

\vspace{0.1cm}
\paragraph{$\bullet$ The quantum Turing machine}
\hspace{10cm}

\vspace{0.1cm}
\noindent
The notion of \emph{quantum TM} (QTM) was seminaly introduced in Deutsch's 1985 paper, `Quantum theory, the Church-Turing principle and the universal quantum computer', see \url{http://old.ceid.upatras.gr/tech_news/papers/quantum_theory.pdf}. The notion of QTM enables, in particular, to capture \emph{quantum algorithms} (see below) in action tables, and may shed light on the implementation of quantum algorithms by an actual quantum computer. 

\iffalse
Let us however recall that explicitly recasting a (quantum) algorithm as a (quantum) TM, is usually not necessary, since current algorithms, either classical or quantum, fit well enough to the mathematical definition of an algorithm which was provided by the seminal definition of TMs. Current classical algorithms may be designed according to the computational model which fits best to the architecture of the computer on which one wishes to implement it.
\fi

\vspace{-0.3cm}
\paragraph{$\bullet$ Quantum algorithms}
\hspace{10cm}

\vspace{0.1cm}
\noindent
As in the classical case, quantum algorithms may be designed according to other \emph{models of quantum computation} than QTMs (see below). That being said, all quantum algorithms share the following characteristics.
Instead of manipulating \emph{binary data}, they manipulate what we may call \emph{quantum binary data}, i.e. (i) the \emph{(logical\textsuperscript{}\footnote{To be opposed to \emph{physical} qubits. We will drop the precision `logical' from now on in this paragraph, but it is implicit since we are in an abstract context.}\hspace{-0.1cm}) bits} are replaced by \emph{qubits}, which can be in any superposition of zeros and ones, and which we could call  \emph{superposed bits}, (ii) the \emph{bit strings} are replaced by \emph{entangled qubit strings} and (iii) the \emph{classical gates} are replaced by \emph{quantum gates}.

Two important quantum algorithms, developed in the 90's, are Shor's \cite{Shor94,Shor97}, for prime-number factorization, and Grover's \cite{Grover96}, for database search.

\subsubsection{Universal abstract quantum computer}
 
 \vspace{0.1cm}
\paragraph{$\bullet$ The universal quantum Turing machine}
\hspace{10cm}

\vspace{0.1cm}
\noindent
The notion of \emph{universal QTM} was, together with that of QTM, also naturally introduced in Deutsch's seminal work (see above). Note that this is three years after Feynman's seminal 1982 paper \cite{Feynman1982}, which was about universal quantum \emph{simulation} -- of (local) physics.

\vspace{-0.3cm}
\paragraph{$\bullet$ Other~models~of~universal~quantum~computation}
\hspace{3cm}

\vspace{0.1cm}
\noindent
Quantum algorithmics has, from Deutsch's seminal work, developed enough for several realistic theoretical models of \emph{universal} quantum computation to be suggested within the last 10 years, other than universal QTMs. Examples of such models are that by Childs, based on CTQWs \cite{Childs2009,CGW13}, or that by Lovett et al., based on DTQWs \cite{Lovett2010}, which is the DTQW version of the first paper Childs \cite{Childs2009}.

There is also a quantum counterpart to classical circuit families (CFs), namely \emph{quantum CFs} (QCFs).
The following review work on the ``computational equivalence between QTMs and QCFs'', \url{http://www.math.ku.dk/~moller/students/chrW.pdf}, ends with the following conclusion: ``Finitely-generated QCFs are thus an intermediate of polynomial-time uniformly generated QCFs and uniform QCFs, such that their computational power exactly matches that of polynomial-time QTMs.'' This makes me believe that the computational power of QCFs relative to QTMs is lower than that of CFs relative to TMs, since uniform CFs are equivalent to (polynomial-time) TMs. That being said, I am currently not enough qualified to be certain about such a statement, and even if this statement is true, I currently do not master well enough the various concepts I have introduced to speak further about this.

\subsection{Actual quantum computer}

\subsubsection{Specific-task actual quantum computer}

These computers are currently developing. They work at best with, in average, a few tens of physical qubits. The company IBM has given online free access to a five-qubit quantum computer, see \url{https://quantumexperience.ng.bluemix.net/qx/editor}, and it also offers online access to a 16-qubit one, for the use of which one has to pay.

\subsubsection{Universal actual quantum computer}

The current difficulty in achieving the construction of a universal quantum computer is essentially a technical one, namely, scalability. Scalability is the scale-up potential of a device. The scalability of current quantum-computing devices is poor, if one aims at beating classical computers. This is essentially due to the difficulty of increasing the size of memories, i.e. the difficulty to manipulate a high number of qubits coherently.

Standard (personal) current (universal) classical computers, use several Gigabytes of RAM, i.e. tens of millions of physical bits to store programs to be executed, let us say around 100 million physical bits.

\newpage
The following lines have been produced essentially after reading the broad-audience untechnical article `Quantum Computing: How Close Are We?', by E. Cartlidge [Cart], see \url{https://www.osa-opn.org/home/articles/volume_27/october_2016/features/quantum_computing_how_close_are_we/}. With current quantum gate fidelity and quantum error tolerance (see below), Martinis et al. \cite{Fowler2012} estimate that around 100 million physical \emph{qubits}, manipulated coherently, would be needed to carry out Shor's factoring algorithm on a 2000-bit number in around a day, which is a task ``far beyond the reach of today's best supercomputers'' [Cart]. 

Several technological candidates currently compete in the design of future universal quantum computers (the physical architecture of such machines may involve several of these techniques, but there will likely be a predominant one). Four of such candidates are summed-up in [Cart]: spin qubits, superconducting circuits, ion traps, and integrated photonic circuits. Cold atoms in optical lattices, which are not in the list, have also been considered for large-scale quantum computation, e.g. by Inaba et al. in 2014 \cite{Inaba2014} (see also the larger-audience paper at \url{https://phys.org/news/2014-03-entangling-atoms-optical-lattice-quantum.html}).

   %%%%%%%%%%%%%%%%%%%%%%%%%%%%%%
%%  Alias bibliographiques  %%
%%%%%%%%%%%%%%%%%%%%%%%%%%%%%%
\defcitealias{NRFortran}{Numerical Recipes}
\defcitealias{NRF77}{Numerical Recipes}
\defcitealias{NRF90}{Numerical Recipes}
\defcitealias{IUPAC}{IUPAC}%International Union of Pure and Applied Chemistry

%\chapter{\textsc{Compléments mathématiques}\label{Ann:Maths}}% Intermezzo mathématique

\chapter{On so-called quantum non-locality} \label{app:on_quantum_non_locality}

\emph{We remind the reader that the doubled quotation marks are used to quote other authors. These quotes are always followed by the reference from which they have been extracted.} \\

\noindent
In their well-known 1935 paper, Einstein, Podolsky and Rosen (EPR) \cite{Einstein1935} exposed a certain quantum-mechanical thought experiment that they considered as paradoxical, the so-called EPR paradox. According to them, this paradox implied the incompleteness of quantum mechanics (QM) as a theoretical tool to predict results about Nature. By incompleteness, they explicitly meant that there are ``elements of the physical reality'' \cite{Einstein1935} which have no ``counterpart'' \cite{Einstein1935} in the theory, i.e. to which no ``physical quantity'' \cite{Einstein1935} is associated in the theory. It is suitable for the purpose of the present discussion to rephrase EPR's statement as follows:  there is information in Nature that QM is unable to capture, \emph{while it obviously should}. The current overwhelmingly-dominant vision, that progressively emerged out of the analysis of many experiments of increasing both subtlety and precision, is that this obviousness was simply not, and that QM is an \emph{ontologic revolution}, in the following sense: a particle does not have both a definite position and momentum before we perform a measurement of one of these two quantities; in other words: prior to measurement, position and momentum \emph{are not determined}, or, in stronger words, \emph{do not exist} as classically understood\footnote{{\bf \emph{A `preview' to what follows in this appendix.}} This ontologic revolution is associated to the fact that some information is sometimes not accessible locally, where the word `locally' is understood in a sense which is \emph{stronger} than the relativistic one, and which is that of `local measurement', seminally defined by local hidden variables, see further down, and later extensively discussed \cite{Peres2004,book_Barnett}. Indeed, if we set aside universe's expansion, the information we are speaking about is \emph{always} locally accessible in a relativistic sense, since during the entangled-state preparation, each of the two subsystems must be within the light cone of the other, so that this remains true afterwards. Although, for the sake of simplicity, we just identified \emph{quantum non-locality}, which I may also call \emph{non-localizability}, to entanglement, there are actually some non-entangled states that can also encode non-local information. Conversely, one can show that maximal entanglement does not always provide maximal quantum non-locality \cite{Junge2011,Vidick2011,Liang2011}.}.

When introducing EPR's paradox, the above-mentioned obviousness is often presented, following EPR's line of argument, as grounded on a certain conception of ``reality'' \cite{Einstein1935}. Indeed, EPR's paper seems to hope to convince the reader by, essentially, elaborating on a ``reasonable'' \cite{Einstein1935} definition of reality, that they attempt to give\footnote{Bohr criticized the vagueness of the concepts developed by EPR.}. Throughout the paper, their concept of reality is basically used as a physical  characteristic that can take two values, either `real' or `unreal'. That being said, such a presentation of EPR's paradox, which strictly follows EPR's formal presentation, structured around the notion of reality by an extensive use of logical inferences, tends to overlook the background on which the paper was conceived, which is quite enlightening.

One of the key aspects of EPR's issue is that the axioms of quantum mechanics give no information about the physical quantity under consideration before it is measured. Let use call this epistemic characteristic  \emph{(pre-measurement) undetermination}. We shall come back to it further down. Another key aspect of EPR's issue is the axiom of quantum mechanics which states that, instantaneously or extremely quickly after measuring the system, the state of the system changes and becomes that of the outcome. Let us call this axiom the \emph{state-reduction axiom}. In the Copenhagen interpretation of QM, largely conceived by Bohr and Heisenberg around 1930, this axiom is viewed as a ``kind of action'' \cite{book_Heisenberg}, i.e. as accounting for a physical process, the so-called \emph{wavepacket collapse}, which must be superluminal, although Heisenberg also quickly pointed out that this kind of action ``can never be utilized for the transmission of signals so that it is not in conflict with the postulates of the theory of relativity'' \cite{book_Heisenberg}. Einstein criticized such an interpretation an referred to it as a ``spooky action at a distance'', reminiscent of the old gravitational and electromagnetic actions at a distance. 

Now, I still need to investigate about the following two historical possibilities. Either, first possibility, EPR were actually rather convinced by Heisenberg's no-signaling (or no-communication) claim, and were already advocating, in their paper, for the so-called \emph{local realism}\footnote{Note that this terminology is not used in their paper.}; in that case, their main concern would indeed have been pre-measurement undetermination. Or, second possibility, EPR were not convinced by no-signaling claims because no fully convincing proof of this result had been developed yet; in that second case, QM was, in EPR's thought experiment, breaking causality essentially \emph{because it was breaking relativistic locality}, where, here, `relativistic locality' stands at least for slower-than-light propagation of energy, if not for Lorentz invariance\footnote{This meaning will be used throughout the rest of this appendix}.  Both EPR's paper and the way the paradox is presented mainly suggests the first historical possibility, but EPR's line of argument is both subtle and vague enough to suggest that both concerns might have intertwined more than often presented.

Whatever concern was the dominant one to EPR, one can for sure say that the incompleteness of QM was to them unavoidable if one wished to preserve causality, whatever the meaning we assign to the word causality. 

Let me again follow the second historical possibility. It might be that incompleteness was EPR's answer to a result that they view as paradoxical, not that much because of their vision of reality, but rather because of their vision of what a physical theory must be\footnote{Regarding this matter, EPR's concept of completeness is precisely developed in the perspective of setting constraints on what a physical theory should be.}. Indeed, it might be that EPR could not admit that the state-reduction axiom was not physical in the usual sense, as Heisenberg suggested: because, to EPR, if one was willing to introduce this axiom, it had to be physical in the usual sense, then it was, in some situations, such as that they present in their paper, actually breaking relativistic locality and could not be a good or at least sufficient axiom. I follow this maybe unconventional line of thought because one might say that EPR's conception of reality was, essentially, merely that of a \emph{causal reality ensured by relativistic locality}. The no-signaling theorem ensures that this vision \emph{is still in order nowadays}. But rigorous proofs of the no-signaling theorem are rather subtle, and this result might not have been as firmly grounded at EPR's time, so that they could not be convinced by it. To prove the no-signaling theorem, indeed, one needs to fundamentally reconsider what a measurement actually is; it demands to realize that it involves both `microscopical' and `macroscopical' or `topological' aspects \cite{Peres2004, book_Barnett}.

In his well-known 1964 paper\footnote{A scan of the paper is available at \url{http://www.drchinese.com/David/Bell_Compact.pdf}.} [Bell1964], i.e. 29 years after EPR's paper, Bell still concluded that there is some signal that ``must propagate instantaneously'' [Bell1964], if his thought experiment happened to be verified, as he believed it would. Bell also added that this would imply that QM ``could not be Lorentz invariant'' [Bell1964].  The above-mentioned local-realist vision was developed between EPR and Bell's papers. A \emph{local realist theory} is, technically, a \emph{strict synonym} for a \emph{local hidden-variables theory}\footnote{Local hidden variables were mathematically defined by Bell, and somehow tailored to prove its well-known inequalities. Bell theoretically proved that no local hidden-variable theory can reproduce the full range of QM predictions.}. It is a \emph{much stronger} requirement than the sole respect of relativistic locality. At that time, however, it might be that this requirement was seen as, if not equivalent, at least not that stronger than the requirement of relativistic locality. For sure, the local-realist requirement was seen as needed to preserve slower-than-light propagation \emph{at least for this exotic action at a distance}\footnote{I need to investigate on how grounded was the no-signaling theorem at that time. If it was firmly grounded, then local realism was seen as needed \emph{only} for the exotic action at a distance of the wavepacket collapse to be subluminal. If not, then local realism was, \emph{more largely, seen as actually needed to firmly ground no-signaling}. This historical question is not obvious. Indeed, while there were for sure already several claims for no-signaling, why then, if this no-signaling was already firmly grounded, insisting on introcuding an exotic signal that conveys no information?}.

It seems, as the introduction of Bell's paper suggests, that one of the important triggers for this paper was Bell's realization that one could introduce hidden variables similar to those informally suggested by EPR, as shown by Bohm \cite{Bohm1952I,Bohm1952II}, but that, instead of reducing EPR's concern, this seemed to actually `naturally' augment what was viewed as a tangible ``gross'' [Bell1964] non-locality that violated slower-than-light propagation. We shall clarify these concepts and arguments below.

Several experiments are viewed as demonstrating, with an increasing precision, the violation of Bell's inequalities\footnote{In his paper, Bell underlined that it was ``crucial'' [Bell1964] to realize ``experiments of the type proposed by Bohm and Aharonov'' [Bell1964], ``in which the settings [of the measuring devices] are changed during the flight of the particles'' [Bell1964]. The first experiment of this kind was realized by Aspect et al. in 1982 \cite{Aspect1982}.}. This is viewed as a proof of the so-called \emph{non-locality} of the information carried by (quantum) physical systems, which can be summed-up as followed. Consider a certain physical system $S_1$; then, provided this system interacts \emph{in a certain way} with another one $S_2$ at some instant $t_0$, and provided the quantum coherence of both $S_1$ and $S_2$ is ensured for $t > t_0$, then $S_1$ will, for any $t > t_0$, contain information about $S_2$, \emph{regardless of any future interaction with $S_2$} (and vice versa). This fundamental property disappears with decoherence. Not to confuse it with relativistic locality, we may rather use the terminology \emph{non-localizability} instead of  \emph{quantum non-locality}.

The terminology `quantum non-locality' is sometimes almost used as a synonym for the \emph{non-separability} of entangled systems, although this non-localizability can appear in separable systems\footnote{Regarding these questions, one may first read Wikipedia's dedicated article, Section 2, at \url{https://en.wikipedia.org/wiki/Quantum_nonlocality}}. Quantum non-locality sometimes only refers to the type of non-locality which appears in Bohm's interpretation of quantum mechanics, which is actually rather rare in the Copenhagen interpretation, although it can happen\footnote{See the following discussion: \url{https://physics.stackexchange.com/questions/197530/non-locality-vs-non-realism-arbitrary-choice}.}.

\iffalse
 One can sometimes hear that the experimental disproof of local realism implies that one has to choose between locality and realism. This is a standard logical error: the negation, or, in a set-theory terminology, the complementary set of $A \cap B$ (local and realist) has been taken to be $A \text{or} B$ with an exclusif `or', which is the \emph{complementary set} of $A \cap B$, often denoted by $\complement (A \cap B)$, instead of taking the logical negation of $A \cap B$, denoted by $\bar{A \cap B}$, and which is known to be the union $\bar{A \cup B}$ 
\fi

A trace of no-signaling in QM is that the state evolution is described by a PDE, namely the Schrödinger equation, which by the way had already been given relativistic counterparts when EPR's paper was published, namely the Klein-Gordon and Dirac equations.

In the light of the no-signaling theorem, the Aharonov-Bohm effect can be viewed as a manifestation of the non-localizability of information: within such an interpretation, the electromagnetic potential should not be regarded as `more physical than the electromagnetic field', a picture sometimes conveyed, but as `containing non-local information', in the same way quantum entanglement can do \cite{Vaidman2012}.

\defcitealias{NRFortran}{Numerical Recipes}
\defcitealias{NRF77}{Numerical Recipes}
\defcitealias{NRF90}{Numerical Recipes}
\defcitealias{IUPAC}{IUPAC}%International Union of Pure and Applied Chemistry

%\chapter{\textsc{Compléments mathématiques}\label{Ann:Maths}}% Intermezzo mathématique

\chapter{Aharonov's quantum random walks} \label{app:Aharonovs_scheme}

\section{Generic scheme (or protocol)}

\subsection{Definition (one-step scheme)}

Aharonov's  scheme \cite{ADZ93a} takes as input a ket belonging to the position (or external) Hilbert space, that we will call `external ket' $| \psi_j \rangle_{\text{p}}$, where the subscript `p' is for `position', and outputs, after one time step, another external ket $| \psi_{j+1} \rangle_{\text{p}}$ given by
\begin{equation}
| \psi_{j+1} \rangle_{\text{p}} = \hat{A}^{\pm}_j  | \psi_j \rangle_{\text{p}} \, ,
\end{equation}
where I define the Aharonov operator by
\begin{equation}
\hat{A}^{\pm}_j = \hat{M}^{\pm}_j \hat{W}^{\text{a}}_j \hat{T}_j \, . 
\end{equation}
As formally written, the scheme is the succession of three operations:
\begin{enumerate}

\item The preparation of the spin in a given superposition along some direction. This superposition can be time dependent (this is not the case in the original paper though), which can always be viewed as, \emph{e.g.}, keeping the original superposition but along a time-dependent direction, with a possible overall phase change. This operation corresponds to tensorizing the external input ket with the prepared superposition, and reads:
\begin{equation}
\hat{T}_j = \left( \begin{matrix}
c^+_j \\ c^-_j
\end{matrix} \right) \otimes \, ,
\end{equation}
with the basis
\begin{equation}
|+\rangle_{\text{c}} = \left( \begin{matrix}
1 \\ 0
\end{matrix} \right) \, , \ \ \ \ \
|-\rangle_{\text{c}} = \left( \begin{matrix}
0 \\ 1
\end{matrix} \right) \, ,
\end{equation}
where the subscript `$\text{c}$' indicates that the kets belong to the coin Hilbert space.

\item A DTQW one-step operation, combination of a spin-dependent shift and a coin operation:
\begin{equation}
\hat{W}^{\text{a}}_j = \hat{C}^{\text{a}}_j \hat{S}^{\text{a}} = 
e^{i\omega_j} \underbrace{\begin{bmatrix}
\alpha_j & \beta_j \\ - \beta^{\ast}_j & \alpha^{\ast}_j 
\end{bmatrix}}_{\in \  \mathrm{SU}(2)}
\begin{bmatrix}
e^{-i\hat{P}} & 0 \\ 0 & e^{i\hat{P}}
\end{bmatrix} \, ,
\end{equation}
where $\hat{P}$ is the momentum operator, that acts on the external ket. The superscript `a' is to indicate the use of the `Aharonov' convention, in which, although the shift is applied before the coin operation, as in the convention used in this thesis and on the contrary to the standard convention (see Appendix \ref{app:time_reversal}), the shift however shifts the spin up to the right, and the spin down to the left, as in the standard convention and on the contrary to that used in this thesis. In the original scheme, the coin operation does not depend on time.

The spin-dependent shift operation, $\hat{S}^{\text{a}}$,  entangles the spin and external degrees of freedom of the initially non-entangled input state, given by 
\begin{equation}
| \Psi_j \rangle 
\equiv \hat{T}_j| \psi_j \rangle_{\text{p}} \, .
\end{equation}
Both an upper case psi letter and no subscript associated to the ket mean that the ket belongs to the full Hilbert space, tensorial product of the coin and position Hilbert spaces. One can view generic position-spin entanglement as making a spin state depend on position, which generates two probability amplitudes, that for spin up and that for spin down. The spin-dependent shift of DTQWs performs a particular type of entanglement, where the upper probability amplitude, $|\psi^+\rangle_{\text{p}} \equiv _{\text{c}}\langle + | \Psi \rangle$, is shifted by one space step in a given direction, while the lower, $|\psi^-\rangle_{\text{p}} \equiv _{\text{c}}\langle - | \Psi \rangle$, is shifted in the other direction. 

The coin operation, $\hat{C}_j^{\text{s}}$, superposes the upper and lower spin components of the entangled state, namely $\hat{S}^{\text{s}} | \Psi_j \rangle$.  This coin operation can be viewed as (case 1) one of the two following possibilities (i) and (ii), as well as (case 2) the result of a combination of these two possibilities: (i) a (non-destructive) rotation of the spin, and (ii) a measurement in a different basis (destructive rotation), thus corresponding to the operation $M_j^{\pm'} = M_j^{\pm} \hat{K}_j^{\text{s}}$, where $\hat{K}_j^{\text{s}}$ is either (case 1)  $\hat{C}_j^{\text{s}}$ itself or (case 2) a part of $\hat{C}_j^{\text{s}}$ (imagine we decompose $\hat{C}_j^{\text{s}}$ into two spin rotations, one corresponding to (i) and the other to (ii)).

\item A measurement of the spin, which desuperposes the spin state in some direction, i.e. determines the value of the spin along this direction, either plus or minus, which at the same time disentangles the quantum state, i.e. makes the spin state independent of the position (this is a possible way to view disentanglement). The operation $\hat{M}_j^{\pm}$ thus corresponds to two possible operations, as indicated by the superscript $\pm$,
\begin{equation}
\hat{M}^{\pm}_j =
 \frac{_{\text{c}}\langle \pm |}{\mathcal{N}^{\pm}_j} \, .
\end{equation}
The probabilities of the spin-measurement outcomes are given by the squared position-Hilbert-space norms, 
\begin{equation} \label{eq:plus_minus_probas}
\mathrm{P}^{\pm}_j = (\mathcal{N}^{\pm}_j)^2 \equiv  || {_{\text{c}}\langle \pm | \hat{W}^{\text{a}}_j | \Psi_j \rangle} ||_{\text{p}}^2 = \left( _{\text{c}}\langle \pm | \hat{W}^{\text{a}}_j | \Psi_j \rangle \right)^{\dag}  {_{\text{c}}\langle \pm | \hat{W}^{\text{a}}_j | \Psi_j \rangle} \, ,
\end{equation}
with
\begin{equation}
\mathrm{P}^+_j = 1-\mathrm{P}_j^-=\mathrm{P}_j \, . 
\end{equation}
Again, the subscript `p' indicates that the norm is that of the position Hilbert space. The indicated time dependence of this operation is that associated to the norm, i.e. to the product $W_j^{\text{a}}T_j$, and has nothing to do with the (quantum) randomness of the measurement outcome. As said above, this operation disentangles the state previously entangled by the spin-dependent shift operation, and selects only an external ket. We could define the measurement operation as usual, i.e. with a projector instead of a bra, but then the scheme anyways only keeps the external part of the state (and reinitializes the spin state), so we would have to introduce an operator for this selection, which would precisely be a bra that, applied on the corresponding projector, whould deliver the same bra; we then simply define the measurement by the action of a bra and not a projector.

\end{enumerate}

\subsection{Compact expressions for the one-step scheme}

Now that we have defined each of the three substeps of the one-step scheme, let us describe this one-step scheme compactly.
A more explicit expression of the action of the walk operator is 
\begin{equation}
\hat{W}_j^{\text{s}} | \Psi_j \rangle  = e^{i\omega_j}\left( \begin{matrix}
\alpha_j \,  e^{-i\hat{P}} | \psi^+_j \rangle_{\text{p}} +  \beta_j \, e^{i\hat{P}} | \psi^-_j \rangle_{\text{p}} \\
-\beta^{\ast}_j \,  e^{-i\hat{P}} | \psi^+_j \rangle_{\text{p}} +  \alpha^{\ast}_j \, e^{i\hat{P}} | \psi^-_j \rangle_{\text{p}} 
\end{matrix} \right)
\equiv
\left( \begin{matrix}
_{\text{c}}\langle + | \hat{W}_j^{\text{s}} | \Psi_j \rangle \\
_{\text{c}}\langle - | \hat{W}_j^{\text{s}} | \Psi_j \rangle
\end{matrix} \right) \, ,
\end{equation}
with
\begin{equation}
| \Psi_j \rangle \equiv \left( \begin{matrix}
_{\text{c}}\langle + | \Psi_j \rangle = |\psi^+_j\rangle_{\text{p}} \\
_{\text{c}}\langle - | \Psi_j \rangle = |\psi^-_j\rangle_{\text{p}} 
\end{matrix} \right)   .
\end{equation}
The $+$ and $-$ probabilities, (\ref{eq:plus_minus_probas}), are respectively given by
\begin{align}
&\mathrm{P}^+_j=|| _{\text{c}}\langle + | \hat{W}_j^{\text{s}} | \Psi_j \rangle ||^2_{\text{p}} = |\alpha_j|^2 ||\psi^+_j||^2_{\text{p}} + |\beta_j|^2 ||\psi^-_j||^2_{\text{p}} + 2 \, \text{Re}\left\lbrace \alpha_j \beta^{\ast}_j \ {_{\text{p}}\langle \psi^-_j | e^{-2i\hat{P}} | \psi^+_j \rangle_{\text{p}}} \right\rbrace \nonumber \\
&\mathrm{P}^-_j=|| _{\text{c}}\langle - | \hat{W}_j^{\text{s}} | \Psi_j \rangle ||^2_{\text{p}} = |\beta_j|^2 ||\psi^+_j||^2_{\text{p}} + |\alpha_j|^2 ||\psi^-_j||^2_{\text{p}} - 2 \, \text{Re}\left\lbrace \alpha_j \beta^{\ast}_j \ {_{\text{p}}\langle \psi^-_j | e^{-2i\hat{P}} | \psi^+_j \rangle_{\text{p}}} \right\rbrace \, ,
\end{align}
both quantities being positive by construction (they are norms).
Now, even if we start at $j=0$ with an entangled state $|\Psi_{j=0}\rangle$, instead of a factorized one of the form $\hat{T}_{j=0} |\psi_{j=0 }\rangle_{\text{p}}$, the scheme runs with an external ket, namely $|\psi_{j}\rangle_{\text{p}} \equiv {_{\text{c}}\langle \pm | \hat{W}_{j-1}^{\text{s}} | \Psi_{{j-1}} \rangle / \sqrt{\mathrm{P}^{\pm}_{j-1}}}$, and the Aharonov operators, that act on this external ket,  read
\begin{align} \label{eq:Aharonov_operators}
\hat{A}^+_j &= e^{i\omega_j}\left(\alpha_j c^+_j \,  e^{-i\hat{P}}  +  \beta_j c^-_j \, e^{i\hat{P}}  \right) / {\sqrt{\mathrm{P}^{+}_j}} \nonumber \\
\hat{A}^-_j &= e^{i\omega_j}\left(-\beta^{\ast}_j c^+_j \,  e^{-i\hat{P}}  +  \alpha^{\ast}_j c^-_j \, e^{i\hat{P}}  \right) / \sqrt{\mathrm{P}^{-}_j} \, .
\end{align}
The expressions of the $+$ and $-$ probabilities given above simplify into
\begin{align}
&\mathrm{P}^+_j = |\alpha_j|^2 |c^+_j|^2 + |\beta_j|^2 |c^-_j|^2 + 2 \, \text{Re}\left\lbrace \alpha_j \beta^{\ast}_j c^+_j {c^-_j}^{\ast} \ {_{\text{p}}\langle \psi_j | e^{-2i\hat{P}} | \psi_j \rangle_{\text{p}}} \right\rbrace \nonumber \\
&\mathrm{P}^-_j = |\beta_j|^2 |c^+_j|^2 + |\alpha_j|^2 |c^-_j|^2 - 2 \, \text{Re}\left\lbrace \alpha_j \beta^{\ast}_j c^+_j {c^-_j}^{\ast} \ {_{\text{p}}\langle \psi_j | e^{-2i\hat{P}} | \psi_j \rangle_{\text{p}}} \right\rbrace \, .
\end{align}

\subsection{Explicit expression for the $N$-step outcome}

After $N$ time steps, the resulting state is 
\begin{equation}\label{eq:N_steps}
|\psi_N^{a^k_i} \rangle_{\text{p}} = \underbrace{\hat{A}^+_{N}\hat{A}^+_{N-1}\hat{A}^-_{N-2}\hat{A}^+_{N-3} \  ... \hat{A}^-_{j+2}\hat{A}^-_{j+1}\hat{A}^+_{j}\hat{A}^+_{j-1}\hat{A}^+_{j-2}\hat{A}^-_{j-3} \ ... \ \hat{A}^-_2 \hat{A}^+_1 \hat{A}^+_0}_{N \ \text{elements in the product}} |\psi_0 \rangle_{\text{p}} \, ,
\end{equation}
where $a^k_i$, $i=1, ... , \mathcal{A}^k_N$, is one of the $\mathcal{A}^k_N=k! {{N}\choose{k}}$ possible arrangements of $k$ operators $\hat{A}^+$ among the $N$ possible locations in an ordered product (which is combinatorially equivalent to a list) of $N$ elements, the other locations being occupied by $\hat{A}^-$ operators. 

This state has the following probability to happen:
\begin{equation} \label{eq:given_outcome_proba}
\mathcal{P}( \psi_N^{a^k_i}) = \prod_{m=1}^k \mathrm{P}^{+}_{j = \tau^+(m)} \prod_{n=1}^{N-k} \mathrm{P}^{-}_{j = \tau^-(n)} \, ,
\end{equation}
where $\tau^+(m)$ (resp. $\tau^-(n)$)  is the time at which the $m$-th $\hat{A}^+$ (resp. $n$-th $\hat{A}^-$) happens. The couple of sequences ($\tau^+$,$\tau^-$) is mapped one to one with one of the $a^k_i$'s.
This probability of a given sequence of outcomes, Eq. (\ref{eq:given_outcome_proba}), results from the combination of two effects:
\begin{enumerate}
\item The quantum interferences between the left- and right-going walkers, realised by DTQW part of the Aharonov operators, and which manifest themselves in the one-step probabilities, the $\mathrm{P}_j^+$'s and $\mathrm{P}_j^-$'s.
\item An effectively-classical randomness effect, realised by the measurement part of the Aharonov operator, which is reflected by the product of all the $\mathrm{P}^+_j$'s and  $\mathrm{P}^-_j$'s. To be completely explicit: the probability of obtaining a given sequence of Aharonov operators is that of a given sequence of outcomes in successive Bernouilli experiments, with possibly time-dependent probabilities. In the next subsection, we will show that this connection with effectively-classical randomness effect goes further, and grows with the amount of averaging over several realizations of Aharonov's scheme.
\end{enumerate}

In the case where the walk operator $\hat{W}^{\text{s}}$ and the spin preparation $\hat{T}$ do not depend on time, the Aharonov operators $\hat{A}^{\pm}$ do not depend on time either, and each $a^k_i$ state has the same probability to happen (thus independent of $i$):
\begin{equation}
\mathcal{P}(\psi_N^{a^k_i}) = \mathrm{P}^k(1-\mathrm{P})^{N-k} \, .
\end{equation}
With no further assumptions, Aharonov's scheme has no simpler formal expression than (\ref{eq:N_steps}), with the Aharonov plus and minus operators given by Eqs. (\ref{eq:Aharonov_operators}).

\subsection{Spatial probability distributions}

\subsubsection{Probability distribution of a given outcome: possibly high quantum-interferences effects}

The probability of being at time $j$ on site $p$ having obtained the arrangement $a_i^k$ after $j$ successive spin measurements, is given by
\begin{equation} \label{eq:proba_given_outcome}
\mathcal{P}(j,p|\psi^{a^k_i}_j) \equiv |\psi_{j,p}^{a^k_i}|^2  \, .
\end{equation}
As manifest by looking at Eqs. (\ref{eq:N_steps}) and (\ref{eq:Aharonov_operators}), this spatial probability distribution will be in part the result of quantum inteferences between the left- and right-going walkers.

\subsubsection{Averaging over all possible outcomes suppresses quantum interferences}

Averaging over all possible outcomes yields
\begin{align}
P_{j,p} &= \sum_{\psi_j} \mathcal{P}(j,p|\psi_j) \, \mathcal{P}(\psi_j) 
= \sum_{k=1}^j \sum_{i=1}^{\mathcal{A}^k_j} \, \mathcal{P}(j,p|\psi_j^{a_k^i}) \mathcal{P}( \psi_j^{a_k^i} ) \, ,
\end{align}
where we have used the abbreviation $\psi_j \equiv  \psi_j^{a_k^i}$ in the first equality.

One can also perform this average on the before-last outcomes:
\begin{align} \label{eq:average_before_last}
P_{j+1,p} &= \sum_{\psi_{j}} \mathcal{P}(j+1,p|\psi_{j}) \, \mathcal{P}(\psi_{j}) \, ,
\end{align}
where the probability of being at time $j+1$ on site $p$ having obtained the arrangement $a_{i}^{k}$ after $j-1$ successive spin measurements, is given by
\begin{equation} \label{eq:knowing_before_last}
\mathcal{P}(j+1,p|\psi_{j}) =  \mathrm{P}^+_j \ | \, {_{\text{p}}\langle p | A^+_j | \psi_j \rangle_{\text{p}}} \, |^2 + \mathrm{P}^-_j \ | \, {_{\text{p}}\langle p | A^-_j | \psi_j \rangle_{\text{p}}} \, |^2 \, .
\end{equation}
Now, because
\begin{align}
| \, {_{\text{p}}\langle p | A^+_j | \psi_j \rangle_{\text{p}}} \, |^2 &= \frac{1}{\mathrm{P}^+_j} \left[ |\alpha_j|^2 |c^+_j|^2 |\psi_{j,p-1}|^2 + |\beta_j|^2 |c^-_j|^2 |\psi_{j,p+1}|^2 \right. \nonumber   \\
& \ \ \ \ \ \ \ \ \ \ \left. + 2 \, \text{Re} \left\lbrace \alpha_j \beta^{\ast}_j c^+_j {c^-_j}^{\ast} \psi_{j,p-1} \psi_{j,p+1}^{\ast} \right\rbrace  \right] \nonumber \\
| \, {_{\text{p}}\langle p | A^-_j | \psi_j \rangle_{\text{p}}} \, |^2 &= \frac{1}{\mathrm{P}^-_j} \left[ |\beta_j|^2 |c^+_j|^2 |\psi_{j,p-1}|^2 + |\alpha_j|^2 |c^-_j|^2 |\psi_{j,p+1}|^2 \right. \nonumber   \\
& \ \ \ \ \ \ \ \ \ \ \left. - 2 \, \text{Re} \left\lbrace \alpha_j \beta^{\ast}_j c^+_j {c^-_j}^{\ast} \psi_{j,p-1} \psi_{j,p+1}^{\ast} \right\rbrace  \right] \, ,
\end{align}
the interference terms between the left- and right-going walkers cancel each other in Eq. (\ref{eq:knowing_before_last}), which yields
\begin{equation}
\mathcal{P}(j+1,p|\psi_{j}) = |c_{j}^+|^2 |\psi_{j,p-1}|^2 + |c_{j}^-|^2 |\psi_{j,p+1}|^2 \, .
\end{equation}
Plugging this expression in the average over the before-last outcomes, Eq. (\ref{eq:average_before_last}), delivers
\begin{equation}
P_{j+1,p} = |c_{j-1}^+|^2 P_{j-1,p-1} + |c_{j-1}^-|^2 P_{j-1,p+1} \, .
\end{equation}
The averaged probability ditribution $P_{j,p}$ thus follows a classical random walk with probabilities of going right and left, from time $j$ to time $j+1$, respectively given by $\pi_j=\pi^+_j=|c_{j}^+|^2$ and $\pi^-_j=|c_{j}^-|^2=1-\pi^+_j$. After averaging over all possible outcomes, the quantumness of the system is removed. The average displacement (resp. variance) of $P_{N,p}$ is the sum of the average displacements (resp. variances) between two consecutive time steps, $\sum_{j=0}^N(\pi_{j}^+-\pi_{j}^-)$ (resp. $\sum_{j=0}^N \pi^+_j \pi^-_j$), which results in $N(\pi^+-\pi^-)$ (resp. $N\pi^+\pi^-$) when $\pi_j$ do not depend on the time $j$.

\iffalse

For a classical random walk, the conditional probability is preserved in time, that is,
\begin{equation} \label{eq:classical_conditional_proba}
\mathcal{P}_{\text{class}}(j,p|a^k_i) = \mathcal{P}_{\text{class}}(j=0,p) = |\psi_{j=0,p}|^2 \, .
\end{equation}
Although this is not the case in this quantum walk scheme, the scheme still yields an averaged probability distribution which is effectively classical, i.e. the average make the interference terms cancel out. Indeed, we have
\begin{equation}
P_{j+1,p} = \mathrm{P}^+_j \ | \, {_{\text{p}}\langle p | A^+_j | \psi_j \rangle_{\text{p}}} \, |^2 + \mathrm{P}^-_j \ | \, {_{\text{p}}\langle p | A^-_j | \psi_j \rangle_{\text{p}}} \, |^2 \, ,
\end{equation}
with

The interference terms thus cancel out after performing the average:
\begin{equation}
P_{j+1,p} = \mathrm{P}_j^+  P_{j,p-1} + \mathrm{P}_j^-  P_{j,p+1} \, ,
\end{equation}
and $P_{j,p}$ is the probability distribution of classical random walk with probabilities $\mathrm{P}_j^{\pm}$ of going to the right (plus sign) or to the left (minus sign) at each time step. The initial condition is $P_{j=0,p}=|\psi_{j=0}|^2$. 

\fi

In the continuous-spacetime limit, and in the case where the probability $\pi_j$ does not depend on time, the probability distribution $P(t,x)$ follows a diffusion equation with a bias drift speed  $c=\lim \{(\pi^+-\pi^-) \, l/\Delta t\}$ and a diffusion coefficient $D=\lim \{\pi^+\pi^- \, l^2 /(2 \Delta t)\}$, where the limit corresponds to a space step $l\rightarrow 0$ and a time step $\Delta t \rightarrow 0$. Note that $l^2 /(2 \Delta t)$ must have a finite limit for this continuum limit to exist, and thus that $\pi^+-\pi^-$ must scale as the length step $l$, which makes sense since $P(t,x)l \Delta t$ is the probability  to be between time $t$ and $t+\Delta t$ and position $x$ and $x+l$. In the case where $\pi_j$ does depend on time, one can still define a bias drift speed and a diffusion coefficient, but locally in time.

\section{Coinless scheme: no more quantum interferences on given realizations of the scheme}

In this case $\beta_j=0$, $|\alpha_j|^2=1$, and the outcomes of the measurement operation yield no superposition between left- and right-going states. As a consequence, not only the probability distribution averaged over all possible outcomes follows a (possibly time-dependent) classical random walk, which is the case even in the coined scheme, as shown previously, but also the mean position of a given realization of the quantum random walk, and even more: at each time step, the initial probability distribution of the quantum walker is actually simply shifted right or left by one space step, with probabilities $\pi^+_j$ and $\pi^-_j$, and without deformation.

The quantumness of the scheme still shows up on three aspects:
\begin{enumerate}
\item The scheme transports the initial external ket $|\psi_{j=0}\rangle_{\text{p}}$ up to time $j=N$ without destroying it (along some classical random path determined by the successive outcomes of spin measurement). One can thus use this quantum state at any time and make it interfere with another quantum state.
\item The scheme generically adds a phase information, since at each time step, the walker $|\psi_j \rangle_{\text{p}}$ catches, while being shifted right (resp. left) with probability $\pi^+_j$ (resp. $\pi^-_j$), a phase $c^{\pm}_j/|c^{\pm}_j|$. 
\item The classicality of a given realisation of the coinless quantum random walk is only effective, since what produces this is the intrinsic randomness of quantum measurement, which has nothing to do\footnote{At least this is the current dominant view, since, in particular, the experiments of Alain Aspect from 1980 to 1983, which impose to make a choice between locality and causality in the description of quantum mechanics; see his thesis at \url{https://tel.archives-ouvertes.fr/tel-00011844/document}.} with classical randomness, that models a lack of information of the system under study.
\end{enumerate}

For the sake of completeness, let us give the Aharonov operators in this coinless case:
\begin{equation}
\hat{A}^{\pm} = e^{i\phi^{\pm}_j} e^{\mp i\hat{P}l} \, ,
\end{equation}
where 
\begin{equation}
e^{i\phi^{\pm}_j} \equiv \frac{c^{\pm}_j}{|c^{\pm}_j|} \, ,
\end{equation}
so that the state at time $N$ is given by
\begin{equation}
|\psi_{j=N}^{a^k_i} \rangle_{\text{p}} = \left[ \prod_{m=1}^k \exp\left(i\phi^{+}_{j = \tau^+(m)}\right) \prod_{n=1}^{N-k} \exp\left(i\phi^{-}_{j = \tau^-(n)}\right) \right] |\psi_{j=0}(p_0 + (k - (N-k))) \rangle_{\text{p}} \, .
\end{equation}
where we have noted $\psi_j = \psi_j(p_0)$, $p_0$ being a reference position (typically, the center of a wavepacket), that is shifted left or right by the spin-dependent translation operators. In this coinless case, to each arrangement $a^k_i$ of Aharonov operators, corresponds one of the $\mathcal{A}^k_N$ posssible classical zig-zag paths that the walker follows if we impose it to make $k$ steps to the right and $N-k$ to the left. 

\section{An effect of the quantum interferences realized by the coined scheme: classical-path deformation}

We refer the reader to both the original paper \cite{ADZ93a} and Kempe's review \cite{Kempe_review} for details on this example, which shows one of the possible effects of quantum interferences in the coined scheme. In this example, one considers a wavepacket whose width is kept large with respect to the space step during the time evolution. In this approximation, one can choose the angle of the coin operation in such a way that the one-step scheme is the following: the walker's wave function is either shifted to the left (resp. right) by far more (resp. less) than a lattice step, with a small (resp. high) probability. Because this is an approximation, the wave function is actually slightly deformed, on the contrary to the coinless scheme, but one sees that, apart from this small wave-function deformation, the effect of quantum interferences in this example is a deformation of the classical path: the step taken to the left (resp. right) is superclassical (resp. subclassical), and the corresponding probability is subclassical (resp. superclassical\footnote{By `superclassical (resp. subclassical) step', I mean that the quantum-average displacement of the left-going (resp. right-going) walker is bigger (resp. smaller) than a lattice step. By `superclassical (resp. subclassical) probability', I mean that the quantum walker goes right (resp. left), i.e. we measure a spin up (resp. down), with a probability which is higher (resp. lower) than that of its classical counterpart going right (resp. left).}). As a pedagogical example, the authors choose the strictly-left-going walker, see Fig. \hspace{-0.15cm}1 of \cite{ADZ93a}, whose mean position overcomes by far the distance reached by its classical counterpart. For the wave-function deformation to appear, the final outcome must be computed exactly (without the approximation used to explain the effect, which assumes no deformation); in the case of a strictly-left-going walker, one simply needs to apply, to the initial state, the $N$-th power of the minus Aharonov operator, in order to obtain the state at time $N$, which results in Eq. (4) of \cite{ADZ93a}.

%Although the deformation of the wavepacket can seem small graphically, it plays an important role, as underlined in ... (it is necessary for this picture of classical-path deformation to make sens).

   %\include{Chapters/App_Schrodinger_Heisenberg}
   %%%%%%%%%%%%%%%%%%%%%%%%%%%%%%
%%  Alias bibliographiques  %%
%%%%%%%%%%%%%%%%%%%%%%%%%%%%%%
\defcitealias{NRFortran}{Numerical Recipes}
\defcitealias{NRF77}{Numerical Recipes}
\defcitealias{NRF90}{Numerical Recipes}
\defcitealias{IUPAC}{IUPAC}%International Union of Pure and Applied Chemistry

%\chapter{\textsc{Compléments mathématiques}\label{Ann:Maths}}% Intermezzo mathématique

\chapter{On the momentum and the quasimomentum} \label{app:quasimomentum}

Consider a continuous spacetime. To any classical, i.e.  point particle having momentum $\boldsymbol P$, one can associate a wave having a so-called de Broglie wavevector $\boldsymbol K = \boldsymbol P / \hbar$, where $\hbar$ is the reduced Planck's constant.  It is one of the two components of the correspondence principle of quantum mechanics, the complementary component being the relation between energy and frequency. In this thesis I use dimensionless variables, which is equivalent to work in so-called natural units, with $\hbar=1$, so that the classical momentum and the wavevector have the same value.

Now, the DTQWs are defined on a graph which is often eventually mapped to a spacetime lattice. On a lattice, the choice of the region -- i.e., in the one-dimensional case, of the interval~-- to which the wavevectors belong is not unique\footnote{Only the volume of this region is imposed, but its location in (the unbounded) $\boldsymbol k$-space is arbitrary. This volume is the length of the interval in the one-dimensional case, and the wavevector is a real number,~$k$.}. Hence, referring to $k$ as `the momentum', in the sense this word is used in standard quantum mechanics, can be misleading. Instead, $k$ will be referred to, in a continuous-spacetime perspective, as our preferred \emph{quasimomentum}, and it will be called `the' quasimomentum, although it is only defined modulo the interval length $L$, in the sense that any wavevector equal to a given one modulo $L$ delivers exactly the same results. In the continuum limit considered in this work, this quasimomentum $k$ will be replaced by a certain continuum-limit momentum $K$, i.e. a certain value of \emph{the} quantum-mechanical variable which is canonically conjugate to the position. More precisely, a certain continuum-limit momentum $K$ is typically, roughly speaking,  a certain limit of the ratio between the quasimomentum and the step of the spacetime lattice, $\epsilon$. Indeed, both quantities $k$ and $\epsilon$ must typically go to zero in the continuum limit, and their ratio can have an arbitrary finite limit, so that one actually maps a certain continuum-limit momentum $K$ to a particular \emph{family of quasimomenta  which is indexed by $\epsilon$}, of the form $k=f_{\epsilon}(K)$, where the function $f_{\epsilon}$ depends on the interval chosen for the quasimomenta. The sign of a given quasimomentum is typically preserved, when taking the continuum limit, if this interval is \emph{partitioned into a (strictly) positive and a (strictly) negative subset}. We choose the symmetric interval $[-\pi;\pi[$ to facilitate the use of intuition; with such a symmetric interval, one can in particular merely set $k=K\epsilon$.

Note that, the word `momentum' being simply non-used in a lattice situation, it could actually be appropriate to use it instead of the word  `quasimomentum', typically if (i) this quasimomentum has a standard-momentum counterpart in a certain continuum spacetime,  which is the case in the present work, and (ii) as long as our main intuition of the quasimomentum is not in conflict with that of the continuum-limit standard momentum, which to me is the case with the symmetric-interval choice. To me, indeed, the relation $k=K\epsilon$ makes it rather natural to extend the use of the word `momentum'  to the lattice, because of the following reason: to satisfy condition (ii), one could (ii.a) require that the quasimomenta interval contain $0$, and (ii.b) that it contain $-k$ if it contains $k$, which selects the \emph{unique} symmetric interval. Instead of (ii.b), one could also require that all quasimomenta are positive, which induces a mapping with the standard-momentum norm instead of its component on the momentum axis, and also selects a unique interval.

In solid-state physics, the terminology `quasimomentum' is widely used, and is interchangeable with the terminology `crystal momentum'. In the present work, there is, obviously, a priori no crystal, and the terminology `quasimomentum' is a priori only related to a common mathematical feature of the present work and solid-state physics, that is, the presence of a lattice. While this common mathematical description obviously translates into common effective mathematical features shared by solid-state systems and DTQWs, note that, in the continuum-limit applications of the present work, the spacetime lattice on which the DTQWs are defined have, not only a priori, but indeed nothing to do with the crystal lattice of solid state physics. One could envisage making physical connections between DTQWs and solid-state physics. To make a physical connection between the `quasimomentum' of this thesis and the `crystal momentum' of solid-state physics, one would have typically to find a proper mapping between the overlap integrals of solid-state physics and the parameters that define our DTQWs. The effective mass of solid-state physics, which is a measure of the resistance to transport, increases when the overlap, i.e. the `connectivity' between nodes diminishes, so that I would expect the overlap to be a decreasing function of the mixing angle $\theta$, if a proper mapping to solid-state physics can be found. The solid-state picture could in particular help to develop an intuition on the phenomena arising from DTQWs' connections between nodes.

   %%%%%%%%%%%%%%%%%%%%%%%%%%%%%%
%%  Alias bibliographiques  %%
%%%%%%%%%%%%%%%%%%%%%%%%%%%%%%
\defcitealias{NRFortran}{Numerical Recipes}
\defcitealias{NRF77}{Numerical Recipes}
\defcitealias{NRF90}{Numerical Recipes}
\defcitealias{IUPAC}{IUPAC}%International Union of Pure and Applied Chemistry

%\chapter{\textsc{Compléments mathématiques}\label{Ann:Maths}}% Intermezzo mathématique

\chapter{Unitary group} \label{app:unitary_group}

\section{Dimension $N$} \label{sec:dimN}

The unitary group of dimension $N$, U($N$), is the group of $N \times N$ unitary matrices, i.e. $\text{U}(N)= \{ U \in M_N(\mathbb{C}) \, / \, U^{\dag}U=\mathbf{1}_{N} \}$. It is a compact and (not simply) connected Lie group, so that the exponential map generates the whole group\footnote{See Appendix \ref{app:topology} for a glimpse on topology, in particular on the notions of compactness and connectedness.}. Consider an arbitrary element $U \in \text{U}(N)$. Its determinant $\text{det} \, U = e^{i\omega}$, $\omega \in [0, 2\pi[$, which follows directly from $U^{\dag}U=\mathbf{1}$. We define $\bar{U}=U/\delta$, where $\delta^N=\text{det}\, U$. The matrix $\bar{U}$ is unitary and has unit determinant, i.e. it is an element of $\text{SU}(N)$. Any element of $\text{U}(N)$ can thus be factorized into the product of an element of $\text{U}(1)$ by an element of $\text{SU}(N)$. This factorization is not unique because $\delta$ is not uniquely defined by the equation $\delta^N=\text{det} \, U$. Indeed, this equation has the $N$ distinct solutions $\delta_k= \exp[i(\omega+2k\pi)/N]$, $k=0,...,N-1$, and each solution defines a different factorization. We choose $k=0$, which defines unambiguously a unique factorisation.

\section{Dimension $2$} \label{sec:dim2}

\subsection{A possible parametrization of $\mathrm{U(2)}$} \label{subsec:param}

According to the above explanation, we can factorize uniquely any $U\in \text{U}(2)$ as the product of $e^{i\alpha}$, $\alpha=\omega/2 \in [0, \pi[$, by $\bar{U} \in \text{SU}(2)$. Writing $\bar{U}^{\dag}{\bar{U}}=\mathbf{1}_2$ with
\begin{equation}
\bar{U}=\begin{bmatrix}
a & b \\ c & d
\end{bmatrix} ,
\end{equation}
yields the following equations,
\begin{subequations} \label{eq:u123}
\begin{align}
|a|^2+|c|^2&=1 \label{eq:unitarity_1} \\ 
|b|^2 + |d|^2&=1  \label{eq:unitarity_2} \\
a^{\ast}b+c^{\ast}d&=0 \, , \label{eq:unitarity_3}
\end{align}
\end{subequations}
and $\text{det} \, \bar{U}=1$ reads
\begin{equation}
ad-bc=1 \, .
\end{equation}
If $b=0$, then $ad=1$ and $c^{\ast}d=0$, so that $c=0$ and then $a=d^{\ast}$. If $b\neq 0$, then plugging $a=-c d^{\ast}/b^{\ast}$ into (\ref{eq:unitarity_1}) and using (\ref{eq:unitarity_2}) yields $|b|=|c|$. We thus have $|a|^2+|b|^2=1$, so that there exist $\theta \in \, [-\pi/2, 0[$ such that $|a|=\cos \theta$ and $|b|=-\sin \theta$, and we can actually use $\theta = 0$ to parametrize the case $b=0$ described previously, so that $\theta \in \, [-\pi/2, 0]$. Hence, there exist $(\xi,\zeta) \in \, ]-2\pi, 0] \times [0, 2\pi[$  such that
\begin{subequations} \label{eq:ab}
\begin{align} 
a &=e^{i\xi} \cos \theta  \label{eq:param_a} \\
b &=e^{i\zeta} \sin \theta \, . \label{eq:param_b} 
\end{align}
\end{subequations}
From Eq. (\ref{eq:unitarity_2}), there also exist $\rho \in [0, 2\pi[$ such that
\begin{equation} \label{eq:param_d}
d=e^{i\rho} \cos \theta \, ,
\end{equation}
and then plugging (\ref{eq:param_a}), (\ref{eq:param_b}), and (\ref{eq:param_d}) in (\ref{eq:unitarity_3}) yields  $c=-e^{i(\xi-\zeta+\rho)} \sin \theta$. Writing then $ad-bc=1$, one deduces that $\xi=-\rho$. Any $\bar{U} \in \text{SU}(2)$ has thus a unique writing of the following form,
\begin{subequations}
\begin{align}
\bar{U}&=
\begin{bmatrix}
e^{i\xi} \cos \theta & e^{i\zeta} \sin\theta \\ -e^{-i\zeta} \sin\theta  & e^{-i\xi} \cos \theta 
\end{bmatrix} \nonumber \\
&=e^{i\frac{\xi+\zeta}{2} \sigma_3} e^{i\theta \sigma_2} e^{i\frac{\xi-\zeta}{2} \sigma_3} \, , \label{eq:euler_angles_rot}
\end{align}
\end{subequations}
as recalled in Eq. (\ref{eq:M_matrix}), with the parameter ranges specified by the set (\ref{eq:mapping_set_2}).

\subsection{Link with the Euler angles of $\mathrm{SO(3)}$ for an active rotation} \label{subsec:link_Euler}

We define
\begin{align}
\psi &=  \zeta - \xi  \nonumber \\
\Theta & = -2 \theta \\
\phi &= -\zeta - \xi  \, ,
\end{align}
so that $\psi \in [0, 4\pi[$ and $\phi \in \, ]-2\pi, 2\pi[$. It is actually enough to restrict $\phi$ to $[0, 2\pi[$ to map the whole SU(2) group\footnote{If $\phi \in ]-2\pi,0[$, then either (i) $\psi \in [0,2\pi[$, in which case choosing $\phi+2\pi \in [0,2\pi[$ and $\psi+2\pi \in [0,4\pi[$ yields the same matrix, or (ii) $\psi \in [2\pi,4\pi[$, in which case choosing $\phi+2\pi \in [0,2\pi[$ and $\psi-2\pi \in [0,4\pi[$ also yields the same  matrix.}. The SU(2) matrix $\bar{U}$ thus reads
\begin{equation}
\bar{U}=e^{-i \phi (\sigma_3/2)} e^{-i\Theta (\sigma_2/2)} e^{-i \psi (\sigma_3/2)} \, ,
\end{equation}
and the mapping of SU(2) is one to one (if $ab\neq 0$) with the following ranges:
\begin{equation}
(\psi,\Theta,\phi) \in [0, 4\pi[ \times [0, \pi] \times \, [0, 2\pi[ \, .
\end{equation}
Now, the angles $\psi$, $\Theta$ and $\phi$ are the Euler angles of an active rotation of SO(3): $\psi$ is the so called \emph{rotation} angle, $\Theta$ the \emph{nutation} angle, and $\phi$ the \emph{precession} angle. To map SO(3) one to one with these Euler angles, one has to restrict $\psi$ to $[0, 2\pi[$ (SU(2) covers SO(3) twice).  Any matrix of SO(3) can be parametrized with these Euler angles: indeed, to define any rotation, one can first perform a rotation of \emph{arbitrary} angle $\psi \in [0, 2\pi[$ around $\boldsymbol{e}_z$, and then change the rotation unit vector from $\boldsymbol{e}_z$ to an \emph{arbitrary} direction $\boldsymbol{e}'_z$ obtained by reorienting $\boldsymbol{e}_z$ (unchanged after the rotation above itself by $\psi$) in the direction specified by the arbitrary spherical-coordinates angles $(\Theta,\phi) \in [0, \pi] \times [0, 2\pi[$, above $\boldsymbol{e}_y$ and `the original' $\boldsymbol{e}_z$, respectively\footnote{A good way for me to see this is the following. Let $J_1, J_2, J_3$ be the generators of SO(3).   The coordinates of the rotated vector in the original basis are $v' = e^{-i\phi J_3}e^{-i\Theta J_2} e^{-i\psi J_3} v$. We know that its coordinates in the basis of spherical coordinates $(\Theta,\phi)$ are given by $e^{i \Theta J_2} e^{i\phi J_3} v'$, which is equal to $e^{-i\psi J_3} v$: in this new basis, we have thus indeed just performed a rotation of angle $\psi$ above the vertical direction, which is $\boldsymbol{e}'_z$.}.   To a matrix of SO(3) having some $\psi_0 \in [0, 2\pi[$ corresponds 2 matrices of SU(2): that with $\psi_0$ and that with $\psi_0+2\pi$, which is the opposite of the first\footnote{For extensive details and proofs on this SU(2)/SO(3) mapping, see, e.g., Ref. \url{http://www.hep.caltech.edu/~fcp/physics/quantumMechanics/angularMomentum/angularMomentum.pdf}, in particular Eqs. (52) to (54) and pages 11 to 14.}. When $\Theta=0$ we are at the north pole of the Bloch sphere, and when $\Theta=\pi$, we are at the south pole (spin flipping). Fore more details, see the following links\footnote{The convention for the Euler angles may change from one document to the other.}: 
\begin{enumerate}
\item \url{http://www.hep.caltech.edu/~fcp/physics/quantumMechanics/angularMomentum/angularMomentum.pdf},
\item \url{http://www.phys.nthu.edu.tw/~class/Group_theory/Chap\%207.pdf}, 
\item \url{http://www.mat.univie.ac.at/~westra/so3su2.pdf},
\item \url{https://www.google.es/url?sa=t&rct=j&q=&esrc=s&source=web&cd=1&ved=0ahUKEwjSlMXSr9fVAhXJhrQKHc9RAWIQFggrMAA&url=https\%3A\%2F\%2Fwww.math.ens.fr\%2Fenseignement\%2Ftelecharger_fichier.php\%3Ffichier\%3D588&usg=AFQjCNH-G8XJeMZdqmhzTg4mazqc19Hvjw}.
\end{enumerate}

\iffalse
\pablo{Write something about active and passive view points. I think $\theta$, $\xi$ and $\zeta$ can be  considered as Euler Angles in either the active of the passive view points, and the the other triplet is for the reciprocal viewpoint.}
\fi
   %%%%%%%%%%%%%%%%%%%%%%%%%%%%%%
%%  Alias bibliographiques  %%
%%%%%%%%%%%%%%%%%%%%%%%%%%%%%%
\defcitealias{NRFortran}{Numerical Recipes}
\defcitealias{NRF77}{Numerical Recipes}
\defcitealias{NRF90}{Numerical Recipes}
\defcitealias{IUPAC}{IUPAC}%International Union of Pure and Applied Chemistry

%\chapter{\textsc{Compléments mathématiques}\label{Ann:Maths}}% Intermezzo mathématique

\chapter{Connecting two conventions for the DTQW on the line} \label{app:time_reversal}

%\emph{These lines came to my mind while reading the lectures notes of the Quantum Field Theory course given by Adel Bilal to the second-year students of the ICFP Master at the \'Ecole Normale %Supérieure in Paris.} \\

In our definition of the walk, we choose to apply first the shift $\hat{S}$  and then the coin operation $\hat{U}$ for practical reasons:  in the sections concerning non-homogeneous walks, the Taylor expansion is simpler with this order in the operations because the values of the spacetime-dependent angles defining the coin operation are not shifted from the running spacetime-lattice point, so that we don't have to Taylor expand them. The most-used convention is, however, the inverse one, namely applying first the coin operation $\hat{U}^{\text{s}}$ and then the shift $\hat{S}^{\text{s}}$, yielding a walk operator $\hat{W}^{\text{s}}= \hat{S}^{\text{s}} \hat{U}^{\text{s}}$, the superscript `s' being for `standard'.  We are going to show that both conventions map one to one through a time-reversal operation.

Let us first consider an arbitrary DTQW with the convention used in this thesis, i.e. let us consider the evolution given by Eq. (\ref{eq:protocol}) for time $j \in [ \! [ 0, j_{\text{max}}  ] \! ]$. The state of the walker at time $j_{\max}$ is given by
\begin{equation} \label{eq:main_}
|\Psi_{j_{\max}}\rangle = \hat{E}_{j_{\text{max}}} | \Psi_{0}\rangle \ ,
\end{equation}
where the $j_{\text{max}}$-step evolution operator  reads
\begin{equation}
\hat{E}_{j_{\text{max}}} = \prod_{j=0}^{j_{\text{max}}-1} \hat{U}_j \hat{S} \, ,
\end{equation}
the product being growingly time ordered from right to left.
One can rewrite Eq. (\ref{eq:main_}) as
\begin{equation} \label{eq:QW_ours_inverted}
 |\Psi_{0}\rangle = \hat{E}_{j_{\text{max}}}^{-1} |\Psi_{j_{\text{max}}} \rangle \, ,
 \end{equation}
where the inverse $j_{\text{max}}$-step evolution operator reads
 \begin{equation}
\hat{E}^{-1}_{j_{\text{max}}} = \prod_{j=1}^{j_{\text{max}}}  \hat{S}^{-1} (\hat{U}_{j_{\text{max}}-j})^{-1}  \, .
\end{equation}

Let us now consider the one-step evolution of a generic standard-convention DTQW:
\begin{equation} 
|\Phi_{r+1}\rangle = \hat{W}^{\text{s}}_r |\Phi_{r}\rangle \ .
\end{equation}
The label $r \in [ \! [ 0,j_{\text{max}}  ] \! ]$ indicates discrete time on another timeline. 
The state of the walker at time $r=j_{\text{max}}$ is thus given by
\begin{equation} \label{eq:QW_standard}
|\Phi_{r=j_{\text{max}}} \rangle = \hat{E}^{\text{s}}_{r=j_{\text{max}}} |\Phi_{r=0}\rangle \ ,
\end{equation}
where the $j_{\text{max}}$-step evolution operator reads, in this standard convention,
\begin{align} 
\hat{E}^{\text{s}}_{r=j_{\text{max}}} = \prod_{r=0}^{j_{\text{max}}-1} \hat{S}^{\text{s}} \hat{U}_r^{\text{s}}  \ .
\end{align}
 
Hence, if we choose 
\begin{align} \label{eq:time_rev}
\hat{S}^{\text{s}} &\equiv \hat{S}^{-1} \nonumber \\
\hat{U}_r^{\text{s}} &\equiv (\hat{U}_{j_{\text{max}}-j})^{-1} = (\hat{U}_{j_{\text{max}}-j})^{\dag} \ ,
\end{align}
that is to say, 
\begin{equation} \label{eq:ev}
\hat{E}^{\text{s}}_{r=j_{\text{max}}} \equiv \hat{E}^{-1}_{j_{\text{max}}} \, ,
\end{equation}
then Eq. (\ref{eq:QW_standard}) is the same as Eq. (\ref{eq:QW_ours_inverted}), provided that we also choose
\begin{align}
|\Phi_{r=0}\rangle = |\Psi_{j=j_\text{max}}\rangle \, ,
\end{align}
which imposes
\begin{equation}
|\Phi_{r=j_\text{max}}\rangle = |\Psi_{j=0}\rangle  \, .
\end{equation}

\begin{framed}
Any evolution operator written in our convention, $\hat{E}_{j_{\text{max}}}$, taking and initial state $|\Psi_{j=0}\rangle$  to a final state $|\Psi_{j=j_{\text{max}}}\rangle$, thus  always coincides with an inverse, i.e. time-reversed evolution operator of the standard convention, namely $(\hat{E}_{r=j_{\text{max}}}^{\text{s}})^{-1}  \equiv \hat{E}_{j_{\text{max}}}$, and we can write $r=j_{\text{max}}- j$.
\end{framed}

The converse result obviously also holds. In particular, the PDE obtained in the continuum limit of a standard-convention DTQW having an evolution operator $\hat{E}^{\text{s}}_{r_{\text{max}}}$, satisfying appropriate zeroth-order constraints but otherwise \emph{arbitrary}, is the same as that obtained for the DTQW of our convention which is defined by the evolution operator $\hat{E}_{j=r_{\text{max}}} \equiv (\hat{E}^{\text{s}}_{r_{\text{max}}})^{-1}$. 

Actually, if the aim is simply to determine the continuum limit of some DTQW written in the standard convention, regardless of any initial and final states, i.e. regardless of connecting the two different timelines, it is enough to work on the one-step evolution equation: take the inverse of some $S^{\text{s}}U^{\text{s}}_j$ originally defined, i.e. simply compute $(U^{\text{s}}_j)^{\dag}$, apply any continuum limit results to the DTQW written in our convention with coin operation $U_j = (U^{\text{s}}_j)^{\dag}$, and then perform the substitution $\partial_{t} \rightarrow - \partial_{t}$\footnote{The (continuous) variable $t$ corresponds to the (continuous-time limit of the discrete) time of our (resp. the standard) convention before (resp. after) the substitution.} in the final PDE to obtain the desired result.

Note that the standard-convention shift operation makes the upper spin component go right in time, and the lower go left, whereas this is reversed in our protocol. 

   %%%%%%%%%%%%%%%%%%%%%%%%%%%%%%
%%  Alias bibliographiques  %%
%%%%%%%%%%%%%%%%%%%%%%%%%%%%%%
\defcitealias{NRFortran}{Numerical Recipes}
\defcitealias{NRF77}{Numerical Recipes}
\defcitealias{NRF90}{Numerical Recipes}
\defcitealias{IUPAC}{IUPAC}%International Union of Pure and Applied Chemistry

%\chapter{\textsc{Compléments mathématiques}\label{Ann:Maths}}% Intermezzo mathématique

\chapter{Geometry of field theories} \label{app:geometry}

\section{Fiber bundles}

\subsection{Definition}

To each element $x$ of a so-called base space $B$, we attach a copy of a so-called fiber space $F$, which is then noted $F_x$, the fiber at $x$. The collection of all the fibers, $E_F \equiv \bigcup_{x \in B} F_x$, is called the total space. We also introduce a surjection $\pi:E_F \rightarrow B$, called projection on the base, such that $\pi^{-1}(x) = F_x$. Now, for the total space to be a so-called fiber bundle, it must be locally (topologically) trivial, i.e. locally (homeomorphically mappable to) the simplest space one can build out of two spaces $F$ and $B$, which is their cartesian product. Formally, this means that for any $x \in B$, there exist a neighborhood $U$ of $x$ such that the inverse projection of $U$ is homeomorphic to $U \times F$. This homeomorphism $\phi$ is like a  chart (function) of the subset $\pi^{-1}(U)$ of $E_F$ (and thus a \emph{local} chart of $E_F$), $\phi: U \times F \rightarrow  \pi^{-1}(U)$, called local trivialization of $E_F$ (it is a trivialization of $\pi^{-1}(U)$)\footnote{The notion of chart we use here is more general than the `standard' one used for manifolds (that is why we rather say `trivialization' than `chart'). Indeed, the codomain $U \times F$ cannot be homeomorphised to a Euclidean space $\mathbb{R}^n$ if $U$ and $F$ are not themselves homeomorphic to some Euclidean space. Moreover, we are not, at this stage, interested in localising absolute positions (which would require a metric, while the notions we define here do not need any metric): the role of $\phi$ is simply to indicate that the local topology is trivial, i.e. is reduced to that of the objects we take as inputs, i.e. the base and the fiber, to build a new one, i.e. the fiber bundle; we will see below that this implies that we do not need to specify the trivializing functions out of the `reconnection domains' (see below). As a sum-up: locally, there is no additional topological information in the object we build than that of the input objects. Note however that, if we introduce a metric, there can be additional metric information even locally, i.e.  in $\pi^{-1}(U)$ with respect to that contained in $U$ and $F$. 
\iffalse
One can view this by introducing a metric (in finite dimensions, all norms are equivalent) and a curvature. by going from the domain $\pi^{-1}(U)$ to the codomain $U \times F$, we get rid of a possible non-trivial topology induced by the way we glue the fibers together.); there can be a remaining curvature, but only coming from either $B$, $F$, or an homeomorphic image of the global cartesian product $B\times F$ (while there is no additional curvature if the codomain is Euclidean).
\fi 

Note that `standard' charts on manifolds are rather defined from the manifold domain to the Euclidean codomain. Here the word `trivialization' should also rather be used for the inverse function $\phi^{-1}$, but it is common, in this fiber-bundle context, to define $\phi$ from $U \times F$ to $\pi^{-1}(U)$, because it is practical to track the elements of the fiber $F$ as arguments of the function we define, instead of images.}. This means that the additional (topological) information on $E_F$, apart from the that of the fiber and the base, is contained in the global structure of $E_F$; locally, $E_F$ carries no more (topological) information than that contained in the fiber and the base.

This condition of local (topological) triviality thus defines fiber bundles. Now, how to describe the variety of these objects? The question is how to characterise a given fiber bundle, how to construct it, how to attach the fibers to the base in order to build a globally (topologically) non-trivial space out of the fiber and the base? The most common way of characterising/building the fiber bundle is by specifying how a so-called structure group $G$ acts on the fiber.

Before explaining what the action of a group $G$ is, and how such an action on the fiber is related to the topological structure of the fiber bundle $E_F$, we must introduce the standard object used to describe the topology of a space, that is, an atlas\footnote{Again, the notion of `atlas' used here is more general than that used in the study of manifolds, because there is no metric for now, only topology, while there is metricity by definition in the study of manifolds, since we refer to Euclidean spaces.}. Of course, it would not be very clever to introduce an atlas of $E_F$ ignoring that it is a fiber bundle. Let us introduce a suitable atlas. Let $(U_i)_{i \in I}$ be an open cover of the base, and $\phi_i: U_i \times F \rightarrow  \pi^{-1}(U_i)$ the trivialization of $\pi^{-1}(U_i)$. The particular suitable atlas of $E_F$ I was announcing is thus simply $(\pi^{-1}(U_i),\phi_i)_{i \in I}$. Of course, we can end up with a `standard' Euclidean atlas if we can provide Euclidean atlases for $U$ and $F$. The global topology of the space $E_F$ is encoded in the so-called transition functions defined by $t_{i,j}(x)=\phi_{i,x} \circ \phi_{x,j}: F \rightarrow F$, where $\phi_{i,x}(f)\equiv \phi_i(x,f)$, and the domain of definition of $t_{i,j}$ is the intersection $U_i \cap U_j$. Since we already know the local topology of $E_F$, it is (topologically) useless to specify the trivializing functions out of the intersections\footnote{Again, this is why we use the word `trivialization' rather than `chart', which would imply that we want to localise an absolute position, which can only be done if we introduce a metric, while this is not necessary for our purpose right now.}, but we do need to specify the transition functions.

Let us now define the (left) action $\varphi$ of some group $G$ on some space $S$ which is then called (left) $G$-set. Have in mind, as a broader motivation for such a definition, that we often find the situation where some family of internal transformations in some space $S$ happen to form a group in the mathematical sense\footnote{The minimal non-trivial internal transformation in some space $S$ is the exchange between the (generically abstract) positions, i.e. the labels, of two elements of $S$, called a transposition. The set of such an exchange transformation and the identity is the simplest example of non-trivial group we can give.}. We thus define
\begin{align}
\varphi: \, &  G \times S \rightarrow S \\
& (g,s) \mapsto \varphi(g,s) \equiv g \cdot s \nonumber \, ,
\end{align}
which must satisfy the two following conditions, for all $x \in S$:
\begin{equation}
\begin{array}{ll}
\text{identity condition:} & e \cdot x = x \\
\text{compatibility condition:}  & (gh) \cdot x = g \cdot (h \cdot x) \, ,
\end{array}
\end{equation}
where $(g,h) \in G^2$ and $e$ is the neutral element of $G$.

Now, as announced, we require that some group $G$ structure the fiber, which means the following: the (image of some $x\in U_i \cap U_j$ by the) transition function is constrained to be an element of $G$, i.e. $t_{i,j}: U_i \cap U_j \rightarrow G$. The fiber bundle is then called fiber $G$-bundle, and the atlas a $G$-atlas. The fiber bundle is said trivial when $G$ is reduced to its neutral element.

\subsection{Examples}

\subsubsection{Sphere bundles}

The lines correspond to the base space, and the columns to the fiber. \\

\noindent
{\small
\begin{tabular}{|c||c|c|c|}
\hline  
 \multicolumn{4}{|c|}{$G=\{e\}$: trivial bundle}  \\
 
\hline
\hline
 & segment & circle & sphere   \\

\hline
\hline
segment & filled rectangle & cylinder & filled torus if segment length smaller than diameter, ball otherwise\\

\hline
circle & cylinder & torus & filled torus \\
 
\hline
sphere & 3D annulus & 3D annulus & 3D annulus  \\
     
\hline
\end{tabular} \\
\vspace{0.3cm}

\noindent
\begin{tabular}{|c||c|c|c|}
\hline  
\multicolumn{4}{|c|}{$G=\{e,r\}$, $r$=reflexion}  \\
 
\hline
\hline
& segment & circle & sphere   \\

\hline
\hline
segment & filled rectangle & cylinder & \\

\hline
circle & M\"{o}bius strip & Klein bottle &  \\
 
\hline
sphere &  &  &   \\
     
\hline
\end{tabular}
\vspace{0.3cm}

\noindent
\begin{tabular}{|c||c|c|c|}
\hline  
 \multicolumn{4}{|c|}{$G=U(1)$=circle}  \\
 
\hline
\hline
& $G$ with right action (or $S^1$ embeded in $S^3$ (in a fibration view)) &  & \\

\hline
\hline
$S^3$ (or $S^2$) & Hopf fibration &  &  \\

\hline
&  &  &  \\
 
\hline
&  &  &   \\
     
\hline
\end{tabular}

}

\subsubsection{Vector bundles}

The fiber is a vectorial space $V$.

\vspace{-0.3cm}
\paragraph{{$\bullet$ Tangent bundle}}\hspace{1cm}

\vspace{0.1cm}

\noindent
The fiber at $x$ is the tangent space $T_x M$ of some base manifold\footnote{To obtain a manifold from a topological space, we must endow it with a differential structure, i.e. we must be able to do differential calculus on it, although this is not that easy when the manifold is not flat, and is best approached, e.g., with Cartan's exterior calculus.} $B=M$ at $x$. The so-called tangent bundle is then denoted $T M$.

\vspace{-0.3cm}
\paragraph{{$\bullet$ Cotangent bundle}} \hspace{1cm}

\vspace{0.1cm}

\noindent
It is useful to define the cotangent space to do calculus on the manifold. Let us use Dirac notations, and let $\ket{v_x}$ be a vector of $T_xM$. The cotangent space at $x$ is that generated by all the bras $\langle v_x |$. More generically, given a vector space $\{\ket{v}\}$, we define the co-space, or dual space $\{ \langle v |\}$. The cotangent space at $x$ is denoted $T^{\ast}_xM$.

\vspace{-0.3cm}
\paragraph{{$\bullet$ Vector bundle of a representation-space generated by linear representations of groups}} \hspace{1cm}

\vspace{0.1cm}

\noindent
I find Wikipedia's ``Group representation'' web page very good as a general (very) short presentation of this wide subject.
Consider a given group $G$ acting on several spaces $S_{\alpha}$'s (say, from the left):
\begin{align}
\varphi_{\alpha}: \, &  G \times S_{\alpha} \rightarrow S_{\alpha} \\
& (g,s) \mapsto \varphi_{\alpha}(g,s) \nonumber \, .
\end{align}
If $S_{\alpha}$ is a vectorial space $S_{\alpha}=V_{\alpha}$, it is natural to be interested in linear actions of $G$ on the $S_{\alpha}$'s, in which case one would use a notation such as $\varphi_{\alpha}(g,s) \equiv \rho_{\alpha}(g) \cdot s = \rho_{\alpha}(g) s$. The function $\rho_{\alpha}$ is called linear representation of $G$, $\rho_{\alpha}(G)$ is the representation image, and $V_{\alpha}$ the representation space.
The main part (this is a euphemism) of the mathematical litterature on group representation theory  deals with linear representations\footnote{Some references on non-linear representations of groups are \url{https://en.wikipedia.org/wiki/Nonlinear_realization}, \url{http://www.numdam.org/article/ASENS_1977_4_10_3_405_0.pdf}, \url{http://www.ams.org/journals/bull/1984-11-02/S0273-0979-1984-15290-X/S0273-0979-1984-15290-X.pdf}, \url{https://www.researchgate.net/profile/Daniel_Sternheimer/publication/225993353_Nonlinear_group_representations_and_evolution_equations/links/0deec52b064c7c2508000000/Nonlinear-group-representations-and-evolution-equations.pdf},\url{https://www.researchgate.net/profile/Daniel_Sternheimer/publication/225993353_Nonlinear_group_representations_and_evolution_equations/links/0deec52b064c7c2508000000/Nonlinear-group-representations-and-evolution-equations.pdf}. In the work of DAriano, there is also a non-linear representation \cite{DAriano2016}.}.

\subsubsection{Principal bundles}

The fiber is the structure group $G$ itself (endowed with right action). The fiber bundle is then called principal $G$-bundle.

\iffalse

\paragraph{{$\bullet$ Frame bundle associated to a vector bundle}} \hspace{1cm}

\vspace{0.3cm}

\section{Connections on fiber bundles}

As we have mentioned, doing calculus on a generic manifold is not that easy. Let us be more precise.

\section{Exterior derivative, Cartan's calculus}

\section{Applications to general relativity}

\section{Applications to gauge theories}

\fi

   %%%%%%%%%%%%%%%%%%%%%%%%%%%%%%
%%  Alias bibliographiques  %%
%%%%%%%%%%%%%%%%%%%%%%%%%%%%%%
\defcitealias{NRFortran}{Numerical Recipes}
\defcitealias{NRF77}{Numerical Recipes}
\defcitealias{NRF90}{Numerical Recipes}
\defcitealias{IUPAC}{IUPAC}%International Union of Pure and Applied Chemistry

%\chapter{\textsc{Compléments mathématiques}\label{Ann:Maths}}% Intermezzo mathématique

\chapter{Gauge-invariant generator of a symmetry of the electromagnetic field (Hamiltonian mechanics)} \label{app:gauge_invariant_gen}

\section{Introduction: electromagnetic field and potential}

Consider a particle of charge $q$ and mass $m$ in an \emph{electromagnetic field} $(\mathbf{E}(\mathbf{r},t),\mathbf{B}(\mathbf{r},t))$. To analyse this system in the framework of analytical (and hence quantum) mechanics, we must describe the electromagnetic field with an \emph{electromagnetic potential} $(\phi_{g}(\mathbf{r},t),\mathbf{A}_{g}(\mathbf{r},t))$, satisfying
\begin{align}
\mathbf{B} & = \boldsymbol{\nabla} \wedge \mathbf{A}_{g}\ , \\
\mathbf{E} & = - \boldsymbol{\nabla} \phi_{g} - \frac{\partial \mathbf{A}_{g} }{\partial t} 
\ .
\end{align}
\noindent
The existence of such a potential is a consequence of the homogeneous Maxwell's equations. This potential is not unique, and the subscript $g$ denotes the chosen \emph{gauge}, which is at this stage arbitrary. 

\section{Symmetry of the electromagnetic field and associated conserved quantity (Lagrangian formalism)}

A possible \emph{Lagrangian} for the previous system is
\begin{equation}
L_g(\mathbf{r},\mathbf{v},t) = \frac{1}{2} m \mathbf{v}^2 
- q\left[\phi_{g}(\mathbf{r},t) - \mathbf{v} \cdot \mathbf{A}_{g}(\mathbf{r},t)\right] \ .
\end{equation}
Suppose the electromagnetic field is invariant under some \emph{continuous symmetry} $\mathcal{S}$.
This implies that there exist a gauge $g_o$ in which the electromagnetic potential $(\phi_{g_o}(\mathbf{r},t),\mathbf{A}_{g_o}(\mathbf{r},t))$, and hence the Lagrangian $L_{g_o}(\mathbf{r},\mathbf{v},t)$, are invariant under $\mathcal{S}$ (otherwise the \emph{action}, which is function only of the fields and not the potentials if one neglects the boundary terms, could not be invariant under $\mathcal{S}$). Noether's theorem then exhibits a \emph{conserved quantity} $Q^{\mathcal{S}}(\mathbf{r},\mathbf{v},t)$ associated to this symmetry $\mathcal{S}$, in terms of a derivative of $L_{g_o}(\mathbf{r},\mathbf{v},t)$. The conservation equation reads:
\begin{equation} \label{conservation_eq}
\frac{\mathrm{d}}{\mathrm{d}t} Q^{\mathcal{S}}(\mathbf{r}(t),\mathbf{v}(t),t) = 0 \ .
\end{equation}

\section{Generator of the symmetry (Hamiltonian formalism)}

The \emph{Hamiltonian} of the system reads
\begin{equation}
H_{g}(\mathbf{r},\mathbf{p}_{g},t) = \frac{\left( \mathbf{p}_{g} - q \mathbf{A}_{g}(\mathbf{r},t) \right)^2}{2m} + q \phi_{g}(\mathbf{r},t) \ ,
\end{equation}
\noindent
where the \emph{canonical momentum} is defined by
\begin{equation} \label{canonical_momentum}
\mathbf{p}_{g} \equiv \frac{\partial L_g}{\partial {\mathbf{v}}}  = m \mathbf{v} + q \mathbf{A}_{g} \ ,
\end{equation}
and depends manifestly on the gauge $g$\footnote{Indeed, the speed of the particle, $\mathbf{v}$, does not depend on the gauge, since it is determined by Newton's law, where only the electromagnetic field plays a role (through the Lorentz force), and not the potential.}.
The invariance of $L_{g_o}(\mathbf{r},\mathbf{v},t)$ under $\mathcal{S}$ implies the invariance of $H_{g_o}(\mathbf{r},\mathbf{p}_{g_o},t)$ under $\mathcal{S}$.

We rewrite the conserved quantity as a function of $(\mathbf{r},\mathbf{p}_{g_o},t)$, using (\ref{canonical_momentum}) for $g = g_o$ :
\begin{equation}
G^{\mathcal{S}}_{g_o}(\mathbf{r},\mathbf{p}_{g_o},t) \equiv Q^{\mathcal{S}}\left(\mathbf{r},\frac{\mathbf{p}_{g_o} - q \mathbf{A}_{g_o}(\mathbf{r},t)}{m},t\right) \ .
\end{equation}
As a function of its variables, $G^{\mathcal{S}}_{g_o}$ is called \emph{generator} of $\mathcal{S}$. This function $G^{\mathcal{S}}_{g_o}$ depends explicitly on $g_o$ through $\mathbf{A}_{g_o}$ (hence the subscript $g_o$ in $G^{\mathcal{S}}_{g_o}$). This means that this function is not gauge invariant in the following sense: if we consider a generic gauge $g$, the quantity $G^{\mathcal{S}}_{g_o}(\mathbf{r},\mathbf{p}_{g},t)$ is not a conserved quantity, and this simply because the canonical momentum is not gauge invariant. 

Recall  that total time-derivative of some function $f(\mathbf{r},\mathbf{p}_{g},t)$ is given by:
\begin{equation}
\frac{\mathrm{d}f}{\mathrm{d}t} = \{f,H_g\}_g + \frac{\partial f}{\partial t} \ ,
\end{equation}
where the $g$- \emph{Poisson brackets} of two functions $A$ and $B$ of the variables $(\mathbf{r},\mathbf{p}_{g},t)$ is defined by
\begin{equation}
\{A,B\}_g = \sum_{i=1}^{3} \frac{\partial A}{\partial x^i} \frac{\partial B}{\partial p_g^i} - \frac{\partial B}{\partial x^i} \frac{\partial A}{\partial p_g^i} \ ,
\end{equation}
so that the conservation equation, (\ref{conservation_eq}), reads:
\begin{equation} \label{conservation_eq_bis}
\{G^{\mathcal{S}}_{g_o},H_{g_o}\}_{g_o} + \frac{\partial G^{\mathcal{S}}_{g_o}}{\partial t} = 0 \ .
\end{equation}
When the generator does not depend explicitly on time, which is equivalent to state that the Hamiltonian $H_{g_o}(\mathbf{r},\mathbf{p}_{g_o},t)$ does not depend explicitly on time, its conservation is then simply expressed by its commutation (in the sense of Poisson brackets) with this Hamiltonian. Let us rephrase what we have seen in the former paragraph: if we write the Hamiltonian in an arbitrary gauge $g$, the function of $(\mathbf{r},\mathbf{p}_{g},t)$ associated to $\mathcal{S}$, {\sl i.e.} that will commute with the Hamiltonian in the sense of (\ref{conservation_eq_bis}), is not $G^{\mathcal{S}}_{g_o}$. We are looking for the expression of such a function, that we will call \emph{gauge-invariant generator}, as a function of the usual generator $G^{\mathcal{S}}_{g_o}$.

\section{Gauge-invariant generator}

Gauges $g$ and $g_o$ are generically linked by:
\begin{align}
\phi_g(\mathbf{r},t) &= \phi_{g_o}(\mathbf{r},t) - \frac{\partial}{\partial t} \chi_g({\mathbf{r},t}) \ , \\
\mathbf{A}_g(\mathbf{r},t) &= \mathbf{A}_{g_o}(\mathbf{r},t) + \boldsymbol{\nabla} \chi_g({\mathbf{r},t}) \ ,
\end{align}
so that the canonical momenta in the two gauges are linked by
\begin{equation}
\mathbf{p}_g = \mathbf{p}_{g_o} + q \boldsymbol{\nabla} \chi_g \ .
\end{equation}
We can then express $G^{\mathcal{S}}_{g_o}(\mathbf{r},\mathbf{p}_{g_o},t)$ as a function of the momentum in a generic gauge, namely  $\mathbf{p}_g$:
\begin{equation} \label{solution}
\boxed{
G^{\mathcal{S}}_g(\mathbf{r},\mathbf{p}_{g},t)
\equiv G^{\mathcal{S}}_{g_o}(\mathbf{r},\mathbf{p}_{g} - q \boldsymbol{\nabla} \chi_g(\mathbf{r},t),t) } \ . 
\end{equation} 
This expression is by construction a conserved quantity, and it is gauge invariant in the following sense: for any gauges $g$ and $g'$, we have
\begin{equation}
G^{\mathcal{S}}_g(\mathbf{r},\mathbf{p}_{g},t) = G^{\mathcal{S}}_{g'}(\mathbf{r},\mathbf{p}_{g'},t) = Q^{\mathcal{S}}(\mathbf{r},\mathbf{v},t) \ .
\end{equation}
Thus, whatever gauge $g$ we choose to write the Hamiltonian of the system, be it non invariant under $\mathcal{S}$, we can always, from the knowledge of the generator $G^{\mathcal{S}}_{g_o}$, find a function of $\mathbf{p}_{g}$, namely the gauge-invariant generator $G^{\mathcal{S}}_g$, which is conserved in time, that is:
\begin{equation} \label{conservation_eq_bis_bis}
\{G^{\mathcal{S}}_{g},H_{g}\}_{g} + \frac{\partial G^{\mathcal{S}}_{g}}{\partial t} = 0 \ .
\end{equation}
   %%%%%%%%%%%%%%%%%%%%%%%%%%%%%%
%%  Alias bibliographiques  %%
%%%%%%%%%%%%%%%%%%%%%%%%%%%%%%
\defcitealias{NRFortran}{Numerical Recipes}
\defcitealias{NRF77}{Numerical Recipes}
\defcitealias{NRF90}{Numerical Recipes}
\defcitealias{IUPAC}{IUPAC}%International Union of Pure and Applied Chemistry

%\chapter{\textsc{Compléments mathématiques}\label{Ann:Maths}}% Intermezzo mathématique

\chapter{On the isospin} \label{app:isospin}

The mass of the proton and that of the neutron differ approximately by 0.1\% from each other. At sufficiently low energy\footnote{Typically, at or below nuclear energies.}, one can thus not detect the difference between both masses, so that the proton and neutron differ only in (i) their electric charge, $e$ or $0$,  and (ii) their so-called (nuclear) isospin (contraction of `isotopic spin'), $+1/2$ or $-1/2$.

The isospin internal degree of freedom (d.o.f.) is that associated to the nuclear force, as the electromagnetic internal d.o.f., i.e. the complex phase of the considered matter field, is associated to the electromagnetic force, which was found out by Weyl, Fock and London, and widespread by Pauli in 1941 \cite{Pauli1941}.

The nuclear-isospin d.o.f. is however more complex than the electromagnetic phase. It is named that way because it is described by the spin-1/2 formalism: indeed, it belongs to a two-dimensional complex Hilbert space and transforms under its associated (global) gauge transformations as the spin transforms under spatial rotations. The projection of the isospin d.o.f. on some arbitrary axis is $+1/2$ for a proton, and $-1/2$ for a neutron. Note two things: (i) this has nothing to do with measurement or physical angular momentum, only the mathematics are the same, and (ii) a rotation of the isospin d.o.f. from projection $+1/2$ to $-1/2$ is enough to transform a proton into a neutron (or vice versa),  \emph{only} if one can neglect the electromagnetic interaction.

The idea of the isospin, i.e. that the neutron and the proton are two manifestations of the same nucleon, was introduced by Heisenberg in 1932 \cite{Heisenberg1932I,  Heisenberg1932II, Heisenberg1932III}, but it was named that way only in 1937, by Wigner, who underlined the aforementioned link with the spin (see the introduction of \cite{Yang1954} for historical details).

In 1954, Yang and Mills proposed, extending Weyl's gauge principle for quantum electromagnetism to the nuclear interaction, to make the invariance of nuclear interactions under global isospin rotations, a local invariance. Since the SU$(2)$ group of isospin gauge transformations is non-Abelian, on the contrary to the U$(1)$ group of electromagnetic gauge transformations, the isospin gauge field is called a non-Abelian gauge field. Yang and Mills did not reach a conclusion regarding the mass of this nuclear non-Abelian gauge field, which mediates the nuclear force, as the electromagnetic field mediates the electromagnetic force. This mass had to be computed and compared with that of pions, the mesons that mediate the nuclear interaction, as described by Yukawa's meson theory; in his paper introducing the now so-called Yukawa's potential  for the nuclear interaction \cite{Yukawa1935}, Yukawa gave a prediction for the mass of the mediator, which was later recognized to be the pi meson, i.e. the pion, discovered in 1947, see \url{https://www.nature.com/physics/looking-back/lattes/lattes.pdf} for the original paper and \url{http://fafnir.phyast.pitt.edu/particles/pion.html} for 1997 comments from the CERN Courier. This picture of the nuclear interaction between nucleons as mediated by pions was later understood as a residual effect of the strong interaction between quarks inside and between nucleons.

   %%%%%%%%%%%%%%%%%%%%%%%%%%%%%%
%%  Alias bibliographiques  %%
%%%%%%%%%%%%%%%%%%%%%%%%%%%%%%
\defcitealias{NRFortran}{Numerical Recipes}
\defcitealias{NRF77}{Numerical Recipes}
\defcitealias{NRF90}{Numerical Recipes}
\defcitealias{IUPAC}{IUPAC}%International Union of Pure and Applied Chemistry

%\chapter{\textsc{Compléments mathématiques}\label{Ann:Maths}}% Intermezzo mathématique

\chapter{Classical fields and spin}  \label{app:classical_fields_and_spin}

We can point out two notable differences between half-integer- and integer-spin (classical i.e. first-quantized) fields, in the way they couple to gauge fields: (i) the Noether current $J$ (associated to the gauge-group charge conservation) of half-integer-spin fields must be written with gamma matrices (trace of the spinorial (i.e. half-integer-spin) nature of the field in the Noether current), and (ii), apart from (i), the coupling between the gauge field and the Noether current is the same for both field natures, in the sense that it takes the form (in the Lagrangian or the Hamiltonian) $J^{\mu} A_{\mu}$, with $J_{\mu} \sim \Psi \partial_{\mu} \Psi$ for a half-integer-spin field $\Psi$ and  $J \sim \Phi \partial_{\mu} \Phi$ for an integer-spin field $\Phi$, \emph{but} for integer-spin fields, there is an additional mass-like coupling $(\Phi A_{\mu})^2$ (in the Lagrangian or the Hamiltonian), which is involved in pair creation; an application to the case of the Higgs mechanism is detailed on the following slides by Englert: \url{https://indico.lal.in2p3.fr/event/2187/contribution/0/material/slides/0.pdf}.

Apart from these differences regarding their coupling to gauge fields, half-integer- and integer-spin fields differ in their free dynamics: the matter free term in the Lagrangian takes the schematic form $\Psi  \partial_{\mu}\Psi$ for a half-integer-spin field, and $(\partial_{\mu}\Phi)^2$ for an integer-spin field.

I wonder how (if) we can single out in a simple picture the phenomenal properties of the Noether quantum coupling between matter and gauge fields, i.e. that of the form $J A_{\mu}$, which are independent of the matter-field spin.

In the non-relativistic limit, these phenomenal properties result, at least for unstructured particles (i.e., in a non-quantum framework, point particles) from the interplay between the usual features of quantum mechanics and classical electrodynamics, which are formally captured by the Hamiltonian of an unstructured particle coupled to a gauge field (consider minimal coupling), if we forget about the half-integer-spin additional term that couples to the magnetic field (in the case of an Abelian gauge field); in other words, in such a non-relativistic framework, the essence of (first-quantized) quantum electrodynamics is contained in the correspondance principle applied to the electrodynamical classical Hamiltonian. The following papers are interesting: \url{https://www.lorentz.leidenuniv.nl/IL-publications/sources/Kramers_50.pdf},  \url{http://link.springer.com/chapter/10.1007\%2F978-3-662-04360-8_42}, \url{https://www.math.toronto.edu/sigal/publications/53.pdf}. There is also a book by Healey entitled \emph{Non-relativistic Quantum Electrodynamics}. Here are some notes on the scattering theory of non-relativistic quantum electrodynamics: \url{http://iopscience.iop.org/article/10.1088/0305-4470/16/1/014/meta}.

In a relativistic context, the idea would be to determine the phenomenal role played by the trace of the spinorial nature of the field (i.e. the presence of gamma matrices) in the current $J$, and the relativistic phenomenal component which is independent from the spinorial nature: is this phenomenal component `merely' contained in the relativistic dispersion relation, which is the same whether  the field has half-integer or integer spin?
   %%%%%%%%%%%%%%%%%%%%%%%%%%%%%%
%%  Alias bibliographiques  %%
%%%%%%%%%%%%%%%%%%%%%%%%%%%%%%
\defcitealias{NRFortran}{Numerical Recipes}
\defcitealias{NRF77}{Numerical Recipes}
\defcitealias{NRF90}{Numerical Recipes}
\defcitealias{IUPAC}{IUPAC}%International Union of Pure and Applied Chemistry

%\chapter{\textsc{Compléments mathématiques}\label{Ann:Maths}}% Intermezzo mathématique

\chapter{Effect of linear plane gravitational waves on a DTQW matter distribution located in the polarization plane of the wave} \label{app:GWs}

\section{Discussion without equations}

Suppose we are interested in how a GW influences an initial matter distribution (or density), located in some plane transverse to this GW, which is stationary in free space i.e. when there is no GW.

Since the free-space $(\xi=0)$ walk operator is local in the momentum, any initial state whose spatial part (as opposed to the spin part) is a single Fourier mode (this is of course an ideal, non-physical situation), is stationary in free space (its density is homogeneous at any time), i.e. the free-space walk operator can only change the polarization of this Fourier mode. Now, among these states, we choose plane waves, i.e. eigenstates of the free-space walk operator, whose polarization is not changed by the time evolution, in order to capture only the density change induced by the GW and get rid of that induced by the free-space polarization change; One could think of considering such a single free-space plane wave as initial state. However, the density of such a state will not be modified by the GW, because the walk operator is local in the momentum even in the presence of a GW $(\xi\neq0)$.

One must then consider a superposition of Fourier modes as initial state. The minimal superposition is that of two modes, with generic wavevectors $\boldsymbol{k}^1$ and $\boldsymbol{k}^2$. As said above, we want this state to be a free-space eigenstate, so that these two Fourier modes must have the same free-space eigen-energy. Their respective wavevectors must then have the same norm; and both polarizations must correspond to same-sign, say positive energies. The initial condition used in Subsection VI B of the paper is constructed in that way. For simplicity purposes, the retained wavevectors $\boldsymbol{k}^1$ and $\boldsymbol{k}^2$ are chosen to be orthogonal, respectively along the $X$ and the $Y$ axis.

In addition to (i) its easy technical tractability and (ii) its pedagogical interest (minimal free-space eigenstate having a density modified by the GW), this situation is interesting because this initial double-peaked momentum distribution yields an interference pattern in the physical-space density of the walker (even in the absence of GW), and this interference pattern facilitates the detection of the deformation induced by the GW, because of at least two reasons: (i) the high contrast between low- and high-density regions, and (ii) the fact that a measurement on a sample displaying a pattern predicted periodic by theory is more precise than if the pattern is non-periodic, because one can make (spatial) statistics thanks to these replications.

\section{Formal material}

If $\lambda_{\pm}(\xi,G(T);q_X,q_Y)$ are the eigenvalues of the walk operator in Fourier space,  namely, $W(\xi,G(T);q_X,q_Y)$, we define the eigen-energies by 
\begin{equation}
\lambda_{\pm}(\xi,G(T);q_X,q_Y) = 
\exp [-i E_{\pm}(\xi,G(T);q_X,q_Y) ] \, .
\end{equation}
In Section V of the paper, we have computed the walk operator in Fourier space at first order in $\xi$; it reads
\begin{align}
&W(\xi,G(T);q_X,q_Y) \\
& \ \ \ \ \simeq \mathcal{W}^{(1)}(\xi,G(T);q_X,q_Y) \nonumber \\
& \ \ \ \ = W^{(0)}(q_X,q_Y) + \xi G(T) \, W^{(1)}(q_X,q_Y) \, , \nonumber
\end{align}
where the superscript in $\mathcal{W}^{(1)}$ means a computation at first order in $\xi$, and $W^{(0)}$ and $W^{(1)}$ are given by Eqs. (28) and (29) of the paper, respectively.
If we work at first order in $\xi$, it only makes sens to search for eigen-elements of the walk operator at first order in $\xi$.  
We thus look for eigen-energies and eigenvectors of the form
\begin{subequations}
\begin{align}
&E_{\pm}(\xi,G(T);q_X,q_Y) \\
&\ \ \ \ \ \  \ \simeq \mathcal{E}_{\pm}^{(1)} (\xi,G(T);q_X,q_Y) \nonumber \\
& \ \ \ \ \ \  \ = E_{\pm}^{(0)}(q_X,q_Y)
+ \xi G(T) \, E^{(1)}(q_X,q_Y)  \, ,\nonumber \\ 
&V_{\pm}(\xi,G(T);q_X,q_Y) \\
&\ \ \ \ \ \  \ \simeq \mathcal{V}_{\pm}^{(1)}(\xi,G(T);q_X,q_Y) \nonumber \\
& \ \ \ \ \ \  \ = V_{\pm}^{(0)}(q_X,q_Y)
+ \xi G(T) \, V^{(1)}_{\pm}(q_X,q_Y) \ , \nonumber 
\end{align}
\end{subequations}
assuming no degeneracy, and where $\mathcal{E}_{\pm}^{(r)}(\xi,G(T);q_X,q_Y)$ and $\mathcal{V}_{\pm}^{(r)}(\xi,G(T);q_X,q_Y)$ are respectively the eigen-energies and eigenvectors computed at order $r$ in $\xi$.
%(of course, the zeroth orders, given by the substitution $\xi=0$, coincide with $E_{\pm}^{(0)}(q_X,q_Y)$ and $V_{\pm}^{(0)}(q_X,q_Y)$). 
A small calculation (that of standard perturbation theory) would then gives us the corrections as functions of the zeroth-order eigen-elements and of the walk perturbation $W^{(1)}$. 
In Subsection VI A, we have computed these eigen-elements, not only at first order in $\xi$, but also at first order in $q_X$ and $q_Y$, i.e. only for large spatial scales, which are those relevant to the continuum limit.

We want to go beyond. We are interested in how a gravitational wave, described by the walk operator at first order in $\xi$, influences an initial matter distribution (in the transverse $(X,Y)$ plane) which is stationary in free space (i.e. if there is no wave). Elementary stationary distributions are generated by free walkers of given energy, i.e. eigenvectors of the real-space version of $W^{(0)}(q_X,q_Y)$\footnote{As mentioned in the previous section, there \emph{are} stationary distributions which are \emph{not} eigenvectors of (the real-space version of) $W^{(0)}(q_X,q_Y)$: these correspond to Fourier modes whose polarization is \emph{not} an eigenpolarization of the zeroth-order walk operator}. As mentioned, we want to study not only large scales, i.e. not only the continuum-limit situation, which is that of a Dirac fermion in curved space, but also small scales comparable to the lattice spacing, where the DTQW evolution differs from that of a Dirac fermion in curved space; we will thus make \emph{no} Taylor developments in $q_X$ and $q_Y$.

 In terms of perturbation theory, the term accounting for such a modification of the distribution is the walk perturbation $W^{(1)}$ applied to a zeroth-order eigenvector, namely, $\xi G(T) \, W^{(1)} V_{\pm}^{(0)}(q_X,q_Y)$. For the initial distribution to be stationary in the presence of the wave, at first order in $\xi$, one would need to add the term $\xi G(T) \, W^{(0)} V_{\pm}^{(1)}(q_X,q_Y)$. We do not add this term, and thus the initial distribution is indeed modified by the wave.

Generically, any initial walker $\tilde{\Psi}_0(k_X,k_Y)$ can be decomposed as a superposition of the plus and minus eigenvectors, since these are a basis of the Hilbert space; such a decomposition can be written
\begin{align}
&\tilde{\Psi}_0(k_X,k_Y) \\
& \ \ \ \ \ = \ \ \alpha_+(k_X,k_Y) \frac{V_+^{(0)}(q_X,q_Y)}{||V_+^{(0)}(q_X,q_Y) ||_{\text{spin}}} \nonumber \\
& \ \ \ \ \ \ \ + \alpha_-(k_X,k_Y) \frac{V_-^{(0)}(q_X,q_Y)}{||V_-^{(0)}(q_X,q_Y) ||_{\text{spin}}} \, , \nonumber
\end{align}
where $||\cdot||_{\text{spin}}$ indicates the norm in spin space, i.e. the square-rooted density. This decomposition must be normalised, i.e.
\begin{align}
&\int \! \! \!\int \tilde{\Psi}^{\dag}_0(k_X,k_Y)\tilde{\Psi}_0(k_X,k_Y) dk_Xdk_Y \nonumber \\
& \ \ \ \ \ =\int \! \! \!\int ( |\alpha_+(k_X,k_Y)|^2 + |\alpha_-(k_X,k_Y)|^2) dk_Xdk_Y \nonumber \\
& \ \ \ \ \ = 1 \, ,
\end{align}
where
\begin{equation}
\int \! \! \! \int = \frac{1}{\sqrt{2\pi}^2}\int_{k_X=-\pi}^{\pi} \int_{k_Y=-\pi}^{\pi} \, .
\end{equation}
In order for the initial distribution to be stationary under $W^{(0)}(q_X,q_Y)$, it has to be an eigenvector of this operator (as already mentioned), say $V_+^{(0)}(q_X,q_Y)$, which means choosing $\alpha_-=0$ and $\alpha_+$ such that
\begin{equation}
\frac{1}{\sqrt{2\pi}^2}\int_{k_X=-\pi}^{\pi} \int_{k_Y=-\pi}^{\pi} |\alpha_+(k_x,k_Y)|^2 dk_Xdk_Y = 1 \, .
\end{equation}

One could consider $|\alpha_+(k_x,k_Y)|^2$ to be a Gaussian centered around some wavevector ${\boldsymbol k}_0$, that is, classically, a walker going at some given average classical speed ${\boldsymbol k}_0$. One would thus have, to go to physical space, to integrate the derivative of a Gaussian (product of $V_+$, which contains terms proportional to the wavevector, by a Gaussian), which would yield a Gaussian. If the Gaussian in momentum space is very peaked, the Gaussian in physical space would have a very large spread, i.e. it would be almost a uniform distribution in some finite region centered around its maximum. If the Gaussian in momentum space has a large spread, this would yield a walker localised in physical space. One would then see how the uniform distribution or the localised one are deformed by the wave.

Another interesting situation to look at is that of a momentum distribution containing two peaks, one at some ${\boldsymbol k}^1$ and another at some ${\boldsymbol k}^2$, because then this yields, in physical space, an interference pattern in the $(X,Y)$ plane, on which the deformation induced by the wave could be more detectable because of high contrast between low- and high-density regions. These two wavevectors, ${\boldsymbol k}^1$ and ${\boldsymbol k}^2$, must have the same norm if we want the distribution to be stationary under $W^{(0)}(q_X,q_Y)$ in physical space; indeed, the two expontentials containing the time dependency must coincide, so that we can factor this exponential out of the momentum-space integral. This corresponds to choosing 
\begin{equation}
|\alpha_+(k_x,k_Y)|^2 = c^1 G({\boldsymbol k}-{\boldsymbol k}^1) + c^2 G({\boldsymbol k}-{\boldsymbol k}^2) \, ,
\end{equation}
where $G({\boldsymbol k}-{\boldsymbol k}_0)$ is a Gaussian centered around ${\boldsymbol k}_0$. We choose to normalise $G({\boldsymbol k})$ to $1/2$, so that giving the same weight to the two Gaussians imposes, for a proper normalisation, to choose $c^1=c^2=1$.
Now this implies we can choose
\begin{align}
\alpha_+(k_x,k_Y) &= \sqrt{ G({\boldsymbol k}-{\boldsymbol k}^1) +  G({\boldsymbol k}-{\boldsymbol k}^2)} \nonumber \\
& \simeq \sqrt{ G({\boldsymbol k}-{\boldsymbol k}^1)} +  \sqrt{G({\boldsymbol k}-{\boldsymbol k}^2)} \, ,
\end{align}
where the approximated equality holds as long as the Gaussians are peaked enough.

 Now, to simplify this exemplifying study, we will work in the limit where the square-rooted Gaussians tend to delta functions, i.e.
\begin{equation}
\alpha_+(k_x,k_Y) = \delta^{(2)}({\boldsymbol k}-{\boldsymbol k}^1) + \delta^{(2)}({\boldsymbol k}-{\boldsymbol k}^2) \, ,
\end{equation}
which yields in physical space the interference between two plane waves corresponding to the selected modes, ${\boldsymbol k}^1$ and ${\boldsymbol k}^2$. Again, for the sake of simplicity, we choose ${\boldsymbol k}^1=(k,0)$ and  ${\boldsymbol k}^2=(0,k)$, where $k=q/2$, so that
\begin{equation}
\alpha_+(k_x,k_Y) = \delta(k_X-k)\delta(k_Y) + \delta(k_X)\delta(k_Y-k) \, .
\end{equation}
The initial wave function then reads
\begin{equation}
{\Psi}_0(X,Y) = {\Psi}^1 e^{i q X} + {\Psi}^2 e^{i q Y} \, ,
\end{equation}
where $\Psi^1$ and $\Psi^2$ are given by Eqs. (36) of the paper. This initial state yields the density of Eq. (38).

   %\include{Chapters/App_A_digression}
   %%%%%%%%%%%%%%%%%%%%%%%%%%%%%%
%%  Alias bibliographiques  %%
%%%%%%%%%%%%%%%%%%%%%%%%%%%%%%
\defcitealias{NRFortran}{Numerical Recipes}
\defcitealias{NRF77}{Numerical Recipes}
\defcitealias{NRF90}{Numerical Recipes}
\defcitealias{IUPAC}{IUPAC}%International Union of Pure and Applied Chemistry

%\chapter{\textsc{Compléments mathématiques}\label{Ann:Maths}}% Intermezzo mathématique

\chapter{Dirac equation in curved spacetime} \label{app:spinors}

In this Appendix, all physical quantities related to spin, such as the spinors $\Phi$ or $\Psi$, or the gamma matrices $\gamma^{\mu}$ or $\gamma^a$, are abstract quantities in the spin space, while in Appendix B of Publication 4, i.e. Ref \cite{AD17}, we have chosen to use respectively the ket and the hat notations to designate such an abstract nature, despite the cumbersomness, in order to be coherent with the main matter of the paper, which already uses non-ket and non-hat notations for a \emph{particular representation} of the Clifford algebra, i.e. for quantities written in a particular basis of the spin space.

\section{The Dirac equation in a generic curved spacetime}

The Dirac equation in an $(N=1+n)$-dimensional curved spacetime with metric $g_{\mu\nu}$ reads
\begin{equation}
\left( i \gamma^{\mu}D_{\mu} - m \right) \Phi = 0 \ .
\end{equation}
The wavefunction $\Phi$ is a $d$-component spinor, where the smallest possible $d$ is $d=2^{N/2}$ if $N$ is even and $d=2^{(N-1)/2}$ if $N$ is odd. The $\gamma^{\mu}$ (greek indices) satisfy the $g_{\mu\nu}$-Clifford algebra, $\{\gamma^{\mu},\gamma^{\nu}\}=g^{\mu\nu}$. Eventually, $D$ is the spin-space covariant derivative, given by
\begin{equation}
D_{\mu} = \partial_{\mu} + \Gamma_{\mu} \ ,
\end{equation}
where the spinorial connection $\Gamma_{\mu}$ is of the form
\begin{equation}
\Gamma_{\mu} = \frac{1}{2} \omega_{ab\mu}S^{ab} \ .
\end{equation}
In this expression, $S^{ab}$ are the generators of the $d$-component spinorial representation of SO($n,1$), and one can prove that they are given by
\begin{equation}
S^{ab} = \frac{1}{4} [ \gamma^{a},\gamma^{b} ] \ ,
\end{equation}
where the $\gamma^{a}$ (latin indices) satisfy the $\eta_{ab}$-Clifford algebra.  The $\omega_{ab\mu}$ are the coefficients of the spinorial connection, and one can show that they are given, in the $N$-ads formalism, by
\begin{equation}
\omega_{ab\mu} = g_{\alpha\beta} e_{a}^{\beta} \nabla_{\mu} e_{b}^{\alpha}  \ .
\end{equation}
In this expression, the $e_{a}^{\mu}$ is the inverse $N$-bein field, {\sl i.e.} the contravariant components of the $N$-ads $(e_a)_{a=1,N}$ on the coordinate basis $(e_{\mu})_{\mu=1,N}$. Eventually, $\nabla$ is the physical-space covariant derivative, defined for any vector $V$ by
\begin{equation}
\nabla_{\mu} V^{\alpha} = \partial_{\mu} V^{\alpha} + \Gamma^{\alpha}_{\mu\nu} V^{\nu} \ ,
\end{equation}
where the $\Gamma^{\alpha}_{\mu\nu}$ are the coefficients of the vectorial connection. An $N$-ad is generically a basis of spacetime, but we often use this term to refer to a fixed (and hence generally non-coordinate) basis, which is in the present context chosen orthonormal, {\sl i.e.} corresponding to a Minkowskian metric: $e_a\cdot e_b = \eta_{ab}$. We can go from the Minkowskian (or Lorentzian) components $V^{a}$ to the coordinate components $V^{\mu}$, and inversely, by
\begin{equation}
V^{\mu} = e_{a}^{\mu} V^{a} \ , \ \ \ \ \ \ \ V^{a} = e_{\mu}^{a} V^{\mu} \ ,
\end{equation}
where $e_{\mu}^{a}$ is the $N$-bein field, {\sl i.e.} the contravariant components of the coordinate-basis vectors on the $N$-ad. These relations can also be used for the $\gamma$ matrices, just replace $V$ by $\gamma$ (don't forget however that $\gamma$ is not a tensor by any means).  Here is an important practical remark: to take the derivative of a spinor in curved spacetime with this $N$-ad formalism, we don't need to compute the $\Gamma^{\alpha}_{\mu\nu}$, as the following computation shows:
\begin{align}
\Gamma_{\mu} & = \frac{1}{2} \omega_{ab\mu} S^{ab} \\
& =  \frac{1}{2} g_{\alpha\beta} e_{a}^{\beta} \left( \partial_{\mu} e^{\alpha}_{b} + \Gamma^{\alpha}_{\mu\nu} e^{\nu}_{b} \right) S^{ab} \\
& = \frac{1}{2} g_{\alpha\beta} e_{a}^{\beta} \left( \partial_{\mu} e^{\alpha}_{b} \right) S^{ab} \ ,
\end{align}
where, in going from the second to the third line, the second term of the sum vanishes because $S^{ab}$ is antisymmetric while $\Gamma^{\alpha}_{\mu\nu} g_{\alpha\beta} e_{a}^{\beta} e^{\nu}_{b}$ is symmetric. A similar remark enables us to write this other equivalent expression for the spinorial connection:
\begin{equation} \label{spin_connection}
\Gamma_{\mu} = \frac{1}{2} e_{a}^{\alpha} \ \! \partial_{\mu} \! \left(g_{\alpha\beta} e_{b}^{\beta} \right)  S^{ab} = \frac{1}{2} e_{a}^{\alpha} \, .
\end{equation}
In the end, the Dirac equation in curved spacetime can be written more explicitly as
\begin{equation} \label{eq:Dirac_curved}
\left[ i \gamma^a e_a^{\mu} \left\{ \partial_{\mu} + \frac{1}{2} e_a^{\beta} \ \! \partial_{\mu} \! \left( g_{\alpha\beta} e_b^{\alpha} \right) S^{ab} \right\} - m \right]\Phi = 0 \ .
\end{equation}
Here come three final remarks. The first remark is that $e^{\mu}_{a}e^a_{\nu} = \delta^{\mu}_{\nu}$ and $e^a_{\mu}e^{\mu}_{b} = \delta^a_b$, so that $[e^{\mu}_{a}] = [e^{a}_{\mu}]^{-1}$, where the upper index always corresponds to rows and the lower index to columns; however, since we always choose a symmetric metric, we can always choose a symetric $N$-bein, in which case rows and columns can be inverted.  The second remark is that denoting the $N$-bein field and its inverse with the same letter $e$ is justified by the fact that $g_{\mu\nu}e^{\nu}_{a} = \eta_{ab}e^{b}_{\mu}$, which can be noted either $e_{a\mu}$ or $e_{\mu a}$; however, we can use the letter $E$ to refer to, say, the inverse $N$-bein, so as not to confuse, for example, $E^1_1$ with $e^1_1$, or use round brackets around the latin indices. The third remark is that we have used the $N$-ad formalism because (i) it is the simplest way to obtain an expression for the coefficients of the spinorial connection (CSC), since it is easier to look for (and classify) the solutions to the Clifford algebra in \emph{flat} spacetime, and (ii) in the end, we don't even need the CSC to covariantly derive a spinor in curved space, but only the $N$-bein field, see expression (\ref{spin_connection}).

\section{The Dirac equation in particular (1+2)D spacetimes}

\subsection{Choice of metric}

We will from now on work in (1+2)D spacetime with dimensionless coordinates $X^0 = T, X^1 = X$ and $X^2 = Y$. The metric is noted $G_{\mu\nu}$, with a capital letter to remember that it is function of dimensionless coordinates. To simplify the problem, we assume that
\begin{enumerate}
\item $G_{00}=1$ 
\item (ii) the length element $dS^2$ is invariant under $X^0 \rightarrow -X^0$, {\sl i.e.} all the components $G_{0i}$, $i \neq 0$, vanish.
\iffalse
(Physical meaning ? Stronger than a time-reversible metric...)
\fi
\end{enumerate}

\noindent
The metric then has the form
\begin{equation}
[G_{\mu\nu}] = \left[
\begin{array}{c c c}
1 & 0 & 0 \\
0 & G_{XX} & G_{XY} \\
0 & G_{XY} & G_{YY}
\end{array} 
\right] \ ,
\end{equation}
where $G_{XX}$, $G_{YY}$ and $G_{XY}$ are three arbitrary functions of $(T,X,Y)$. Such a metric is much simpler than a generic one, but still quite general, and it enables to treat many interesting physical situations.

\subsection{Relation between the metric and the 3-bein}

Consider now an orthonormal triad associated to the Minkowskian coordinates $(\tilde{T},\tilde{X},\tilde{Y})$. This triad generates a $3$-bein field $e_{\mu}^{a} = \partial \tilde{X}^a / \partial X^{\mu}$ (its inverse is noted $E_{a}^{\mu}$). At any point, the length element $dS^2$ can be expressed in terms of the global coordinates $X^{\mu}$ but also in terms of the Minkowskian coordinates $\tilde{X}^a$, because it is a local quantity:
\begin{align}  \label{length_element}
dS^2 & = dT^2 + G_{XX} dX^2 + 2 G_{XY} dXdY + G_{YY} dY^2 \nonumber \\
& = d\tilde{T}^2 - d\tilde{X}^2 - d\tilde{Y}^2 \ . 
\end{align}
We choose $\tilde{T} = T$, so that $e_0^0 = 1$ and $e^0_i = e^i_0 =0$ for $i \in \{X,Y\}$. Hence, we can write the following decomposition:
\begin{equation}
d\tilde{X} = \frac{\partial \tilde{X}}{\partial X} dX + \frac{\partial \tilde{X}}{\partial Y} dY \ , \ \ \ \ \ \ \ d\tilde{Y} = \frac{\partial \tilde{Y}}{\partial X} dX + \frac{\partial \tilde{Y}}{\partial Y} dY \ .
\end{equation}

\subsubsection{Case of a symmetric 3-bein}

I wonder if the fact that the metric is symmetric  implies that we can choose a symmetric 3-bein \emph{without loosing any generality}.
Anyways, we can still look for a symmetric $3$-bein, {\sl i.e.} such that
\begin{equation}
e_{2}^{1} = e_{1}^{2} \ , \ \ \  \textsl{i.e.}  \ \ \ \
\frac{\partial \tilde{Y}}{\partial X} = \frac{\partial \tilde{X}}{\partial Y} \ .
\end{equation}
With these choices, Eq. (\ref{length_element}) leads to the following set of equations:
\begin{equation}
-G_{XY}  = e_{2}^{1} \left( e_{1}^{1} + e_{2}^{2} \right) \ , \ \ \ \ \ \ \ \  \ \ \ 
-G_{XX}  = (e_{1}^{1})^2 + (e_{2}^{1})^2 \ , \ \ \ \ \ \ \ \  \ \ \ 
-G_{YY}  = (e_{2}^{2})^2 + (e_{2}^{1})^2 \ .
\end{equation}
A solution to this system is given by 
\begin{equation}
e_{1}^{1} = \frac{-G_{XX} + \sqrt{G}}{\sqrt{2\sqrt{G} - \Sigma}} \ , \ \ \ \ \ \ \ \ \
e_{2}^{2} = \frac{-G_{YY} + \sqrt{G}}{\sqrt{2\sqrt{G} - \Sigma}} \ , \ \ \ \ \ \ \ \ \
e_{2}^{1} = \frac{-G_{XY}}{\sqrt{2\sqrt{G} - \Sigma}} \ ,
\end{equation}
with $G = \mathrm{det} [G_{\mu\nu}] = G_{XX}G_{YY} - G_{XY}^2$ and $\Sigma = G_{XX} + G_{YY}$. This solution can straightfowardly be inverted into
\begin{equation}
\varone = E_{1}^{1} = \frac{1}{\sqrt{G}}\frac{-G_{YY} + \sqrt{G}}{\sqrt{2\sqrt{G} - \Sigma}} \ , \ \ \ \ \ \ \ \ \ \
\vartwo = E_{2}^{2} = \frac{1}{\sqrt{G}} \frac{-G_{XX} + \sqrt{G}}{\sqrt{2\sqrt{G} - \Sigma}} \ , \ \ \ \ \ \ \ \ \ \
\varthree = E_{2}^{1} = \frac{1}{\sqrt{G}} \frac{G_{XY}}{\sqrt{2\sqrt{G} - \Sigma}} \ .
\end{equation}

\subsubsection{`General' case}

\subsection{Working out the Dirac equation in our particular spacetime}

The Dirac equation in the (1+2)D spacetime considered previously can be expanded into
\begin{equation}
\partial_0 \Phi = \left( -L_1 \partial_1  - L_2 \partial_2 + K \right) \Phi \ ,
\end{equation}
where the operators $L_1$, $L_2$ and $K$ are given by
\begin{align}
L_1 & = \varone \gamma^0\gamma^1 + \varthree \gamma^0\gamma^2 \nonumber \\
L_2 & = \varthree \gamma^0\gamma^1 + \vartwo \gamma^0\gamma^2 \\
K & = -\Gamma_0 - L_1 \Gamma_1 - L_2 \Gamma_2 - i \gamma^0 m \ .\nonumber
\end{align}
In (1+2)D, the smallest dimension for the spin Hilbert space is $d=2$. Let $(b_-,b_+)$ be a basis of this two-dimensional Hilbert space, {\sl i.e.} $\Phi = \phi^- b_- + \phi^+ b_+$. We can choose the following representation for the $\gamma$ matrices,
\begin{equation}
\left[{\left(\gamma^0\right)^u}_v\right]
%= \gamma^T 
= \sigma_1 = \left[
    \begin{array}{cc}
       0 & 1 \\
       1 & 0 
    \end{array}
  \right], \ \ \ \ \ \ \ \ \left[{\left(\gamma^1\right)^u}_v\right] = i \sigma_2 = \left[
    \begin{array}{cc}
       0 & 1 \\
       -1 & 0 
    \end{array}
  \right], \ \ \ \ \ \ \ \ \left[{\left(\gamma^2\right)^u}_v\right] = i \sigma_3 =  \left[
    \begin{array}{cc}
       i & 0 \\
       0 & -i
    \end{array}
  \right] ,
\end{equation}
with $(u,v) \in \{-,+\}^2$. In this representation, the Dirac equation reads:
\begin{equation} \label{eq:non_diagonal_metric}
\partial_T [\Phi^u] = \Big\{ \left( \varone \sigma_3 -  \varthree \sigma_2 \right) \partial_X +  \left( \varthree \sigma_3 - \vartwo \sigma_2 \right) \partial_Y + [{K^u}_v] \ \! \Big\} \ \! [\Phi^u] \ .
\end{equation}
If the metric is diagonal, then $\varthree=0$, and the previous equation get simplified into:
\begin{equation}\label{eq:diagonal_metric}
\partial_T [ \Phi^u ] = \big\{ \varone \sigma_3 \partial_X -  \vartwo \sigma_2 \partial_Y + [{K^u}_v] \big\} [ \Phi^u ] \ ,
\end{equation}
where we can choose $\varone = 1/\sqrt{-G_{XX}}$ and $\vartwo = 1/\sqrt{-G_{YY}}$.

\section{2D DTQWs as Dirac fermions in curved spacetime with diagonal metric}

\subsection{Recovering the Dirac dynamics from a DTQW}

We want to find a walk $\Psi_{j+1} = W_{j} \Psi_{j}$ whose continuous-limit dynamics correspond to Eq. (\ref{eq:diagonal_metric}). We must first notice that the continuous-limit dynamics of such a walk will always take the generic form
\begin{equation} \label{eq:generic_CL}
\partial_T [ \Psi^h ] = \left( -M_1 \partial_X - M_2 \partial_Y + Q \right) [ \Psi^h ] \ ,
\end{equation}
with $h \in \{L,R\}$, where $(b_L,b_R)$ is some basis of the spin Hilbert space. 

We set $P_1 = M_1 / \varone$ and $P_2 = M_2 / \vartwo$, so that the previous equation becomes
\begin{equation} \label{eq:generic_CL2}
\partial_T [ \Psi^h ] = \left( -\varone P_1 \partial_X - \vartwo P_2 \partial_Y + Q \right) [ \Psi^h ] \ .
\end{equation}

Suppose we find a walk $W$ such that the passage matrix $\mathcal{P}$ from basis $(b_L,b_R)$ to basis $(b_-,b_+)$, defined by $[ \Psi^h ] = \mathcal{P}[ \Psi^u ]$, satisfies:
\begin{align}
\mathcal{P}^{-1} P_1 \mathcal{P} & = - \sigma_3  \label{eq:C1} \\
\mathcal{P}^{-1} P_2 \mathcal{P} & =  \sigma_2 \ . \label{eq:C2}
\end{align}
In basis $(b_-,b_+)$, Eq. (\ref{eq:generic_CL2}) will then read:
\begin{equation} \label{eq:generic_CL3}
\partial_T [ \Psi^u ] = \left( \varone \sigma_3 \partial_X - \vartwo \sigma_2 \partial_Y + R \right) [ \Psi^u ] \ , \ \ \ \ \ \ \ \ \ \text{with} \ \ \ \ \ R = \mathcal{P}^{-1}(\partial_X + \partial_Y - \partial_T + Q)\mathcal{P} \ .
\end{equation}

Now, to link (\ref{eq:generic_CL3}) to (\ref{eq:diagonal_metric}), we have to match the normalization  conditions. On the one hand, $\Phi$ in Eq. (\ref{eq:diagonal_metric}) is normalized with respect to the covariant volume element $dV = \sqrt{\text{det}G} \, dX dY$, with $\sqrt{\text{det}G} = (\varone \vartwo)^{-1}$:
\begin{equation}
\int dV \left( |\phi^-|^2 + |\phi^+|^2 \right) = 1 \ .
\end{equation}
On the other hand, $\Psi$ in (\ref{eq:generic_CL3}) is normalized with respect to $dX dY$:
\begin{equation}
\int dXdY \left( |\psi^-|^2 + |\psi^+|^2 \right) = 1 \ .
\end{equation}
Hence, if, starting from (\ref{eq:generic_CL3}), we set
\begin{align}
\Phi & \equiv (\text{det}G)^{-1/4} \Psi \\
& = (\varone \vartwo)^{1/2} \Psi \ ,
\end{align}
then (\ref{eq:generic_CL3}) can be put in the form of Eq. (\ref{eq:diagonal_metric}), provided that
\begin{equation}
[{K^u}_v] = \left( R + \varone \sigma_3 \partial_X - \vartwo \sigma_2 \partial_Y - \partial_T \right) (\varone \vartwo)^{-1/2} \ .
\end{equation}

\subsection{Constraints on $P_1$ and the passage matrix $\mathcal{P}$}

Let us now examine the constraints on $P_1$ and $\mathcal{P}$ which are imposed by Eq. (\ref{eq:C1}). Generically, $P_1$ is a $2 \times 2$ matrix with complex entries:
\begin{equation}
P_1 = \left[
\begin{array}{c c}
a & b \\
c & d
\end{array}
\right] \ , \ \ \ \ \ \ \ \ (a,b,c,d) \in \mathbb{C}^4 \ .
\end{equation}
Now, condition (\ref{eq:C1}) imposes:
\begin{equation} \label{eq:form_imposed_to_P1}
P_1 = \left[
\begin{array}{c c}
a & b \\
\frac{1- a^2}{b} & -a
\end{array}
\right] \ ,
\end{equation}
together with the following form for the passage matrix:
\begin{equation}
\mathcal{P} = \left[
\begin{array}{c c}
x_- & x_+ \\
y_- & y_+
\end{array}
\right] \ ,
\end{equation}
with
\begin{align}
b \, y_- & = (-1-a) \, x_- \nonumber \\
b \, y_+ & = (1-a) \, x_+ \ .
\end{align}

\defcitealias{NRFortran}{Numerical Recipes}
\defcitealias{NRF77}{Numerical Recipes}
\defcitealias{NRF90}{Numerical Recipes}
\defcitealias{IUPAC}{IUPAC}%International Union of Pure and Applied Chemistry

%\chapter{\textsc{Compléments mathématiques}\label{Ann:Maths}}% Intermezzo mathématique

\chapter{General ideas on (perturbative) renormalization} \label{app:renormalization}

\emph{Reading that I am far from being an expert -- which is more than a euphemism -- on perturbative renormalization, is obviously unnecessary for any reader that will go through the following lines. I thus ask for his indulgence and suggestions to improve this appendix. I am longing to work, not only on this particular subject, but also many others, and eventually connect them to quantum walks.} \\ \\

\noindent
Perturbative renormalization is a computational procedure, i.e. a series of computational rules, which, to first speak in epistemological terms -- with no need to elaborate on either the technical details or the physical nature of such a procedure --, eventually enabled to `extract' predictive results out of the `first' computations carried out in the history of QFT, which are still standard and useful precisely when perturbative renormalization works. These standard computations are grounded on a perturbation series whose convergence radius actually vanishes in general, which makes them a priori unable to predict any physical result\footnote{This series, which does generally not converge, is essentially the exponential of the interaction term of the considered model. Note that, (i) in a Hilbert-space framework, this simply means that the evolution operator is actually a mathematically ill-defined object, i.e. that one cannot define the exponential of the Hamiltonian operator -- indeed, there is no reason that the free part of the Hamiltonian compensate the interaction part, because these two terms are completely different by nature --, while (ii) in a path-integral framework, this means the path integral is a divergent object.}. 

Indeed, while one is allowed to add an arbitrary (but finite!) number of terms, one a priori expects that the numerical result provided by such a sum will simply be of \emph{no} predictive value if the predictive hope of this sum relies on viewing it as the truncation of a series which does not converge. While this is in general true, i.e. these finite sums give wrong numerical predictions, it is actually possible to produce correct numerical predictions by, as we say, \emph{renormalizing}, or \emph{resuming} this sum, i.e. by modifying it according to some rules. It is at this point necessary to give more technical details. The aforementioned sum, as we said, contains a finite number of terms; however, for these terms, which are integrals, to be finite themselves, i.e. for the integrals to converge, one first has to introduce a so-called ultraviolet  \emph{regulator}, i.e. an upper-bound, or cutoff, for the accessible energies\footnote{This is how perturbative renormalization was historically developed: most of the lowest-orders Feynman diagrams have an ultraviolet divergence, so that an ultraviolet cutoff was introduced. Actually, one realizes after developing the renormalization method, that there are also infrared divergences in the theory, which was quickly pointed out by Landau, in the 50's.}.

Now, the guiding idea underlying the perturbative renormalization `recipe', is the following: while the regulator makes the integrals involved in the sum be well-defined, i.e. convergent, this sum gives wrong predictions `because' the regulation modifies the physical theory (which is anyways mathematically ill-defined); to recover the `good physical theory' (although mathematically ill-defined), one must figure out how to modify these finite integrals so that their limit with infinitely-big regulator, i.e. in the original pre-regularization situation, is finite. 

Eventually, the physics behind these mathematics holds in the following lines. The coupling constant of an interaction, formally introduced as a constant often denoted by $g$ in the interaction Lagrangian, is of no physical significance, and must be, as we say, renormalized, i.e. made function of the typical energy scale $\mu$ at which the interaction occurs; this is achieved by the  computational procedure detailed above, i.e. it is the physical way of viewing such computations. This renormalized coupling constant $g_R(\mu)$ is that which is measured in experiments. The energy scale is simply the input energy if the system can be considered isolated. In particle physics, $\mu$ is thus given by the kinetic energy communicated to the particles before making them collide.

The regulator can be viewed as a mere intermediate tool which is in the end replaced by a renormalization-flow picture, which defines $g_R(\mu)$ by relating the couples $(g_R^i,\mu^i)$ labeled by $i$, which simply indicates that $g_R^i$ is the good numerical value of the measured coupling constant if we are at energy scale $\mu^i$. This connects with the non-perturbative methods based on the renormalization group, see further down.

Let us stress that, while such a perturbative renormalization happens to produce accurate predictions, it still does not solve the mathematical problem of ill-definition of the theory, in the sense that the mathematical idea underlying the addition of the terms in the \emph{renormalized} sum is that this sum should \emph{still} be the truncation of some series which, as the first `bare' series, is standardly seen as not converging either. This is not surprising since perturbative renormalization `simply' modifies the value of the coupling constant, which will never be enough to make a series with radius \emph{zero} be defined, since the litteral model in such a picture is viewed as unchanged\footnote{Perturbative renormalization should thus be regarded as a series of computational rules which, although developed upon a mathematically ill-defined framework, produce predictive results. This subject illustrates particularly beautifully, I believe, that physics need, if not often, at least sometimes, not be proper from the mathematical point of view, but that mathematics are rather used as either (i) a practical tool, namely mathematical \emph{computations}, or (ii) an inspirational tool, thanks to their formal, i.e. \emph{symbolization power}. Indeed, formalism can help to conceptualize useful \emph{physical} concepts, regardless of the mathematical well-definition of the formal objects one introduces; the only thing that physics requires from some formalism is that the \emph{numerical} computations this formalism suggest make predictive results.}\textsuperscript{,}\footnote{I need to investigate more about this matter: it may be that some constructive approaches (see below) are actually able to view the renormalized sum as the truncation of a series which, this time, does converge, which would make sense.}.

The branch of mathematical physics aiming at solving the problem of ill mathematical definition of QFT is generically called \emph{axiomatic QFT}. In an epistemological picture, the so-called \emph{constructive QFT} program, born in the 60's, can be viewed as a desire to connect, as much as possible, the mathematical developments of a given axiomatic QFT with the succesful predictions of renormalization theory; this is done by lowering the standard mathematical `desire of generality' and constructing a series of concrete examples of reference, i.e., a series of Lagrangians, in which one can precisely define \emph{(rather meta-mathematical) translation rules} to relate the mathematically ill-defined, but physically predictive original Lagrangian-based QFT, to such a mathematically well-defined axiomatic QFT.

For more details, see the following introductory review written by Jaffe in 2000, \url{http://www.arthurjaffe.com/Assets/pdf/CQFT.pdf}, the following mathematical-physics Master's thesis of 2013, by Sheikh, 
\sloppy
\burl{https://esc.fnwi.uva.nl/thesis/centraal/files/f1519507583.pdf}, and the following epistemological work by Li \url{http://jamesowenweatherall.com/wp-content/uploads/2015/10/irvine-draft.pdf}, which provides a rough diagram of the links between the different axiomatic and constructive QFT approaches existing in current litterature. Rivasseau is one of the leading figures in these rigorous approaches to QFT, and his books are standards.

In a path-integral framework, axiomatic QFT is grounded on a spacetime-lattice formulation of QFT, the so-called \emph{lattice gauge theories (LGTs)}. While, in a constructivist perspective, LGTs can, on the one hand, sometimes provide a way to derive the perturbative renormalization rules from the lattice situation, e.g., to express the perturbatively-grounded finite sums, and in particular the counterterms added to the regularized finite sums\footnote{Recall that these counterterms are added in order to make each of the terms of the sum, which is an integral, converge when the ultraviolet regulator goes to infinity.}, as functions of the quantities defined on the lattice, LGTs can, on the other hand, be used without explictly referring to the resuming, through so-called non-perturbative methods, which are all based, in LGTs, on the \emph{renormalization group}, which is a non-perturbative way of viewing the perturbative renormalization of the coupling constant according to typical energies involved in the interaction. The renormalization-group picture provides, in particular, a criterion for the predictive power of QFT, which, since non-perturbatively grounded, does not require the perturbative renormalizability of the theory, i.e. the possibility to resum the regularized sums, which would be too strong of a criterion; such a predictive, although non perturbatively-renormalizable theory is said to be \emph{asymptotically safe}, see the following FAQs (frequently asked questions), \url{http://www.percacci.it/roberto/physics/as/faq.html}. Several other non-perturbative methods exist.

    %%%%%%%%%%%%%%%%%%%%%%%%%%%%%%
%%  Alias bibliographiques  %%
%%%%%%%%%%%%%%%%%%%%%%%%%%%%%%
\defcitealias{NRFortran}{Numerical Recipes}
\defcitealias{NRF77}{Numerical Recipes}
\defcitealias{NRF90}{Numerical Recipes}
\defcitealias{IUPAC}{IUPAC}%International Union of Pure and Applied Chemistry

%\chapter{\textsc{Compléments mathématiques}\label{Ann:Maths}}% Intermezzo mathématique

\chapter{Unitarity of time evolution in quantum mechanics and quantum field theory} \label{app:unitarity}

\section{Unitarity in standard lattice gauge theories}

Unitarity is a priori not ensured by Wilson's original lattice discretization \cite{Wilson74}. Indeed, Wilson used the (Euclidean) path-integral quantization, which is directly mapped to the Lagrangian formalism through a Wick rotation. Within such a framework, proving unitarity essentially always demands to map this Lagrangian formalism to a Hamiltonian one\footnote{Regarding this matter, let me quote Weinberg, \emph{The Quantum Theory of Fields I} \cite{Weinberg_QFT1}, Chapter 9, 4th paragraph: ``At this point the reader may be wondering why if the path-integral method is so convenient we bothered in Chapter 7 to introduce the canonical formalism. Indeed, Feynman seems at first to have thought ofhis path-integral approach as a substitute for the ordinary canonical formulation of quantum mechanics. There are two reasons for starting with the canonical formalism. The first is a point of principle: although the path-integral formalism provides us with manifestly Lorentz-invariant diagrammatic rules, it does not make clear why the $S$-matrix calculated in this way is unitary. As far as I know, the only way to show that the path-integral formalism yields a unitary $S$-matrix is to use it to reconstruct the canonical formalism, in which unitarity is obvious. There is a kind of conservation of trouble here; we can use the canonical approach, in which unitarity is obvious and Lorentz invariance obscure, or the path-integral approach, which is manifestly Lorentz-invariant but far from manisfestly unitary. Since the path-integral approach is here dervied from the canonical approach, we know that the two approaches yield the same $S$-matrix, so that the $S$-matrix must indeed be both Lorentz-invariant and unitary.''}. In the standard continuum-spacetime situation, methods are known to do this for relatively general Lagrangians, which essentially ensures unitarity in most practical situations\footnote{For example, in axiomatic, and thus rigorous formulations of QFTs, the time-evolution unitarity is ensured, starting from a Euclidean path-integral, by demanding that this Euclidean path-integral satisfy a property called \emph{reflection positivity}, see \url{https://ncatlab.org/nlab/show/Osterwalder-Schrader+theorem}.}. Hence, discretizing the Lagrangian, as originally done by Wilson, will a priori yield issues regarding unitarity, because then standard continuum-spacetime methods mapping this Lagrangian formalism to the Hamiltonian one have to be reconsidered.

Now, one month after Wilson's original paper, Kogut and Susskind \cite{KogSuss75a} developed a Hamiltonian formulation of lattice gauge theories, which uses the canonical formalism instead of the path-integral one. In this formulation, only the spatial lattice is discretized, i.e. time is kept continuous, and unitarity is ensured by defining an appropriate Hermitian Hamiltonian on the spatial lattice.

%\section{A broader view}

    %%%%%%%%%%%%%%%%%%%%%%%%%%%%%%
%%  Alias bibliographiques  %%
%%%%%%%%%%%%%%%%%%%%%%%%%%%%%%
\defcitealias{NRFortran}{Numerical Recipes}
\defcitealias{NRF77}{Numerical Recipes}
\defcitealias{NRF90}{Numerical Recipes}
\defcitealias{IUPAC}{IUPAC}%International Union of Pure and Applied Chemistry

%\chapter{\textsc{Compléments mathématiques}\label{Ann:Maths}}% Intermezzo mathématique

\chapter{On topology} \label{app:topology}

\emph{This appendix is aimed at giving some intuition of the elementary notions and tools used in this branch of mathematics that topology is. The style of the presentation is often informal.}

\section{Introduction}

\hspace{0.585cm}$\bullet$ A \emph{space} is a \emph{set} endowed with a \emph{structure}. It is customary to distinguish between two main kinds of structures: \emph{algebraic} and \emph{analytic} structures. \\

$\bullet$ An algebraic structure is the given of one or several \emph{binary operations}, which may be simply called \emph{operations}, between \emph{elements} of the set. A binary operation takes as input two \emph{elements} of the set, and produces as output another element of the set. Examples of such operations are the addition or the multiplication. The algebraic structure tells us how the elements of the set are related between each other \emph{in terms of equalities between these elements}. An algebraic structure is thus in particular able to generate, from a well-chosen starting subset, called a \emph{generator set}, the original set, by producing all elements of the original set with inputs from the generator set\footnote{A simple example of algebraic structure is $\mathbb{N}^{\ast}$ endowed with the addition operation. The smallest generating set is $\{1\}$.}. \\

$\bullet$ An analytic structure is the given of, again, one or several binary `operations', but, this time, between \emph{subsets} of the set (hence the quotation marks); these binary `operations' are usually the \emph{union} and the \emph{intersection}. The analytic structure tells us how the elements of the set are related between each other \emph{in terms of groupings of these elements}, through the notions of \emph{neighborhood} and of \emph{open set}. An analytic structure can thus not generate a set\footnote{It can however generate a set of subsets of this set, namely, the set of all possible groupings induced by the considered analytic structure.}, but is aimed at analysing a preexisting set in terms of groupings of its elements. Endowing a set with an analytic structure is a first step to define, very generally, notions such as \emph{connectedness}, \emph{continuity} or \emph{compactness}. \\

$\bullet$ On the one hand, examples of spaces endowed with an algebraic structure are \emph{groups}, or \emph{linear spaces}, also called \emph{vector spaces}. A group is a set (i) left invariant, i.e. stable, under some binary operation between its elements, (ii) which contains a \emph{neutral element}, and (iii) in which all elements have an \emph{inverse}\footnote{A simple example of (non-finite) group is $\mathbb{Z}$ endowed with the addition operation. The smallest generating set is $\{1,-1\}$.}. A vector space is, essentially, a set stable under linear combinations of its elements\footnote{A vector space is thus in particular a (non-finite) group with respect to the addition operation, and it has an additional homothety stability.}. \\

$\bullet$ On the other hand, \emph{topological spaces} are a very general type of spaces endowed with an analytic structure.
\iffalse
\footnote{I don't use the terminology `analytic space' since it is already used for a quite specific type of topological space.}.
\fi

\section{Topological spaces and homeomorphisms}

\hspace{0.585cm}$\bullet$ A topological space can be defined as a couple ($X,\tau$) of a set $X$ and the following structure, which is of analytic kind: a collection of subsets $\tau$ satisfying the following axioms:
\vspace{-0.5cm}
\begin{enumerate}
\item {\bf \emph{axiom} `$\varnothing$ \& $X$':} The empty set $\varnothing$ and the total space $X$ itself belong to $\tau$.
\vspace{-0.3cm}
\item {\bf \emph{axiom} `$\cup$ \& $\cup_{\infty}$':} Any \emph{finite} or \emph{infinite} union of members of $\tau$ belongs to $\tau$.  
\vspace{-0.3cm}
\item {\bf \emph{axiom} `$\cap$':} Any \emph{finite}\footnote{If we allow for infinite intersections, then we can exhibit examples where some elements of $\tau$ are \emph{closed} from the metric point of view, while we want the present axioms to induce a definition of an open set, which eventually coincides with the metric definition of such objects, see the following discussion: \url{https://math.stackexchange.com/questions/284970/in-a-topological-space-why-the-intersection-only-has-to-be-finite}.} intersection of members of  $\tau$ belongs to $\tau$.
\end{enumerate}
\vspace{-0.1cm}
The elements of $\tau$ are called \emph{open sets} and the collection $\tau$ is called a \emph{topology on $X^{\,}$}\footnote{There is an alternative way of defining a topological space, through the notion of neighborhood. This definition is formally more cumbersome, but the notion of neighborhood is very intuitive, and so is the notion of open set that ensues from it: a subset of some set is said open if it is a neighborhood of all the elements it contains. One may first consult Wikipedia on this subject: \url{https://en.wikipedia.org/wiki/Topological_space}.}. One may first look at elementary examples and non-examples on Wikipedia: \url{https://en.wikipedia.org/wiki/Topological_space}. \\

$\bullet$ A function $f:X \rightarrow Y$ is said \emph{continuous} if:
\begin{equation}
\forall x \in X, \forall \ \text{neighborhood} \ N \ \text{of} \ f(x),  \exists \ \text{neighborhood} \ M \ \text{of} \ x \ \text{such that} \ f(M) \subseteq N \, .
\end{equation}
This definition generalizes that used in \emph{real analysis}, where $f: \mathbb{R} \rightarrow \mathbb{R} \, .$  \\

$\bullet$ A \emph{homeomorphism} is a continuous bijection whose inverse is also continuous. \\

$\bullet$ Two topological spaces are said \emph{homeomorphic} if they are related by a homeomorphism. Such two spaces are said to have the \emph{same topology}, or to be \emph{topologically equal}. Note that if the set $X$ is endowed with nothing else than a topology, it can actually not be distinguished, \emph{in terms of relations between its elements}\footnote{This precision can actually be omitted, since the nature of the elements regardless of any relations between these elements is actually not a mathematical question, unless this nature can itself be described by a mathematical sub-structure for each element of the set, which again will deal, by nature, with \emph{relations between the sub-elements of which each element is maid}.}, from a homeomorphic partner, and both sets are strictly \emph{equal up to a homeomorphism}.

\section{Connectedness and homotopy equivalence}

\subsection{Connectedness and path-connectedness}

\hspace{0.585cm}$\bullet$ A space is said \emph{connected} if it cannot be represented as the union of two or more disjoint non-empty open subsets. \\

$\bullet$ A \emph{continuous path}, or simply \emph{path} in $X$, is a continuous function $f:[0,1] \rightarrow X$. The terminology `path' is sometimes used for $f([0,1])$ rather than $f$, since this matches with the intuition behind the word `path'. One should normally be able to determine, from the context, whether the word `path' refers to $f$ or to $f([0,1])$. \\

$\bullet$ A space $X$ is said \emph{path-connected} if any two points $x$ and $y$ in $X$ can be connected by a continuous path, i.e. if:
\begin{equation}
\exists \ \text{continuous} \ f:[0,1] \rightarrow X \ \ \text{such that} \ \ 
\left\{
\begin{array}{l}
f(0) = x \\
f(1) = y
\end{array}\right. .
\end{equation}
\vspace{0.01cm}
 \\

$\bullet$ For example, the Lorentz Lie group $\text{O}(1,3)$ is, as a manifold, disconnected, but has four $\text{path-connected}$ components (which are subgroups): the proper (i.e. special) and orthochronous (denoted by $+$), $\text{SO}^+(1,3)$, the proper and non-orthochronous, the improper and orthochronous, and the improper and non-orthochronous.

\subsection{Homotopy and homotopy equivalence }

\hspace{0.585cm}$\bullet$ In the definition of a homeomorphism, bijectivity is necessary for the topology of the \emph{total} space to be preserved\footnote{This is by definition of `topology preservation', but this definition is obviously set to first match with (and then generalize) intuitive results.}. Relaxing (i) bijectivity while keeping (ii) continuity enables to track (i) \emph{differences} between spaces which are (ii) strictly topological. \\

$\bullet$ Two continuous functions $f, g : X \rightarrow Y$ are said \emph{homotopic} if: 
\begin{equation}
\exists \ \text{continuous} \ H : X \times [0,1] \rightarrow Y \ \ \text{such that} \ \ x \in X \Rightarrow \left\{
\begin{array}{l}
H(x,0) = f(x) \\
H(x,1) = g(x)
\end{array}\right. .
\end{equation}
$H$ is called a \emph{homotopy}, and the relation `$f$ homotopic to $g$' can be noted $f \sim g$. \\

$\bullet$ It is easy to prove that the relation of homotopy between paths \emph{having fixed end points} is an \emph{equivalence relation}, i.e. a relation which is \emph{binary}, \emph{reflexive}, \emph{symmetric} and \emph{transitive}\footnote{For a proof, see, e.g., \url{https://www.math.cornell.edu/~hatcher/AT/ATch1.pdf}, Proposition 1.2.}. It is called \emph{relative homotopy equivalence}, where ``relative'' means `relative to chosen fixed end points'.  \\

$\bullet$ A more general notion can be defined, which does not assume fixed end points, and is, this time, simply called \emph{homotopy equivalence}: $X$ and $Y$ are said homotopically equivalent, or of the same homotopy type, if 
\begin{equation}
\exists \ \text{continuous} 
\left\{
\begin{array}{l}
f : X \rightarrow Y  \\
g : Y \rightarrow X
\end{array} \right. 
\ \ \text{such that} \ \
\left\{
\begin{array}{l}
g \circ f  \sim \mathrm{Id}_X \\
f \circ g  \sim \mathrm{Id}_Y
\end{array}\right. .
\end{equation}
Comparing this definition with that of a homeomorphism reveals the formal procedure to, e.g., relax bijectivity and `transform' a homeomorphism into a the more general notion of homotopy equivalence. \\

\iffalse
$\bullet$ An \emph{elementary homotopy class} is a homotopy class a representative of which cannot be mapped to a representative of another class by a strict surjection, i.e. this can only be done through bijections.
\fi

$\bullet$ In the following, we will be interested homotopy classes of paths, i.e. sets of paths which are all homotopic, since, from the topological point of view, two homotopic paths are not distinguishable.  We will, in particular, be interested in a particular type of paths, that is, \emph{closed paths}, also called \emph{loops}, namely continuous functions of the form $f : S^1 \rightarrow X$. Also, we will consider homotopy classes of loops \emph{having a common fixed point} -- which is both the starting and the end point of the loop path -- since this is necessary for a group structure to be defined, see below. A loop that is homotopic to a point is said \emph{contractible}.

\section{Fundamental group}

\subsection{Genus in one and two dimensions}

\hspace{0.585cm} $\bullet$ The most natural way of introducing the \emph{fundamental group} of a topological space is, to me -- as well as to many others -- through the \emph{genus} $g$ of 1D or 2D real manifolds, which correspond, essentially, to the intuitive notion of curve and surface, respectively\footnote{Don't forget however that these 1D or 2D real manifolds can be made of several disconnected components.}. The genus is intuitively the numbers of \emph{holes} of a manifold, or the number of \emph{handles}. \\

\iffalse
\footnote{In dimension 2, the question of whether the manifold is orientable or not arises.}
\footnote{One can also define the genus of non-orientable manifolds, which are rather seen as less intuitive than orientable ones}
\fi

$\bullet$ For a 1D (non-knoted) real manifold, the genus is the maximum number of cuttings without rendering the resultant manifold disconnected, i.e. non-path-connected.  \\

$\bullet$ The circle, i.e. the 1D sphere, or $1$-sphere, $S^1$, has one hole, i.e. genus $g=1$. Indeed, it can be cut at most once -- at any of its points -- without making it disconnected. \\

$\bullet$ For a 2D (non-knoted) orientable real manifold, the genus is the maximum number of cuttings along non-intersecting loops without rendering the resultant manifold disconnected\footnote{I have rewritten Wikipedia's definition, see \url{https://en.wikipedia.org/wiki/Genus_(mathematics)\#Orientable_surface}.}\textsuperscript{,}\footnote{One can also define the genus of non-orientable 2D real manifolds, such as the Moebius strip or the Klein bottle, which are rather seen as less intuitive than orientable ones.}. \\

\iffalse
The condition ``non-intersecting'' implies, in particular, that all loops that are not elementary, i.e. that cover other loops more than once, are excluded, i.e. do not contribute to the genus. Note that this does not imply that all elementary loops do contribute to the genus, see below the example of the torus.  
\fi

$\bullet$ The standard, i.e. 2D sphere, or $2$-sphere,  $S^2$, has no hole, i.e. genus $g=0$. Indeed, it contains no loop along which one can cut it without making it disconnected. The standard, i.e. 2D torus, or $2$-torus, $T^2$, also called (empty\footnote{We will omit this precision afterwards.}) donut, has one hole, i.e. genus $g=1$. Indeed, consider the donut as a fiber bundle of fiber $S^1$ over the base $S^1$. One can, without making it disconnected, cut it at most once along, e.g., either (i) any loop homotopic to the base, i.e. winding around the handle, which yields a cylinder\footnote{Having as a base the \emph{base} of the original donut fiber bundle.}, or (ii) any loop homotopic to the fiber, i.e. winding around the hole, which yields a 2D annulus\footnote{Which is actually, topologically, also a cylinder, but having as a base the \emph{fiber} of the original donut fiber bundle.}, but not both -- although the resulting manifold would still be connected (it would be a rectangle) -- since loops of type (i) intersect those of type (ii). \\

\iffalse
Note that any loop corresponding to a fiber does not contribute to the genus because cutting along this loop disconnects the torus. \\
\fi

\iffalse
Note that the elementary loop corresponding to a fiber does not contribute to the genus because cutting along this loop disconnects the torus. 
\fi

$\bullet$ The genus $g$ is related to -- and can thus also be defined through -- the Euler characteristic $\chi$. For an orientable surface, the relation reads
\begin{equation}
\left\{
\begin{array}{l l}
\chi = 2 - 2g  \ & \text{closed surface} \\
\chi = 2 - 2g - b \ & \text{surface with $b$ boundary components.}
\end{array}
\right.
\end{equation}
\vspace{0.1cm}

$\bullet$ Endowing these manifolds with more properties, such as, typically, differentiability or, even more, analyticity, may simplifiy the topological study. For example, the genus can be defined in a more computational manner for \emph{algebraic curves}\footnote{An algebraic curve is a set of points on the Euclidean plane whose coordinates are zeros of some polynomial in two variables.} and \emph{Riemman surfaces}\footnote{A Riemman surface is a \emph{1D complex manifold}, i.e., essentially, a 2D real manifolds endowed with analyticity.}. If one can then show that the obtained results do not depend on these extra properties, i.e. that the latter have merely been used as tools but are not necessary assumptions, then one can extend the results to all homeomorphically equivalent spaces lacking those properties.

\subsection{Homotopy classes of loops, fundamental group}

\hspace{0.585cm} $\bullet$ From the previous examples, the intuition that the reader may have of the mathematical notion of loop defined above may be that of a loop `that winds once', be it around a hole (e.g. hole of a donut) or not (e.g. handle of the donut). Actually, as mathematically defined above, a loop can, in some cases, wind several times. \\

\vspace{-0.17cm}
$\bullet$ Note that in dimension 1, if the loop $f$ is, e.g., a loop winding twice around a circle, then $f([0,1])$ is necessarily strictly surjective, more precisely, we must `pass by' \emph{each} point of the circle twice, which does not prevent $f$ from being continuous. In dimension 2, instead, it is possible to draw a closed curve $f([0,1])$ on the surface of the donut, that winds twice around the hole of the donut, but still with a bijective $f$. For loops that wind several times around the handle of the donut, $f$ must be strictly surjective, but it can be -- and is naturally -- so only over a discrete set of points\footnote{In contrast with loops winding several times around the hole of a circle, for which $f$ must be surjective over \emph{a continuous interval}.}. \\

\vspace{-0.17cm}
$\bullet$ One can show that all loops that wind the same number of times around the same \emph{topological barrier} -- e.g., a hole, or a handle -- belong to the same homotopy class, while this is not so otherwise. The homotopy class of loops that wind once say, clockwise, around some topological barrier, will be denoted by $1$, that of loops winding twice anticlockwise will be denoted by $-2$, etc. \\

\vspace{-0.17cm}
$\bullet$ We denote by $\pi_1(X)$ the set of all homotopy classes of loops of the topological space $X$, relative to a given fixed point $x_0$ which is omitted from the notation. The subscript `$1$' refers to the fact that the tools we are using to track the topology are loops, i.e. $1$-dimensional closed manifolds. \\

\vspace{-0.17cm}
$\bullet$ One can show that $\pi_1(X)$ has a \emph{group structure} with respect to a certain composition operation between (homotopy classes of) loops, that we may call \emph{loop product} (or \emph{addition}). This operation corresponds to the more general \emph{composition of paths}, also called \emph{path product}, when applied to loops\footnote{For a definition of the path product, see, e.g., \url{https://www.math.cornell.edu/~hatcher/AT/ATch1.pdf}, page 26.}. The path product enables to generate, from two paths belonging to two given classes, respectively, a new path belonging to another given class. The group $\pi_1(X)$ is called the fundamental group (FG) of $X$. The generators of this group are the (homotopy classes of) loops that wind only once around a given topological barrier, together with the respective inverses. \\

\vspace{-0.17cm}
$\bullet$ The FG of the circle $S^1$ is isomorphic to  $\mathbb{Z}$ equipped with the addition operation, which we simply note $\pi_1(S^1)=\mathbb{Z}$, where the equality sign means `up to a group isomorphism'. The neutral element of $\pi_1(S^1)$ is the class of loops contractible into the fixed point, i.e. $0$. \\

$\bullet$ The FG contains more information than the genus. A first example for this is that given above: the circle $S^1$ has genus $1$, but its FG is $\pi_1(S^1)=\mathbb{Z}$\footnote{I'm looking for an example in which a loop can only wind a finite number of times, i.e. in which it is not possible to find, as continuous path, a loop that winds more than a finite number of times.}. Another example is the donut $T^2$, which also has genus $1$,  but whose FG is $\pi_1(T^2)=\mathbb{Z}^2$, which traces, in particular, the fact that the manifold is two-dimensional, through the power $2$ in $\mathbb{Z}^2$.

\iffalse
\footnote{}
\fi
\iffalse
It is useful to have in mind the 2D-manifold hole picture in the following.
\fi

\iffalse
\footnote{For \emph{orientable manifolds} (i.e. no Moëbius-strip-like manifold), one could first use the even simpler example of 1D curves, but in this case the homotopy classes are trivial i.e. reduced to a single representative, unless one allows for continuous deformations of the total space itself, which is indeed usually implied in topology. Indeed, if we do not allow for, e.g., a circle to be continuously deformed, each homotopy class has a single representative, namely  the union of a given number of copies of this circle -- together with an orientation, $\pm$, which gives information about whether the representative is followed clockwise or anticlockwise.}\footnote{The genus can be generalized to higher dimensions. This was first done, \emph{arithmetically}, by Severi in 1909, see \url{https://link.springer.com/chapter/10.1007/978-3-319-42312-8_37/fulltext.html}.}.
\fi

\section{Simple connectedness, weaker and higher forms of connectedness, i.e., respectively, $k$-uple connectedness and $n$-connectedness}

\hspace{0.585cm}$\bullet$ A topological space $X$ is said \emph{simply connected} when its FG is trivial, i.e. equal to the neutral element, which we denote by $\pi_1(X)=0$. For example, $\pi_1(S^N)=0$ for $N\geq 2$. When the FG is not trivial, the space is not simply connected. The space is said \emph{$k$-uply connected} if, roughly speaking, it has $k$ holes, or equivalently, $k$ handles. \\

$\bullet$  The FG does not contain all the topological information about a topological space. It only records, roughly speaking, information about the holes of a manifold, which are the coarser empty spaces around which, purposely vaguely speaking, the manifold `winds' or `wraps'\footnote{That being said, recall that the FG still records more than the number of holes, i.e. the genus, as we have seen above by comparing the FGs of two equal-genus but topologically different manifolds, namely the circle, with FG $\mathbb{Z}$, and the torus, with FG $\mathbb{Z}^2$. The FG also records information about how the simply-connected sub-components glue together, i.e., in particular, information about the edges of the manifold, or its orientability -- although not always, e.g. the Moebius strip, which is a non-orientable surface, has the same FG as the 2D annulus and the circle. The FG can record information about the structure group of a fiber bundle. For instance, consider the donut and the Klein bottle, which are two fiber bundles of $S^1$ over $S^1$, but with different structure groups: the torus has a trivial structure group, i.e. reduced to a neutral element, while the Klein bottle has the structure group $\mathbb{Z}^2$ acting by reflection on $S^1 \subset \mathbb{R}^2$, see \url{https://books.google.fr/books?id=wuUlBQAAQBAJ\&pg=PA111\&lpg=PA111\&dq=klein+bottle+fiber+bundle+structure+group\&source=bl\&ots=LFjgyl1NKm\&sig=T60iDmPFUmjDUADczRVvTBqkhaM\&hl=fr\&sa=X&ved=0ahUKEwiUnvyotbfVAhUjK8AKHTO-Dv4Q6AEISjAD\#v=onepage\&q=klein\%20bottle\%20fiber\%20bundle\%20structure\%20group\&f=false}. Well, the FG of the Klein bottle is different from that of the donut.}. \\

$\bullet$ There are higher-order homotopy groups, which record information about the topology of the space at a more refined level. Let us give an intuitive picture of this. Consider the standard, i.e. 3D ball, or $3$-sphere, also called 3D sphere, $S^3$. One can pierce it entirely with a cylinder, which creates a hole and makes it become, topologically, a full donut, $D^2 \times S^1$, where $D^2$ is the standard disk. But one can also imagine that the ball has some empty space inside, as if some nuclueus was absent; such a space is called a 3D annulus, let us denote it by $A^3$. Now, loops are not able to track this empty space, that is: from the point of view of loops, $A^3$ is trivial, since any loop inside it is contractible, i.e. $\pi_1(A^3)=0$, i.e. $A^3$ is simply connected (with respect to loops\footnote{This precision is always omitted in the standard terminology.}). But one intuitively feels that $A^3$ is not homeomorphic to $S^3$. Indeed, $A^3$ is not trivial if we track its topology with $2$-spheres. From the point of view of such spheres, $A^3$ is doubly connected, i.e. it has one `spherical hole': a sphere that wraps around the `spherical hole' is not contractible. Moreover, one can show (annulus theorem) that $A^3$ is homeomorphic to $S^2 \times [0,1]$, so that its homotopy group with respect to spheres, that we note $\pi_2(A^3)$, satisfies $\pi_2(A^3) = \pi_2(S^2) = \mathbb{Z}$. \\

$\bullet$ More generally, the set $\pi_n(X)$ of all homotopy classes of continuous tracking functions of the form $f : S^n \rightarrow X$ (with common fixed point omitted from the notation), has a group structure with respect to the composition of $n$-spheres, and is called called $n$-th homotopy group of $X$. The space $X$ is said \emph{$n$-connected} if its homotopy groups are all trivial up to order $n$\footnote{$n$-connectedness must not be confused with $k$-uple connectedness, defined above, $k$ being essentially the number of holes of the space. In particular, the first notion is stronger than simple connectedness, while the second is weaker.}, i.e. 
\begin{equation}
\forall j \in [ \! [1, n] \! ], \ \pi_j(X)=0\, .
\end{equation}

\iffalse
, i.e. more refined -- that is, lower-dimensional -- empty spaces winded by the topological space)
\fi

\section{Examples:  $\mathrm{SU}(N)$, and $\mathrm{SO}(N)$ and its \emph{universal cover}, $\mathrm{Spin}(N)$.}

\hspace{0.585cm}$\bullet$ The group $\text{SU}(N)$ is simply connected for all $N$, i.e. it has no hole. \\

$\bullet$ $\text{SO}(N)$ has a \emph{double cover} called the spin group $\text{Spin}(N)$. \\

$\bullet$ $N=1$: $\text{SO}(1) = 1$, and $\text{Spin}(1)=\{-1,1\}=O(1)=\{\text{reflexion}, \text{identity}\}$, which is not connected. \\

$\bullet$ $N=2$: $\text{Spin}(2) = U(1)$, which is homeomorphic to both $SO(2)$ itself and $S^1$, i.e. the circle, and is thus manifestly (i.e. through its homeomorphism to $S^1$) connected, but doubly, and not simply, i.e. it has exactly one hole. \\

$\bullet$ $N \geq 3$: \ \ \hspace{1cm} \begin{enumerate}
\item[$\star$] $\text{SO}(N)$ is doubly connected.
\item[$\star$] Its double cover $\text{Spin}(N)$ is simply connected and thus coincides, by definition, with the universal covering group of $\text{SO}(N)$.
\end{enumerate}

An extensive list of examples of homotopy groups useful in physics is available at \url{http://felix.physics.sunysb.edu/~abanov/Teaching/Spring2009/Notes/abanov-cpA1-upload.pdf}.

\section{Compactness}

\hspace{0.585cm}$\bullet$  Compactness is a notion that can be defined given an arbitrary topological space, that generalizes the notion of a subset of Euclidean space being both \emph{closed} (that is, containing all its limit points) and \emph{bounded} (that is, having all its points lie within some fixed distance of each other).

$\bullet$ A topological space $X$ is said \emph{compact} if each of its open covers has a finite subcover. That is, $X$ is compact if for every collection $C$ of open subsets of $X$ such that
\begin{equation}
X = \bigcup_{x \in C} x \, ,
\end{equation}
there is a finite subset $F$ of $C$ such that
\begin{equation}
X = \bigcup_{x \in F} x \, .
\end{equation}

$\bullet$  The fact that we are here indeed dealing with a generalization of the notion of close and bounded subset of an Euclidean space is indicated by two facts: first, this notion can be defined in non-metric spaces; second, the Borel-Lebesgue theorem states that any closed and bounded subset of $\mathbb{R}$ is compact\footnote{The theorem is the proof of the former implying the latter. The latter implying the former is straightforward to prove.}.

\end{appendices}

%\cleartoleftpage

%\bibliographystyle{apsrev4-1}
%\nocite{apsrev41Control}

%\bibliography{bibli.bib,revtex-custom}
%\printbibliography

%merlin.mbs apsrev4-1.bst 2010-07-25 4.21a (PWD, AO, DPC) hacked
%Control: key (0)
%Control: author (72) initials jnrlst
%Control: editor formatted (1) identically to author
%Control: production of article title (0) allowed
%Control: page (1) range
%Control: year (0) verbatim
%Control: production of eprint (0) enabled
%

\newpage\null\thispagestyle{empty}\newpage
\newpage\null\thispagestyle{empty}\newpage

%{\footnotesize\bibliography{These}}

\begin{thebibliography}{296}%
\makeatletter
\providecommand \@ifxundefined [1]{%
 \@ifx{#1\undefined}
}%
\providecommand \@ifnum [1]{%
 \ifnum #1\expandafter \@firstoftwo
 \else \expandafter \@secondoftwo
 \fi
}%
\providecommand \@ifx [1]{%
 \ifx #1\expandafter \@firstoftwo
 \else \expandafter \@secondoftwo
 \fi
}%
\providecommand \natexlab [1]{#1}%
\providecommand \enquote  [1]{``#1''}%
\providecommand \bibnamefont  [1]{#1}%
\providecommand \bibfnamefont [1]{#1}%
\providecommand \citenamefont [1]{#1}%
\providecommand \href@noop [0]{\@secondoftwo}%
\providecommand \href [0]{\begingroup \@sanitize@url \@href}%
\providecommand \@href[1]{\@@startlink{#1}\@@href}%
\providecommand \@@href[1]{\endgroup#1\@@endlink}%
\providecommand \@sanitize@url [0]{\catcode `\\12\catcode `\$12\catcode
  `\&12\catcode `\#12\catcode `\^12\catcode `\_12\catcode `\%12\relax}%
\providecommand \@@startlink[1]{}%
\providecommand \@@endlink[0]{}%
\providecommand \url  [0]{\begingroup\@sanitize@url \@url }%
\providecommand \@url [1]{\endgroup\@href {#1}{\urlprefix }}%
\providecommand \urlprefix  [0]{URL }%
\providecommand \Eprint [0]{\href }%
\providecommand \doibase [0]{http://dx.doi.org/}%
\providecommand \selectlanguage [0]{\@gobble}%
\providecommand \bibinfo  [0]{\@secondoftwo}%
\providecommand \bibfield  [0]{\@secondoftwo}%
\providecommand \translation [1]{[#1]}%
\providecommand \BibitemOpen [0]{}%
\providecommand \bibitemStop [0]{}%
\providecommand \bibitemNoStop [0]{.\EOS\space}%
\providecommand \EOS [0]{\spacefactor3000\relax}%
\providecommand \BibitemShut  [1]{\csname bibitem#1\endcsname}%
\let\auto@bib@innerbib\@empty
%</preamble>
\bibitem [{\citenamefont {Arnault}\ and\ \citenamefont
  {Debbasch}(2015)}]{AD15}%
  \BibitemOpen
  \bibfield  {author} {\bibinfo {author} {\bibfnamefont {P.}~\bibnamefont
  {Arnault}}\ and\ \bibinfo {author} {\bibfnamefont {F.}~\bibnamefont
  {Debbasch}},\ }\bibfield  {title} {\bibinfo {title} {Landau levels for
  discrete-time quantum walks in artificial magnetic fields},\ }\href
  {http://www.sciencedirect.com/science/article/pii/S0378437115006664}
  {\bibfield  {journal} {\bibinfo  {journal} {Physica A}\ }\textbf {\bibinfo
  {volume} {443}},\ \bibinfo {pages} {179--191} (\bibinfo {year}
  {2015})}\BibitemShut {NoStop}%
\bibitem [{\citenamefont {Arnault}\ and\ \citenamefont
  {Debbasch}(2016)}]{AD16}%
  \BibitemOpen
  \bibfield  {author} {\bibinfo {author} {\bibfnamefont {P.}~\bibnamefont
  {Arnault}}\ and\ \bibinfo {author} {\bibfnamefont {F.}~\bibnamefont
  {Debbasch}},\ }\bibfield  {title} {\bibinfo {title} {Quantum walks and
  discrete gauge theories},\ }\href
  {https://link.aps.org/doi/10.1103/PhysRevA.93.052301} {\bibfield  {journal}
  {\bibinfo  {journal} {Phys. Rev. A}\ }\textbf {\bibinfo {volume} {93}},\
  \bibinfo {pages} {052301} (\bibinfo {year} {2016})}\BibitemShut {NoStop}%
\bibitem [{\citenamefont {Arnault}\ \emph {et~al.}(2016)\citenamefont
  {Arnault}, \citenamefont {{Di Molfetta}}, \citenamefont {Brachet},\ and\
  \citenamefont {Debbasch}}]{ADMDB16}%
  \BibitemOpen
  \bibfield  {author} {\bibinfo {author} {\bibfnamefont {P.}~\bibnamefont
  {Arnault}}, \bibinfo {author} {\bibfnamefont {G.}~\bibnamefont {{Di
  Molfetta}}}, \bibinfo {author} {\bibfnamefont {M.}~\bibnamefont {Brachet}}, \
  and\ \bibinfo {author} {\bibfnamefont {F.}~\bibnamefont {Debbasch}},\
  }\bibfield  {title} {\bibinfo {title} {Quantum walks and non-{A}belian
  discrete gauge theory},\ }\href {\doibase 10.1103/physreva.94.012335}
  {\bibfield  {journal} {\bibinfo  {journal} {Phys. Rev. A}\ }\textbf {\bibinfo
  {volume} {94}},\ \bibinfo {pages} {012335} (\bibinfo {year}
  {2016})}\BibitemShut {NoStop}%
\bibitem [{\citenamefont {Arnault}\ and\ \citenamefont
  {Debbasch}(2017)}]{AD17}%
  \BibitemOpen
  \bibfield  {author} {\bibinfo {author} {\bibfnamefont {P.}~\bibnamefont
  {Arnault}}\ and\ \bibinfo {author} {\bibfnamefont {F.}~\bibnamefont
  {Debbasch}},\ }\bibfield  {title} {\bibinfo {title} {Quantum walks and
  gravitational waves},\ }\href
  {http://www.sciencedirect.com/science/article/pii/S0003491617301094}
  {\bibfield  {journal} {\bibinfo  {journal} {Ann. Physics}\ }\textbf {\bibinfo
  {volume} {383}},\ \bibinfo {pages} {645--661} (\bibinfo {year}
  {2017})}\BibitemShut {NoStop}%
\bibitem [{\citenamefont {Trabesinger}(2012)}]{Trabesinger2012}%
  \BibitemOpen
  \bibfield  {author} {\bibinfo {author} {\bibfnamefont {A.}~\bibnamefont
  {Trabesinger}},\ }\bibfield  {title} {\bibinfo {title} {Quantum simulation},\
  }\href {\doibase 10.1038/nphys2258} {\bibfield  {journal} {\bibinfo
  {journal} {Nat. Phys.}\ }\textbf {\bibinfo {volume} {8}},\ \bibinfo {pages}
  {263--263} (\bibinfo {year} {2012})}\BibitemShut {NoStop}%
\bibitem [{\citenamefont {Schaetz}\ \emph {et~al.}(2013)\citenamefont
  {Schaetz}, \citenamefont {Monroe},\ and\ \citenamefont
  {Esslinger}}]{Schaetz2013}%
  \BibitemOpen
  \bibfield  {author} {\bibinfo {author} {\bibfnamefont {T.}~\bibnamefont
  {Schaetz}}, \bibinfo {author} {\bibfnamefont {C.~R.}\ \bibnamefont {Monroe}},
  \ and\ \bibinfo {author} {\bibfnamefont {T.}~\bibnamefont {Esslinger}},\
  }\bibfield  {title} {\bibinfo {title} {Focus on quantum simulation},\ }\href
  {\doibase 10.1088/1367-2630/15/8/085009} {\bibfield  {journal} {\bibinfo
  {journal} {New J. Phys.}\ }\textbf {\bibinfo {volume} {15}},\ \bibinfo
  {pages} {085009} (\bibinfo {year} {2013})}\BibitemShut {NoStop}%
\bibitem [{\citenamefont {Georgescu}\ \emph {et~al.}(2014)\citenamefont
  {Georgescu}, \citenamefont {Ashhab},\ and\ \citenamefont
  {Nori}}]{Georgescu2014}%
  \BibitemOpen
  \bibfield  {author} {\bibinfo {author} {\bibfnamefont {I.~M.}\ \bibnamefont
  {Georgescu}}, \bibinfo {author} {\bibfnamefont {S.}~\bibnamefont {Ashhab}}, \
  and\ \bibinfo {author} {\bibfnamefont {F.}~\bibnamefont {Nori}},\ }\bibfield
  {title} {\bibinfo {title} {Quantum simulation},\ }\href
  {https://link.aps.org/doi/10.1103/RevModPhys.86.153} {\bibfield  {journal}
  {\bibinfo  {journal} {Rev. Mod. Phys.}\ }\textbf {\bibinfo {volume} {86}},\
  \bibinfo {pages} {153--185} (\bibinfo {year} {2014})}\BibitemShut {NoStop}%
\bibitem [{\citenamefont {Johnson}\ \emph {et~al.}(2014)\citenamefont
  {Johnson}, \citenamefont {Clark},\ and\ \citenamefont
  {Jaksch}}]{Johnson2014}%
  \BibitemOpen
  \bibfield  {author} {\bibinfo {author} {\bibfnamefont {T.~H.}\ \bibnamefont
  {Johnson}}, \bibinfo {author} {\bibfnamefont {S.~R.}\ \bibnamefont {Clark}},
  \ and\ \bibinfo {author} {\bibfnamefont {D.}~\bibnamefont {Jaksch}},\
  }\bibfield  {title} {\bibinfo {title} {What is a quantum simulator?}\ }\href
  {\doibase 10.1140/epjqt10} {\bibfield  {journal} {\bibinfo  {journal} {{EPJ}
  Quantum Technol.}\ }\textbf {\bibinfo {volume} {1}},\ \bibinfo {pages} {10}
  (\bibinfo {year} {2014})}\BibitemShut {NoStop}%
\bibitem [{\citenamefont {Childs}(2009{\natexlab{a}})}]{Childs2009}%
  \BibitemOpen
  \bibfield  {author} {\bibinfo {author} {\bibfnamefont {A.~M.}\ \bibnamefont
  {Childs}},\ }\bibfield  {title} {\bibinfo {title} {Universal computation by
  quantum walk},\ }\href {\doibase 10.1103/PhysRevLett.102.180501} {\bibfield
  {journal} {\bibinfo  {journal} {Phys. Rev. Lett.}\ }\textbf {\bibinfo
  {volume} {102}},\ \bibinfo {pages} {180501} (\bibinfo {year}
  {2009}{\natexlab{a}})}\BibitemShut {NoStop}%
\bibitem [{\citenamefont {Childs}\ \emph {et~al.}(2013)\citenamefont {Childs},
  \citenamefont {Gosset},\ and\ \citenamefont {Webb}}]{CGW13}%
  \BibitemOpen
  \bibfield  {author} {\bibinfo {author} {\bibfnamefont {A.~M.}\ \bibnamefont
  {Childs}}, \bibinfo {author} {\bibfnamefont {D.}~\bibnamefont {Gosset}}, \
  and\ \bibinfo {author} {\bibfnamefont {Z.}~\bibnamefont {Webb}},\ }\bibfield
  {title} {\bibinfo {title} {Universal computation by multiparticle quantum
  walk},\ }\href {\doibase 10.1126/science.1229957} {\bibfield  {journal}
  {\bibinfo  {journal} {Science}\ }\textbf {\bibinfo {volume} {339}},\ \bibinfo
  {pages} {791--794} (\bibinfo {year} {2013})}\BibitemShut {NoStop}%
\bibitem [{\citenamefont {Arrighi}\ and\ \citenamefont
  {Grattage}(2011)}]{Arrighi2011}%
  \BibitemOpen
  \bibfield  {author} {\bibinfo {author} {\bibfnamefont {P.}~\bibnamefont
  {Arrighi}}\ and\ \bibinfo {author} {\bibfnamefont {J.}~\bibnamefont
  {Grattage}},\ }\bibfield  {title} {\bibinfo {title} {Partitioned quantum
  cellular automata are intrinsically universal},\ }\href
  {https://doi.org/10.1007/s11047-011-9277-6} {\bibfield  {journal} {\bibinfo
  {journal} {Nat. Comput.}\ }\textbf {\bibinfo {volume} {11}},\ \bibinfo
  {pages} {13--22} (\bibinfo {year} {2011})}\BibitemShut {NoStop}%
\bibitem [{\citenamefont {Arrighi}\ and\ \citenamefont
  {Grattage}(2012)}]{Arrighi2012}%
  \BibitemOpen
  \bibfield  {author} {\bibinfo {author} {\bibfnamefont {P.}~\bibnamefont
  {Arrighi}}\ and\ \bibinfo {author} {\bibfnamefont {J.}~\bibnamefont
  {Grattage}},\ }\bibfield  {title} {\bibinfo {title} {Intrinsically universal
  n-dimensional quantum cellular automata},\ }\href
  {https://doi.org/10.1016/j.jcss.2011.12.008} {\bibfield  {journal} {\bibinfo
  {journal} {J. Comput. Sys. Sci.}\ }\textbf {\bibinfo {volume} {78}},\
  \bibinfo {pages} {1883--1898} (\bibinfo {year} {2012})}\BibitemShut {NoStop}%
\bibitem [{\citenamefont {Wiese}(2013)}]{Wiese2013}%
  \BibitemOpen
  \bibfield  {author} {\bibinfo {author} {\bibfnamefont {U.-J.}\ \bibnamefont
  {Wiese}},\ }\bibfield  {title} {\bibinfo {title} {Ultracold quantum gases and
  lattice systems: quantum simulation of lattice gauge theories},\ }\href
  {\doibase 10.1002/andp.201300104} {\bibfield  {journal} {\bibinfo  {journal}
  {Ann. Physik (Leipzig)}\ }\textbf {\bibinfo {volume} {525}},\ \bibinfo
  {pages} {777--796} (\bibinfo {year} {2013})}\BibitemShut {NoStop}%
\bibitem [{\citenamefont {Zohar}\ \emph {et~al.}(2013)\citenamefont {Zohar},
  \citenamefont {Cirac},\ and\ \citenamefont {Reznik}}]{Zohar2013}%
  \BibitemOpen
  \bibfield  {author} {\bibinfo {author} {\bibfnamefont {E.}~\bibnamefont
  {Zohar}}, \bibinfo {author} {\bibfnamefont {J.~I.}\ \bibnamefont {Cirac}}, \
  and\ \bibinfo {author} {\bibfnamefont {B.}~\bibnamefont {Reznik}},\
  }\bibfield  {title} {\bibinfo {title} {Cold-atom quantum simulator for
  {S{U}}(2) {Y}ang-{M}ills lattice gauge theory},\ }\href
  {https://doi.org/10.1103%2Fphysrevlett.110.125304} {\bibfield  {journal}
  {\bibinfo  {journal} {Phys. Rev. Lett.}\ }\textbf {\bibinfo {volume} {110}},\
  \bibinfo {pages} {125304} (\bibinfo {year} {2013})}\BibitemShut {NoStop}%
\bibitem [{\citenamefont {Banerjee}\ \emph {et~al.}(2013)\citenamefont
  {Banerjee}, \citenamefont {B\"{o}gli}, \citenamefont {Dalmonte},
  \citenamefont {Rico}, \citenamefont {Stebler}, \citenamefont {Wiese},\ and\
  \citenamefont {Zoller}}]{Banerjee2013}%
  \BibitemOpen
  \bibfield  {author} {\bibinfo {author} {\bibfnamefont {D.}~\bibnamefont
  {Banerjee}}, \bibinfo {author} {\bibfnamefont {M.}~\bibnamefont {B\"{o}gli}},
  \bibinfo {author} {\bibfnamefont {M.}~\bibnamefont {Dalmonte}}, \bibinfo
  {author} {\bibfnamefont {E.}~\bibnamefont {Rico}}, \bibinfo {author}
  {\bibfnamefont {P.}~\bibnamefont {Stebler}}, \bibinfo {author} {\bibfnamefont
  {U.-J.}\ \bibnamefont {Wiese}}, \ and\ \bibinfo {author} {\bibfnamefont
  {P.}~\bibnamefont {Zoller}},\ }\bibfield  {title} {\bibinfo {title} {Atomic
  quantum simulation of {U}(${N}$) and {S{U}}(${N}$) non-{A}belian lattice
  gauge theories},\ }\href {\doibase 10.1103/physrevlett.110.125303} {\bibfield
   {journal} {\bibinfo  {journal} {Phys. Rev. Lett.}\ }\textbf {\bibinfo
  {volume} {110}},\ \bibinfo {pages} {125303} (\bibinfo {year}
  {2013})}\BibitemShut {NoStop}%
\bibitem [{\citenamefont {Rico}\ \emph {et~al.}(2014)\citenamefont {Rico},
  \citenamefont {Pichler}, \citenamefont {Dalmonte}, \citenamefont {Zoller},\
  and\ \citenamefont {Montangero}}]{Rico2014}%
  \BibitemOpen
  \bibfield  {author} {\bibinfo {author} {\bibfnamefont {E.}~\bibnamefont
  {Rico}}, \bibinfo {author} {\bibfnamefont {T.}~\bibnamefont {Pichler}},
  \bibinfo {author} {\bibfnamefont {M.}~\bibnamefont {Dalmonte}}, \bibinfo
  {author} {\bibfnamefont {P.}~\bibnamefont {Zoller}}, \ and\ \bibinfo {author}
  {\bibfnamefont {S.}~\bibnamefont {Montangero}},\ }\bibfield  {title}
  {\bibinfo {title} {Tensor networks for lattice gauge theories and atomic
  quantum simulation},\ }\href {\doibase 10.1103/physrevlett.112.201601}
  {\bibfield  {journal} {\bibinfo  {journal} {Phys. Rev. Lett.}\ }\textbf
  {\bibinfo {volume} {112}},\ \bibinfo {pages} {201601} (\bibinfo {year}
  {2014})}\BibitemShut {NoStop}%
\bibitem [{\citenamefont {{Di Molfetta}}(2015)}]{dimolfetta:tel-01230891}%
  \BibitemOpen
  \bibfield  {author} {\bibinfo {author} {\bibfnamefont {G.}~\bibnamefont {{Di
  Molfetta}}},\ }\emph {\bibinfo {title} {{Discrete time quantum walks: from
  synthetic gauge fields to spontaneous equilibration}}},\ \href
  {https://tel.archives-ouvertes.fr/tel-01230891} {Ph.D. thesis},\ \bibinfo
  {school} {{Universit{\'e} Pierre et Marie Curie - Paris VI}} (\bibinfo {year}
  {2015})\BibitemShut {NoStop}%
\bibitem [{\citenamefont {Succi}\ \emph {et~al.}(2015)\citenamefont {Succi},
  \citenamefont {Fillion-Gourdeau},\ and\ \citenamefont
  {Palpacelli}}]{Succi2015}%
  \BibitemOpen
  \bibfield  {author} {\bibinfo {author} {\bibfnamefont {S.}~\bibnamefont
  {Succi}}, \bibinfo {author} {\bibfnamefont {F.}~\bibnamefont
  {Fillion-Gourdeau}}, \ and\ \bibinfo {author} {\bibfnamefont
  {S.}~\bibnamefont {Palpacelli}},\ }\bibfield  {title} {\bibinfo {title}
  {Quantum lattice {B}oltzmann is a quantum walk},\ }\href
  {https://link.springer.com/article/10.1140/epjqt/s40507-015-0025-1}
  {\bibfield  {journal} {\bibinfo  {journal} {{EPJ} Quantum Technol.}\ }\textbf
  {\bibinfo {volume} {2}},\ \bibinfo {pages} {12} (\bibinfo {year}
  {2015})}\BibitemShut {NoStop}%
\bibitem [{\citenamefont {Dalibard}(2015)}]{Dalibar2015}%
  \BibitemOpen
  \bibfield  {author} {\bibinfo {author} {\bibfnamefont {J.}~\bibnamefont
  {Dalibard}},\ }\bibfield  {title} {\bibinfo {title} {Introduction to the
  physics of artificial gauge fields},\ }\href
  {https://arxiv.org/abs/1504.05520} {\bibfield  {journal} {\bibinfo  {journal}
  {arXiv:1504.05520}\ } (\bibinfo {year} {2015})}\BibitemShut {NoStop}%
\bibitem [{\citenamefont {Hamilton}\ \emph {et~al.}(2016)\citenamefont
  {Hamilton}, \citenamefont {Barkhofen}, \citenamefont {Sansoni}, \citenamefont
  {Jex},\ and\ \citenamefont {Silberhorn}}]{Hamilton2016}%
  \BibitemOpen
  \bibfield  {author} {\bibinfo {author} {\bibfnamefont {C.~S.}\ \bibnamefont
  {Hamilton}}, \bibinfo {author} {\bibfnamefont {S.}~\bibnamefont {Barkhofen}},
  \bibinfo {author} {\bibfnamefont {L.}~\bibnamefont {Sansoni}}, \bibinfo
  {author} {\bibfnamefont {I.}~\bibnamefont {Jex}}, \ and\ \bibinfo {author}
  {\bibfnamefont {C.}~\bibnamefont {Silberhorn}},\ }\bibfield  {title}
  {\bibinfo {title} {Driven discrete time quantum walks},\ }\href
  {https://doi.org/10.1088/1367-2630/18/7/073008} {\bibfield  {journal}
  {\bibinfo  {journal} {New J. Phys.}\ }\textbf {\bibinfo {volume} {18}},\
  \bibinfo {pages} {073008} (\bibinfo {year} {2016})}\BibitemShut {NoStop}%
\bibitem [{\citenamefont {Bloch}\ \emph {et~al.}(2012)\citenamefont {Bloch},
  \citenamefont {Dalibard},\ and\ \citenamefont
  {Nascimb{\`{e}}ne}}]{Bloch2012}%
  \BibitemOpen
  \bibfield  {author} {\bibinfo {author} {\bibfnamefont {I.}~\bibnamefont
  {Bloch}}, \bibinfo {author} {\bibfnamefont {J.}~\bibnamefont {Dalibard}}, \
  and\ \bibinfo {author} {\bibfnamefont {S.}~\bibnamefont {Nascimb{\`{e}}ne}},\
  }\bibfield  {title} {\bibinfo {title} {Quantum simulations with ultracold
  quantum gases},\ }\href {\doibase 10.1038/nphys2259} {\bibfield  {journal}
  {\bibinfo  {journal} {Nat. Phys.}\ }\textbf {\bibinfo {volume} {8}},\
  \bibinfo {pages} {267--276} (\bibinfo {year} {2012})}\BibitemShut {NoStop}%
\bibitem [{\citenamefont {Scholl}(2014)}]{scholl:tel-01165961}%
  \BibitemOpen
  \bibfield  {author} {\bibinfo {author} {\bibfnamefont {M.}~\bibnamefont
  {Scholl}},\ }\emph {\bibinfo {title} {{Probing an ytterbium Bose-Einstein
  condensate using an ultranarrow optical line: towards artificial gauge fields
  in optical lattices}}},\ \href
  {https://tel.archives-ouvertes.fr/tel-01165961} {Ph.D. thesis},\ \bibinfo
  {school} {{Universit{\'e} Pierre et Marie Curie - Paris VI}} (\bibinfo {year}
  {2014})\BibitemShut {NoStop}%
\bibitem [{\citenamefont {Aidelsburger}(2016)}]{Aidelsburger2016}%
  \BibitemOpen
  \bibfield  {author} {\bibinfo {author} {\bibfnamefont {M.}~\bibnamefont
  {Aidelsburger}},\ }\href {\doibase 10.1007/978-3-319-25829-4} {\emph
  {\bibinfo {title} {Artificial Gauge Fields with Ultracold Atoms in Optical
  Lattices}}}\ (\bibinfo  {publisher} {Springer},\ \bibinfo {year} {2016})\
  \bibinfo {note} {({P}h.{D}. thesis, Ludwig-Maximilians-Universit{\"a}t
  M{\"u}nchen)}\BibitemShut {NoStop}%
\bibitem [{\citenamefont {Martinez}\ \emph {et~al.}(2016)\citenamefont
  {Martinez}, \citenamefont {Muschik}, \citenamefont {Schindler}, \citenamefont
  {Nigg}, \citenamefont {Erhard}, \citenamefont {Heyl}, \citenamefont {Hauke},
  \citenamefont {Dalmonte}, \citenamefont {Monz}, \citenamefont {Zoller},\ and\
  \citenamefont {Blatt}}]{Martinez16}%
  \BibitemOpen
  \bibfield  {author} {\bibinfo {author} {\bibfnamefont {E.~A.}\ \bibnamefont
  {Martinez}}, \bibinfo {author} {\bibfnamefont {C.~A.}\ \bibnamefont
  {Muschik}}, \bibinfo {author} {\bibfnamefont {P.}~\bibnamefont {Schindler}},
  \bibinfo {author} {\bibfnamefont {D.}~\bibnamefont {Nigg}}, \bibinfo {author}
  {\bibfnamefont {A.}~\bibnamefont {Erhard}}, \bibinfo {author} {\bibfnamefont
  {M.}~\bibnamefont {Heyl}}, \bibinfo {author} {\bibfnamefont {P.}~\bibnamefont
  {Hauke}}, \bibinfo {author} {\bibfnamefont {M.}~\bibnamefont {Dalmonte}},
  \bibinfo {author} {\bibfnamefont {T.}~\bibnamefont {Monz}}, \bibinfo {author}
  {\bibfnamefont {P.}~\bibnamefont {Zoller}}, \ and\ \bibinfo {author}
  {\bibfnamefont {R.}~\bibnamefont {Blatt}},\ }\bibfield  {title} {\bibinfo
  {title} {Real-time dynamics of lattice gauge theories with a few-qubit
  quantum computer},\ }\href
  {https://www.nature.com/nature/journal/v534/n7608/full/nature18318.html}
  {\bibfield  {journal} {\bibinfo  {journal} {Nature}\ }\textbf {\bibinfo
  {volume} {534}},\ \bibinfo {pages} {516–519} (\bibinfo {year}
  {2016})}\BibitemShut {NoStop}%
\bibitem [{\citenamefont {O'Malley}\ \emph {et~al.}(2016)\citenamefont
  {O'Malley}, \citenamefont {Babbush}, \citenamefont {Kivlichan}, \citenamefont
  {Romero}, \citenamefont {McClean}, \citenamefont {Barends}, \citenamefont
  {Kelly}, \citenamefont {Roushan}, \citenamefont {Tranter}, \citenamefont
  {Ding}, \citenamefont {Campbell}, \citenamefont {Chen}, \citenamefont {Chen},
  \citenamefont {Chiaro}, \citenamefont {Dunsworth}, \citenamefont {Fowler},
  \citenamefont {Jeffrey}, \citenamefont {Lucero}, \citenamefont {Megrant},
  \citenamefont {Mutus}, \citenamefont {Neeley}, \citenamefont {Neill},
  \citenamefont {Quintana}, \citenamefont {Sank}, \citenamefont {Vainsencher},
  \citenamefont {Wenner}, \citenamefont {White}, \citenamefont {Coveney},
  \citenamefont {Love}, \citenamefont {Neven}, \citenamefont {Aspuru-Guzik},\
  and\ \citenamefont {Martinis}}]{OMalley2016}%
  \BibitemOpen
  \bibfield  {author} {\bibinfo {author} {\bibfnamefont {P.~J.~J.}\
  \bibnamefont {O'Malley}}, \bibinfo {author} {\bibfnamefont {R.}~\bibnamefont
  {Babbush}}, \bibinfo {author} {\bibfnamefont {I.~D.}\ \bibnamefont
  {Kivlichan}}, \bibinfo {author} {\bibfnamefont {J.}~\bibnamefont {Romero}},
  \bibinfo {author} {\bibfnamefont {J.~R.}\ \bibnamefont {McClean}}, \bibinfo
  {author} {\bibfnamefont {R.}~\bibnamefont {Barends}}, \bibinfo {author}
  {\bibfnamefont {J.}~\bibnamefont {Kelly}}, \bibinfo {author} {\bibfnamefont
  {P.}~\bibnamefont {Roushan}}, \bibinfo {author} {\bibfnamefont
  {A.}~\bibnamefont {Tranter}}, \bibinfo {author} {\bibfnamefont
  {N.}~\bibnamefont {Ding}}, \bibinfo {author} {\bibfnamefont {B.}~\bibnamefont
  {Campbell}}, \bibinfo {author} {\bibfnamefont {Y.}~\bibnamefont {Chen}},
  \bibinfo {author} {\bibfnamefont {Z.}~\bibnamefont {Chen}}, \bibinfo {author}
  {\bibfnamefont {B.}~\bibnamefont {Chiaro}}, \bibinfo {author} {\bibfnamefont
  {A.}~\bibnamefont {Dunsworth}}, \bibinfo {author} {\bibfnamefont {A.~G.}\
  \bibnamefont {Fowler}}, \bibinfo {author} {\bibfnamefont {E.}~\bibnamefont
  {Jeffrey}}, \bibinfo {author} {\bibfnamefont {E.}~\bibnamefont {Lucero}},
  \bibinfo {author} {\bibfnamefont {A.}~\bibnamefont {Megrant}}, \bibinfo
  {author} {\bibfnamefont {J.~Y.}\ \bibnamefont {Mutus}}, \bibinfo {author}
  {\bibfnamefont {M.}~\bibnamefont {Neeley}}, \bibinfo {author} {\bibfnamefont
  {C.}~\bibnamefont {Neill}}, \bibinfo {author} {\bibfnamefont
  {C.}~\bibnamefont {Quintana}}, \bibinfo {author} {\bibfnamefont
  {D.}~\bibnamefont {Sank}}, \bibinfo {author} {\bibfnamefont {A.}~\bibnamefont
  {Vainsencher}}, \bibinfo {author} {\bibfnamefont {J.}~\bibnamefont {Wenner}},
  \bibinfo {author} {\bibfnamefont {T.~C.}\ \bibnamefont {White}}, \bibinfo
  {author} {\bibfnamefont {P.~V.}\ \bibnamefont {Coveney}}, \bibinfo {author}
  {\bibfnamefont {P.~J.}\ \bibnamefont {Love}}, \bibinfo {author}
  {\bibfnamefont {H.}~\bibnamefont {Neven}}, \bibinfo {author} {\bibfnamefont
  {A.}~\bibnamefont {Aspuru-Guzik}}, \ and\ \bibinfo {author} {\bibfnamefont
  {J.~M.}\ \bibnamefont {Martinis}},\ }\bibfield  {title} {\bibinfo {title}
  {Scalable quantum simulation of molecular energies},\ }\href
  {https://doi.org/10.1103/physrevx.6.031007} {\bibfield  {journal} {\bibinfo
  {journal} {Phys. Rev. X}\ }\textbf {\bibinfo {volume} {6}},\ \bibinfo {pages}
  {031007} (\bibinfo {year} {2016})}\BibitemShut {NoStop}%
\bibitem [{\citenamefont {Tame}\ \emph {et~al.}(2014)\citenamefont {Tame},
  \citenamefont {Bell}, \citenamefont {Franco}, \citenamefont {Wadsworth},\
  and\ \citenamefont {Rarity}}]{Tame2014}%
  \BibitemOpen
  \bibfield  {author} {\bibinfo {author} {\bibfnamefont {M.~S.}\ \bibnamefont
  {Tame}}, \bibinfo {author} {\bibfnamefont {B.~A.}\ \bibnamefont {Bell}},
  \bibinfo {author} {\bibfnamefont {C.~D.}\ \bibnamefont {Franco}}, \bibinfo
  {author} {\bibfnamefont {W.~J.}\ \bibnamefont {Wadsworth}}, \ and\ \bibinfo
  {author} {\bibfnamefont {J.~G.}\ \bibnamefont {Rarity}},\ }\bibfield  {title}
  {\bibinfo {title} {Experimental realization of a one-way quantum computer
  algorithm solving {S}imon's problem},\ }\href
  {https://doi.org/10.1103/physrevlett.113.200501} {\bibfield  {journal}
  {\bibinfo  {journal} {Phys. Rev. Lett.}\ }\textbf {\bibinfo {volume} {113}},\
  \bibinfo {pages} {200501} (\bibinfo {year} {2014})}\BibitemShut {NoStop}%
\bibitem [{\citenamefont {Wilson}(1974)}]{Wilson74}%
  \BibitemOpen
  \bibfield  {author} {\bibinfo {author} {\bibfnamefont {K.~G.}\ \bibnamefont
  {Wilson}},\ }\bibfield  {title} {\bibinfo {title} {Confinement of quarks},\
  }\href {https://link.aps.org/doi/10.1103/PhysRevD.10.2445} {\bibfield
  {journal} {\bibinfo  {journal} {Phys. Rev. D}\ }\textbf {\bibinfo {volume}
  {10}},\ \bibinfo {pages} {2445} (\bibinfo {year} {1974})}\BibitemShut
  {NoStop}%
\bibitem [{\citenamefont {Rothe}(2012)}]{book_Rothe}%
  \BibitemOpen
  \bibfield  {author} {\bibinfo {author} {\bibfnamefont {H.~J.}\ \bibnamefont
  {Rothe}},\ }\href {http://www.worldscientific.com/worldscibooks/10.1142/8229}
  {\emph {\bibinfo {title} {Lattice Gauge Theories: An Introduction{,} 4th
  {E}dition}}}\ (\bibinfo  {publisher} {World Scientific},\ \bibinfo {year}
  {2012})\BibitemShut {NoStop}%
\bibitem [{\citenamefont {Schweber}(1986)}]{Schweber86a}%
  \BibitemOpen
  \bibfield  {author} {\bibinfo {author} {\bibfnamefont {S.~S.}\ \bibnamefont
  {Schweber}},\ }\bibfield  {title} {\bibinfo {title} {Feynman and the
  visualization of space-time processes},\ }\href
  {https://link.aps.org/doi/10.1103/RevModPhys.58.449} {\bibfield  {journal}
  {\bibinfo  {journal} {Rev. Mod. Phys.}\ }\textbf {\bibinfo {volume} {58}},\
  \bibinfo {pages} {449} (\bibinfo {year} {1986})}\BibitemShut {NoStop}%
\bibitem [{\citenamefont {Berezin}(1966)}]{book_Berezin}%
  \BibitemOpen
  \bibfield  {author} {\bibinfo {author} {\bibfnamefont {F.~A.}\ \bibnamefont
  {Berezin}},\ }\href@noop {} {\emph {\bibinfo {title} {The {M}ethod of
  {S}econd {Q}uantization}}}\ (\bibinfo  {publisher} {Academic Press, United
  States},\ \bibinfo {year} {1966})\BibitemShut {NoStop}%
\bibitem [{\citenamefont {Weinberg}(1995)}]{Weinberg_QFT1}%
  \BibitemOpen
  \bibfield  {author} {\bibinfo {author} {\bibfnamefont {S.}~\bibnamefont
  {Weinberg}},\ }\href
  {http://www.cambridge.org/catalogue/catalogue.asp?isbn=9780521670531} {\emph
  {\bibinfo {title} {The Quantum Theory of Fields I}}}\ (\bibinfo  {publisher}
  {Cambridge {U}niversity {P}ress},\ \bibinfo {year} {1995})\BibitemShut
  {NoStop}%
\bibitem [{\citenamefont {Aharonov}\ \emph {et~al.}(1993)\citenamefont
  {Aharonov}, \citenamefont {Davidovich},\ and\ \citenamefont
  {Zagury}}]{ADZ93a}%
  \BibitemOpen
  \bibfield  {author} {\bibinfo {author} {\bibfnamefont {Y.}~\bibnamefont
  {Aharonov}}, \bibinfo {author} {\bibfnamefont {L.}~\bibnamefont
  {Davidovich}}, \ and\ \bibinfo {author} {\bibfnamefont {N.}~\bibnamefont
  {Zagury}},\ }\bibfield  {title} {\bibinfo {title} {Quantum random walks},\
  }\href {https://link.aps.org/doi/10.1103/PhysRevA.48.1687} {\bibfield
  {journal} {\bibinfo  {journal} {Phys. {R}ev. A}\ }\textbf {\bibinfo {volume}
  {48}},\ \bibinfo {pages} {1687--1690} (\bibinfo {year} {1993})}\BibitemShut
  {NoStop}%
\bibitem [{\citenamefont {Meyer}(1996)}]{Meyer96a}%
  \BibitemOpen
  \bibfield  {author} {\bibinfo {author} {\bibfnamefont {D.~A.}\ \bibnamefont
  {Meyer}},\ }\bibfield  {title} {\bibinfo {title} {From quantum cellular
  automata to quantum lattice gases},\ }\href
  {https://link.springer.com/article/10.1007/BF02199356} {\bibfield  {journal}
  {\bibinfo  {journal} {{J}. {S}tat. {P}hys.}\ }\textbf {\bibinfo {volume}
  {85}},\ \bibinfo {pages} {551--574} (\bibinfo {year} {1996})}\BibitemShut
  {NoStop}%
\bibitem [{\citenamefont {Farhi}\ and\ \citenamefont {Gutmann}(1998)}]{FG98a}%
  \BibitemOpen
  \bibfield  {author} {\bibinfo {author} {\bibfnamefont {E.}~\bibnamefont
  {Farhi}}\ and\ \bibinfo {author} {\bibfnamefont {S.}~\bibnamefont
  {Gutmann}},\ }\bibfield  {title} {\bibinfo {title} {Quantum computation and
  decision trees},\ }\href {https://link.aps.org/doi/10.1103/PhysRevA.58.915}
  {\bibfield  {journal} {\bibinfo  {journal} {Phys. {R}ev. A}\ }\textbf
  {\bibinfo {volume} {58}},\ \bibinfo {pages} {915--928} (\bibinfo {year}
  {1998})}\BibitemShut {NoStop}%
\bibitem [{\citenamefont {Strauch}(2006{\natexlab{a}})}]{Strauch06b}%
  \BibitemOpen
  \bibfield  {author} {\bibinfo {author} {\bibfnamefont {F.~W.}\ \bibnamefont
  {Strauch}},\ }\bibfield  {title} {\bibinfo {title} {Connecting the discrete-
  and continuous-time quantum walks},\ }\href
  {https://link.aps.org/doi/10.1103/PhysRevA.74.030301} {\bibfield  {journal}
  {\bibinfo  {journal} {Phys. {R}ev. A}\ }\textbf {\bibinfo {volume} {74}},\
  \bibinfo {pages} {030301} (\bibinfo {year} {2006}{\natexlab{a}})}\BibitemShut
  {NoStop}%
\bibitem [{\citenamefont {Shor}(1994)}]{Shor94}%
  \BibitemOpen
  \bibfield  {author} {\bibinfo {author} {\bibfnamefont {P.~W.}\ \bibnamefont
  {Shor}},\ }\bibfield  {title} {\bibinfo {title} {Algorithms for quantum
  computation: discrete logarithms and factoring},\ }in\ \href
  {http://ieeexplore.ieee.org/document/365700/} {\emph {\bibinfo {booktitle}
  {Proceedings of the 35th annual IEEE Symposium on Foundations of Computer
  Science}}}\ (\bibinfo {year} {1994})\BibitemShut {NoStop}%
\bibitem [{\citenamefont {Shor}(1997)}]{Shor97}%
  \BibitemOpen
  \bibfield  {author} {\bibinfo {author} {\bibfnamefont {P.~W.}\ \bibnamefont
  {Shor}},\ }\bibfield  {title} {\bibinfo {title} {Polynomial-time algorithms
  for prime factorization and discrete logarithms on a quantum computer},\
  }\href {http://epubs.siam.org/doi/10.1137/S0097539795293172} {\bibfield
  {journal} {\bibinfo  {journal} {SIAM J. Comput.}\ }\textbf {\bibinfo {volume}
  {26}},\ \bibinfo {pages} {1484--1509} (\bibinfo {year} {1997})}\BibitemShut
  {NoStop}%
\bibitem [{\citenamefont {Grover}(1996)}]{Grover96}%
  \BibitemOpen
  \bibfield  {author} {\bibinfo {author} {\bibfnamefont {L.~K.}\ \bibnamefont
  {Grover}},\ }\bibfield  {title} {\bibinfo {title} {A fast quantum mechanical
  algorithm for database search},\ }in\ \href
  {http://dl.acm.org/citation.cfm?doid=237814.237866} {\emph {\bibinfo
  {booktitle} {Proceedings of the 28th annual ACM Symposium on Theory of
  Computing - STOC96}}}\ (\bibinfo {year} {1996})\BibitemShut {NoStop}%
\bibitem [{\citenamefont {Ambainis}\ \emph {et~al.}(2001)\citenamefont
  {Ambainis}, \citenamefont {Bach}, \citenamefont {Nayak}, \citenamefont
  {Vishwanath},\ and\ \citenamefont {Watrous}}]{Ambainis2001}%
  \BibitemOpen
  \bibfield  {author} {\bibinfo {author} {\bibfnamefont {A.}~\bibnamefont
  {Ambainis}}, \bibinfo {author} {\bibfnamefont {E.}~\bibnamefont {Bach}},
  \bibinfo {author} {\bibfnamefont {A.}~\bibnamefont {Nayak}}, \bibinfo
  {author} {\bibfnamefont {A.}~\bibnamefont {Vishwanath}}, \ and\ \bibinfo
  {author} {\bibfnamefont {J.}~\bibnamefont {Watrous}},\ }\bibfield  {title}
  {\bibinfo {title} {One-dimensional quantum walks},\ }in\ \href {\doibase
  10.1145/380752.380757} {\emph {\bibinfo {booktitle} {Proceedings of the 33rd
  annual {ACM} Symposium on Theory of Computing - {STOC}01}}}\ (\bibinfo {year}
  {2001})\BibitemShut {NoStop}%
\bibitem [{\citenamefont {A.~Ambainis}(2005)}]{AKR2005}%
  \BibitemOpen
  \bibfield  {author} {\bibinfo {author} {\bibfnamefont {A.~R.}\ \bibnamefont
  {A.~Ambainis}, \bibfnamefont {J.~Kempe}},\ }\bibfield  {title} {\bibinfo
  {title} {Coins make quantum walks faster},\ }in\ \href
  {http://dl.acm.org/citation.cfm?id=1070432.1070590} {\emph {\bibinfo
  {booktitle} {Proceedings of the 16th annual ACM-SIAM Symposium on Discrete
  Algorithms - SODA05}}}\ (\bibinfo {year} {2005})\BibitemShut {NoStop}%
\bibitem [{\citenamefont {Ambainis}(2007)}]{Amb07a}%
  \BibitemOpen
  \bibfield  {author} {\bibinfo {author} {\bibfnamefont {A.}~\bibnamefont
  {Ambainis}},\ }\bibfield  {title} {\bibinfo {title} {Quantum walk algorithm
  for element distinctness},\ }\href
  {http://epubs.siam.org/doi/10.1137/S0097539705447311} {\bibfield  {journal}
  {\bibinfo  {journal} {SIAM J. Comput.}\ }\textbf {\bibinfo {volume} {37}},\
  \bibinfo {pages} {210--239} (\bibinfo {year} {2007})}\BibitemShut {NoStop}%
\bibitem [{\citenamefont {Feynman}(1982)}]{Feynman1982}%
  \BibitemOpen
  \bibfield  {author} {\bibinfo {author} {\bibfnamefont {R.~P.}\ \bibnamefont
  {Feynman}},\ }\bibfield  {title} {\bibinfo {title} {Simulating physics with
  computers},\ }\href {\doibase 10.1007/bf02650179} {\bibfield  {journal}
  {\bibinfo  {journal} {Int. J. Theor. Phys.}\ }\textbf {\bibinfo {volume}
  {21}},\ \bibinfo {pages} {467--488} (\bibinfo {year} {1982})}\BibitemShut
  {NoStop}%
\bibitem [{\citenamefont {Lloyd}(1996)}]{Lloyd1996}%
  \BibitemOpen
  \bibfield  {author} {\bibinfo {author} {\bibfnamefont {S.}~\bibnamefont
  {Lloyd}},\ }\bibfield  {title} {\bibinfo {title} {Universal quantum
  simulators},\ }\href {https://doi.org/10.1126/science.273.5278.1073}
  {\bibfield  {journal} {\bibinfo  {journal} {Science}\ }\textbf {\bibinfo
  {volume} {273}},\ \bibinfo {pages} {1073--1078} (\bibinfo {year}
  {1996})}\BibitemShut {NoStop}%
\bibitem [{\citenamefont {Farrelly}()}]{Farrelly15}%
  \BibitemOpen
  \bibfield  {author} {\bibinfo {author} {\bibfnamefont {T.~C.}\ \bibnamefont
  {Farrelly}},\ }\bibfield  {title} {\bibinfo {title} {\emph{Insights from
  quantum information into fundamental physics}},\ }\href@noop {} {\bibinfo
  {journal} {Ph.D. thesis, {U}niversity of {C}ambridge (2015),
  \href{https://arxiv.org/abs/1708.08897}{arXiv:1708.08897} \hspace{-0.25cm}}\
  }\BibitemShut {NoStop}%
\bibitem [{\citenamefont {Navier}(1822)}]{N1822}%
  \BibitemOpen
\bibfield  {journal} {  }\bibfield  {author} {\bibinfo {author} {\bibfnamefont
  {H.}~\bibnamefont {Navier}},\ }\bibfield  {title} {\bibinfo {title}
  {M\'{e}moire sur les lois du mouvement des fluides},\ }\href
  {http://gallica.bnf.fr/ark:/12148/bpt6k3221x/f577.image} {\bibfield
  {journal} {\bibinfo  {journal} {M\'{e}moires Acad. Roy. Sci.}\ }\textbf
  {\bibinfo {volume} {6}} (\bibinfo {year} {1822})}\BibitemShut {NoStop}%
\bibitem [{\citenamefont {Stokes}(1845)}]{S1845}%
  \BibitemOpen
  \bibfield  {author} {\bibinfo {author} {\bibfnamefont {G.~G.}\ \bibnamefont
  {Stokes}},\ }\bibfield  {title} {\bibinfo {title} {On the theories of the
  internal friction of fluids in motion},\ }\href
  {https://archive.org/details/cbarchive_39179_onthetheoriesoftheinternalfric1849}
  {\bibfield  {journal} {\bibinfo  {journal} {Trans. Cambridge Philos. Soc.}\
  }\textbf {\bibinfo {volume} {8}},\ \bibinfo {pages} {287--305} (\bibinfo
  {year} {1845})}\BibitemShut {NoStop}%
\bibitem [{\citenamefont {Debbasch}\ and\ \citenamefont {Rivet}(1998)}]{DR98}%
  \BibitemOpen
  \bibfield  {author} {\bibinfo {author} {\bibfnamefont {F.}~\bibnamefont
  {Debbasch}}\ and\ \bibinfo {author} {\bibfnamefont {J.-P.}\ \bibnamefont
  {Rivet}},\ }\bibfield  {title} {\bibinfo {title} {A diffusion equation from
  the relativistic {O}rnstein-{U}hlenbeck process},\ }\href
  {https://link.springer.com/article/10.1023/A:1023275210656} {\bibfield
  {journal} {\bibinfo  {journal} {J. Stat. Phys.}\ }\textbf {\bibinfo {volume}
  {90}},\ \bibinfo {pages} {1179--1199} (\bibinfo {year} {1998})}\BibitemShut
  {NoStop}%
\bibitem [{\citenamefont {Ivancevic}\ and\ \citenamefont
  {Ivancevic}(2013)}]{II2013}%
  \BibitemOpen
  \bibfield  {author} {\bibinfo {author} {\bibfnamefont {V.~G.}\ \bibnamefont
  {Ivancevic}}\ and\ \bibinfo {author} {\bibfnamefont {T.~T.}\ \bibnamefont
  {Ivancevic}},\ }\bibfield  {title} {\bibinfo {title} {Sine-{G}ordon solitons,
  kinks and breathers as physical models of nonlinear excitations in living
  cellular structures},\ }\href {ttps://arxiv.org/abs/1305.0613} {\bibfield
  {journal} {\bibinfo  {journal} {arXiv:1305.0613}\ } (\bibinfo {year}
  {2013})}\BibitemShut {NoStop}%
\bibitem [{\citenamefont {Debbasch}\ \emph {et~al.}(1997)\citenamefont
  {Debbasch}, \citenamefont {Mallick},\ and\ \citenamefont {Rivet}}]{DMR97}%
  \BibitemOpen
  \bibfield  {author} {\bibinfo {author} {\bibfnamefont {F.}~\bibnamefont
  {Debbasch}}, \bibinfo {author} {\bibfnamefont {K.}~\bibnamefont {Mallick}}, \
  and\ \bibinfo {author} {\bibfnamefont {J.-P.}\ \bibnamefont {Rivet}},\
  }\bibfield  {title} {\bibinfo {title} {Relativistic {O}rnstein-{U}hlenbeck
  process},\ }\href
  {https://link.springer.com/article/10.1023/B:JOSS.0000015180.16261.53}
  {\bibfield  {journal} {\bibinfo  {journal} {J. Stat. Phys.}\ }\textbf
  {\bibinfo {volume} {88}},\ \bibinfo {pages} {945--966} (\bibinfo {year}
  {1997})}\BibitemShut {NoStop}%
\bibitem [{\citenamefont {Freist\"{u}hler}\ and\ \citenamefont
  {Temple}(2017)}]{Freisthler2017}%
  \BibitemOpen
  \bibfield  {author} {\bibinfo {author} {\bibfnamefont {H.}~\bibnamefont
  {Freist\"{u}hler}}\ and\ \bibinfo {author} {\bibfnamefont {B.}~\bibnamefont
  {Temple}},\ }\bibfield  {title} {\bibinfo {title} {Causal dissipation for the
  relativistic dynamics of ideal gases},\ }\href {\doibase
  10.1098/rspa.2016.0729} {\bibfield  {journal} {\bibinfo  {journal} {Proc.
  Roy. Soc. A}\ }\textbf {\bibinfo {volume} {473}},\ \bibinfo {pages}
  {20160729} (\bibinfo {year} {2017})}\BibitemShut {NoStop}%
\bibitem [{\citenamefont {Cercignani}\ and\ \citenamefont
  {Kremer}(2002)}]{CK2002}%
  \BibitemOpen
  \bibfield  {author} {\bibinfo {author} {\bibfnamefont {C.}~\bibnamefont
  {Cercignani}}\ and\ \bibinfo {author} {\bibfnamefont {G.~M.}\ \bibnamefont
  {Kremer}},\ }\href {http://www.springer.com/gp/book/9783764366933} {\emph
  {\bibinfo {title} {The Relativistic Boltzmann Equation: Theory and
  Applications}}}\ (\bibinfo  {publisher} {Springer},\ \bibinfo {year}
  {2002})\BibitemShut {NoStop}%
\bibitem [{\citenamefont {Landauer}(1991)}]{Landauer1991}%
  \BibitemOpen
  \bibfield  {author} {\bibinfo {author} {\bibfnamefont {R.}~\bibnamefont
  {Landauer}},\ }\bibfield  {title} {\bibinfo {title} {Information is
  physical},\ }\href {https://doi.org/10.1063/1.881299} {\bibfield  {journal}
  {\bibinfo  {journal} {Phys. Today}\ }\textbf {\bibinfo {volume} {44}},\
  \bibinfo {pages} {23--29} (\bibinfo {year} {1991})}\BibitemShut {NoStop}%
\bibitem [{\citenamefont {Peres}\ and\ \citenamefont
  {Terno}(2004)}]{Peres2004}%
  \BibitemOpen
  \bibfield  {author} {\bibinfo {author} {\bibfnamefont {A.}~\bibnamefont
  {Peres}}\ and\ \bibinfo {author} {\bibfnamefont {D.~R.}\ \bibnamefont
  {Terno}},\ }\bibfield  {title} {\bibinfo {title} {Quantum information and
  relativity theory},\ }\href {\doibase 10.1103/revmodphys.76.93} {\bibfield
  {journal} {\bibinfo  {journal} {Rev. Mod. Phys.}\ }\textbf {\bibinfo {volume}
  {76}},\ \bibinfo {pages} {93--123} (\bibinfo {year} {2004})}\BibitemShut
  {NoStop}%
\bibitem [{\citenamefont {Roug\'e}(2002)}]{Rouge}%
  \BibitemOpen
  \bibfield  {author} {\bibinfo {author} {\bibfnamefont {A.}~\bibnamefont
  {Roug\'e}},\ }\href@noop {} {\emph {\bibinfo {title} {Introduction \`a la
  relativit\'e}}}\ (\bibinfo  {publisher} {\'Editions de l'\'Ecole
  Polytechnique},\ \bibinfo {year} {2002})\BibitemShut {NoStop}%
\bibitem [{\citenamefont {Hardy}\ \emph {et~al.}(1976)\citenamefont {Hardy},
  \citenamefont {de~Pazzis},\ and\ \citenamefont {Pomeau}}]{Hardy1976}%
  \BibitemOpen
  \bibfield  {author} {\bibinfo {author} {\bibfnamefont {J.}~\bibnamefont
  {Hardy}}, \bibinfo {author} {\bibfnamefont {O.}~\bibnamefont {de~Pazzis}}, \
  and\ \bibinfo {author} {\bibfnamefont {Y.}~\bibnamefont {Pomeau}},\
  }\bibfield  {title} {\bibinfo {title} {Molecular dynamics of a classical
  lattice gas: transport properties and time correlation functions},\ }\href
  {\doibase 10.1103/physreva.13.1949} {\bibfield  {journal} {\bibinfo
  {journal} {Phys. Rev. A}\ }\textbf {\bibinfo {volume} {13}},\ \bibinfo
  {pages} {1949--1961} (\bibinfo {year} {1976})}\BibitemShut {NoStop}%
\bibitem [{\citenamefont {Frisch}\ \emph {et~al.}(1986)\citenamefont {Frisch},
  \citenamefont {Hasslacher},\ and\ \citenamefont {Pomeau}}]{FHP86}%
  \BibitemOpen
  \bibfield  {author} {\bibinfo {author} {\bibfnamefont {U.}~\bibnamefont
  {Frisch}}, \bibinfo {author} {\bibfnamefont {B.}~\bibnamefont {Hasslacher}},
  \ and\ \bibinfo {author} {\bibfnamefont {Y.}~\bibnamefont {Pomeau}},\
  }\bibfield  {title} {\bibinfo {title} {Lattice-gas automata for the
  {N}avier-{S}tokes equation},\ }\href
  {https://doi.org/10.1103/physrevlett.56.1505} {\bibfield  {journal} {\bibinfo
   {journal} {Phys. Rev. Lett.}\ }\textbf {\bibinfo {volume} {56}},\ \bibinfo
  {pages} {1505--1508} (\bibinfo {year} {1986})}\BibitemShut {NoStop}%
\bibitem [{\citenamefont {Frisch}\ \emph {et~al.}(1987)\citenamefont {Frisch},
  \citenamefont {D'Humi\`{e}res}, \citenamefont {Hasslacher}, \citenamefont
  {Lallemand}, \citenamefont {Pomeau},\ and\ \citenamefont {Rivet}}]{FHHLPR87}%
  \BibitemOpen
  \bibfield  {author} {\bibinfo {author} {\bibfnamefont {U.}~\bibnamefont
  {Frisch}}, \bibinfo {author} {\bibfnamefont {D.}~\bibnamefont
  {D'Humi\`{e}res}}, \bibinfo {author} {\bibfnamefont {B.}~\bibnamefont
  {Hasslacher}}, \bibinfo {author} {\bibfnamefont {P.}~\bibnamefont
  {Lallemand}}, \bibinfo {author} {\bibfnamefont {Y.}~\bibnamefont {Pomeau}}, \
  and\ \bibinfo {author} {\bibfnamefont {J.-P.}\ \bibnamefont {Rivet}},\
  }\bibfield  {title} {\bibinfo {title} {Lattice gas hydrodynamics in two and
  three dimensions},\ }\href {http://www.complex-systems.com/pdf/01-4-7.pdf}
  {\bibfield  {journal} {\bibinfo  {journal} {Complex Systems}\ }\textbf
  {\bibinfo {volume} {1}},\ \bibinfo {pages} {649--707} (\bibinfo {year}
  {1987})}\BibitemShut {NoStop}%
\bibitem [{\citenamefont {Yepez}(2016)}]{Yepez2016}%
  \BibitemOpen
  \bibfield  {author} {\bibinfo {author} {\bibfnamefont {J.}~\bibnamefont
  {Yepez}},\ }\bibfield  {title} {\bibinfo {title} {Quantum lattice gas
  algorithmic representation of gauge field theory},\ }in\ \href {\doibase
  10.1117/12.2246702} {\emph {\bibinfo {booktitle} {Quantum Inf. Sci. Technol.
  {II}}}}\ (\bibinfo  {publisher} {{SPIE}-Intl Soc. Optical Eng.},\ \bibinfo
  {year} {2016})\BibitemShut {NoStop}%
\bibitem [{\citenamefont {D'Ariano}(2016)}]{DAriano2016}%
  \BibitemOpen
  \bibfield  {author} {\bibinfo {author} {\bibfnamefont {G.~M.}\ \bibnamefont
  {D'Ariano}},\ }\bibfield  {title} {\bibinfo {title} {Physics without
  physics},\ }\href {\doibase 10.1007/s10773-016-3172-y} {\bibfield  {journal}
  {\bibinfo  {journal} {Int. J. Theor. Phys.}\ }\textbf {\bibinfo {volume}
  {56}},\ \bibinfo {pages} {97--128} (\bibinfo {year} {2016})}\BibitemShut
  {NoStop}%
\bibitem [{\citenamefont {Arrighi}\ \emph
  {et~al.}(2014{\natexlab{a}})\citenamefont {Arrighi}, \citenamefont
  {Facchini},\ and\ \citenamefont {Forets}}]{Arrighi2014}%
  \BibitemOpen
  \bibfield  {author} {\bibinfo {author} {\bibfnamefont {P.}~\bibnamefont
  {Arrighi}}, \bibinfo {author} {\bibfnamefont {S.}~\bibnamefont {Facchini}}, \
  and\ \bibinfo {author} {\bibfnamefont {M.}~\bibnamefont {Forets}},\
  }\bibfield  {title} {\bibinfo {title} {Discrete {L}orentz covariance for
  quantum walks and quantum cellular automata},\ }\href
  {https://doi.org/10.1088%2F1367-2630%2F16%2F9%2F093007} {\bibfield  {journal}
  {\bibinfo  {journal} {New J. Phys.}\ }\textbf {\bibinfo {volume} {16}},\
  \bibinfo {pages} {093007} (\bibinfo {year} {2014}{\natexlab{a}})}\BibitemShut
  {NoStop}%
\bibitem [{\citenamefont {Grössing}\ and\ \citenamefont
  {Zeilinger}(1988)}]{Grossing_Zeilinger_88}%
  \BibitemOpen
  \bibfield  {author} {\bibinfo {author} {\bibfnamefont {G.}~\bibnamefont
  {Grössing}}\ and\ \bibinfo {author} {\bibfnamefont {A.}~\bibnamefont
  {Zeilinger}},\ }\bibfield  {title} {\bibinfo {title} {Quantum cellular
  automata},\ }\href
  {http://www.complex-systems.com/abstracts/v02_i02_a04.html} {\bibfield
  {journal} {\bibinfo  {journal} {Complex Sys.}\ }\textbf {\bibinfo {volume}
  {2}},\ \bibinfo {pages} {197--208} (\bibinfo {year} {1988})}\BibitemShut
  {NoStop}%
\bibitem [{\citenamefont {Huberman}(2017)}]{Huberman2017}%
  \BibitemOpen
  \bibfield  {author} {\bibinfo {author} {\bibfnamefont {S.}~\bibnamefont
  {Huberman}},\ }\bibfield  {title} {\bibinfo {title} {Two-particle quantum
  walks with topological phases},\ }\href {https://arxiv.org/abs/1702.00821}
  {\bibfield  {journal} {\bibinfo  {journal} {arXiv:1702.00821}\ } (\bibinfo
  {year} {2017})}\BibitemShut {NoStop}%
\bibitem [{\citenamefont {Berry}\ and\ \citenamefont {Wang}(2011)}]{Berry2011}%
  \BibitemOpen
  \bibfield  {author} {\bibinfo {author} {\bibfnamefont {S.~D.}\ \bibnamefont
  {Berry}}\ and\ \bibinfo {author} {\bibfnamefont {J.~B.}\ \bibnamefont
  {Wang}},\ }\bibfield  {title} {\bibinfo {title} {Two-particle quantum walks:
  entanglement and graph isomorphism testing},\ }\href
  {https://doi.org/10.1103%2Fphysreva.83.042317} {\bibfield  {journal}
  {\bibinfo  {journal} {Phys. Rev. A}\ }\textbf {\bibinfo {volume} {83}},\
  \bibinfo {pages} {042317} (\bibinfo {year} {2011})}\BibitemShut {NoStop}%
\bibitem [{\citenamefont {Carson}\ \emph {et~al.}(2015)\citenamefont {Carson},
  \citenamefont {Loke},\ and\ \citenamefont {Wang}}]{Carson2015}%
  \BibitemOpen
  \bibfield  {author} {\bibinfo {author} {\bibfnamefont {G.~R.}\ \bibnamefont
  {Carson}}, \bibinfo {author} {\bibfnamefont {T.}~\bibnamefont {Loke}}, \ and\
  \bibinfo {author} {\bibfnamefont {J.~B.}\ \bibnamefont {Wang}},\ }\bibfield
  {title} {\bibinfo {title} {Entanglement dynamics of two-particle quantum
  walks},\ }\href {\doibase 10.1007/s11128-015-1047-4} {\bibfield  {journal}
  {\bibinfo  {journal} {Quantum Inf. Process.}\ }\textbf {\bibinfo {volume}
  {14}},\ \bibinfo {pages} {3193--3210} (\bibinfo {year} {2015})}\BibitemShut
  {NoStop}%
\bibitem [{\citenamefont {Sansoni}\ \emph {et~al.}(2012)\citenamefont
  {Sansoni}, \citenamefont {Sciarrino}, \citenamefont {Vallone}, \citenamefont
  {Mataloni}, \citenamefont {Crespi}, \citenamefont {Ramponi},\ and\
  \citenamefont {Osellame}}]{Sansoni11a}%
  \BibitemOpen
  \bibfield  {author} {\bibinfo {author} {\bibfnamefont {L.}~\bibnamefont
  {Sansoni}}, \bibinfo {author} {\bibfnamefont {F.}~\bibnamefont {Sciarrino}},
  \bibinfo {author} {\bibfnamefont {G.}~\bibnamefont {Vallone}}, \bibinfo
  {author} {\bibfnamefont {P.}~\bibnamefont {Mataloni}}, \bibinfo {author}
  {\bibfnamefont {A.}~\bibnamefont {Crespi}}, \bibinfo {author} {\bibfnamefont
  {R.}~\bibnamefont {Ramponi}}, \ and\ \bibinfo {author} {\bibfnamefont
  {R.}~\bibnamefont {Osellame}},\ }\bibfield  {title} {\bibinfo {title}
  {Two-particle bosonic-fermionic quantum walk via 3{D} integrated photonics},\
  }\href {https://link.aps.org/doi/10.1103/PhysRevLett.108.010502} {\bibfield
  {journal} {\bibinfo  {journal} {Phys. Rev. Lett.}\ }\textbf {\bibinfo
  {volume} {108}},\ \bibinfo {pages} {010502} (\bibinfo {year}
  {2012})}\BibitemShut {NoStop}%
\bibitem [{\citenamefont {Mahan}(2000)}]{Mahan2000}%
  \BibitemOpen
  \bibfield  {author} {\bibinfo {author} {\bibfnamefont {G.~D.}\ \bibnamefont
  {Mahan}},\ }\href {\doibase 10.1007/978-1-4757-5714-9} {\emph {\bibinfo
  {title} {Many-Particle Physics}}}\ (\bibinfo  {publisher} {Springer},\
  \bibinfo {year} {2000})\BibitemShut {NoStop}%
\bibitem [{\citenamefont {Boettcher}\ \emph {et~al.}(2013)\citenamefont
  {Boettcher}, \citenamefont {Falkner},\ and\ \citenamefont
  {Portugal}}]{Boettcher2013}%
  \BibitemOpen
  \bibfield  {author} {\bibinfo {author} {\bibfnamefont {S.}~\bibnamefont
  {Boettcher}}, \bibinfo {author} {\bibfnamefont {S.}~\bibnamefont {Falkner}},
  \ and\ \bibinfo {author} {\bibfnamefont {R.}~\bibnamefont {Portugal}},\
  }\bibfield  {title} {\bibinfo {title} {Renormalization group for quantum
  walks},\ }\href {\doibase 10.1088/1742-6596/473/1/012018} {\bibfield
  {journal} {\bibinfo  {journal} {J. Phys. Conf. Ser.}\ }\textbf {\bibinfo
  {volume} {473}},\ \bibinfo {pages} {012018} (\bibinfo {year}
  {2013})}\BibitemShut {NoStop}%
\bibitem [{\citenamefont {Boettcher}\ \emph {et~al.}(2017)\citenamefont
  {Boettcher}, \citenamefont {Li},\ and\ \citenamefont
  {Portugal}}]{Boettcher2016}%
  \BibitemOpen
  \bibfield  {author} {\bibinfo {author} {\bibfnamefont {S.}~\bibnamefont
  {Boettcher}}, \bibinfo {author} {\bibfnamefont {S.}~\bibnamefont {Li}}, \
  and\ \bibinfo {author} {\bibfnamefont {R.}~\bibnamefont {Portugal}},\
  }\bibfield  {title} {\bibinfo {title} {Renormalization of the unitary
  evolution equation for coined quantum walks},\ }\href
  {http://stacks.iop.org/1751-8121/50/i=12/a=125302} {\bibfield  {journal}
  {\bibinfo  {journal} {J. Phys. A}\ }\textbf {\bibinfo {volume} {50}},\
  \bibinfo {pages} {125302} (\bibinfo {year} {2017})}\BibitemShut {NoStop}%
\bibitem [{\citenamefont {Arrighi}\ and\ \citenamefont
  {Grattage}(2009)}]{AG09}%
  \BibitemOpen
  \bibfield  {author} {\bibinfo {author} {\bibfnamefont {P.}~\bibnamefont
  {Arrighi}}\ and\ \bibinfo {author} {\bibfnamefont {J.}~\bibnamefont
  {Grattage}},\ }\bibfield  {title} {\bibinfo {title} {Intrinsically universal
  n-dimensional quantum cellular automata},\ }\href
  {https://arxiv.org/abs/0907.3827} {\bibfield  {journal} {\bibinfo  {journal}
  {arXiv:0907.3827}\ } (\bibinfo {year} {2009})}\BibitemShut {NoStop}%
\bibitem [{\citenamefont {Turing}(1937)}]{Turing1937}%
  \BibitemOpen
  \bibfield  {author} {\bibinfo {author} {\bibfnamefont {A.~M.}\ \bibnamefont
  {Turing}},\ }\bibfield  {title} {\bibinfo {title} {On computable numbers,
  with an application to the {E}ntscheidungsproblem},\ }\href
  {https://doi.org/10.1112/plms/s2-42.1.230} {\bibfield  {journal} {\bibinfo
  {journal} {Proceedings of the London Mathematical Society}\ }\textbf
  {\bibinfo {volume} {s2-42}},\ \bibinfo {pages} {230--265} (\bibinfo {year}
  {1937})}\BibitemShut {NoStop}%
\bibitem [{\citenamefont {Turing}(1938)}]{Turing1938}%
  \BibitemOpen
  \bibfield  {author} {\bibinfo {author} {\bibfnamefont {A.~M.}\ \bibnamefont
  {Turing}},\ }\bibfield  {title} {\bibinfo {title} {On computable numbers,
  with an application to the {E}ntscheidungsproblem; a correction},\ }\href
  {\doibase 10.1112/plms/s2-43.6.544} {\bibfield  {journal} {\bibinfo
  {journal} {Proceedings of the London Mathematical Society}\ }\textbf
  {\bibinfo {volume} {s2-43}},\ \bibinfo {pages} {544--546} (\bibinfo {year}
  {1938})}\BibitemShut {NoStop}%
\bibitem [{\citenamefont {Giulini}\ \emph {et~al.}(1996)\citenamefont
  {Giulini}, \citenamefont {Joos}, \citenamefont {Kiefer}, \citenamefont
  {Kupsch}, \citenamefont {Stamatescu},\ and\ \citenamefont {Zeh}}]{var96a}%
  \BibitemOpen
  \bibfield  {author} {\bibinfo {author} {\bibfnamefont {D.}~\bibnamefont
  {Giulini}}, \bibinfo {author} {\bibfnamefont {E.}~\bibnamefont {Joos}},
  \bibinfo {author} {\bibfnamefont {C.}~\bibnamefont {Kiefer}}, \bibinfo
  {author} {\bibfnamefont {J.}~\bibnamefont {Kupsch}}, \bibinfo {author}
  {\bibfnamefont {I.-O.}\ \bibnamefont {Stamatescu}}, \ and\ \bibinfo {author}
  {\bibfnamefont {H.~D.}\ \bibnamefont {Zeh}},\ }\href
  {http://www.springer.com/in/book/9783540003908} {\emph {\bibinfo {title}
  {Decoherence and the Appearance of a Classical World in Quantum Theory}}}\
  (\bibinfo  {publisher} {Springer},\ \bibinfo {year} {1996})\BibitemShut
  {NoStop}%
\bibitem [{\citenamefont {Schlosshauer}(2007)}]{book_Schlosshauer}%
  \BibitemOpen
  \bibfield  {author} {\bibinfo {author} {\bibfnamefont {M.~A.}\ \bibnamefont
  {Schlosshauer}},\ }\href {http://www.springer.com/gp/book/9783540357735}
  {\emph {\bibinfo {title} {Decoherence and the Quantum-To-Classical
  Transition}}}\ (\bibinfo  {publisher} {Springer},\ \bibinfo {year}
  {2007})\BibitemShut {NoStop}%
\bibitem [{\citenamefont {Arrighi}\ \emph
  {et~al.}(2014{\natexlab{b}})\citenamefont {Arrighi}, \citenamefont {Nesme},\
  and\ \citenamefont {Forets}}]{Arrighi_higher_dim_2014}%
  \BibitemOpen
  \bibfield  {author} {\bibinfo {author} {\bibfnamefont {P.}~\bibnamefont
  {Arrighi}}, \bibinfo {author} {\bibfnamefont {V.}~\bibnamefont {Nesme}}, \
  and\ \bibinfo {author} {\bibfnamefont {M.}~\bibnamefont {Forets}},\
  }\bibfield  {title} {\bibinfo {title} {The {D}irac equation as a quantum
  walk: higher dimensions, observational convergence},\ }\href
  {https://doi.org/10.1088/1751-8113/47/46/465302} {\bibfield  {journal}
  {\bibinfo  {journal} {J. Phys. A}\ }\textbf {\bibinfo {volume} {47}},\
  \bibinfo {pages} {465302} (\bibinfo {year} {2014}{\natexlab{b}})}\BibitemShut
  {NoStop}%
\bibitem [{\citenamefont {Tagliacozzo}\ \emph {et~al.}(2014)\citenamefont
  {Tagliacozzo}, \citenamefont {Celi},\ and\ \citenamefont
  {Lewenstein}}]{Tagliacozzo2014}%
  \BibitemOpen
  \bibfield  {author} {\bibinfo {author} {\bibfnamefont {L.}~\bibnamefont
  {Tagliacozzo}}, \bibinfo {author} {\bibfnamefont {A.}~\bibnamefont {Celi}}, \
  and\ \bibinfo {author} {\bibfnamefont {M.}~\bibnamefont {Lewenstein}},\
  }\bibfield  {title} {\bibinfo {title} {Tensor networks for lattice gauge
  theories with continuous groups},\ }\href
  {https://doi.org/10.1103%2Fphysrevx.4.041024} {\bibfield  {journal} {\bibinfo
   {journal} {Phys. Rev. X}\ }\textbf {\bibinfo {volume} {4}},\ \bibinfo
  {pages} {041024} (\bibinfo {year} {2014})}\BibitemShut {NoStop}%
\bibitem [{\citenamefont {Bru}\ \emph {et~al.}(2016{\natexlab{a}})\citenamefont
  {Bru}, \citenamefont {de~Valc{\'{a}}rcel}, \citenamefont {{Di Molfetta}},
  \citenamefont {P{\'{e}}rez}, \citenamefont {Rold{\'{a}}n},\ and\
  \citenamefont {Silva}}]{Bru2016b}%
  \BibitemOpen
  \bibfield  {author} {\bibinfo {author} {\bibfnamefont {L.~A.}\ \bibnamefont
  {Bru}}, \bibinfo {author} {\bibfnamefont {G.~J.}\ \bibnamefont
  {de~Valc{\'{a}}rcel}}, \bibinfo {author} {\bibfnamefont {G.}~\bibnamefont
  {{Di Molfetta}}}, \bibinfo {author} {\bibfnamefont {A.}~\bibnamefont
  {P{\'{e}}rez}}, \bibinfo {author} {\bibfnamefont {E.}~\bibnamefont
  {Rold{\'{a}}n}}, \ and\ \bibinfo {author} {\bibfnamefont {F.}~\bibnamefont
  {Silva}},\ }\bibfield  {title} {\bibinfo {title} {Quantum walk on a
  cylinder},\ }\href {https://doi.org/10.1103%2Fphysreva.94.032328} {\bibfield
  {journal} {\bibinfo  {journal} {Phys. Rev. A}\ }\textbf {\bibinfo {volume}
  {94}},\ \bibinfo {pages} {032328} (\bibinfo {year}
  {2016}{\natexlab{a}})}\BibitemShut {NoStop}%
\bibitem [{\citenamefont {{Di Molfetta}}\ and\ \citenamefont
  {P{\'{e}}rez}(2016)}]{Molfetta2016}%
  \BibitemOpen
  \bibfield  {author} {\bibinfo {author} {\bibfnamefont {G.}~\bibnamefont {{Di
  Molfetta}}}\ and\ \bibinfo {author} {\bibfnamefont {A.}~\bibnamefont
  {P{\'{e}}rez}},\ }\bibfield  {title} {\bibinfo {title} {Quantum walks as
  simulators of neutrino oscillations in a vacuum and matter},\ }\href
  {\doibase 10.1088/1367-2630/18/10/103038} {\bibfield  {journal} {\bibinfo
  {journal} {New J. Phys.}\ }\textbf {\bibinfo {volume} {18}},\ \bibinfo
  {pages} {103038} (\bibinfo {year} {2016})}\BibitemShut {NoStop}%
\bibitem [{\citenamefont {M{\'{a}}rquez-Mart{\'{\i}}n}\ \emph
  {et~al.}(2017)\citenamefont {M{\'{a}}rquez-Mart{\'{\i}}n}, \citenamefont {{Di
  Molfetta}},\ and\ \citenamefont {P{\'{e}}rez}}]{MrquezMartn2017}%
  \BibitemOpen
  \bibfield  {author} {\bibinfo {author} {\bibfnamefont {I.}~\bibnamefont
  {M{\'{a}}rquez-Mart{\'{\i}}n}}, \bibinfo {author} {\bibfnamefont
  {G.}~\bibnamefont {{Di Molfetta}}}, \ and\ \bibinfo {author} {\bibfnamefont
  {A.}~\bibnamefont {P{\'{e}}rez}},\ }\bibfield  {title} {\bibinfo {title}
  {Fermion confinement via quantum walks in (2+1)-dimensional and
  (3+1)-dimensional space-time},\ }\href {\doibase 10.1103/physreva.95.042112}
  {\bibfield  {journal} {\bibinfo  {journal} {Phys. Rev. A}\ }\textbf {\bibinfo
  {volume} {95}},\ \bibinfo {pages} {042112} (\bibinfo {year}
  {2017})}\BibitemShut {NoStop}%
\bibitem [{\citenamefont {Bru}\ \emph {et~al.}(2016{\natexlab{b}})\citenamefont
  {Bru}, \citenamefont {Hinarejos}, \citenamefont {Silva}, \citenamefont
  {de~Valc{\'{a}}rcel},\ and\ \citenamefont {Rold{\'{a}}n}}]{Bru2016}%
  \BibitemOpen
  \bibfield  {author} {\bibinfo {author} {\bibfnamefont {L.~A.}\ \bibnamefont
  {Bru}}, \bibinfo {author} {\bibfnamefont {M.}~\bibnamefont {Hinarejos}},
  \bibinfo {author} {\bibfnamefont {F.}~\bibnamefont {Silva}}, \bibinfo
  {author} {\bibfnamefont {G.~J.}\ \bibnamefont {de~Valc{\'{a}}rcel}}, \ and\
  \bibinfo {author} {\bibfnamefont {E.}~\bibnamefont {Rold{\'{a}}n}},\
  }\bibfield  {title} {\bibinfo {title} {Electric quantum walks in two
  dimensions},\ }\href {https://doi.org/10.1103%2Fphysreva.93.032333}
  {\bibfield  {journal} {\bibinfo  {journal} {Phys. Rev. A}\ }\textbf {\bibinfo
  {volume} {93}},\ \bibinfo {pages} {032333} (\bibinfo {year}
  {2016}{\natexlab{b}})}\BibitemShut {NoStop}%
\bibitem [{\citenamefont {Dalibard}\ \emph {et~al.}(2010)\citenamefont
  {Dalibard}, \citenamefont {Gerbier}, \citenamefont {Juzeli{\=u}nas},\ and\
  \citenamefont {{\"O}hberg}}]{dalibard10a}%
  \BibitemOpen
  \bibfield  {author} {\bibinfo {author} {\bibfnamefont {J.}~\bibnamefont
  {Dalibard}}, \bibinfo {author} {\bibfnamefont {F.}~\bibnamefont {Gerbier}},
  \bibinfo {author} {\bibfnamefont {G.}~\bibnamefont {Juzeli{\=u}nas}}, \ and\
  \bibinfo {author} {\bibfnamefont {P.}~\bibnamefont {{\"O}hberg}},\ }\bibfield
   {title} {\bibinfo {title} {Artificial gauge potential for neutral atoms},\
  }\href {https://link.aps.org/doi/10.1103/RevModPhys.83.1523} {\bibfield
  {journal} {\bibinfo  {journal} {{R}ev. Mod. Phys.}\ }\textbf {\bibinfo
  {volume} {83}},\ \bibinfo {pages} {1523} (\bibinfo {year}
  {2010})}\BibitemShut {NoStop}%
\bibitem [{\citenamefont {Goldman}\ and\ \citenamefont
  {Dalibard}(2014)}]{Goldman2014}%
  \BibitemOpen
  \bibfield  {author} {\bibinfo {author} {\bibfnamefont {N.}~\bibnamefont
  {Goldman}}\ and\ \bibinfo {author} {\bibfnamefont {J.}~\bibnamefont
  {Dalibard}},\ }\bibfield  {title} {\bibinfo {title} {Periodically driven
  quantum systems: effective {H}amiltonians and engineered gauge fields},\
  }\href {https://doi.org/10.1103%2Fphysrevx.4.031027} {\bibfield  {journal}
  {\bibinfo  {journal} {Phys. Rev. X}\ }\textbf {\bibinfo {volume} {4}},\
  \bibinfo {pages} {031027} (\bibinfo {year} {2014})}\BibitemShut {NoStop}%
\bibitem [{\citenamefont {Jackson}(1998)}]{book_Jackson}%
  \BibitemOpen
  \bibfield  {author} {\bibinfo {author} {\bibfnamefont {J.~D.}\ \bibnamefont
  {Jackson}},\ }\href
  {http://eu.wiley.com/WileyCDA/WileyTitle/productCd-047130932X.html} {\emph
  {\bibinfo {title} {Classical {E}lectrodynamics{,} 3rd {E}dition}}}\ (\bibinfo
   {publisher} {Wiley},\ \bibinfo {year} {1998})\BibitemShut {NoStop}%
\bibitem [{\citenamefont {Kolovsky}\ and\ \citenamefont
  {Mantica}(2010)}]{Kolovsky10}%
  \BibitemOpen
  \bibfield  {author} {\bibinfo {author} {\bibfnamefont {A.~R.}\ \bibnamefont
  {Kolovsky}}\ and\ \bibinfo {author} {\bibfnamefont {G.}~\bibnamefont
  {Mantica}},\ }\bibfield  {title} {\bibinfo {title} {Cyclotron-{B}loch
  dynamics of a quantum particle in a 2{D} lattice},\ }\href
  {https://arxiv.org/abs/1012.3041} {\bibfield  {journal} {\bibinfo  {journal}
  {arXiv:1012.3041}\ } (\bibinfo {year} {2010})}\BibitemShut {NoStop}%
\bibitem [{\citenamefont {Kolovsky}\ and\ \citenamefont
  {Mantica}(2014)}]{KolovskyMantica14}%
  \BibitemOpen
  \bibfield  {author} {\bibinfo {author} {\bibfnamefont {A.~R.}\ \bibnamefont
  {Kolovsky}}\ and\ \bibinfo {author} {\bibfnamefont {G.}~\bibnamefont
  {Mantica}},\ }\bibfield  {title} {\bibinfo {title} {Landau-{S}tark states and
  cyclotron-{B}loch oscillations of a quantum particle},\ }\href
  {https://arxiv.org/abs/1406.0276v1} {\bibfield  {journal} {\bibinfo
  {journal} {arXiv:1406.0276}\ } (\bibinfo {year} {2014})}\BibitemShut
  {NoStop}%
\bibitem [{\citenamefont {Hamilton}\ \emph {et~al.}(2014)\citenamefont
  {Hamilton}, \citenamefont {Kruse}, \citenamefont {Sansoni}, \citenamefont
  {Silberhorn},\ and\ \citenamefont {Jex}}]{Hamilton2014}%
  \BibitemOpen
  \bibfield  {author} {\bibinfo {author} {\bibfnamefont {C.~S.}\ \bibnamefont
  {Hamilton}}, \bibinfo {author} {\bibfnamefont {R.}~\bibnamefont {Kruse}},
  \bibinfo {author} {\bibfnamefont {L.}~\bibnamefont {Sansoni}}, \bibinfo
  {author} {\bibfnamefont {C.}~\bibnamefont {Silberhorn}}, \ and\ \bibinfo
  {author} {\bibfnamefont {I.}~\bibnamefont {Jex}},\ }\bibfield  {title}
  {\bibinfo {title} {Driven quantum walks},\ }\href
  {https://doi.org/10.1103%2Fphysrevlett.113.083602} {\bibfield  {journal}
  {\bibinfo  {journal} {Phys. Rev. Lett.}\ }\textbf {\bibinfo {volume} {113}},\
  \bibinfo {pages} {083602} (\bibinfo {year} {2014})}\BibitemShut {NoStop}%
\bibitem [{\citenamefont {Khemani}\ \emph {et~al.}(2016)\citenamefont
  {Khemani}, \citenamefont {Lazarides}, \citenamefont {Moessner},\ and\
  \citenamefont {Sondhi}}]{Khemani2016}%
  \BibitemOpen
  \bibfield  {author} {\bibinfo {author} {\bibfnamefont {V.}~\bibnamefont
  {Khemani}}, \bibinfo {author} {\bibfnamefont {A.}~\bibnamefont {Lazarides}},
  \bibinfo {author} {\bibfnamefont {R.}~\bibnamefont {Moessner}}, \ and\
  \bibinfo {author} {\bibfnamefont {S.~L.}\ \bibnamefont {Sondhi}},\ }\bibfield
   {title} {\bibinfo {title} {Phase structure of driven quantum systems},\
  }\href {\doibase 10.1103/physrevlett.116.250401} {\bibfield  {journal}
  {\bibinfo  {journal} {Phys. Rev. Lett.}\ }\textbf {\bibinfo {volume} {116}},\
  \bibinfo {pages} {250401} (\bibinfo {year} {2016})}\BibitemShut {NoStop}%
\bibitem [{\citenamefont {Genske}\ \emph {et~al.}(2013)\citenamefont {Genske},
  \citenamefont {Alt}, \citenamefont {Steffen}, \citenamefont {Werner},
  \citenamefont {Werner}, \citenamefont {Meschede},\ and\ \citenamefont
  {Alberti}}]{mesch13a}%
  \BibitemOpen
  \bibfield  {author} {\bibinfo {author} {\bibfnamefont {M.}~\bibnamefont
  {Genske}}, \bibinfo {author} {\bibfnamefont {W.}~\bibnamefont {Alt}},
  \bibinfo {author} {\bibfnamefont {A.}~\bibnamefont {Steffen}}, \bibinfo
  {author} {\bibfnamefont {A.~H.}\ \bibnamefont {Werner}}, \bibinfo {author}
  {\bibfnamefont {R.~F.}\ \bibnamefont {Werner}}, \bibinfo {author}
  {\bibfnamefont {D.}~\bibnamefont {Meschede}}, \ and\ \bibinfo {author}
  {\bibfnamefont {A.}~\bibnamefont {Alberti}},\ }\bibfield  {title} {\bibinfo
  {title} {Electric quantum walks with individual atoms},\ }\href
  {https://link.aps.org/doi/10.1103/PhysRevLett.110.190601} {\bibfield
  {journal} {\bibinfo  {journal} {Phys. {R}ev. Lett.}\ }\textbf {\bibinfo
  {volume} {110}},\ \bibinfo {pages} {190601} (\bibinfo {year}
  {2013})}\BibitemShut {NoStop}%
\bibitem [{\citenamefont {Cedzich}\ \emph {et~al.}(2013)\citenamefont
  {Cedzich}, \citenamefont {Ryb{\'a}r}, \citenamefont {Werner}, \citenamefont
  {Alberti}, \citenamefont {Genske},\ and\ \citenamefont {Werner}}]{ced13}%
  \BibitemOpen
  \bibfield  {author} {\bibinfo {author} {\bibfnamefont {C.}~\bibnamefont
  {Cedzich}}, \bibinfo {author} {\bibfnamefont {T.}~\bibnamefont {Ryb{\'a}r}},
  \bibinfo {author} {\bibfnamefont {A.~H.}\ \bibnamefont {Werner}}, \bibinfo
  {author} {\bibfnamefont {A.}~\bibnamefont {Alberti}}, \bibinfo {author}
  {\bibfnamefont {M.}~\bibnamefont {Genske}}, \ and\ \bibinfo {author}
  {\bibfnamefont {R.~F.}\ \bibnamefont {Werner}},\ }\bibfield  {title}
  {\bibinfo {title} {Propagation of quantum walks in electric fields},\ }\href
  {https://link.aps.org/doi/10.1103/PhysRevLett.111.160601} {\bibfield
  {journal} {\bibinfo  {journal} {Phys. Rev. Lett.}\ }\textbf {\bibinfo
  {volume} {111}},\ \bibinfo {pages} {160601} (\bibinfo {year}
  {2013})}\BibitemShut {NoStop}%
\bibitem [{\citenamefont {Yal{\c{c}}{\i}nkaya}\ and\ \citenamefont
  {Gedik}(2015)}]{Yalcinkaya2015}%
  \BibitemOpen
  \bibfield  {author} {\bibinfo {author} {\bibfnamefont {{\.{I}}.}~\bibnamefont
  {Yal{\c{c}}{\i}nkaya}}\ and\ \bibinfo {author} {\bibfnamefont
  {Z.}~\bibnamefont {Gedik}},\ }\bibfield  {title} {\bibinfo {title}
  {Two-dimensional quantum walk under artificial magnetic field},\ }\href
  {\doibase 10.1103/physreva.92.042324} {\bibfield  {journal} {\bibinfo
  {journal} {Phys. Rev. A}\ }\textbf {\bibinfo {volume} {92}},\ \bibinfo
  {pages} {042324} (\bibinfo {year} {2015})}\BibitemShut {NoStop}%
\bibitem [{\citenamefont {{Di Molfetta}}\ \emph {et~al.}(2014)\citenamefont
  {{Di Molfetta}}, \citenamefont {Debbasch},\ and\ \citenamefont
  {Brachet}}]{DMD14}%
  \BibitemOpen
  \bibfield  {author} {\bibinfo {author} {\bibfnamefont {G.}~\bibnamefont {{Di
  Molfetta}}}, \bibinfo {author} {\bibfnamefont {F.}~\bibnamefont {Debbasch}},
  \ and\ \bibinfo {author} {\bibfnamefont {M.}~\bibnamefont {Brachet}},\
  }\bibfield  {title} {\bibinfo {title} {Quantum walks in artificial electric
  and gravitational fields},\ }\href
  {http://www.sciencedirect.com/science/article/pii/S0378437113011059}
  {\bibfield  {journal} {\bibinfo  {journal} {Physica A}\ }\textbf {\bibinfo
  {volume} {397}},\ \bibinfo {pages} {157--168} (\bibinfo {year}
  {2014})}\BibitemShut {NoStop}%
\bibitem [{\citenamefont {{Di Molfetta}}\ \emph {et~al.}(2013)\citenamefont
  {{Di Molfetta}}, \citenamefont {Debbasch},\ and\ \citenamefont
  {Brachet}}]{DMD13b}%
  \BibitemOpen
  \bibfield  {author} {\bibinfo {author} {\bibfnamefont {G.}~\bibnamefont {{Di
  Molfetta}}}, \bibinfo {author} {\bibfnamefont {F.}~\bibnamefont {Debbasch}},
  \ and\ \bibinfo {author} {\bibfnamefont {M.}~\bibnamefont {Brachet}},\
  }\bibfield  {title} {\bibinfo {title} {Quantum walks as massless {D}irac
  fermions in curved space},\ }\href
  {https://link.aps.org/doi/10.1103/PhysRevA.88.042301} {\bibfield  {journal}
  {\bibinfo  {journal} {Phys. Rev. A}\ }\textbf {\bibinfo {volume} {88}}
  (\bibinfo {year} {2013})}\BibitemShut {NoStop}%
\bibitem [{\citenamefont {{Di Molfetta}}\ and\ \citenamefont
  {Debbasch}(2012)}]{DMD12a}%
  \BibitemOpen
  \bibfield  {author} {\bibinfo {author} {\bibfnamefont {G.}~\bibnamefont {{Di
  Molfetta}}}\ and\ \bibinfo {author} {\bibfnamefont {F.}~\bibnamefont
  {Debbasch}},\ }\bibfield  {title} {\bibinfo {title} {Discrete-time quantum
  walks: continuous limit and symmetries},\ }\href
  {https://doi.org/10.1063/1.4764876} {\bibfield  {journal} {\bibinfo
  {journal} {J. {M}ath. {P}hys.}\ }\textbf {\bibinfo {volume} {53}},\ \bibinfo
  {pages} {123302} (\bibinfo {year} {2012})}\BibitemShut {NoStop}%
\bibitem [{\citenamefont {Debbasch}\ \emph {et~al.}(2012)\citenamefont
  {Debbasch}, \citenamefont {{Di Molfetta}}, \citenamefont {Espaze},\ and\
  \citenamefont {Foulonneau}}]{DDMEF12a}%
  \BibitemOpen
  \bibfield  {author} {\bibinfo {author} {\bibfnamefont {F.}~\bibnamefont
  {Debbasch}}, \bibinfo {author} {\bibfnamefont {G.}~\bibnamefont {{Di
  Molfetta}}}, \bibinfo {author} {\bibfnamefont {D.}~\bibnamefont {Espaze}}, \
  and\ \bibinfo {author} {\bibfnamefont {V.}~\bibnamefont {Foulonneau}},\
  }\bibfield  {title} {\bibinfo {title} {Propagation in quantum walks and
  relativistic diffusions},\ }\href
  {https://doi.org/10.1088/0031-8949/2012/t151/014044} {\bibfield  {journal}
  {\bibinfo  {journal} {Phys. Scripta}\ }\textbf {\bibinfo {volume} {151}},\
  \bibinfo {pages} {014044} (\bibinfo {year} {2012})}\BibitemShut {NoStop}%
\bibitem [{\citenamefont {Venegas-Andraca}(2012)}]{Venegas_review}%
  \BibitemOpen
  \bibfield  {author} {\bibinfo {author} {\bibfnamefont {S.~E.}\ \bibnamefont
  {Venegas-Andraca}},\ }\bibfield  {title} {\bibinfo {title} {Quantum walks: a
  comprehensive review},\ }\href {http://dx.doi.org/10.1007/s11128-012-0432-5}
  {\bibfield  {journal} {\bibinfo  {journal} {Quantum Inf. Process.}\ }\textbf
  {\bibinfo {volume} {11}},\ \bibinfo {pages} {1015--1106} (\bibinfo {year}
  {2012})}\BibitemShut {NoStop}%
\bibitem [{\citenamefont {Patel}\ \emph {et~al.}(2005)\citenamefont {Patel},
  \citenamefont {Raghunathan},\ and\ \citenamefont {Rungta}}]{Patel2005}%
  \BibitemOpen
  \bibfield  {author} {\bibinfo {author} {\bibfnamefont {A.}~\bibnamefont
  {Patel}}, \bibinfo {author} {\bibfnamefont {K.~S.}\ \bibnamefont
  {Raghunathan}}, \ and\ \bibinfo {author} {\bibfnamefont {P.}~\bibnamefont
  {Rungta}},\ }\bibfield  {title} {\bibinfo {title} {Quantum random walks do
  not need a coin toss},\ }\href {https://doi.org/10.1103/physreva.71.032347}
  {\bibfield  {journal} {\bibinfo  {journal} {Phys. Rev. A}\ }\textbf {\bibinfo
  {volume} {71}} (\bibinfo {year} {2005})}\BibitemShut {NoStop}%
\bibitem [{\citenamefont {Weyl}(1929)}]{Weyl1929}%
  \BibitemOpen
  \bibfield  {author} {\bibinfo {author} {\bibfnamefont {H.}~\bibnamefont
  {Weyl}},\ }\bibfield  {title} {\bibinfo {title} {Gravitation and the
  electron},\ }\href {\doibase 10.1073/pnas.15.4.323} {\bibfield  {journal}
  {\bibinfo  {journal} {Proc. Natl. Acad. Sci. U.S.A.}\ }\textbf {\bibinfo
  {volume} {15}},\ \bibinfo {pages} {323--334} (\bibinfo {year}
  {1929})}\BibitemShut {NoStop}%
\bibitem [{\citenamefont {Feynman}\ and\ \citenamefont
  {Hibbs}(1965)}]{FeynHibbs65a}%
  \BibitemOpen
  \bibfield  {author} {\bibinfo {author} {\bibfnamefont {R.~P.}\ \bibnamefont
  {Feynman}}\ and\ \bibinfo {author} {\bibfnamefont {A.~R.}\ \bibnamefont
  {Hibbs}},\ }\href@noop {} {\emph {\bibinfo {title} {Quantum Mechanics and
  Path Integrals}}}\ (\bibinfo  {publisher} {International Series in Pure and
  Applied Physics, McGraw-Hill Book Company},\ \bibinfo {year}
  {1965})\BibitemShut {NoStop}%
\bibitem [{\citenamefont {Narlikar}(1972)}]{Narlikar71}%
  \BibitemOpen
  \bibfield  {author} {\bibinfo {author} {\bibfnamefont {J.~V.}\ \bibnamefont
  {Narlikar}},\ }\bibfield  {title} {\bibinfo {title} {Path amplitudes for
  {D}irac particles},\ }\href {http://hdl.handle.net/11007/1064} {\bibfield
  {journal} {\bibinfo  {journal} {J. Indian Math. Soc.}\ }\textbf {\bibinfo
  {volume} {36}},\ \bibinfo {pages} {9--32} (\bibinfo {year}
  {1972})}\BibitemShut {NoStop}%
\bibitem [{\citenamefont {Gersch}(1981)}]{Gersch1981}%
  \BibitemOpen
  \bibfield  {author} {\bibinfo {author} {\bibfnamefont {H.~A.}\ \bibnamefont
  {Gersch}},\ }\bibfield  {title} {\bibinfo {title} {Feynman's relativistic
  chessboard as an {I}sing model},\ }\href {\doibase 10.1007/bf00669436}
  {\bibfield  {journal} {\bibinfo  {journal} {Int. J. Theor. Phys.}\ }\textbf
  {\bibinfo {volume} {20}},\ \bibinfo {pages} {491--501} (\bibinfo {year}
  {1981})}\BibitemShut {NoStop}%
\bibitem [{\citenamefont {Jacobson}\ and\ \citenamefont
  {Schulman}(1984)}]{JS84}%
  \BibitemOpen
  \bibfield  {author} {\bibinfo {author} {\bibfnamefont {T.}~\bibnamefont
  {Jacobson}}\ and\ \bibinfo {author} {\bibfnamefont {L.~S.}\ \bibnamefont
  {Schulman}},\ }\bibfield  {title} {\bibinfo {title} {Quantum stochastics: the
  passage from a relativistic to a non-relativistic path integral},\ }\href
  {http://stacks.iop.org/0305-4470/17/i=2/a=023} {\bibfield  {journal}
  {\bibinfo  {journal} {J. Phys. A}\ }\textbf {\bibinfo {volume} {17}},\
  \bibinfo {pages} {375--383} (\bibinfo {year} {1984})}\BibitemShut {NoStop}%
\bibitem [{\citenamefont {Jacobson}(1985)}]{Jacobson1985review}%
  \BibitemOpen
  \bibfield  {author} {\bibinfo {author} {\bibfnamefont {T.}~\bibnamefont
  {Jacobson}},\ }\bibfield  {title} {\bibinfo {title} {Feynman's checkerboard
  and other games},\ }in\ \href {https://doi.org/10.1007%2F3-540-15213-x_88}
  {\emph {\bibinfo {booktitle} {Non-Linear Equations in Classical and Quantum
  Field Theory: Proceedings of a Seminar Series Held at DAPHE, Observatoire de
  Meudon, and LPTHE, Universit{\'e}Pierre et Marie Curie, Paris, Between
  October 1983 and October 1984}}}\ (\bibinfo  {publisher} {Springer Berlin
  Heidelberg},\ \bibinfo {year} {1985})\BibitemShut {NoStop}%
\bibitem [{\citenamefont {Kauffman}\ and\ \citenamefont {Noyes}(1996)}]{KN96}%
  \BibitemOpen
  \bibfield  {author} {\bibinfo {author} {\bibfnamefont {L.~H.}\ \bibnamefont
  {Kauffman}}\ and\ \bibinfo {author} {\bibfnamefont {H.~P.}\ \bibnamefont
  {Noyes}},\ }\bibfield  {title} {\bibinfo {title} {Discrete physics and the
  {D}irac equation},\ }\href
  {http://www.sciencedirect.com/science/article/pii/0375960196004367}
  {\bibfield  {journal} {\bibinfo  {journal} {Phys. Lett. A}\ }\textbf
  {\bibinfo {volume} {218}},\ \bibinfo {pages} {139--146} (\bibinfo {year}
  {1996})}\BibitemShut {NoStop}%
\bibitem [{\citenamefont {Kull}\ and\ \citenamefont {Treumann}(1999)}]{KT99}%
  \BibitemOpen
  \bibfield  {author} {\bibinfo {author} {\bibfnamefont {A.}~\bibnamefont
  {Kull}}\ and\ \bibinfo {author} {\bibfnamefont {R.~A.}\ \bibnamefont
  {Treumann}},\ }\bibfield  {title} {\bibinfo {title} {On the path integral of
  the relativistic electron},\ }\href
  {http://dx.doi.org/10.1023/A:1026637015146} {\bibfield  {journal} {\bibinfo
  {journal} {Int. J. Theor. Phys.}\ }\textbf {\bibinfo {volume} {38}},\
  \bibinfo {pages} {1423--1428} (\bibinfo {year} {1999})}\BibitemShut {NoStop}%
\bibitem [{\citenamefont {Earle}(2010)}]{E08}%
  \BibitemOpen
  \bibfield  {author} {\bibinfo {author} {\bibfnamefont {K.~A.}\ \bibnamefont
  {Earle}},\ }\bibfield  {title} {\bibinfo {title} {Notes on the {F}eynman
  checkerboard problem},\ }\href {https://arxiv.org/abs/1012.1564} {\bibfield
  {journal} {\bibinfo  {journal} {arXiv:1012.1564}\ } (\bibinfo {year}
  {2010})}\BibitemShut {NoStop}%
\bibitem [{\citenamefont {Jacobson}(1984)}]{J84}%
  \BibitemOpen
  \bibfield  {author} {\bibinfo {author} {\bibfnamefont {T.}~\bibnamefont
  {Jacobson}},\ }\bibfield  {title} {\bibinfo {title} {Spinor chain path
  integral for the {D}irac equation},\ }\href
  {http://stacks.iop.org/0305-4470/17/i=12/a=015} {\bibfield  {journal}
  {\bibinfo  {journal} {J. Phys. A}\ }\textbf {\bibinfo {volume} {17}},\
  \bibinfo {pages} {2433--2451} (\bibinfo {year} {1984})}\BibitemShut {NoStop}%
\bibitem [{\citenamefont {Foster}\ and\ \citenamefont
  {Jacobson}(2016)}]{Foster2016}%
  \BibitemOpen
  \bibfield  {author} {\bibinfo {author} {\bibfnamefont {B.~Z.}\ \bibnamefont
  {Foster}}\ and\ \bibinfo {author} {\bibfnamefont {T.}~\bibnamefont
  {Jacobson}},\ }\bibfield  {title} {\bibinfo {title} {Spin on a 4{D} {F}eynman
  checkerboard},\ }\href {\doibase 10.1007/s10773-016-3170-0} {\bibfield
  {journal} {\bibinfo  {journal} {Int. J. Theor. Phys.}\ }\textbf {\bibinfo
  {volume} {56}},\ \bibinfo {pages} {129--144} (\bibinfo {year}
  {2016})}\BibitemShut {NoStop}%
\bibitem [{\citenamefont {Ichinose}(2014)}]{Ichinose2014}%
  \BibitemOpen
  \bibfield  {author} {\bibinfo {author} {\bibfnamefont {W.}~\bibnamefont
  {Ichinose}},\ }\bibfield  {title} {\bibinfo {title} {On the {F}eynman path
  integral for the {D}irac equation in the general dimensional spacetime},\
  }\href {\doibase 10.1007/s00220-014-1997-x} {\bibfield  {journal} {\bibinfo
  {journal} {Commun. Math. Phys.}\ }\textbf {\bibinfo {volume} {329}},\
  \bibinfo {pages} {483--508} (\bibinfo {year} {2014})}\BibitemShut {NoStop}%
\bibitem [{\citenamefont {Feynman}(1985)}]{Feynman1985book}%
  \BibitemOpen
  \bibfield  {author} {\bibinfo {author} {\bibfnamefont {R.~P.}\ \bibnamefont
  {Feynman}},\ }\href {http://press.princeton.edu/titles/8169.html} {\emph
  {\bibinfo {title} {QED: The {S}trange {T}heory of {L}ight and {M}atter}}}\
  (\bibinfo  {publisher} {Princeton University Press},\ \bibinfo {year}
  {1985})\BibitemShut {NoStop}%
\bibitem [{\citenamefont {Ichinose}(2005)}]{Ichinose2005}%
  \BibitemOpen
  \bibfield  {author} {\bibinfo {author} {\bibfnamefont {T.}~\bibnamefont
  {Ichinose}},\ }\bibfield  {title} {\bibinfo {title} {Path integral for the
  radial {D}irac equation},\ }\href {\doibase 10.1063/1.1829374} {\bibfield
  {journal} {\bibinfo  {journal} {J. Math. Phys.}\ }\textbf {\bibinfo {volume}
  {46}},\ \bibinfo {pages} {022103} (\bibinfo {year} {2005})}\BibitemShut
  {NoStop}%
\bibitem [{\citenamefont {Rosen}(1983)}]{Rosen1983}%
  \BibitemOpen
  \bibfield  {author} {\bibinfo {author} {\bibfnamefont {G.}~\bibnamefont
  {Rosen}},\ }\bibfield  {title} {\bibinfo {title} {Feynman path summation for
  the {D}irac equation: an underlying one-dimensional aspect of relativistic
  particle motion},\ }\href {\doibase 10.1103/physreva.28.1139} {\bibfield
  {journal} {\bibinfo  {journal} {Phys. Rev. A}\ }\textbf {\bibinfo {volume}
  {28}},\ \bibinfo {pages} {1139--1140} (\bibinfo {year} {1983})}\BibitemShut
  {NoStop}%
\bibitem [{\citenamefont {Ord}\ and\ \citenamefont {Mckeon}(1993)}]{Ord1993}%
  \BibitemOpen
  \bibfield  {author} {\bibinfo {author} {\bibfnamefont {G.~N.}\ \bibnamefont
  {Ord}}\ and\ \bibinfo {author} {\bibfnamefont {D.~G.~C.}\ \bibnamefont
  {Mckeon}},\ }\bibfield  {title} {\bibinfo {title} {On the {D}irac equation in
  3 + 1 dimensions},\ }\href {\doibase 10.1006/aphy.1993.1022} {\bibfield
  {journal} {\bibinfo  {journal} {Ann. Physics}\ }\textbf {\bibinfo {volume}
  {222}},\ \bibinfo {pages} {244--253} (\bibinfo {year} {1993})}\BibitemShut
  {NoStop}%
\bibitem [{\citenamefont {{Smith Jr}}(1995)}]{SmithJr95}%
  \BibitemOpen
  \bibfield  {author} {\bibinfo {author} {\bibfnamefont {F.~D.}\ \bibnamefont
  {{Smith Jr}}},\ }\bibfield  {title} {\bibinfo {title} {Hyper{D}iamond
  {F}eynman checkerboard in 4-dimensional spacetime},\ }\href
  {https://arxiv.org/abs/quant-ph/9503015} {\bibfield  {journal} {\bibinfo
  {journal} {arXiv:quant-ph/9503015}\ } (\bibinfo {year} {1995})}\BibitemShut
  {NoStop}%
\bibitem [{\citenamefont {Riazanov}(1958)}]{Riazanov58}%
  \BibitemOpen
  \bibfield  {author} {\bibinfo {author} {\bibfnamefont {G.~V.}\ \bibnamefont
  {Riazanov}},\ }\bibfield  {title} {\bibinfo {title} {The {F}eynman path
  integral for the {D}irac equation},\ }\href
  {http://jetp.ac.ru/cgi-bin/dn/e_006_06_1107.pdf} {\bibfield  {journal}
  {\bibinfo  {journal} {J. Exp. Theor. Phys.}\ }\textbf {\bibinfo {volume}
  {33}},\ \bibinfo {pages} {1437--1444} (\bibinfo {year} {1958})}\BibitemShut
  {NoStop}%
\bibitem [{\citenamefont {Gaveau}\ \emph {et~al.}(1984)\citenamefont {Gaveau},
  \citenamefont {Jacobson}, \citenamefont {Kac},\ and\ \citenamefont
  {Schulman}}]{Gaveau1984}%
  \BibitemOpen
  \bibfield  {author} {\bibinfo {author} {\bibfnamefont {B.}~\bibnamefont
  {Gaveau}}, \bibinfo {author} {\bibfnamefont {T.}~\bibnamefont {Jacobson}},
  \bibinfo {author} {\bibfnamefont {M.}~\bibnamefont {Kac}}, \ and\ \bibinfo
  {author} {\bibfnamefont {L.~S.}\ \bibnamefont {Schulman}},\ }\bibfield
  {title} {\bibinfo {title} {Relativistic extension of the analogy between
  quantum mechanics and {B}rownian motion},\ }\href
  {https://doi.org/10.1103%2Fphysrevlett.53.419} {\bibfield  {journal}
  {\bibinfo  {journal} {Phys. Rev. Lett.}\ }\textbf {\bibinfo {volume} {53}},\
  \bibinfo {pages} {419--422} (\bibinfo {year} {1984})}\BibitemShut {NoStop}%
\bibitem [{\citenamefont {Karmanov}(1993)}]{Karmanov1993}%
  \BibitemOpen
  \bibfield  {author} {\bibinfo {author} {\bibfnamefont {V.~A.}\ \bibnamefont
  {Karmanov}},\ }\bibfield  {title} {\bibinfo {title} {On the derivation of the
  electron propagator from a random walk},\ }\href
  {https://doi.org/10.1016%2F0375-9601%2893%2990193-4} {\bibfield  {journal}
  {\bibinfo  {journal} {Phys. Lett. A}\ }\textbf {\bibinfo {volume} {174}},\
  \bibinfo {pages} {371--376} (\bibinfo {year} {1993})}\BibitemShut {NoStop}%
\bibitem [{\citenamefont {Kempe}(2003{\natexlab{a}})}]{Kempe_review}%
  \BibitemOpen
  \bibfield  {author} {\bibinfo {author} {\bibfnamefont {J.}~\bibnamefont
  {Kempe}},\ }\bibfield  {title} {\bibinfo {title} {Quantum random walks - an
  introductory overview},\ }\href {https://arxiv.org/abs/quant-ph/0303081}
  {\bibfield  {journal} {\bibinfo  {journal} {arXiv:quant-ph/0303081}\ }
  (\bibinfo {year} {2003}{\natexlab{a}})}\BibitemShut {NoStop}%
\bibitem [{\citenamefont {Schr\"{o}dinger}(1926)}]{Schrodinger1926}%
  \BibitemOpen
  \bibfield  {author} {\bibinfo {author} {\bibfnamefont {E.}~\bibnamefont
  {Schr\"{o}dinger}},\ }\bibfield  {title} {\bibinfo {title} {Quantisierung als
  {E}igenwertproblem},\ }\href {\doibase 10.1002/andp.19263840404} {\bibfield
  {journal} {\bibinfo  {journal} {Ann. Physik}\ }\textbf {\bibinfo {volume}
  {384}},\ \bibinfo {pages} {361--376} (\bibinfo {year} {1926})}\BibitemShut
  {NoStop}%
\bibitem [{\citenamefont {Kac}(1951)}]{kac1951}%
  \BibitemOpen
  \bibfield  {author} {\bibinfo {author} {\bibfnamefont {M.}~\bibnamefont
  {Kac}},\ }\bibfield  {title} {\bibinfo {title} {On some connections between
  probability theory and differential and integral equations},\ }in\ \href
  {http://projecteuclid.org/euclid.bsmsp/1200500229} {\emph {\bibinfo
  {booktitle} {Proceedings of the 2nd Berkeley Symposium on Mathematical
  Statistics and Probability}}}\ (\bibinfo  {publisher} {University of
  California Press},\ \bibinfo {year} {1951})\BibitemShut {NoStop}%
\bibitem [{\citenamefont {Fick}(1855{\natexlab{a}})}]{Fick1855Allemand}%
  \BibitemOpen
  \bibfield  {author} {\bibinfo {author} {\bibfnamefont {A.}~\bibnamefont
  {Fick}},\ }\bibfield  {title} {\bibinfo {title} {\"{U}ber diffusion},\ }\href
  {http://onlinelibrary.wiley.com/doi/10.1002/andp.18551700105/abstract}
  {\bibfield  {journal} {\bibinfo  {journal} {Ann. Physik und Chemie}\ }\textbf
  {\bibinfo {volume} {94}},\ \bibinfo {pages} {59--86} (\bibinfo {year}
  {1855}{\natexlab{a}})}\BibitemShut {NoStop}%
\bibitem [{\citenamefont {Fick}(1855{\natexlab{b}})}]{Fick1855}%
  \BibitemOpen
  \bibfield  {author} {\bibinfo {author} {\bibfnamefont {A.}~\bibnamefont
  {Fick}},\ }\bibfield  {title} {\bibinfo {title} {On liquid diffusion},\
  }\href {http://www.tandfonline.com/doi/abs/10.1080/14786445508641925}
  {\bibfield  {journal} {\bibinfo  {journal} {Philos. Mag. J. Sci.}\ }\textbf
  {\bibinfo {volume} {10}},\ \bibinfo {pages} {30--39} (\bibinfo {year}
  {1855}{\natexlab{b}})}\BibitemShut {NoStop}%
\bibitem [{\citenamefont {Einstein}(1905{\natexlab{a}})}]{Einstein1905}%
  \BibitemOpen
  \bibfield  {author} {\bibinfo {author} {\bibfnamefont {A.}~\bibnamefont
  {Einstein}},\ }\bibfield  {title} {\bibinfo {title} {\"{U}ber die von der
  molekularkinetischen {T}heorie der {W}\"{a}rme geforderte {B}ewegung von in
  ruhenden {F}l\"{u}ssigkeiten suspendierten {T}eilchen},\ }\href
  {https://doi.org/10.1002%2Fandp.19053220806} {\bibfield  {journal} {\bibinfo
  {journal} {Ann. Physik}\ }\textbf {\bibinfo {volume} {322}},\ \bibinfo
  {pages} {549--560} (\bibinfo {year} {1905}{\natexlab{a}})}\BibitemShut
  {NoStop}%
\bibitem [{\citenamefont {Wick}(1954)}]{Wick1954}%
  \BibitemOpen
  \bibfield  {author} {\bibinfo {author} {\bibfnamefont {G.~C.}\ \bibnamefont
  {Wick}},\ }\bibfield  {title} {\bibinfo {title} {Properties of
  {B}ethe-{S}alpeter wave functions},\ }\href
  {https://link.aps.org/doi/10.1103/PhysRev.96.1124} {\bibfield  {journal}
  {\bibinfo  {journal} {Phys. Rev.}\ }\textbf {\bibinfo {volume} {96}},\
  \bibinfo {pages} {1124--1134} (\bibinfo {year} {1954})}\BibitemShut {NoStop}%
\bibitem [{\citenamefont {Kac}(1974)}]{Kac74}%
  \BibitemOpen
  \bibfield  {author} {\bibinfo {author} {\bibfnamefont {M.}~\bibnamefont
  {Kac}},\ }\bibfield  {title} {\bibinfo {title} {A stochastic model to the
  telegrapher's equation},\ }\href
  {https://projecteuclid.org/euclid.rmjm/1250130879} {\bibfield  {journal}
  {\bibinfo  {journal} {Rocky Mountain J. Math.}\ }\textbf {\bibinfo {volume}
  {4}},\ \bibinfo {pages} {497--509} (\bibinfo {year} {1974})},\ \bibinfo
  {note} {reprinted from Magnolia Petroleum Company Colloquium Lectures in the
  Pure and Applied Sciences, No. 2, 1956, ``Some stochastic problems in physics
  and mathematics''}\BibitemShut {NoStop}%
\bibitem [{\citenamefont {Strauch}(2006{\natexlab{b}})}]{Strauch06a}%
  \BibitemOpen
  \bibfield  {author} {\bibinfo {author} {\bibfnamefont {F.~W.}\ \bibnamefont
  {Strauch}},\ }\bibfield  {title} {\bibinfo {title} {Relativistic quantum
  walks},\ }\href {https://link.aps.org/doi/10.1103/PhysRevA.73.054302}
  {\bibfield  {journal} {\bibinfo  {journal} {Phys. {R}ev. A}\ }\textbf
  {\bibinfo {volume} {73}},\ \bibinfo {pages} {054302} (\bibinfo {year}
  {2006}{\natexlab{b}})}\BibitemShut {NoStop}%
\bibitem [{\citenamefont {Wiener}(1923)}]{Wiener1923}%
  \BibitemOpen
  \bibfield  {author} {\bibinfo {author} {\bibfnamefont {N.}~\bibnamefont
  {Wiener}},\ }\bibfield  {title} {\bibinfo {title} {Differential space},\
  }\href {https://doi.org/10.1002/sapm192321131} {\bibfield  {journal}
  {\bibinfo  {journal} {J. Math. \& Phys.}\ }\textbf {\bibinfo {volume} {2}},\
  \bibinfo {pages} {132--174} (\bibinfo {year} {1923})}\BibitemShut {NoStop}%
\bibitem [{\citenamefont {Huepe}\ \emph {et~al.}(2000)\citenamefont {Huepe},
  \citenamefont {Brachet},\ and\ \citenamefont {Debbasch}}]{Huepe2000}%
  \BibitemOpen
  \bibfield  {author} {\bibinfo {author} {\bibfnamefont {C.}~\bibnamefont
  {Huepe}}, \bibinfo {author} {\bibfnamefont {M.-E.}\ \bibnamefont {Brachet}},
  \ and\ \bibinfo {author} {\bibfnamefont {F.}~\bibnamefont {Debbasch}},\
  }\bibfield  {title} {\bibinfo {title} {Generic inflationary and
  non-inflationary behavior in toy-cosmology},\ }\href
  {https://doi.org/10.1016%2Fs0167-2789%2800%2900032-4} {\bibfield  {journal}
  {\bibinfo  {journal} {Physica D}\ }\textbf {\bibinfo {volume} {144}},\
  \bibinfo {pages} {20--36} (\bibinfo {year} {2000})}\BibitemShut {NoStop}%
\bibitem [{\citenamefont {Feynman}(1948)}]{Feynman1948}%
  \BibitemOpen
  \bibfield  {author} {\bibinfo {author} {\bibfnamefont {R.~P.}\ \bibnamefont
  {Feynman}},\ }\bibfield  {title} {\bibinfo {title} {Space-time approach to
  non-relativistic quantum mechanics},\ }\href
  {https://doi.org/10.1103%2Frevmodphys.20.367} {\bibfield  {journal} {\bibinfo
   {journal} {Rev. Mod. Phys.}\ }\textbf {\bibinfo {volume} {20}},\ \bibinfo
  {pages} {367--387} (\bibinfo {year} {1948})}\BibitemShut {NoStop}%
\bibitem [{\citenamefont {Candlin}(1956)}]{Candlin1956}%
  \BibitemOpen
  \bibfield  {author} {\bibinfo {author} {\bibfnamefont {D.~J.}\ \bibnamefont
  {Candlin}},\ }\bibfield  {title} {\bibinfo {title} {On sums over trajectories
  for systems with {F}ermi statistics},\ }\href {\doibase 10.1007/bf02745446}
  {\bibfield  {journal} {\bibinfo  {journal} {Il Nuovo Cimento}\ }\textbf
  {\bibinfo {volume} {4}},\ \bibinfo {pages} {231--239} (\bibinfo {year}
  {1956})}\BibitemShut {NoStop}%
\bibitem [{\citenamefont {DeWitt}(1984)}]{DeWitt}%
  \BibitemOpen
  \bibfield  {author} {\bibinfo {author} {\bibfnamefont {B.}~\bibnamefont
  {DeWitt}},\ }\href@noop {} {\emph {\bibinfo {title} {Supermanifolds}}}\
  (\bibinfo  {publisher} {Cambridge University Press},\ \bibinfo {year}
  {1984})\BibitemShut {NoStop}%
\bibitem [{\citenamefont {Gaveau}\ and\ \citenamefont {Schulman}(1989)}]{GS89}%
  \BibitemOpen
  \bibfield  {author} {\bibinfo {author} {\bibfnamefont {B.}~\bibnamefont
  {Gaveau}}\ and\ \bibinfo {author} {\bibfnamefont {L.~S.}\ \bibnamefont
  {Schulman}},\ }\bibfield  {title} {\bibinfo {title} {Dirac-equation path
  integral: interpreting the {G}rassman variables},\ }\href
  {http://dx.doi.org/10.1007/BF02450232} {\bibfield  {journal} {\bibinfo
  {journal} {Il Nuovo Cimento D}\ }\textbf {\bibinfo {volume} {11}},\ \bibinfo
  {pages} {31} (\bibinfo {year} {1989})}\BibitemShut {NoStop}%
\bibitem [{\citenamefont {Neumann}\ and\ \citenamefont
  {Burks}(1966)}]{VonNeumann66}%
  \BibitemOpen
  \bibfield  {author} {\bibinfo {author} {\bibfnamefont {J.~V.}\ \bibnamefont
  {Neumann}}\ and\ \bibinfo {author} {\bibfnamefont {A.~W.}\ \bibnamefont
  {Burks}},\ }\href@noop {} {\emph {\bibinfo {title} {Theory of
  Self-Reproducing Automata}}}\ (\bibinfo  {publisher} {University of Illinois
  Press},\ \bibinfo {year} {1966})\BibitemShut {NoStop}%
\bibitem [{\citenamefont {Deutsch}(1985)}]{Deutsch1985}%
  \BibitemOpen
  \bibfield  {author} {\bibinfo {author} {\bibfnamefont {D.}~\bibnamefont
  {Deutsch}},\ }\bibfield  {title} {\bibinfo {title} {Quantum theory, the
  {C}hurch-{T}uring principle and the universal quantum computer},\ }\href
  {\doibase 10.1098/rspa.1985.0070} {\bibfield  {journal} {\bibinfo  {journal}
  {Proc. R. Soc. London A}\ }\textbf {\bibinfo {volume} {400}},\ \bibinfo
  {pages} {97--117} (\bibinfo {year} {1985})}\BibitemShut {NoStop}%
\bibitem [{\citenamefont {Aharonov}\ \emph {et~al.}(2001)\citenamefont
  {Aharonov}, \citenamefont {Ambainis}, \citenamefont {Kempe},\ and\
  \citenamefont {Vazirani}}]{Aharonov2001}%
  \BibitemOpen
  \bibfield  {author} {\bibinfo {author} {\bibfnamefont {D.}~\bibnamefont
  {Aharonov}}, \bibinfo {author} {\bibfnamefont {A.}~\bibnamefont {Ambainis}},
  \bibinfo {author} {\bibfnamefont {J.}~\bibnamefont {Kempe}}, \ and\ \bibinfo
  {author} {\bibfnamefont {U.}~\bibnamefont {Vazirani}},\ }\bibfield  {title}
  {\bibinfo {title} {Quantum walks on graphs},\ }in\ \href {\doibase
  10.1145/380752.380758} {\emph {\bibinfo {booktitle} {Proceedings of the 33rd
  annual {ACM} Symposium on Theory of Computing - {STOC}01}}}\ (\bibinfo {year}
  {2001})\BibitemShut {NoStop}%
\bibitem [{\citenamefont {Moore}\ and\ \citenamefont
  {Russell}(2002)}]{Moore2002}%
  \BibitemOpen
  \bibfield  {author} {\bibinfo {author} {\bibfnamefont {C.}~\bibnamefont
  {Moore}}\ and\ \bibinfo {author} {\bibfnamefont {A.}~\bibnamefont
  {Russell}},\ }\bibfield  {title} {\bibinfo {title} {Quantum walks on the
  hypercube},\ }in\ \href {\doibase 10.1007/3-540-45726-7_14} {\emph {\bibinfo
  {booktitle} {Randomization and Approximation Techniques in Computer
  Science}}}\ (\bibinfo  {publisher} {Springer Nature},\ \bibinfo {year}
  {2002})\ pp.\ \bibinfo {pages} {164--178}\BibitemShut {NoStop}%
\bibitem [{\citenamefont {Kempe}(2003{\natexlab{b}})}]{Kempe2003}%
  \BibitemOpen
  \bibfield  {author} {\bibinfo {author} {\bibfnamefont {J.}~\bibnamefont
  {Kempe}},\ }\bibfield  {title} {\bibinfo {title} {Discrete quantum walks hit
  exponentially faster},\ }in\ \href {\doibase 10.1007/978-3-540-45198-3_30}
  {\emph {\bibinfo {booktitle} {Approximation, Randomization, and Combinatorial
  Optimization. Algorithms and Techniques.}}}\ (\bibinfo  {publisher} {Springer
  Nature},\ \bibinfo {year} {2003})\ pp.\ \bibinfo {pages}
  {354--369}\BibitemShut {NoStop}%
\bibitem [{\citenamefont {Childs}\ \emph {et~al.}(2002)\citenamefont {Childs},
  \citenamefont {Farhi},\ and\ \citenamefont {Gutmann}}]{Childs2002}%
  \BibitemOpen
  \bibfield  {author} {\bibinfo {author} {\bibfnamefont {A.~M.}\ \bibnamefont
  {Childs}}, \bibinfo {author} {\bibfnamefont {E.}~\bibnamefont {Farhi}}, \
  and\ \bibinfo {author} {\bibfnamefont {S.}~\bibnamefont {Gutmann}},\ }\href
  {\doibase 10.1023/a:1019609420309} {\bibfield  {journal} {\bibinfo  {journal}
  {Quantum Inf. Proc.}\ }\textbf {\bibinfo {volume} {1}},\ \bibinfo {pages}
  {35--43} (\bibinfo {year} {2002})}\BibitemShut {NoStop}%
\bibitem [{\citenamefont {Mackay}\ \emph
  {et~al.}(2002{\natexlab{a}})\citenamefont {Mackay}, \citenamefont {Bartlett},
  \citenamefont {Stephenson},\ and\ \citenamefont {Sanders}}]{Mackay2002}%
  \BibitemOpen
  \bibfield  {author} {\bibinfo {author} {\bibfnamefont {T.~D.}\ \bibnamefont
  {Mackay}}, \bibinfo {author} {\bibfnamefont {S.~D.}\ \bibnamefont
  {Bartlett}}, \bibinfo {author} {\bibfnamefont {L.~T.}\ \bibnamefont
  {Stephenson}}, \ and\ \bibinfo {author} {\bibfnamefont {B.~C.}\ \bibnamefont
  {Sanders}},\ }\bibfield  {title} {\bibinfo {title} {Quantum walks in higher
  dimensions},\ }\href {\doibase 10.1088/0305-4470/35/12/304} {\bibfield
  {journal} {\bibinfo  {journal} {J. Phys. A}\ }\textbf {\bibinfo {volume}
  {35}},\ \bibinfo {pages} {2745--2753} (\bibinfo {year}
  {2002}{\natexlab{a}})}\BibitemShut {NoStop}%
\bibitem [{\citenamefont {Tregenna}\ \emph {et~al.}(2003)\citenamefont
  {Tregenna}, \citenamefont {Flanagan}, \citenamefont {Maile},\ and\
  \citenamefont {Kendon}}]{Tregenna2003}%
  \BibitemOpen
  \bibfield  {author} {\bibinfo {author} {\bibfnamefont {B.}~\bibnamefont
  {Tregenna}}, \bibinfo {author} {\bibfnamefont {W.}~\bibnamefont {Flanagan}},
  \bibinfo {author} {\bibfnamefont {R.}~\bibnamefont {Maile}}, \ and\ \bibinfo
  {author} {\bibfnamefont {V.}~\bibnamefont {Kendon}},\ }\bibfield  {title}
  {\bibinfo {title} {Controlling discrete quantum walks: coins and initial
  states},\ }\href {\doibase 10.1088/1367-2630/5/1/383} {\bibfield  {journal}
  {\bibinfo  {journal} {New J. Phys.}\ }\textbf {\bibinfo {volume} {5}},\
  \bibinfo {pages} {83--83} (\bibinfo {year} {2003})}\BibitemShut {NoStop}%
\bibitem [{\citenamefont {Giri}\ and\ \citenamefont {Korepin}(2016)}]{GK16}%
  \BibitemOpen
  \bibfield  {author} {\bibinfo {author} {\bibfnamefont {P.~R.}\ \bibnamefont
  {Giri}}\ and\ \bibinfo {author} {\bibfnamefont {V.~E.}\ \bibnamefont
  {Korepin}},\ }\bibfield  {title} {\bibinfo {title} {A review on quantum
  search algorithms},\ }\href@noop {} {\bibfield  {journal} {\bibinfo
  {journal} {arXiv:1602.02730v1}\ } (\bibinfo {year} {2016})}\BibitemShut
  {NoStop}%
\bibitem [{\citenamefont {Gordon}(1993)}]{Gordon1993}%
  \BibitemOpen
  \bibfield  {author} {\bibinfo {author} {\bibfnamefont {D.~M.}\ \bibnamefont
  {Gordon}},\ }\bibfield  {title} {\bibinfo {title} {Discrete logarithms in
  ${G}{F}(p)$ using the number field sieve},\ }\href {\doibase 10.1137/0406010}
  {\bibfield  {journal} {\bibinfo  {journal} {{SIAM} J. Discrete Math.}\
  }\textbf {\bibinfo {volume} {6}},\ \bibinfo {pages} {124--138} (\bibinfo
  {year} {1993})}\BibitemShut {NoStop}%
\bibitem [{\citenamefont {Bennett}\ \emph {et~al.}(1997)\citenamefont
  {Bennett}, \citenamefont {Bernstein}, \citenamefont {Brassard},\ and\
  \citenamefont {Vazirani}}]{Bennett1997}%
  \BibitemOpen
  \bibfield  {author} {\bibinfo {author} {\bibfnamefont {C.~H.}\ \bibnamefont
  {Bennett}}, \bibinfo {author} {\bibfnamefont {E.}~\bibnamefont {Bernstein}},
  \bibinfo {author} {\bibfnamefont {G.}~\bibnamefont {Brassard}}, \ and\
  \bibinfo {author} {\bibfnamefont {U.}~\bibnamefont {Vazirani}},\ }\bibfield
  {title} {\bibinfo {title} {Strengths and weaknesses of quantum computing},\
  }\href {\doibase 10.1137/s0097539796300933} {\bibfield  {journal} {\bibinfo
  {journal} {{SIAM} J. Comput.}\ }\textbf {\bibinfo {volume} {26}},\ \bibinfo
  {pages} {1510--1523} (\bibinfo {year} {1997})}\BibitemShut {NoStop}%
\bibitem [{\citenamefont {Boyer}\ \emph {et~al.}(1998)\citenamefont {Boyer},
  \citenamefont {Brassard}, \citenamefont {Hoeyer},\ and\ \citenamefont
  {Tapp}}]{Boyer1998}%
  \BibitemOpen
  \bibfield  {author} {\bibinfo {author} {\bibfnamefont {M.}~\bibnamefont
  {Boyer}}, \bibinfo {author} {\bibfnamefont {G.}~\bibnamefont {Brassard}},
  \bibinfo {author} {\bibfnamefont {P.}~\bibnamefont {Hoeyer}}, \ and\ \bibinfo
  {author} {\bibfnamefont {A.}~\bibnamefont {Tapp}},\ }\bibfield  {title}
  {\bibinfo {title} {Tight bounds on quantum searching},\ }\href
  {http://onlinelibrary.wiley.com/doi/10.1002/(SICI)1521-3978(199806)46:4/5%3C493::AID-PROP493%3E3.0.CO;2-P/abstract}
  {\bibfield  {journal} {\bibinfo  {journal} {Fortschr. Phys.}\ }\textbf
  {\bibinfo {volume} {46}},\ \bibinfo {pages} {493--505} (\bibinfo {year}
  {1998})}\BibitemShut {NoStop}%
\bibitem [{\citenamefont {Zalka}(1999)}]{Zalka1999}%
  \BibitemOpen
  \bibfield  {author} {\bibinfo {author} {\bibfnamefont {C.}~\bibnamefont
  {Zalka}},\ }\bibfield  {title} {\bibinfo {title} {Grover's quantum searching
  algorithm is optimal},\ }\href {\doibase 10.1103/physreva.60.2746} {\bibfield
   {journal} {\bibinfo  {journal} {Phys. Rev. A}\ }\textbf {\bibinfo {volume}
  {60}},\ \bibinfo {pages} {2746--2751} (\bibinfo {year} {1999})}\BibitemShut
  {NoStop}%
\bibitem [{\citenamefont {Shenvi}\ \emph {et~al.}(2003)\citenamefont {Shenvi},
  \citenamefont {Kempe},\ and\ \citenamefont {Whaley}}]{Shenvi2003}%
  \BibitemOpen
  \bibfield  {author} {\bibinfo {author} {\bibfnamefont {N.}~\bibnamefont
  {Shenvi}}, \bibinfo {author} {\bibfnamefont {J.}~\bibnamefont {Kempe}}, \
  and\ \bibinfo {author} {\bibfnamefont {K.~B.}\ \bibnamefont {Whaley}},\
  }\bibfield  {title} {\bibinfo {title} {Quantum random-walk search
  algorithm},\ }\href {\doibase 10.1103/physreva.67.052307} {\bibfield
  {journal} {\bibinfo  {journal} {Phys. Rev. A}\ }\textbf {\bibinfo {volume}
  {67}} (\bibinfo {year} {2003}),\ 10.1103/physreva.67.052307}\BibitemShut
  {NoStop}%
\bibitem [{\citenamefont {Aaronson}\ and\ \citenamefont
  {Ambainis}(2005)}]{Aaronson2005}%
  \BibitemOpen
  \bibfield  {author} {\bibinfo {author} {\bibfnamefont {S.}~\bibnamefont
  {Aaronson}}\ and\ \bibinfo {author} {\bibfnamefont {A.}~\bibnamefont
  {Ambainis}},\ }\bibfield  {title} {\bibinfo {title} {Quantum search of
  spatial regions},\ }\href {\doibase 10.4086/toc.2005.v001a004} {\bibfield
  {journal} {\bibinfo  {journal} {Theor. Comput.}\ }\textbf {\bibinfo {volume}
  {1}},\ \bibinfo {pages} {47--79} (\bibinfo {year} {2005})}\BibitemShut
  {NoStop}%
\bibitem [{\citenamefont {Benioff}(2002)}]{Benioff2002}%
  \BibitemOpen
  \bibfield  {author} {\bibinfo {author} {\bibfnamefont {P.}~\bibnamefont
  {Benioff}},\ }\href {\doibase 10.1090/conm/305/05212} {\bibinfo {title}
  {Space searches with a quantum robot},\ } (\bibinfo {year}
  {2002})\BibitemShut {NoStop}%
\bibitem [{\citenamefont {Szegedy}()}]{Szegedy}%
  \BibitemOpen
  \bibfield  {author} {\bibinfo {author} {\bibfnamefont {M.}~\bibnamefont
  {Szegedy}},\ }\bibfield  {title} {\bibinfo {title} {Quantum speed-up of
  markov chain based algorithms},\ }in\ \href {\doibase 10.1109/focs.2004.53}
  {\emph {\bibinfo {booktitle} {Proceedings of the 45th annual IEEE Symposium
  on Foundations of Computer Science}}}\ (\bibinfo  {publisher} {Institute of
  Electrical and Electronics Engineers ({IEEE})})\BibitemShut {NoStop}%
\bibitem [{\citenamefont {Magniez}\ \emph {et~al.}(2007)\citenamefont
  {Magniez}, \citenamefont {Nayak}, \citenamefont {Roland},\ and\ \citenamefont
  {Santha}}]{Magniez2007}%
  \BibitemOpen
  \bibfield  {author} {\bibinfo {author} {\bibfnamefont {F.}~\bibnamefont
  {Magniez}}, \bibinfo {author} {\bibfnamefont {A.}~\bibnamefont {Nayak}},
  \bibinfo {author} {\bibfnamefont {J.}~\bibnamefont {Roland}}, \ and\ \bibinfo
  {author} {\bibfnamefont {M.}~\bibnamefont {Santha}},\ }\bibfield  {title}
  {\bibinfo {title} {Search via quantum walk},\ }in\ \href {\doibase
  10.1145/1250790.1250874} {\emph {\bibinfo {booktitle} {Proceedings of the
  39th annual {ACM} Symposium on Theory of Computing - {STOC}07}}}\ (\bibinfo
  {publisher} {Association for Computing Machinery ({ACM})},\ \bibinfo {year}
  {2007})\BibitemShut {NoStop}%
\bibitem [{\citenamefont {Santha}(2008)}]{Santha08quantumwalk}%
  \BibitemOpen
  \bibfield  {author} {\bibinfo {author} {\bibfnamefont {M.}~\bibnamefont
  {Santha}},\ }\bibfield  {title} {\bibinfo {title} {Quantum walk based search
  algorithms},\ }in\ \href@noop {} {\emph {\bibinfo {booktitle} {In Proceedings
  of 5th TAMC}}}\ (\bibinfo {year} {2008})\ pp.\ \bibinfo {pages}
  {31--46}\BibitemShut {NoStop}%
\bibitem [{\citenamefont {Magniez}\ \emph {et~al.}(2011)\citenamefont
  {Magniez}, \citenamefont {Nayak}, \citenamefont {Roland},\ and\ \citenamefont
  {Santha}}]{Magniez2011}%
  \BibitemOpen
  \bibfield  {author} {\bibinfo {author} {\bibfnamefont {F.}~\bibnamefont
  {Magniez}}, \bibinfo {author} {\bibfnamefont {A.}~\bibnamefont {Nayak}},
  \bibinfo {author} {\bibfnamefont {J.}~\bibnamefont {Roland}}, \ and\ \bibinfo
  {author} {\bibfnamefont {M.}~\bibnamefont {Santha}},\ }\bibfield  {title}
  {\bibinfo {title} {Search via quantum walk},\ }\href {\doibase
  10.1137/090745854} {\bibfield  {journal} {\bibinfo  {journal} {{SIAM} J.
  Comput.}\ }\textbf {\bibinfo {volume} {40}},\ \bibinfo {pages} {142--164}
  (\bibinfo {year} {2011})}\BibitemShut {NoStop}%
\bibitem [{\citenamefont {Poto{\v{c}}ek}\ \emph {et~al.}(2009)\citenamefont
  {Poto{\v{c}}ek}, \citenamefont {G{\'{a}}bris}, \citenamefont {Kiss},\ and\
  \citenamefont {Jex}}]{Potoek2009}%
  \BibitemOpen
  \bibfield  {author} {\bibinfo {author} {\bibfnamefont {V.}~\bibnamefont
  {Poto{\v{c}}ek}}, \bibinfo {author} {\bibfnamefont {A.}~\bibnamefont
  {G{\'{a}}bris}}, \bibinfo {author} {\bibfnamefont {T.}~\bibnamefont {Kiss}},
  \ and\ \bibinfo {author} {\bibfnamefont {I.}~\bibnamefont {Jex}},\ }\bibfield
   {title} {\bibinfo {title} {Optimized quantum random-walk search algorithms
  on the hypercube},\ }\href {\doibase 10.1103/physreva.79.012325} {\bibfield
  {journal} {\bibinfo  {journal} {Phys. Rev. A}\ }\textbf {\bibinfo {volume}
  {79}},\ \bibinfo {pages} {012325} (\bibinfo {year} {2009})}\BibitemShut
  {NoStop}%
\bibitem [{\citenamefont {Ambainis}\ \emph {et~al.}(2013)\citenamefont
  {Ambainis}, \citenamefont {Ba{\v{c}}kurs}, \citenamefont {Nahimovs},
  \citenamefont {Ozols},\ and\ \citenamefont {Rivosh}}]{Ambainis2013}%
  \BibitemOpen
  \bibfield  {author} {\bibinfo {author} {\bibfnamefont {A.}~\bibnamefont
  {Ambainis}}, \bibinfo {author} {\bibfnamefont {A.}~\bibnamefont
  {Ba{\v{c}}kurs}}, \bibinfo {author} {\bibfnamefont {N.}~\bibnamefont
  {Nahimovs}}, \bibinfo {author} {\bibfnamefont {R.}~\bibnamefont {Ozols}}, \
  and\ \bibinfo {author} {\bibfnamefont {A.}~\bibnamefont {Rivosh}},\
  }\bibfield  {title} {\bibinfo {title} {Search by quantum walks on
  two-dimensional grid without amplitude amplification},\ }in\ \href {\doibase
  10.1007/978-3-642-35656-8_7} {\emph {\bibinfo {booktitle} {Theory of Quantum
  Computation, Communication, and Cryptography}}}\ (\bibinfo  {publisher}
  {Springer Nature},\ \bibinfo {year} {2013})\ pp.\ \bibinfo {pages}
  {87--97}\BibitemShut {NoStop}%
\bibitem [{\citenamefont {Tulsi}(2008)}]{Tulsi2008}%
  \BibitemOpen
  \bibfield  {author} {\bibinfo {author} {\bibfnamefont {A.}~\bibnamefont
  {Tulsi}},\ }\bibfield  {title} {\bibinfo {title} {Faster quantum-walk
  algorithm for the two-dimensional spatial search},\ }\href {\doibase
  10.1103/physreva.78.012310} {\bibfield  {journal} {\bibinfo  {journal} {Phys.
  Rev. A}\ }\textbf {\bibinfo {volume} {78}},\ \bibinfo {pages} {012310}
  (\bibinfo {year} {2008})}\BibitemShut {NoStop}%
\bibitem [{\citenamefont {Krovi}\ \emph {et~al.}(2010)\citenamefont {Krovi},
  \citenamefont {Magniez}, \citenamefont {Ozols},\ and\ \citenamefont
  {Roland}}]{Krovi2010}%
  \BibitemOpen
  \bibfield  {author} {\bibinfo {author} {\bibfnamefont {H.}~\bibnamefont
  {Krovi}}, \bibinfo {author} {\bibfnamefont {F.}~\bibnamefont {Magniez}},
  \bibinfo {author} {\bibfnamefont {M.}~\bibnamefont {Ozols}}, \ and\ \bibinfo
  {author} {\bibfnamefont {J.}~\bibnamefont {Roland}},\ }\bibfield  {title}
  {\bibinfo {title} {Finding is as easy as detecting for quantum walks},\ }in\
  \href {\doibase 10.1007/978-3-642-14165-2_46} {\emph {\bibinfo {booktitle}
  {Automata, Languages and Programming}}}\ (\bibinfo  {publisher} {Springer
  Nature},\ \bibinfo {year} {2010})\ pp.\ \bibinfo {pages}
  {540--551}\BibitemShut {NoStop}%
\bibitem [{\citenamefont {Childs}\ \emph {et~al.}(2003)\citenamefont {Childs},
  \citenamefont {Cleve}, \citenamefont {Deotto}, \citenamefont {Farhi},
  \citenamefont {Gutmann},\ and\ \citenamefont {Spielman}}]{Childs2003}%
  \BibitemOpen
  \bibfield  {author} {\bibinfo {author} {\bibfnamefont {A.~M.}\ \bibnamefont
  {Childs}}, \bibinfo {author} {\bibfnamefont {R.}~\bibnamefont {Cleve}},
  \bibinfo {author} {\bibfnamefont {E.}~\bibnamefont {Deotto}}, \bibinfo
  {author} {\bibfnamefont {E.}~\bibnamefont {Farhi}}, \bibinfo {author}
  {\bibfnamefont {S.}~\bibnamefont {Gutmann}}, \ and\ \bibinfo {author}
  {\bibfnamefont {D.~A.}\ \bibnamefont {Spielman}},\ }\bibfield  {title}
  {\bibinfo {title} {Exponential algorithmic speedup by a quantum walk},\ }in\
  \href {\doibase 10.1145/780542.780552} {\emph {\bibinfo {booktitle}
  {Proceedings of the 35th annual {ACM} Symposium on Theory of Computing -
  {STOC}03}}}\ (\bibinfo  {publisher} {Association for Computing Machinery
  ({ACM})},\ \bibinfo {year} {2003})\BibitemShut {NoStop}%
\bibitem [{\citenamefont {Childs}\ and\ \citenamefont
  {Goldstone}(2004)}]{Childs2004}%
  \BibitemOpen
  \bibfield  {author} {\bibinfo {author} {\bibfnamefont {A.~M.}\ \bibnamefont
  {Childs}}\ and\ \bibinfo {author} {\bibfnamefont {J.}~\bibnamefont
  {Goldstone}},\ }\bibfield  {title} {\bibinfo {title} {Spatial search by
  quantum walk},\ }\href {\doibase 10.1103/physreva.70.022314} {\bibfield
  {journal} {\bibinfo  {journal} {Phys. Rev. A}\ }\textbf {\bibinfo {volume}
  {70}},\ \bibinfo {pages} {022314} (\bibinfo {year} {2004})}\BibitemShut
  {NoStop}%
\bibitem [{\citenamefont {Childs}(2009{\natexlab{b}})}]{ChildsCD2009}%
  \BibitemOpen
  \bibfield  {author} {\bibinfo {author} {\bibfnamefont {A.~M.}\ \bibnamefont
  {Childs}},\ }\bibfield  {title} {\bibinfo {title} {On the relationship
  between continuous- and discrete-time quantum walk},\ }\href
  {https://link.springer.com/article/10.1007/s00220-009-0930-1} {\bibfield
  {journal} {\bibinfo  {journal} {Commun. Math. Phys.}\ }\textbf {\bibinfo
  {volume} {294}},\ \bibinfo {pages} {581--603} (\bibinfo {year}
  {2009}{\natexlab{b}})}\BibitemShut {NoStop}%
\bibitem [{\citenamefont {Lovett}\ \emph {et~al.}(2010)\citenamefont {Lovett},
  \citenamefont {Cooper}, \citenamefont {Everitt}, \citenamefont {Trevers},\
  and\ \citenamefont {Kendon}}]{Lovett2010}%
  \BibitemOpen
  \bibfield  {author} {\bibinfo {author} {\bibfnamefont {N.~B.}\ \bibnamefont
  {Lovett}}, \bibinfo {author} {\bibfnamefont {S.}~\bibnamefont {Cooper}},
  \bibinfo {author} {\bibfnamefont {M.}~\bibnamefont {Everitt}}, \bibinfo
  {author} {\bibfnamefont {M.}~\bibnamefont {Trevers}}, \ and\ \bibinfo
  {author} {\bibfnamefont {V.}~\bibnamefont {Kendon}},\ }\bibfield  {title}
  {\bibinfo {title} {Universal quantum computation using the discrete-time
  quantum walk},\ }\href {https://doi.org/10.1103/physreva.81.042330}
  {\bibfield  {journal} {\bibinfo  {journal} {Phys. Rev. A}\ }\textbf {\bibinfo
  {volume} {81}} (\bibinfo {year} {2010})}\BibitemShut {NoStop}%
\bibitem [{\citenamefont {Foulger}\ \emph {et~al.}(2015)\citenamefont
  {Foulger}, \citenamefont {Gnutzmann},\ and\ \citenamefont {Tanner}}]{FG2014}%
  \BibitemOpen
  \bibfield  {author} {\bibinfo {author} {\bibfnamefont {I.}~\bibnamefont
  {Foulger}}, \bibinfo {author} {\bibfnamefont {S.}~\bibnamefont {Gnutzmann}},
  \ and\ \bibinfo {author} {\bibfnamefont {G.}~\bibnamefont {Tanner}},\
  }\bibfield  {title} {\bibinfo {title} {Quantum walks and quantum search on
  graphene lattices},\ }\href {\doibase 10.1103/PhysRevA.91.062323} {\bibfield
  {journal} {\bibinfo  {journal} {Phys. Rev. A}\ }\textbf {\bibinfo {volume}
  {91}},\ \bibinfo {pages} {062323} (\bibinfo {year} {2015})}\BibitemShut
  {NoStop}%
\bibitem [{\citenamefont {Foulger}(2014)}]{Foulger_thesis_2014}%
  \BibitemOpen
  \bibfield  {author} {\bibinfo {author} {\bibfnamefont {I.}~\bibnamefont
  {Foulger}},\ }\emph {\bibinfo {title} {{Quantum walks and quantum search on
  graphene lattices}}},\ \href
  {http://eprints.nottingham.ac.uk/27717/1/IainFoulger_thesis.pdf} {Ph.D.
  thesis},\ \bibinfo  {school} {University of Nottingham} (\bibinfo {year}
  {2014})\BibitemShut {NoStop}%
\bibitem [{\citenamefont {Chakraborty}\ \emph {et~al.}(2016)\citenamefont
  {Chakraborty}, \citenamefont {Novo}, \citenamefont {Ambainis},\ and\
  \citenamefont {Omar}}]{Chakraborty2016}%
  \BibitemOpen
  \bibfield  {author} {\bibinfo {author} {\bibfnamefont {S.}~\bibnamefont
  {Chakraborty}}, \bibinfo {author} {\bibfnamefont {L.}~\bibnamefont {Novo}},
  \bibinfo {author} {\bibfnamefont {A.}~\bibnamefont {Ambainis}}, \ and\
  \bibinfo {author} {\bibfnamefont {Y.}~\bibnamefont {Omar}},\ }\bibfield
  {title} {\bibinfo {title} {Spatial search by quantum walk is optimal for
  almost all graphs},\ }\href {\doibase 10.1103/physrevlett.116.100501}
  {\bibfield  {journal} {\bibinfo  {journal} {Phys. Rev. Lett.}\ }\textbf
  {\bibinfo {volume} {116}},\ \bibinfo {pages} {100501} (\bibinfo {year}
  {2016})}\BibitemShut {NoStop}%
\bibitem [{\citenamefont {Wong}(2016)}]{Wong2016}%
  \BibitemOpen
  \bibfield  {author} {\bibinfo {author} {\bibfnamefont {T.~G.}\ \bibnamefont
  {Wong}},\ }\bibfield  {title} {\bibinfo {title} {Quantum walk search on
  {J}ohnson graphs},\ }\href {\doibase 10.1088/1751-8113/49/19/195303}
  {\bibfield  {journal} {\bibinfo  {journal} {J. Phys. A}\ }\textbf {\bibinfo
  {volume} {49}},\ \bibinfo {pages} {195303} (\bibinfo {year}
  {2016})}\BibitemShut {NoStop}%
\bibitem [{\citenamefont {Ambainis}(2004)}]{Ambainis04quantumwalk}%
  \BibitemOpen
  \bibfield  {author} {\bibinfo {author} {\bibfnamefont {A.}~\bibnamefont
  {Ambainis}},\ }\bibfield  {title} {\bibinfo {title} {Quantum walk algorithms
  for element distinctness},\ }in\ \href
  {http://ieeexplore.ieee.org/stamp/stamp.jsp?arnumber=1366221} {\emph
  {\bibinfo {booktitle} {Proceedings of the 45th annual IEEE Symposium on
  Foundations of Computer Science}}}\ (\bibinfo {year} {2004})\BibitemShut
  {NoStop}%
\bibitem [{\citenamefont {Ambainis}(2014)}]{AmbainisElementDistinctness14}%
  \BibitemOpen
  \bibfield  {author} {\bibinfo {author} {\bibfnamefont {A.}~\bibnamefont
  {Ambainis}},\ }\bibfield  {title} {\bibinfo {title} {Quantum walk algorithms
  for element distinctness},\ }\href {https://arxiv.org/abs/quant-ph/0311001}
  {\bibfield  {journal} {\bibinfo  {journal} {arXiv:quant-ph/0311001}\ }
  (\bibinfo {year} {2014})}\BibitemShut {NoStop}%
\bibitem [{\citenamefont {Frisch}(2008)}]{U08}%
  \BibitemOpen
  \bibfield  {author} {\bibinfo {author} {\bibfnamefont {U.}~\bibnamefont
  {Frisch}},\ }\bibfield  {title} {\bibinfo {title} {Translation of {L}eonhard
  {E}uler's: General principles of the motion of fluids},\ }\href
  {https://arxiv.org/abs/0802.2383} {\bibfield  {journal} {\bibinfo  {journal}
  {arXiv:0802.2383}\ } (\bibinfo {year} {2008})}\BibitemShut {NoStop}%
\bibitem [{\citenamefont {Laplace}(1816)}]{L1816}%
  \BibitemOpen
  \bibfield  {author} {\bibinfo {author} {\bibfnamefont {P.~S.}\ \bibnamefont
  {Laplace}},\ }\bibfield  {title} {\bibinfo {title} {Sur la vitesse du son
  dans l’air et dans l’eau},\ }\href
  {http://gallica.bnf.fr/ark:/12148/bpt6k77603s/f303} {\bibfield  {journal}
  {\bibinfo  {journal} {Ann. de Chimie et de Physique}\ }\textbf {\bibinfo
  {volume} {3}},\ \bibinfo {pages} {297--300} (\bibinfo {year}
  {1816})}\BibitemShut {NoStop}%
\bibitem [{\citenamefont {Christodoulou}\ and\ \citenamefont
  {Mia}(2013)}]{CM13}%
  \BibitemOpen
  \bibfield  {author} {\bibinfo {author} {\bibfnamefont {D.}~\bibnamefont
  {Christodoulou}}\ and\ \bibinfo {author} {\bibfnamefont {S.}~\bibnamefont
  {Mia}},\ }\bibfield  {title} {\bibinfo {title} {Compressible {F}low and
  {E}uler’s {E}quations},\ }\href {https://arxiv.org/abs/1212.2867}
  {\bibfield  {journal} {\bibinfo  {journal} {arXiv:1212.2867v2}\ } (\bibinfo
  {year} {2013})}\BibitemShut {NoStop}%
\bibitem [{\citenamefont {Clausius}(1865)}]{C1865}%
  \BibitemOpen
  \bibfield  {author} {\bibinfo {author} {\bibfnamefont {R.}~\bibnamefont
  {Clausius}},\ }\bibfield  {title} {\bibinfo {title} {Uber verschiedene
  f\"{u}r die {A}nwendung bequeme {F}ormen der {H}auptgleichungen der
  mechanischen {W}\"{a}rmatheorie},\ }\href
  {http://www.ngzh.ch/archiv/1865_10/10_1/10_3.pdf} {\bibfield  {journal}
  {\bibinfo  {journal} {Ann. Physik und Chemie}\ }\textbf {\bibinfo {volume}
  {125}},\ \bibinfo {pages} {353--400} (\bibinfo {year} {1865})}\BibitemShut
  {NoStop}%
\bibitem [{\citenamefont {Boltzmann}(1872)}]{Boltzmann1872}%
  \BibitemOpen
  \bibfield  {author} {\bibinfo {author} {\bibfnamefont {L.}~\bibnamefont
  {Boltzmann}},\ }\bibfield  {title} {\bibinfo {title} {Weitere {S}tudien
  \"{u}ber das {W}\"{a}rmegleichgewicht unter {G}asmolek\"{u}len},\ }\href
  {https://link.springer.com/chapter/10.1007%2F978-3-322-84986-1_3} {\bibfield
  {journal} {\bibinfo  {journal} {Sitzungsberichte der Kaiserlichen Akademie
  der Wissenschaften, Wien}\ }\textbf {\bibinfo {volume} {66}},\ \bibinfo
  {pages} {275--370} (\bibinfo {year} {1872})},\ \bibinfo {note} {{E}nglish
  translation by S. G. Brush in \emph{Kinetic Theory of Gases}, vol. 2, p. 88,
  Imperial College Press, 2003.}\BibitemShut {Stop}%
\bibitem [{\citenamefont {Boltzmann}(2012)}]{Boltzmann2012}%
  \BibitemOpen
  \bibfield  {author} {\bibinfo {author} {\bibfnamefont {L.}~\bibnamefont
  {Boltzmann}},\ }\href {http://store.doverpublications.com/0486684555.html}
  {\emph {\bibinfo {title} {Lectures on {G}as {T}heory}}},\ edited by\ \bibinfo
  {editor} {\bibfnamefont {C.}~\bibnamefont {corporation}}\ (\bibinfo
  {publisher} {Dover publications},\ \bibinfo {year} {2012})\ \bibinfo {note}
  {\hspace{-0.05cm}, reproduction of the {E}nglish translation by S. G. Brush,
  University of California Press, 1964.}\BibitemShut {Stop}%
\bibitem [{\citenamefont {Maxwell}(1890)}]{M1890}%
  \BibitemOpen
  \bibfield  {author} {\bibinfo {author} {\bibfnamefont {J.~C.}\ \bibnamefont
  {Maxwell}},\ }\href
  {http://strangebeautiful.com/other-texts/maxwell-scientificpapers-vol-ii-dover.pdf}
  {\emph {\bibinfo {title} {The Scientific Papers, Vol. 2}}}\ (\bibinfo
  {publisher} {Cambridge University Press},\ \bibinfo {year} {1890})\ \bibinfo
  {note} {\hspace{-0.1cm}, link towards a version published by Dover
  publications}\BibitemShut {NoStop}%
\bibitem [{\citenamefont {Enskog}(1917)}]{E17}%
  \BibitemOpen
  \bibfield  {author} {\bibinfo {author} {\bibfnamefont {D.}~\bibnamefont
  {Enskog}},\ }\emph {\bibinfo {title} {{K}inetische {T}heorie der
  {V}org\"{a}nge in m\"{a}ssig verd\"{u}nnten {G}asen}},\ \href@noop {} {Ph.D.
  thesis},\ \bibinfo  {school} {Uppsala University} (\bibinfo {year}
  {1917})\BibitemShut {NoStop}%
\bibitem [{\citenamefont {Chapman}\ and\ \citenamefont
  {Cowling}(1952)}]{CC52a}%
  \BibitemOpen
  \bibfield  {author} {\bibinfo {author} {\bibfnamefont {S.}~\bibnamefont
  {Chapman}}\ and\ \bibinfo {author} {\bibfnamefont {T.~G.}\ \bibnamefont
  {Cowling}},\ }\href
  {http://www.cambridge.org/catalogue/catalogue.asp?isbn=9780521075770} {\emph
  {\bibinfo {title} {The {M}athematical {T}heory of {N}on-{U}niform {G}ases}}}\
  (\bibinfo  {publisher} {Cambridge University Press},\ \bibinfo {year}
  {1952})\BibitemShut {NoStop}%
\bibitem [{\citenamefont {Saint-Raymond}(2009)}]{L09}%
  \BibitemOpen
  \bibfield  {author} {\bibinfo {author} {\bibfnamefont {L.}~\bibnamefont
  {Saint-Raymond}},\ }\href {http://www.springer.com/gb/book/9783540928461}
  {\emph {\bibinfo {title} {Hydrodynamic {L}imits of the {B}oltzmann
  {E}quation}}}\ (\bibinfo  {publisher} {Springer},\ \bibinfo {year}
  {2009})\BibitemShut {NoStop}%
\bibitem [{\citenamefont {Chow}\ \emph {et~al.}(2016)\citenamefont {Chow},
  \citenamefont {Fan},\ and\ \citenamefont {Yuen}}]{Chow2016}%
  \BibitemOpen
  \bibfield  {author} {\bibinfo {author} {\bibfnamefont {K.}~\bibnamefont
  {Chow}}, \bibinfo {author} {\bibfnamefont {E.}~\bibnamefont {Fan}}, \ and\
  \bibinfo {author} {\bibfnamefont {M.}~\bibnamefont {Yuen}},\ }\bibfield
  {title} {\bibinfo {title} {The analytical solutions for the $n$-dimensional
  damped compressible {E}uler equations},\ }\href {\doibase 10.1111/sapm.12154}
  {\bibfield  {journal} {\bibinfo  {journal} {Studies in Applied Mathematics}\
  }\textbf {\bibinfo {volume} {138}},\ \bibinfo {pages} {294--316} (\bibinfo
  {year} {2016})}\BibitemShut {NoStop}%
\bibitem [{\citenamefont {Hardy}\ \emph {et~al.}(1973)\citenamefont {Hardy},
  \citenamefont {Pomeau},\ and\ \citenamefont {de~Pazzis}}]{Hardy1973}%
  \BibitemOpen
  \bibfield  {author} {\bibinfo {author} {\bibfnamefont {J.}~\bibnamefont
  {Hardy}}, \bibinfo {author} {\bibfnamefont {Y.}~\bibnamefont {Pomeau}}, \
  and\ \bibinfo {author} {\bibfnamefont {O.}~\bibnamefont {de~Pazzis}},\
  }\bibfield  {title} {\bibinfo {title} {Time evolution of a two-dimensional
  classical lattice system},\ }\href {\doibase 10.1103/physrevlett.31.276}
  {\bibfield  {journal} {\bibinfo  {journal} {Phys. Rev. Lett.}\ }\textbf
  {\bibinfo {volume} {31}},\ \bibinfo {pages} {276--279} (\bibinfo {year}
  {1973})}\BibitemShut {NoStop}%
\bibitem [{\citenamefont {Harris}(1966)}]{Harris1966}%
  \BibitemOpen
  \bibfield  {author} {\bibinfo {author} {\bibfnamefont {S.}~\bibnamefont
  {Harris}},\ }\bibfield  {title} {\bibinfo {title} {Approach to equilibrium in
  a moderately dense discrete velocity gas},\ }\href
  {https://doi.org/10.1063%2F1.1761848} {\bibfield  {journal} {\bibinfo
  {journal} {Phys. Fluids}\ }\textbf {\bibinfo {volume} {9}},\ \bibinfo {pages}
  {1328--1332} (\bibinfo {year} {1966})}\BibitemShut {NoStop}%
\bibitem [{\citenamefont {Lakshmi}(1989)}]{RajLakshmi1989}%
  \BibitemOpen
  \bibfield  {author} {\bibinfo {author} {\bibfnamefont {M.~R.}\ \bibnamefont
  {Lakshmi}},\ }\bibfield  {title} {\bibinfo {title} {Cellular automaton fluids
  -- a review},\ }\href {\doibase 10.1007/bf02812024} {\bibfield  {journal}
  {\bibinfo  {journal} {Sadhana}\ }\textbf {\bibinfo {volume} {14}},\ \bibinfo
  {pages} {133--172} (\bibinfo {year} {1989})}\BibitemShut {NoStop}%
\bibitem [{\citenamefont {Succi}(2001)}]{succi2001lattice}%
  \BibitemOpen
  \bibfield  {author} {\bibinfo {author} {\bibfnamefont {S.}~\bibnamefont
  {Succi}},\ }\href {https://books.google.fr/books?id=OC0Sj\_xgnhAC} {\emph
  {\bibinfo {title} {The Lattice {B}oltzmann Equation: For Fluid Dynamics and
  Beyond}}},\ Numerical Mathematics and Scientific Computation\ (\bibinfo
  {publisher} {Clarendon Press},\ \bibinfo {year} {2001})\BibitemShut {NoStop}%
\bibitem [{\citenamefont {Succi}\ and\ \citenamefont
  {Benzi}(1993)}]{Succi1993}%
  \BibitemOpen
  \bibfield  {author} {\bibinfo {author} {\bibfnamefont {S.}~\bibnamefont
  {Succi}}\ and\ \bibinfo {author} {\bibfnamefont {R.}~\bibnamefont {Benzi}},\
  }\bibfield  {title} {\bibinfo {title} {Lattice {B}oltzmann equation for
  quantum mechanics},\ }\href {\doibase 10.1016/0167-2789(93)90096-j}
  {\bibfield  {journal} {\bibinfo  {journal} {Physica D}\ }\textbf {\bibinfo
  {volume} {69}},\ \bibinfo {pages} {327--332} (\bibinfo {year}
  {1993})}\BibitemShut {NoStop}%
\bibitem [{\citenamefont {Mackay}\ \emph
  {et~al.}(2002{\natexlab{b}})\citenamefont {Mackay}, \citenamefont {Bartlett},
  \citenamefont {Stephenson},\ and\ \citenamefont {Sanders}}]{MBSS2002}%
  \BibitemOpen
  \bibfield  {author} {\bibinfo {author} {\bibfnamefont {D.}~\bibnamefont
  {Mackay}}, \bibinfo {author} {\bibfnamefont {S.~D.}\ \bibnamefont
  {Bartlett}}, \bibinfo {author} {\bibfnamefont {L.~T.}\ \bibnamefont
  {Stephenson}}, \ and\ \bibinfo {author} {\bibfnamefont {B.~C.}\ \bibnamefont
  {Sanders}},\ }\bibfield  {title} {\bibinfo {title} {Quantum walks in higher
  dimensions},\ }\href {https://doi.org/10.1088/0305-4470/35/12/304} {\bibfield
   {journal} {\bibinfo  {journal} {J. Phys. A}\ }\textbf {\bibinfo {volume}
  {35}},\ \bibinfo {pages} {0305--4470} (\bibinfo {year}
  {2002}{\natexlab{b}})}\BibitemShut {NoStop}%
\bibitem [{\citenamefont {Wolfram}(2002)}]{Wolfram2002}%
  \BibitemOpen
  \bibfield  {author} {\bibinfo {author} {\bibfnamefont {S.}~\bibnamefont
  {Wolfram}},\ }\href {https://www.wolframscience.com} {\emph {\bibinfo {title}
  {A {N}ew {K}ind of {S}cience}}}\ (\bibinfo  {publisher} {Wolfram Media},\
  \bibinfo {year} {2002})\BibitemShut {NoStop}%
\bibitem [{\citenamefont {Wolfram}(1984)}]{Wolfram1984}%
  \BibitemOpen
  \bibfield  {author} {\bibinfo {author} {\bibfnamefont {S.}~\bibnamefont
  {Wolfram}},\ }\bibfield  {title} {\bibinfo {title} {Cellular automata as
  models of complexity},\ }\href {\doibase 10.1038/311419a0} {\bibfield
  {journal} {\bibinfo  {journal} {Nature}\ }\textbf {\bibinfo {volume} {311}},\
  \bibinfo {pages} {419--424} (\bibinfo {year} {1984})}\BibitemShut {NoStop}%
\bibitem [{\citenamefont {'t~Hooft}(1988)}]{Hooft1988}%
  \BibitemOpen
  \bibfield  {author} {\bibinfo {author} {\bibfnamefont {G.}~\bibnamefont
  {'t~Hooft}},\ }\bibfield  {title} {\bibinfo {title} {Equivalence relations
  between deterministic and quantum mechanical systems},\ }\href
  {https://doi.org/10.1007%2Fbf01011560} {\bibfield  {journal} {\bibinfo
  {journal} {J. Stat. Phys.}\ }\textbf {\bibinfo {volume} {53}},\ \bibinfo
  {pages} {323--344} (\bibinfo {year} {1988})}\BibitemShut {NoStop}%
\bibitem [{\citenamefont {'t~Hooft}(2014)}]{Hooft2014}%
  \BibitemOpen
  \bibfield  {author} {\bibinfo {author} {\bibfnamefont {G.}~\bibnamefont
  {'t~Hooft}},\ }\href {https://arxiv.org/abs/1405.1548} {\emph {\bibinfo
  {title} {The {C}ellular {A}utomaton {I}nterpretation of {Q}uantum
  {M}echanics}}}\ (\bibinfo  {publisher} {arXiv:1405.1548},\ \bibinfo {year}
  {2014})\BibitemShut {NoStop}%
\bibitem [{\citenamefont {Bialynicki-Birula}(1994)}]{BB94a}%
  \BibitemOpen
  \bibfield  {author} {\bibinfo {author} {\bibfnamefont {I.}~\bibnamefont
  {Bialynicki-Birula}},\ }\bibfield  {title} {\bibinfo {title} {Weyl, {D}irac,
  and {M}axwell equations on a lattice as unitary cellular automata},\ }\href
  {https://journals.aps.org/prd/pdf/10.1103/PhysRevD.49.6920} {\bibfield
  {journal} {\bibinfo  {journal} {Phys. Rev. D}\ }\textbf {\bibinfo {volume}
  {49}},\ \bibinfo {pages} {6920--6927} (\bibinfo {year} {1994})}\BibitemShut
  {NoStop}%
\bibitem [{\citenamefont {Ba{\~{n}}uls}\ \emph {et~al.}(2006)\citenamefont
  {Ba{\~{n}}uls}, \citenamefont {Navarrete}, \citenamefont {P{\'{e}}rez},
  \citenamefont {Rold{\'{a}}n},\ and\ \citenamefont {Soriano}}]{Bauls2006}%
  \BibitemOpen
  \bibfield  {author} {\bibinfo {author} {\bibfnamefont {M.~C.}\ \bibnamefont
  {Ba{\~{n}}uls}}, \bibinfo {author} {\bibfnamefont {C.}~\bibnamefont
  {Navarrete}}, \bibinfo {author} {\bibfnamefont {A.}~\bibnamefont
  {P{\'{e}}rez}}, \bibinfo {author} {\bibfnamefont {E.}~\bibnamefont
  {Rold{\'{a}}n}}, \ and\ \bibinfo {author} {\bibfnamefont {J.~C.}\
  \bibnamefont {Soriano}},\ }\bibfield  {title} {\bibinfo {title} {Quantum walk
  with a time-dependent coin},\ }\href {\doibase 10.1103/physreva.73.062304}
  {\bibfield  {journal} {\bibinfo  {journal} {Phys. Rev. A}\ }\textbf {\bibinfo
  {volume} {73}},\ \bibinfo {pages} {062304} (\bibinfo {year}
  {2006})}\BibitemShut {NoStop}%
\bibitem [{\citenamefont {Graf}\ and\ \citenamefont {Vogl}(1995)}]{Graf1995}%
  \BibitemOpen
  \bibfield  {author} {\bibinfo {author} {\bibfnamefont {M.}~\bibnamefont
  {Graf}}\ and\ \bibinfo {author} {\bibfnamefont {P.}~\bibnamefont {Vogl}},\
  }\bibfield  {title} {\bibinfo {title} {Electromagnetic fields and dielectric
  response in empirical tight-binding theory},\ }\href
  {https://doi.org/10.1103%2Fphysrevb.51.4940} {\bibfield  {journal} {\bibinfo
  {journal} {Phys. Rev. B}\ }\textbf {\bibinfo {volume} {51}},\ \bibinfo
  {pages} {4940--4949} (\bibinfo {year} {1995})}\BibitemShut {NoStop}%
\bibitem [{\citenamefont {Jaffe}\ and\ \citenamefont
  {Singh}(1987)}]{Jaffe1987}%
  \BibitemOpen
  \bibfield  {author} {\bibinfo {author} {\bibfnamefont {M.~D.}\ \bibnamefont
  {Jaffe}}\ and\ \bibinfo {author} {\bibfnamefont {J.}~\bibnamefont {Singh}},\
  }\bibfield  {title} {\bibinfo {title} {Inclusion of spin-orbit coupling into
  tight binding bandstructure calculations for bulk and superlattice
  semiconductors},\ }\href {\doibase 10.1016/0038-1098(87)91042-8} {\bibfield
  {journal} {\bibinfo  {journal} {Solid State Commun.}\ }\textbf {\bibinfo
  {volume} {62}},\ \bibinfo {pages} {399--402} (\bibinfo {year}
  {1987})}\BibitemShut {NoStop}%
\bibitem [{\citenamefont {Barreteau}\ \emph {et~al.}(2016)\citenamefont
  {Barreteau}, \citenamefont {Spanjaard},\ and\ \citenamefont
  {Desjonqu{\`{e}}res}}]{Barreteau2016}%
  \BibitemOpen
  \bibfield  {author} {\bibinfo {author} {\bibfnamefont {C.}~\bibnamefont
  {Barreteau}}, \bibinfo {author} {\bibfnamefont {D.}~\bibnamefont
  {Spanjaard}}, \ and\ \bibinfo {author} {\bibfnamefont {M.-C.}\ \bibnamefont
  {Desjonqu{\`{e}}res}},\ }\bibfield  {title} {\bibinfo {title} {An efficient
  magnetic tight-binding method for transition metals and alloys},\ }\href
  {\doibase 10.1016/j.crhy.2015.12.014} {\bibfield  {journal} {\bibinfo
  {journal} {Comptes Rendus Physique}\ }\textbf {\bibinfo {volume} {17}},\
  \bibinfo {pages} {406--429} (\bibinfo {year} {2016})}\BibitemShut {NoStop}%
\bibitem [{\citenamefont {Franco}\ \emph {et~al.}(2011)\citenamefont {Franco},
  \citenamefont {Gettrick},\ and\ \citenamefont {Busch}}]{DiFranco2011}%
  \BibitemOpen
  \bibfield  {author} {\bibinfo {author} {\bibfnamefont {C.~D.}\ \bibnamefont
  {Franco}}, \bibinfo {author} {\bibfnamefont {M.~M.}\ \bibnamefont
  {Gettrick}}, \ and\ \bibinfo {author} {\bibfnamefont {T.}~\bibnamefont
  {Busch}},\ }\bibfield  {title} {\bibinfo {title} {Mimicking the probability
  distribution of a two-dimensional {G}rover walk with a single-qubit coin},\
  }\href {https://doi.org/10.1103%2Fphysrevlett.106.080502} {\bibfield
  {journal} {\bibinfo  {journal} {Phys. Rev. Lett.}\ }\textbf {\bibinfo
  {volume} {106}} (\bibinfo {year} {2011})}\BibitemShut {NoStop}%
\bibitem [{\citenamefont {Yepez}(2002{\natexlab{a}})}]{Yepez2002Dirac}%
  \BibitemOpen
  \bibfield  {author} {\bibinfo {author} {\bibfnamefont {J.}~\bibnamefont
  {Yepez}},\ }\bibfield  {title} {\bibinfo {title} {An efficient and accurate
  quantum algorithm for the {D}irac equation},\ }\href
  {https://arxiv.org/abs/quant-ph/0210093} {\bibfield  {journal} {\bibinfo
  {journal} {arXiv:quant-ph/0210093}\ } (\bibinfo {year}
  {2002}{\natexlab{a}})}\BibitemShut {NoStop}%
\bibitem [{\citenamefont {Arrighi}\ and\ \citenamefont
  {Facchini}(2016)}]{AF16}%
  \BibitemOpen
  \bibfield  {author} {\bibinfo {author} {\bibfnamefont {P.}~\bibnamefont
  {Arrighi}}\ and\ \bibinfo {author} {\bibfnamefont {F.}~\bibnamefont
  {Facchini}},\ }\bibfield  {title} {\bibinfo {title} {Quantum walking in
  curved spacetime: (3+1) dimensions, and beyond},\ }\href
  {https://arxiv.org/abs/1609.00305} {\bibfield  {journal} {\bibinfo  {journal}
  {arXiv:1609.00305}\ } (\bibinfo {year} {2016})}\BibitemShut {NoStop}%
\bibitem [{\citenamefont {Yepez}\ and\ \citenamefont
  {Boghosian}(2002)}]{Yepez2002Schro}%
  \BibitemOpen
  \bibfield  {author} {\bibinfo {author} {\bibfnamefont {J.}~\bibnamefont
  {Yepez}}\ and\ \bibinfo {author} {\bibfnamefont {B.}~\bibnamefont
  {Boghosian}},\ }\bibfield  {title} {\bibinfo {title} {An efficient and
  accurate quantum lattice-gas model for the many-body {S}chr\"{o}dinger wave
  equation},\ }\href {https://doi.org/10.1016%2Fs0010-4655%2802%2900419-8}
  {\bibfield  {journal} {\bibinfo  {journal} {Comput. Phys. Commun.}\ }\textbf
  {\bibinfo {volume} {146}},\ \bibinfo {pages} {280--294} (\bibinfo {year}
  {2002})}\BibitemShut {NoStop}%
\bibitem [{\citenamefont {Yepez}(2002{\natexlab{b}})}]{Yepez2002Burgers}%
  \BibitemOpen
  \bibfield  {author} {\bibinfo {author} {\bibfnamefont {J.}~\bibnamefont
  {Yepez}},\ }\bibfield  {title} {\bibinfo {title} {Quantum lattice-gas model
  for the {B}urgers equation},\ }\href {\doibase 10.1023/a:1014514805610}
  {\bibfield  {journal} {\bibinfo  {journal} {J. Stat. Phys.}\ }\textbf
  {\bibinfo {volume} {107}},\ \bibinfo {pages} {203--224} (\bibinfo {year}
  {2002}{\natexlab{b}})}\BibitemShut {NoStop}%
\bibitem [{\citenamefont {Haugset}\ \emph {et~al.}(1993)\citenamefont
  {Haugset}, \citenamefont {Ruud},\ and\ \citenamefont {Ravndal}}]{HRR93}%
  \BibitemOpen
  \bibfield  {author} {\bibinfo {author} {\bibfnamefont {T.}~\bibnamefont
  {Haugset}}, \bibinfo {author} {\bibfnamefont {J.~A.}\ \bibnamefont {Ruud}}, \
  and\ \bibinfo {author} {\bibfnamefont {F.}~\bibnamefont {Ravndal}},\
  }\bibfield  {title} {\bibinfo {title} {Gauge invariance of {L}andau levels},\
  }\href {https://doi.org/10.1088/0031-8949/47/6/004} {\bibfield  {journal}
  {\bibinfo  {journal} {Phys. Scripta}\ }\textbf {\bibinfo {volume} {47}},\
  \bibinfo {pages} {715--719} (\bibinfo {year} {1993})}\BibitemShut {NoStop}%
\bibitem [{\citenamefont {Kolovsky}\ and\ \citenamefont
  {Korsch}(2003)}]{Kolovski03}%
  \BibitemOpen
  \bibfield  {author} {\bibinfo {author} {\bibfnamefont {A.~R.}\ \bibnamefont
  {Kolovsky}}\ and\ \bibinfo {author} {\bibfnamefont {H.~J.}\ \bibnamefont
  {Korsch}},\ }\bibfield  {title} {\bibinfo {title} {Bloch oscillations of cold
  atoms in two-dimensional optical lattices},\ }\href
  {https://doi.org/10.1103/physreva.67.063601} {\bibfield  {journal} {\bibinfo
  {journal} {Phys. Rev. A}\ }\textbf {\bibinfo {volume} {67}},\ \bibinfo
  {pages} {063601} (\bibinfo {year} {2003})}\BibitemShut {NoStop}%
\bibitem [{\citenamefont {Kolovsky}\ and\ \citenamefont
  {Korsch}(2004)}]{Kolovsky04}%
  \BibitemOpen
  \bibfield  {author} {\bibinfo {author} {\bibfnamefont {A.~R.}\ \bibnamefont
  {Kolovsky}}\ and\ \bibinfo {author} {\bibfnamefont {H.~J.}\ \bibnamefont
  {Korsch}},\ }\bibfield  {title} {\bibinfo {title} {Bloch oscillations of cold
  atoms in optical lattices},\ }\href
  {http://www.worldscientific.com/doi/abs/10.1142/S0217979204024483} {\bibfield
   {journal} {\bibinfo  {journal} {Int. J. Mod. Phys. B}\ }\textbf {\bibinfo
  {volume} {18}},\ \bibinfo {pages} {1235--1260} (\bibinfo {year}
  {2004})}\BibitemShut {NoStop}%
\bibitem [{\citenamefont {Yang}\ and\ \citenamefont {Mills}(1954)}]{Yang1954}%
  \BibitemOpen
  \bibfield  {author} {\bibinfo {author} {\bibfnamefont {C.~N.}\ \bibnamefont
  {Yang}}\ and\ \bibinfo {author} {\bibfnamefont {R.~L.}\ \bibnamefont
  {Mills}},\ }\bibfield  {title} {\bibinfo {title} {Conservation of isotopic
  spin and isotopic gauge invariance},\ }\href
  {https://doi.org/10.1103%2Fphysrev.96.191} {\bibfield  {journal} {\bibinfo
  {journal} {Phys. Rev.}\ }\textbf {\bibinfo {volume} {96}},\ \bibinfo {pages}
  {191--195} (\bibinfo {year} {1954})}\BibitemShut {NoStop}%
\bibitem [{\citenamefont {Boozer}(2011)}]{Boozer2011}%
  \BibitemOpen
  \bibfield  {author} {\bibinfo {author} {\bibfnamefont {A.~D.}\ \bibnamefont
  {Boozer}},\ }\bibfield  {title} {\bibinfo {title} {Classical {Y}ang-{M}ills
  theory},\ }\href {https://doi.org/10.1119%2F1.3606478} {\bibfield  {journal}
  {\bibinfo  {journal} {Am. J. Phys.}\ }\textbf {\bibinfo {volume} {79}},\
  \bibinfo {pages} {925--931} (\bibinfo {year} {2011})}\BibitemShut {NoStop}%
\bibitem [{\citenamefont {'t~Hooft}(1971{\natexlab{a}})}]{tHooft1971}%
  \BibitemOpen
  \bibfield  {author} {\bibinfo {author} {\bibfnamefont {G.}~\bibnamefont
  {'t~Hooft}},\ }\bibfield  {title} {\bibinfo {title} {Renormalization of
  massless {Y}ang-{M}ills fields},\ }\href
  {https://doi.org/10.1016%2F0550-3213%2871%2990395-6} {\bibfield  {journal}
  {\bibinfo  {journal} {Nucl. Phys. B}\ }\textbf {\bibinfo {volume} {33}},\
  \bibinfo {pages} {173--199} (\bibinfo {year}
  {1971}{\natexlab{a}})}\BibitemShut {NoStop}%
\bibitem [{\citenamefont {'t~Hooft}(1971{\natexlab{b}})}]{Hooft1971b}%
  \BibitemOpen
  \bibfield  {author} {\bibinfo {author} {\bibfnamefont {G.}~\bibnamefont
  {'t~Hooft}},\ }\bibfield  {title} {\bibinfo {title} {Renormalizable
  {L}agrangians for massive {Y}ang-{M}ills fields},\ }\href
  {https://doi.org/10.1016%2F0550-3213%2871%2990139-8} {\bibfield  {journal}
  {\bibinfo  {journal} {Nucl. Phys. B}\ }\textbf {\bibinfo {volume} {35}},\
  \bibinfo {pages} {167--188} (\bibinfo {year}
  {1971}{\natexlab{b}})}\BibitemShut {NoStop}%
\bibitem [{\citenamefont {Berche}\ and\ \citenamefont
  {Medina}(2012)}]{Berche2012}%
  \BibitemOpen
  \bibfield  {author} {\bibinfo {author} {\bibfnamefont {B.}~\bibnamefont
  {Berche}}\ and\ \bibinfo {author} {\bibfnamefont {E.}~\bibnamefont
  {Medina}},\ }\bibfield  {title} {\bibinfo {title} {Classical {Y}ang-{M}ills
  theory in condensed-matter physics},\ }\href
  {https://doi.org/10.1088%2F0143-0807%2F34%2F1%2F161} {\bibfield  {journal}
  {\bibinfo  {journal} {Europ. J. Phys.}\ }\textbf {\bibinfo {volume} {34}},\
  \bibinfo {pages} {161--180} (\bibinfo {year} {2012})}\BibitemShut {NoStop}%
\bibitem [{\citenamefont {Maxwell}(1865)}]{Maxwell1865}%
  \BibitemOpen
  \bibfield  {author} {\bibinfo {author} {\bibfnamefont {J.~C.}\ \bibnamefont
  {Maxwell}},\ }\bibfield  {title} {\bibinfo {title} {A dynamical theory of the
  electromagnetic field},\ }\href {https://doi.org/10.1098%2Frstl.1865.0008}
  {\bibfield  {journal} {\bibinfo  {journal} {Philos. Trans. R. Soc. London}\
  }\textbf {\bibinfo {volume} {155}},\ \bibinfo {pages} {459--512} (\bibinfo
  {year} {1865})}\BibitemShut {NoStop}%
\bibitem [{\citenamefont {Einstein}(1905{\natexlab{b}})}]{Einstein1905b}%
  \BibitemOpen
  \bibfield  {author} {\bibinfo {author} {\bibfnamefont {A.}~\bibnamefont
  {Einstein}},\ }\bibfield  {title} {\bibinfo {title} {Zur {E}lektrodynamik
  bewegter {K}{\"o}rper},\ }\href {\doibase 10.1002/andp.19053221004}
  {\bibfield  {journal} {\bibinfo  {journal} {Ann. Physik}\ }\textbf {\bibinfo
  {volume} {322}},\ \bibinfo {pages} {891--921} (\bibinfo {year}
  {1905}{\natexlab{b}})}\BibitemShut {NoStop}%
\bibitem [{\citenamefont {Rutherford}\ and\ \citenamefont
  {Soddy}(1902{\natexlab{a}})}]{Rutherford1902I}%
  \BibitemOpen
  \bibfield  {author} {\bibinfo {author} {\bibfnamefont {E.}~\bibnamefont
  {Rutherford}}\ and\ \bibinfo {author} {\bibfnamefont {F.}~\bibnamefont
  {Soddy}},\ }\bibfield  {title} {\bibinfo {title} {The radioactivity of
  thorium compounds. i. an investigation of the radioactive emanation},\ }\href
  {\doibase 10.1039/ct9028100321} {\bibfield  {journal} {\bibinfo  {journal}
  {Trans. J. Chem. Soc.}\ }\textbf {\bibinfo {volume} {81}},\ \bibinfo {pages}
  {321--350} (\bibinfo {year} {1902}{\natexlab{a}})}\BibitemShut {NoStop}%
\bibitem [{\citenamefont {Rutherford}\ and\ \citenamefont
  {Soddy}(1902{\natexlab{b}})}]{Rutherford1902II}%
  \BibitemOpen
  \bibfield  {author} {\bibinfo {author} {\bibfnamefont {E.}~\bibnamefont
  {Rutherford}}\ and\ \bibinfo {author} {\bibfnamefont {F.}~\bibnamefont
  {Soddy}},\ }\bibfield  {title} {\bibinfo {title} {The radioactivity of
  thorium compounds {II}. the cause and nature of radioactivity},\ }\href
  {\doibase 10.1039/ct9028100837} {\bibfield  {journal} {\bibinfo  {journal}
  {Trans. J. Chem. Soc.}\ }\textbf {\bibinfo {volume} {81}},\ \bibinfo {pages}
  {837--860} (\bibinfo {year} {1902}{\natexlab{b}})}\BibitemShut {NoStop}%
\bibitem [{\citenamefont {Rutherford}\ and\ \citenamefont
  {Soddy}(1902{\natexlab{c}})}]{Rutherford1903}%
  \BibitemOpen
  \bibfield  {author} {\bibinfo {author} {\bibfnamefont {E.}~\bibnamefont
  {Rutherford}}\ and\ \bibinfo {author} {\bibfnamefont {F.}~\bibnamefont
  {Soddy}},\ }\bibfield  {title} {\bibinfo {title} {Radioactive change},\
  }\href {https://doi.org/10.1080/14786440309462960} {\bibfield  {journal}
  {\bibinfo  {journal} {Philos. Mag.}\ }\textbf {\bibinfo {volume} {5}},\
  \bibinfo {pages} {576--591} (\bibinfo {year}
  {1902}{\natexlab{c}})}\BibitemShut {NoStop}%
\bibitem [{\citenamefont {Fermi}(1934)}]{Fermi1934Ger}%
  \BibitemOpen
  \bibfield  {author} {\bibinfo {author} {\bibfnamefont {E.}~\bibnamefont
  {Fermi}},\ }\bibfield  {title} {\bibinfo {title} {Versuch einer {T}heorie der
  $\beta$-{S}trahlen {I}},\ }\href {\doibase 10.1007/bf01351864} {\bibfield
  {journal} {\bibinfo  {journal} {Z. Physik}\ }\textbf {\bibinfo {volume}
  {88}},\ \bibinfo {pages} {161--177} (\bibinfo {year} {1934})}\BibitemShut
  {NoStop}%
\bibitem [{\citenamefont {Fermi}(1968)}]{Fermi1934Eng}%
  \BibitemOpen
  \bibfield  {author} {\bibinfo {author} {\bibfnamefont {E.}~\bibnamefont
  {Fermi}},\ }\bibfield  {title} {\bibinfo {title} {Fermi's theory of beta
  decay},\ }\href {\doibase 10.1119/1.1974382} {\bibfield  {journal} {\bibinfo
  {journal} {Am. J. Phys.}\ }\textbf {\bibinfo {volume} {36}},\ \bibinfo
  {pages} {1150--1160} (\bibinfo {year} {1968})},\ \bibinfo {note} {translation
  of Fermi's original paper, by F. L. Wilson}\BibitemShut {NoStop}%
\bibitem [{\citenamefont {Lee}\ and\ \citenamefont {Yang}(1956)}]{Lee1956}%
  \BibitemOpen
  \bibfield  {author} {\bibinfo {author} {\bibfnamefont {T.~D.}\ \bibnamefont
  {Lee}}\ and\ \bibinfo {author} {\bibfnamefont {C.~N.}\ \bibnamefont {Yang}},\
  }\bibfield  {title} {\bibinfo {title} {Question of parity conservation in
  weak interactions},\ }\href {\doibase 10.1103/physrev.104.254} {\bibfield
  {journal} {\bibinfo  {journal} {Phys. Rev.}\ }\textbf {\bibinfo {volume}
  {104}},\ \bibinfo {pages} {254--258} (\bibinfo {year} {1956})}\BibitemShut
  {NoStop}%
\bibitem [{\citenamefont {Wu}\ \emph {et~al.}(1957)\citenamefont {Wu},
  \citenamefont {Ambler}, \citenamefont {Hayward}, \citenamefont {Hoppes},\
  and\ \citenamefont {Hudson}}]{Wu1957}%
  \BibitemOpen
  \bibfield  {author} {\bibinfo {author} {\bibfnamefont {C.~S.}\ \bibnamefont
  {Wu}}, \bibinfo {author} {\bibfnamefont {E.}~\bibnamefont {Ambler}}, \bibinfo
  {author} {\bibfnamefont {R.~W.}\ \bibnamefont {Hayward}}, \bibinfo {author}
  {\bibfnamefont {D.~D.}\ \bibnamefont {Hoppes}}, \ and\ \bibinfo {author}
  {\bibfnamefont {R.~P.}\ \bibnamefont {Hudson}},\ }\bibfield  {title}
  {\bibinfo {title} {Experimental test of parity conservation in beta decay},\
  }\href {\doibase 10.1103/physrev.105.1413} {\bibfield  {journal} {\bibinfo
  {journal} {Phys. Rev.}\ }\textbf {\bibinfo {volume} {105}},\ \bibinfo {pages}
  {1413--1415} (\bibinfo {year} {1957})}\BibitemShut {NoStop}%
\bibitem [{\citenamefont {Sudarshan}\ and\ \citenamefont
  {Marshak}(1957)}]{Sudarshan1957}%
  \BibitemOpen
  \bibfield  {author} {\bibinfo {author} {\bibfnamefont {E.~C.~G.}\
  \bibnamefont {Sudarshan}}\ and\ \bibinfo {author} {\bibfnamefont {R.~E.}\
  \bibnamefont {Marshak}},\ }\bibfield  {title} {\bibinfo {title} {The nature
  of the four-fermion interaction},\ }in\ \href@noop {} {\emph {\bibinfo
  {booktitle} {Proceedings of the Padua-Venice Conference on Mesons and
  Recently Discovered Particles}}}\ (\bibinfo {year} {1957})\BibitemShut
  {NoStop}%
\bibitem [{\citenamefont {Sudarshan}\ and\ \citenamefont
  {Marshak}(1958)}]{Sudarshan1958}%
  \BibitemOpen
  \bibfield  {author} {\bibinfo {author} {\bibfnamefont {E.~C.~G.}\
  \bibnamefont {Sudarshan}}\ and\ \bibinfo {author} {\bibfnamefont {R.~E.}\
  \bibnamefont {Marshak}},\ }\bibfield  {title} {\bibinfo {title} {Chirality
  invariance and the universal {F}ermi interaction},\ }\href
  {https://doi.org/10.1103%2Fphysrev.109.1860.2} {\bibfield  {journal}
  {\bibinfo  {journal} {Phys. Rev.}\ }\textbf {\bibinfo {volume} {109}},\
  \bibinfo {pages} {1860--1862} (\bibinfo {year} {1958})}\BibitemShut {NoStop}%
\bibitem [{\citenamefont {Feynman}\ and\ \citenamefont
  {Gell-Mann}(1958)}]{Feynman1958}%
  \BibitemOpen
  \bibfield  {author} {\bibinfo {author} {\bibfnamefont {R.~P.}\ \bibnamefont
  {Feynman}}\ and\ \bibinfo {author} {\bibfnamefont {M.}~\bibnamefont
  {Gell-Mann}},\ }\bibfield  {title} {\bibinfo {title} {Theory of the fermi
  interaction},\ }\href {\doibase 10.1103/physrev.109.193} {\bibfield
  {journal} {\bibinfo  {journal} {Phys. Rev.}\ }\textbf {\bibinfo {volume}
  {109}},\ \bibinfo {pages} {193--198} (\bibinfo {year} {1958})}\BibitemShut
  {NoStop}%
\bibitem [{\citenamefont {Landau}(1957)}]{Landau1957}%
  \BibitemOpen
  \bibfield  {author} {\bibinfo {author} {\bibfnamefont {L.}~\bibnamefont
  {Landau}},\ }\bibfield  {title} {\bibinfo {title} {On the conservation laws
  for weak interactions},\ }\href {\doibase 10.1016/0029-5582(57)90061-5}
  {\bibfield  {journal} {\bibinfo  {journal} {Nucl. Phys.}\ }\textbf {\bibinfo
  {volume} {3}},\ \bibinfo {pages} {127--131} (\bibinfo {year}
  {1957})}\BibitemShut {NoStop}%
\bibitem [{\citenamefont {Christenson}\ \emph {et~al.}(1964)\citenamefont
  {Christenson}, \citenamefont {Cronin}, \citenamefont {Fitch},\ and\
  \citenamefont {Turlay}}]{Christenson1964}%
  \BibitemOpen
  \bibfield  {author} {\bibinfo {author} {\bibfnamefont {J.~H.}\ \bibnamefont
  {Christenson}}, \bibinfo {author} {\bibfnamefont {J.~W.}\ \bibnamefont
  {Cronin}}, \bibinfo {author} {\bibfnamefont {V.~L.}\ \bibnamefont {Fitch}}, \
  and\ \bibinfo {author} {\bibfnamefont {R.}~\bibnamefont {Turlay}},\
  }\bibfield  {title} {\bibinfo {title} {Evidence for the $2\pi$ decay of the
  ${K}_2^0$ meson},\ }\href {\doibase 10.1103/physrevlett.13.138} {\bibfield
  {journal} {\bibinfo  {journal} {Phys. Rev. Lett.}\ }\textbf {\bibinfo
  {volume} {13}},\ \bibinfo {pages} {138--140} (\bibinfo {year}
  {1964})}\BibitemShut {NoStop}%
\bibitem [{\citenamefont {Glashow}(1961)}]{Glashow1961}%
  \BibitemOpen
  \bibfield  {author} {\bibinfo {author} {\bibfnamefont {S.~L.}\ \bibnamefont
  {Glashow}},\ }\bibfield  {title} {\bibinfo {title} {Partial-symmetries of
  weak interactions},\ }\href {\doibase 10.1016/0029-5582(61)90469-2}
  {\bibfield  {journal} {\bibinfo  {journal} {Nucl. Phys.}\ }\textbf {\bibinfo
  {volume} {22}},\ \bibinfo {pages} {579--588} (\bibinfo {year}
  {1961})}\BibitemShut {NoStop}%
\bibitem [{\citenamefont {Nambu}(1960)}]{Nambu1960}%
  \BibitemOpen
  \bibfield  {author} {\bibinfo {author} {\bibfnamefont {Y.}~\bibnamefont
  {Nambu}},\ }\bibfield  {title} {\bibinfo {title} {Quasi-particles and gauge
  invariance in the theory of superconductivity},\ }\href
  {https://doi.org/10.1103%2Fphysrev.117.648} {\bibfield  {journal} {\bibinfo
  {journal} {Phys. Rev.}\ }\textbf {\bibinfo {volume} {117}},\ \bibinfo {pages}
  {648--663} (\bibinfo {year} {1960})}\BibitemShut {NoStop}%
\bibitem [{\citenamefont {Nambu}\ and\ \citenamefont
  {Jona-Lasinio}(1961)}]{Nambu1961}%
  \BibitemOpen
  \bibfield  {author} {\bibinfo {author} {\bibfnamefont {Y.}~\bibnamefont
  {Nambu}}\ and\ \bibinfo {author} {\bibfnamefont {G.}~\bibnamefont
  {Jona-Lasinio}},\ }\bibfield  {title} {\bibinfo {title} {Dynamical model of
  elementary particles based on an analogy with superconductivity {I}},\ }\href
  {\doibase 10.1103/physrev.122.345} {\bibfield  {journal} {\bibinfo  {journal}
  {Phys. Rev.}\ }\textbf {\bibinfo {volume} {122}},\ \bibinfo {pages}
  {345--358} (\bibinfo {year} {1961})}\BibitemShut {NoStop}%
\bibitem [{\citenamefont {Goldstone}(1961)}]{Goldstone1961}%
  \BibitemOpen
  \bibfield  {author} {\bibinfo {author} {\bibfnamefont {J.}~\bibnamefont
  {Goldstone}},\ }\bibfield  {title} {\bibinfo {title} {Field theories with
  ``superconductor" solutions},\ }\href {\doibase 10.1007/bf02812722}
  {\bibfield  {journal} {\bibinfo  {journal} {Il Nuovo Cimento}\ }\textbf
  {\bibinfo {volume} {19}},\ \bibinfo {pages} {154--164} (\bibinfo {year}
  {1961})}\BibitemShut {NoStop}%
\bibitem [{\citenamefont {Goldstone}\ \emph {et~al.}(1962)\citenamefont
  {Goldstone}, \citenamefont {Salam},\ and\ \citenamefont
  {Weinberg}}]{Goldstone1962}%
  \BibitemOpen
  \bibfield  {author} {\bibinfo {author} {\bibfnamefont {J.}~\bibnamefont
  {Goldstone}}, \bibinfo {author} {\bibfnamefont {A.}~\bibnamefont {Salam}}, \
  and\ \bibinfo {author} {\bibfnamefont {S.}~\bibnamefont {Weinberg}},\
  }\bibfield  {title} {\bibinfo {title} {Broken symmetries},\ }\href
  {https://doi.org/10.1103%2Fphysrev.127.965} {\bibfield  {journal} {\bibinfo
  {journal} {Phys. Rev.}\ }\textbf {\bibinfo {volume} {127}},\ \bibinfo {pages}
  {965--970} (\bibinfo {year} {1962})}\BibitemShut {NoStop}%
\bibitem [{\citenamefont {Anderson}(1963)}]{Anderson1963}%
  \BibitemOpen
  \bibfield  {author} {\bibinfo {author} {\bibfnamefont {P.~W.}\ \bibnamefont
  {Anderson}},\ }\bibfield  {title} {\bibinfo {title} {Plasmons, gauge
  invariance, and mass},\ }\href {\doibase 10.1103/physrev.130.439} {\bibfield
  {journal} {\bibinfo  {journal} {Phys. Rev.}\ }\textbf {\bibinfo {volume}
  {130}},\ \bibinfo {pages} {439--442} (\bibinfo {year} {1963})}\BibitemShut
  {NoStop}%
\bibitem [{\citenamefont {Englert}\ and\ \citenamefont
  {Brout}(1964)}]{Englert1964}%
  \BibitemOpen
  \bibfield  {author} {\bibinfo {author} {\bibfnamefont {F.}~\bibnamefont
  {Englert}}\ and\ \bibinfo {author} {\bibfnamefont {R.}~\bibnamefont
  {Brout}},\ }\bibfield  {title} {\bibinfo {title} {Broken symmetry and the
  mass of gauge vector mesons},\ }\href {\doibase 10.1103/physrevlett.13.321}
  {\bibfield  {journal} {\bibinfo  {journal} {Phys. Rev. Lett.}\ }\textbf
  {\bibinfo {volume} {13}},\ \bibinfo {pages} {321--323} (\bibinfo {year}
  {1964})}\BibitemShut {NoStop}%
\bibitem [{\citenamefont {Higgs}(1964)}]{Higgs1964}%
  \BibitemOpen
  \bibfield  {author} {\bibinfo {author} {\bibfnamefont {P.~W.}\ \bibnamefont
  {Higgs}},\ }\bibfield  {title} {\bibinfo {title} {Broken symmetries and the
  masses of gauge bosons},\ }\href {\doibase 10.1103/physrevlett.13.508}
  {\bibfield  {journal} {\bibinfo  {journal} {Phys. Rev. Lett.}\ }\textbf
  {\bibinfo {volume} {13}},\ \bibinfo {pages} {508--509} (\bibinfo {year}
  {1964})}\BibitemShut {NoStop}%
\bibitem [{\citenamefont {Guralnik}\ \emph {et~al.}(1964)\citenamefont
  {Guralnik}, \citenamefont {Hagen},\ and\ \citenamefont
  {Kibble}}]{Guralnik1964}%
  \BibitemOpen
  \bibfield  {author} {\bibinfo {author} {\bibfnamefont {G.~S.}\ \bibnamefont
  {Guralnik}}, \bibinfo {author} {\bibfnamefont {C.~R.}\ \bibnamefont {Hagen}},
  \ and\ \bibinfo {author} {\bibfnamefont {T.~W.~B.}\ \bibnamefont {Kibble}},\
  }\bibfield  {title} {\bibinfo {title} {Global conservation laws and massless
  particles},\ }\href {\doibase 10.1103/physrevlett.13.585} {\bibfield
  {journal} {\bibinfo  {journal} {Phys. Rev. Lett.}\ }\textbf {\bibinfo
  {volume} {13}},\ \bibinfo {pages} {585--587} (\bibinfo {year}
  {1964})}\BibitemShut {NoStop}%
\bibitem [{\citenamefont {Weinberg}(1967)}]{Weinberg1967}%
  \BibitemOpen
  \bibfield  {author} {\bibinfo {author} {\bibfnamefont {S.}~\bibnamefont
  {Weinberg}},\ }\bibfield  {title} {\bibinfo {title} {A model of leptons},\
  }\href {\doibase 10.1103/physrevlett.19.1264} {\bibfield  {journal} {\bibinfo
   {journal} {Phys. Rev. Lett.}\ }\textbf {\bibinfo {volume} {19}},\ \bibinfo
  {pages} {1264--1266} (\bibinfo {year} {1967})}\BibitemShut {NoStop}%
\bibitem [{\citenamefont {Salam}\ and\ \citenamefont {Ward}(1959)}]{Salam1959}%
  \BibitemOpen
  \bibfield  {author} {\bibinfo {author} {\bibfnamefont {A.}~\bibnamefont
  {Salam}}\ and\ \bibinfo {author} {\bibfnamefont {J.~C.}\ \bibnamefont
  {Ward}},\ }\bibfield  {title} {\bibinfo {title} {Weak and electromagnetic
  interactions},\ }\href {\doibase 10.1007/BF02726525} {\bibfield  {journal}
  {\bibinfo  {journal} {Il Nuovo Cimento (1955-1965)}\ }\textbf {\bibinfo
  {volume} {11}},\ \bibinfo {pages} {568--577} (\bibinfo {year}
  {1959})}\BibitemShut {NoStop}%
\bibitem [{\citenamefont {Salam}()}]{Salam1968}%
  \BibitemOpen
  \bibfield  {author} {\bibinfo {author} {\bibfnamefont {A.}~\bibnamefont
  {Salam}},\ }\bibfield  {title} {\bibinfo {title} {Weak and electromagnetic
  interactions},\ }in\ \href@noop {} {\emph {\bibinfo {booktitle} {Elementary
  particle theory}}},\ \bibinfo {editor} {edited by\ \bibinfo {editor}
  {\bibfnamefont {N.}~\bibnamefont {Svartholm}}}\ (\bibinfo  {publisher}
  {Almquist \& Wiksell})\ pp.\ \bibinfo {pages} {367--377}\BibitemShut
  {NoStop}%
\bibitem [{\citenamefont {Thomson}(1904)}]{Thomson1904}%
  \BibitemOpen
  \bibfield  {author} {\bibinfo {author} {\bibfnamefont {J.~J.}\ \bibnamefont
  {Thomson}},\ }\bibfield  {title} {\bibinfo {title} {On the structure of the
  atom: an investigation of the stability and periods of oscillation of a
  number of corpuscles arranged at equal intervals around the circumference of
  a circle with application of the results to the theory of atomic structure},\
  }\href {\doibase 10.1080/14786440409463107} {\bibfield  {journal} {\bibinfo
  {journal} {Philos. Mag. Series 6}\ }\textbf {\bibinfo {volume} {7}},\
  \bibinfo {pages} {237--265} (\bibinfo {year} {1904})}\BibitemShut {NoStop}%
\bibitem [{\citenamefont {Bohr}(1913{\natexlab{a}})}]{Bohr1913I}%
  \BibitemOpen
  \bibfield  {author} {\bibinfo {author} {\bibfnamefont {N.}~\bibnamefont
  {Bohr}},\ }\bibfield  {title} {\bibinfo {title} {On the constitution of atoms
  and molecules {I}},\ }\href {\doibase 10.1080/14786441308634955} {\bibfield
  {journal} {\bibinfo  {journal} {Philos. Mag. Series 6}\ }\textbf {\bibinfo
  {volume} {26}},\ \bibinfo {pages} {1--25} (\bibinfo {year}
  {1913}{\natexlab{a}})}\BibitemShut {NoStop}%
\bibitem [{\citenamefont {Larmor}(1897)}]{Larmor1897}%
  \BibitemOpen
  \bibfield  {author} {\bibinfo {author} {\bibfnamefont {J.}~\bibnamefont
  {Larmor}},\ }\bibfield  {title} {\bibinfo {title} {A dynamical theory of the
  electric and luminiferous medium. part {III}. relations with material
  media},\ }\href {\doibase 10.1098/rsta.1897.0020} {\bibfield  {journal}
  {\bibinfo  {journal} {Philos. Trans. R. Soc. A}\ }\textbf {\bibinfo {volume}
  {190}},\ \bibinfo {pages} {205--493} (\bibinfo {year} {1897})}\BibitemShut
  {NoStop}%
\bibitem [{\citenamefont {Rutherford}(1911)}]{Rutherford1911}%
  \BibitemOpen
  \bibfield  {author} {\bibinfo {author} {\bibfnamefont {E.}~\bibnamefont
  {Rutherford}},\ }\bibfield  {title} {\bibinfo {title} {The scattering of
  $\alpha$ and $\beta$ particles by matter and the structure of the atom},\
  }\href {\doibase 10.1080/14786440508637080} {\bibfield  {journal} {\bibinfo
  {journal} {Philos. Mag. Series 6}\ }\textbf {\bibinfo {volume} {21}},\
  \bibinfo {pages} {669--688} (\bibinfo {year} {1911})}\BibitemShut {NoStop}%
\bibitem [{\citenamefont {Bohr}(1913{\natexlab{b}})}]{Bohr1913II}%
  \BibitemOpen
  \bibfield  {author} {\bibinfo {author} {\bibfnamefont {N.}~\bibnamefont
  {Bohr}},\ }\bibfield  {title} {\bibinfo {title} {On the constitution of atoms
  and molecules {II}},\ }\href {\doibase 10.1080/14786441308634993} {\bibfield
  {journal} {\bibinfo  {journal} {Philos. Mag. Series 6}\ }\textbf {\bibinfo
  {volume} {26}},\ \bibinfo {pages} {476--502} (\bibinfo {year}
  {1913}{\natexlab{b}})}\BibitemShut {NoStop}%
\bibitem [{\citenamefont {Yukawa}(1935)}]{Yukawa1935}%
  \BibitemOpen
  \bibfield  {author} {\bibinfo {author} {\bibfnamefont {H.}~\bibnamefont
  {Yukawa}},\ }\bibfield  {title} {\bibinfo {title} {On the interaction of
  elementary particles {I}},\ }\href
  {https://www.jstage.jst.go.jp/article/ppmsj1919/17/0/17_0_48/_article}
  {\bibfield  {journal} {\bibinfo  {journal} {Proc. Phys. Math. Soc. Japan}\
  }\textbf {\bibinfo {volume} {17}},\ \bibinfo {pages} {48} (\bibinfo {year}
  {1935})}\BibitemShut {NoStop}%
\bibitem [{\citenamefont {Nakano}\ and\ \citenamefont
  {Nishijima}(1953)}]{Nakano1953}%
  \BibitemOpen
  \bibfield  {author} {\bibinfo {author} {\bibfnamefont {T.}~\bibnamefont
  {Nakano}}\ and\ \bibinfo {author} {\bibfnamefont {K.}~\bibnamefont
  {Nishijima}},\ }\bibfield  {title} {\bibinfo {title} {Charge independence for
  ${V}$-particles},\ }\href {\doibase 10.1143/ptp.10.581} {\bibfield  {journal}
  {\bibinfo  {journal} {Prog. Theor. Phys.}\ }\textbf {\bibinfo {volume}
  {10}},\ \bibinfo {pages} {581--582} (\bibinfo {year} {1953})}\BibitemShut
  {NoStop}%
\bibitem [{\citenamefont {Nishijima}(1955)}]{Nishijima1955}%
  \BibitemOpen
  \bibfield  {author} {\bibinfo {author} {\bibfnamefont {K.}~\bibnamefont
  {Nishijima}},\ }\bibfield  {title} {\bibinfo {title} {Charge independence
  theory of ${V}$-particles},\ }\href {\doibase 10.1143/ptp.13.285} {\bibfield
  {journal} {\bibinfo  {journal} {Prog. Theor. Phys.}\ }\textbf {\bibinfo
  {volume} {13}},\ \bibinfo {pages} {285--304} (\bibinfo {year}
  {1955})}\BibitemShut {NoStop}%
\bibitem [{\citenamefont {Sakata}(1956)}]{Sakata1956}%
  \BibitemOpen
  \bibfield  {author} {\bibinfo {author} {\bibfnamefont {S.}~\bibnamefont
  {Sakata}},\ }\bibfield  {title} {\bibinfo {title} {On a composite model for
  the new particles},\ }\href {\doibase 10.1143/ptp.16.686} {\bibfield
  {journal} {\bibinfo  {journal} {Prog. Theor. Phys.}\ }\textbf {\bibinfo
  {volume} {16}},\ \bibinfo {pages} {686--688} (\bibinfo {year}
  {1956})}\BibitemShut {NoStop}%
\bibitem [{\citenamefont {Gell-Mann}(1961)}]{GellMann1961}%
  \BibitemOpen
  \bibfield  {author} {\bibinfo {author} {\bibfnamefont {M.}~\bibnamefont
  {Gell-Mann}},\ }\href {\doibase 10.2172/4008239} {\emph {\bibinfo {title}
  {The eightfold way: a theory of strong interaction symmetry}}},\ \bibinfo
  {type} {Tech. Rep.}\ (\bibinfo {year} {1961})\BibitemShut {NoStop}%
\bibitem [{\citenamefont {Gell-Mann}(1962)}]{GellMann1962}%
  \BibitemOpen
  \bibfield  {author} {\bibinfo {author} {\bibfnamefont {M.}~\bibnamefont
  {Gell-Mann}},\ }\bibfield  {title} {\bibinfo {title} {Symmetries of baryons
  and mesons},\ }\href {\doibase 10.1103/physrev.125.1067} {\bibfield
  {journal} {\bibinfo  {journal} {Phys. Rev.}\ }\textbf {\bibinfo {volume}
  {125}},\ \bibinfo {pages} {1067--1084} (\bibinfo {year} {1962})}\BibitemShut
  {NoStop}%
\bibitem [{\citenamefont {Ne'eman}(1961)}]{Neeman1961}%
  \BibitemOpen
  \bibfield  {author} {\bibinfo {author} {\bibfnamefont {Y.}~\bibnamefont
  {Ne'eman}},\ }\bibfield  {title} {\bibinfo {title} {Derivation of strong
  interactions from a gauge invariance},\ }\href
  {https://doi.org/10.1016%2F0029-5582%2861%2990134-1} {\bibfield  {journal}
  {\bibinfo  {journal} {Nucl. Phys.}\ }\textbf {\bibinfo {volume} {26}},\
  \bibinfo {pages} {222--229} (\bibinfo {year} {1961})}\BibitemShut {NoStop}%
\bibitem [{\citenamefont {{Barnes et al}}(1964)}]{Barnes1964}%
  \BibitemOpen
  \bibfield  {author} {\bibinfo {author} {\bibfnamefont {V.~E.}\ \bibnamefont
  {{Barnes et al}}},\ }\bibfield  {title} {\bibinfo {title} {Observation of a
  hyperon with strangeness minus three},\ }\href
  {https://doi.org/10.1103%2Fphysrevlett.12.204} {\bibfield  {journal}
  {\bibinfo  {journal} {Phys. Rev. Lett.}\ }\textbf {\bibinfo {volume} {12}},\
  \bibinfo {pages} {204--206} (\bibinfo {year} {1964})}\BibitemShut {NoStop}%
\bibitem [{\citenamefont {Gell-Mann}(1964)}]{GellMann1964}%
  \BibitemOpen
  \bibfield  {author} {\bibinfo {author} {\bibfnamefont {M.}~\bibnamefont
  {Gell-Mann}},\ }\bibfield  {title} {\bibinfo {title} {A schematic model of
  baryons and mesons},\ }\href {\doibase 10.1016/s0031-9163(64)92001-3}
  {\bibfield  {journal} {\bibinfo  {journal} {Phys. Lett.}\ }\textbf {\bibinfo
  {volume} {8}},\ \bibinfo {pages} {214--215} (\bibinfo {year}
  {1964})}\BibitemShut {NoStop}%
\bibitem [{\citenamefont {Kosmann-Schwarzbach}(2010)}]{KosmannSchwarzbach2010}%
  \BibitemOpen
  \bibfield  {author} {\bibinfo {author} {\bibfnamefont {Y.}~\bibnamefont
  {Kosmann-Schwarzbach}},\ }\href {\doibase 10.1007/978-0-387-78866-1} {\emph
  {\bibinfo {title} {Groups and Symmetries}}}\ (\bibinfo  {publisher} {Springer
  New York},\ \bibinfo {year} {2010})\BibitemShut {NoStop}%
\bibitem [{\citenamefont {Maki}\ and\ \citenamefont {Ohnuki}(1964)}]{Maki1964}%
  \BibitemOpen
  \bibfield  {author} {\bibinfo {author} {\bibfnamefont {Z.}~\bibnamefont
  {Maki}}\ and\ \bibinfo {author} {\bibfnamefont {Y.}~\bibnamefont {Ohnuki}},\
  }\bibfield  {title} {\bibinfo {title} {Quartet scheme for elementary
  particles},\ }\href {\doibase 10.1143/ptp.32.144} {\bibfield  {journal}
  {\bibinfo  {journal} {Prog. Theor. Phys.}\ }\textbf {\bibinfo {volume}
  {32}},\ \bibinfo {pages} {144--158} (\bibinfo {year} {1964})}\BibitemShut
  {NoStop}%
\bibitem [{\citenamefont {Sakata}(1964)}]{Sakita1964}%
  \BibitemOpen
  \bibfield  {author} {\bibinfo {author} {\bibfnamefont {B.}~\bibnamefont
  {Sakata}},\ }\bibfield  {title} {\bibinfo {title} {Supermultiplets of
  elementary particles},\ }\href {\doibase 10.1103/physrev.136.b1756}
  {\bibfield  {journal} {\bibinfo  {journal} {Phys. Rev.}\ }\textbf {\bibinfo
  {volume} {136}},\ \bibinfo {pages} {B1756--B1760} (\bibinfo {year}
  {1964})}\BibitemShut {NoStop}%
\bibitem [{\citenamefont {Kobayashi}\ and\ \citenamefont
  {Maskawa}(1973)}]{Kobayashi1973}%
  \BibitemOpen
  \bibfield  {author} {\bibinfo {author} {\bibfnamefont {M.}~\bibnamefont
  {Kobayashi}}\ and\ \bibinfo {author} {\bibfnamefont {T.}~\bibnamefont
  {Maskawa}},\ }\bibfield  {title} {\bibinfo {title} {{CP}-violation in the
  renormalizable theory of weak interaction},\ }\href
  {https://doi.org/10.1143%2Fptp.49.652} {\bibfield  {journal} {\bibinfo
  {journal} {Prog. Theor. Phys.}\ }\textbf {\bibinfo {volume} {49}},\ \bibinfo
  {pages} {652--657} (\bibinfo {year} {1973})}\BibitemShut {NoStop}%
\bibitem [{\citenamefont {Greenberg}(1964)}]{Greenberg1964}%
  \BibitemOpen
  \bibfield  {author} {\bibinfo {author} {\bibfnamefont {O.~W.}\ \bibnamefont
  {Greenberg}},\ }\bibfield  {title} {\bibinfo {title} {Spin and unitary-spin
  independence in a paraquark model of baryons and mesons},\ }\href
  {https://doi.org/10.1103%2Fphysrevlett.13.598} {\bibfield  {journal}
  {\bibinfo  {journal} {Phys. Rev. Lett.}\ }\textbf {\bibinfo {volume} {13}},\
  \bibinfo {pages} {598--602} (\bibinfo {year} {1964})}\BibitemShut {NoStop}%
\bibitem [{\citenamefont {Han}\ and\ \citenamefont {Nambu}(1965)}]{Han1965}%
  \BibitemOpen
  \bibfield  {author} {\bibinfo {author} {\bibfnamefont {M.~Y.}\ \bibnamefont
  {Han}}\ and\ \bibinfo {author} {\bibfnamefont {Y.}~\bibnamefont {Nambu}},\
  }\bibfield  {title} {\bibinfo {title} {Three-triplet model with double
  {SU}(3) symmetry},\ }\href {\doibase 10.1103/physrev.139.b1006} {\bibfield
  {journal} {\bibinfo  {journal} {Phys. Rev.}\ }\textbf {\bibinfo {volume}
  {139}},\ \bibinfo {pages} {B1006--B1010} (\bibinfo {year}
  {1965})}\BibitemShut {NoStop}%
\bibitem [{\citenamefont {Fritzsch}\ \emph {et~al.}(1973)\citenamefont
  {Fritzsch}, \citenamefont {Gell-Mann},\ and\ \citenamefont
  {Leutwyler}}]{Fritzsch1973}%
  \BibitemOpen
  \bibfield  {author} {\bibinfo {author} {\bibfnamefont {H.}~\bibnamefont
  {Fritzsch}}, \bibinfo {author} {\bibfnamefont {M.}~\bibnamefont {Gell-Mann}},
  \ and\ \bibinfo {author} {\bibfnamefont {H.}~\bibnamefont {Leutwyler}},\
  }\bibfield  {title} {\bibinfo {title} {Advantages of the color octet gluon
  picture},\ }\href {\doibase 10.1016/0370-2693(73)90625-4} {\bibfield
  {journal} {\bibinfo  {journal} {Phys. Lett. B}\ }\textbf {\bibinfo {volume}
  {47}},\ \bibinfo {pages} {365--368} (\bibinfo {year} {1973})}\BibitemShut
  {NoStop}%
\bibitem [{\citenamefont {Gross}\ and\ \citenamefont
  {Wilczek}(1973)}]{Gross1973}%
  \BibitemOpen
  \bibfield  {author} {\bibinfo {author} {\bibfnamefont {D.~J.}\ \bibnamefont
  {Gross}}\ and\ \bibinfo {author} {\bibfnamefont {F.}~\bibnamefont
  {Wilczek}},\ }\bibfield  {title} {\bibinfo {title} {Ultraviolet behavior of
  non-{A}belian gauge theories},\ }\href {\doibase 10.1103/physrevlett.30.1343}
  {\bibfield  {journal} {\bibinfo  {journal} {Phys. Rev. Lett.}\ }\textbf
  {\bibinfo {volume} {30}},\ \bibinfo {pages} {1343--1346} (\bibinfo {year}
  {1973})}\BibitemShut {NoStop}%
\bibitem [{\citenamefont {Politzer}(1973)}]{Politzer1973}%
  \BibitemOpen
  \bibfield  {author} {\bibinfo {author} {\bibfnamefont {H.~D.}\ \bibnamefont
  {Politzer}},\ }\bibfield  {title} {\bibinfo {title} {Reliable perturbative
  results for strong interactions?}\ }\href
  {https://doi.org/10.1103%2Fphysrevlett.30.1346} {\bibfield  {journal}
  {\bibinfo  {journal} {Phys. Rev. Lett.}\ }\textbf {\bibinfo {volume} {30}},\
  \bibinfo {pages} {1346--1349} (\bibinfo {year} {1973})}\BibitemShut {NoStop}%
\bibitem [{\citenamefont {Bertolami}\ and\ \citenamefont
  {Sequeira}(2009)}]{Bertolami2009}%
  \BibitemOpen
  \bibfield  {author} {\bibinfo {author} {\bibfnamefont {O.}~\bibnamefont
  {Bertolami}}\ and\ \bibinfo {author} {\bibfnamefont {M.~C.}\ \bibnamefont
  {Sequeira}},\ }\bibfield  {title} {\bibinfo {title} {Energy conditions and
  stability in $f({R})$ theories of gravity with nonminimal coupling to
  matter},\ }\href {\doibase 10.1103/physrevd.79.104010} {\bibfield  {journal}
  {\bibinfo  {journal} {Phys. Rev. D}\ }\textbf {\bibinfo {volume} {79}},\
  \bibinfo {pages} {104010} (\bibinfo {year} {2009})}\BibitemShut {NoStop}%
\bibitem [{\citenamefont {Wang}\ \emph {et~al.}(2010)\citenamefont {Wang},
  \citenamefont {Wu}, \citenamefont {Guo}, \citenamefont {Yang},\ and\
  \citenamefont {Wang}}]{Wang2010}%
  \BibitemOpen
  \bibfield  {author} {\bibinfo {author} {\bibfnamefont {J.}~\bibnamefont
  {Wang}}, \bibinfo {author} {\bibfnamefont {Y.-B.}\ \bibnamefont {Wu}},
  \bibinfo {author} {\bibfnamefont {Y.-X.}\ \bibnamefont {Guo}}, \bibinfo
  {author} {\bibfnamefont {W.-Q.}\ \bibnamefont {Yang}}, \ and\ \bibinfo
  {author} {\bibfnamefont {L.}~\bibnamefont {Wang}},\ }\bibfield  {title}
  {\bibinfo {title} {Energy conditions and stability in generalized $f({R})$
  gravity with arbitrary coupling between matter and geometry},\ }\href
  {\doibase 10.1016/j.physletb.2010.04.063} {\bibfield  {journal} {\bibinfo
  {journal} {Phys. Lett. B}\ }\textbf {\bibinfo {volume} {689}},\ \bibinfo
  {pages} {133--138} (\bibinfo {year} {2010})}\BibitemShut {NoStop}%
\bibitem [{\citenamefont {Balakin}\ \emph {et~al.}(2010)\citenamefont
  {Balakin}, \citenamefont {Lemos},\ and\ \citenamefont
  {Zayats}}]{Balakin2010}%
  \BibitemOpen
  \bibfield  {author} {\bibinfo {author} {\bibfnamefont {A.~B.}\ \bibnamefont
  {Balakin}}, \bibinfo {author} {\bibfnamefont {J.~P.~S.}\ \bibnamefont
  {Lemos}}, \ and\ \bibinfo {author} {\bibfnamefont {A.~E.}\ \bibnamefont
  {Zayats}},\ }\bibfield  {title} {\bibinfo {title} {Nonminimal coupling for
  the gravitational and electromagnetic fields: traversable electric
  wormholes},\ }\href {\doibase 10.1103/physrevd.81.084015} {\bibfield
  {journal} {\bibinfo  {journal} {Phys. Rev. D}\ }\textbf {\bibinfo {volume}
  {81}},\ \bibinfo {pages} {084015} (\bibinfo {year} {2010})}\BibitemShut
  {NoStop}%
\bibitem [{\citenamefont {Mahajan}(2014)}]{Mahajan2014}%
  \BibitemOpen
  \bibfield  {author} {\bibinfo {author} {\bibfnamefont {N.}~\bibnamefont
  {Mahajan}},\ }\bibfield  {title} {\bibinfo {title} {Some remarks on
  nonminimal coupling of the inflaton},\ }\href
  {https://doi.org/10.1142%2Fs021827181450076x} {\bibfield  {journal} {\bibinfo
   {journal} {Int. J. Mod. Phys. D}\ }\textbf {\bibinfo {volume} {23}},\
  \bibinfo {pages} {1450076} (\bibinfo {year} {2014})}\BibitemShut {NoStop}%
\bibitem [{\citenamefont {Budhi}(2017)}]{Budhi2017}%
  \BibitemOpen
  \bibfield  {author} {\bibinfo {author} {\bibfnamefont {R.~H.~S.}\
  \bibnamefont {Budhi}},\ }\bibfield  {title} {\bibinfo {title} {Inflation due
  to non-minimal coupling of $f({R})$ gravity to a scalar field},\ }\href
  {https://arxiv.org/abs/1701.03814} {\bibfield  {journal} {\bibinfo  {journal}
  {arXiv:1701.03814}\ } (\bibinfo {year} {2017})}\BibitemShut {NoStop}%
\bibitem [{\citenamefont {Harko}\ and\ \citenamefont {Lobo}(2014)}]{Harko2014}%
  \BibitemOpen
  \bibfield  {author} {\bibinfo {author} {\bibfnamefont {T.}~\bibnamefont
  {Harko}}\ and\ \bibinfo {author} {\bibfnamefont {F.}~\bibnamefont {Lobo}},\
  }\bibfield  {title} {\bibinfo {title} {Generalized curvature-matter couplings
  in modified gravity},\ }\href {\doibase 10.3390/galaxies2030410} {\bibfield
  {journal} {\bibinfo  {journal} {Galaxies}\ }\textbf {\bibinfo {volume} {2}},\
  \bibinfo {pages} {410--465} (\bibinfo {year} {2014})}\BibitemShut {NoStop}%
\bibitem [{\citenamefont {Kogut}\ and\ \citenamefont
  {Susskind}(1975)}]{KogSuss75a}%
  \BibitemOpen
  \bibfield  {author} {\bibinfo {author} {\bibfnamefont {J.~B.}\ \bibnamefont
  {Kogut}}\ and\ \bibinfo {author} {\bibfnamefont {L.}~\bibnamefont
  {Susskind}},\ }\bibfield  {title} {\bibinfo {title} {Hamiltonian formulation
  of {W}ilson's lattice gauge theories},\ }\href
  {https://doi.org/10.1103/physrevd.11.395} {\bibfield  {journal} {\bibinfo
  {journal} {Phys. Rev. D}\ }\textbf {\bibinfo {volume} {11}},\ \bibinfo
  {pages} {395--408} (\bibinfo {year} {1975})}\BibitemShut {NoStop}%
\bibitem [{\citenamefont {Byrnes}\ and\ \citenamefont
  {Yamamoto}(2006)}]{Byrnes2006}%
  \BibitemOpen
  \bibfield  {author} {\bibinfo {author} {\bibfnamefont {T.}~\bibnamefont
  {Byrnes}}\ and\ \bibinfo {author} {\bibfnamefont {Y.}~\bibnamefont
  {Yamamoto}},\ }\bibfield  {title} {\bibinfo {title} {Simulating lattice gauge
  theories on a quantum computer},\ }\href
  {https://doi.org/10.1103%2Fphysreva.73.022328} {\bibfield  {journal}
  {\bibinfo  {journal} {Phys. Rev. A}\ }\textbf {\bibinfo {volume} {73}},\
  \bibinfo {pages} {022328} (\bibinfo {year} {2006})}\BibitemShut {NoStop}%
\bibitem [{\citenamefont {Jordan}\ \emph {et~al.}(2012)\citenamefont {Jordan},
  \citenamefont {Lee},\ and\ \citenamefont {Preskill}}]{Jordan2012}%
  \BibitemOpen
  \bibfield  {author} {\bibinfo {author} {\bibfnamefont {S.~P.}\ \bibnamefont
  {Jordan}}, \bibinfo {author} {\bibfnamefont {K.~S.~M.}\ \bibnamefont {Lee}},
  \ and\ \bibinfo {author} {\bibfnamefont {J.}~\bibnamefont {Preskill}},\
  }\bibfield  {title} {\bibinfo {title} {Quantum algorithms for quantum field
  theories},\ }\href {https://doi.org/10.1126/science.1217069} {\bibfield
  {journal} {\bibinfo  {journal} {Science}\ }\textbf {\bibinfo {volume}
  {336}},\ \bibinfo {pages} {1130--1133} (\bibinfo {year} {2012})}\BibitemShut
  {NoStop}%
\bibitem [{\citenamefont {Zohar}\ \emph {et~al.}(2015)\citenamefont {Zohar},
  \citenamefont {Cirac},\ and\ \citenamefont {Reznik}}]{Zohar2015}%
  \BibitemOpen
  \bibfield  {author} {\bibinfo {author} {\bibfnamefont {E.}~\bibnamefont
  {Zohar}}, \bibinfo {author} {\bibfnamefont {J.~I.}\ \bibnamefont {Cirac}}, \
  and\ \bibinfo {author} {\bibfnamefont {B.}~\bibnamefont {Reznik}},\
  }\bibfield  {title} {\bibinfo {title} {Quantum simulations of lattice gauge
  theories using ultracold atoms in optical lattices},\ }\href
  {https://doi.org/10.1088%2F0034-4885%2F79%2F1%2F014401} {\bibfield  {journal}
  {\bibinfo  {journal} {Rep. Prog. Phys.}\ }\textbf {\bibinfo {volume} {79}},\
  \bibinfo {pages} {014401} (\bibinfo {year} {2015})}\BibitemShut {NoStop}%
\bibitem [{\citenamefont {B\"{u}chler}\ \emph {et~al.}(2005)\citenamefont
  {B\"{u}chler}, \citenamefont {Hermele}, \citenamefont {Huber}, \citenamefont
  {Fisher},\ and\ \citenamefont {Zoller}}]{Bchler2005}%
  \BibitemOpen
  \bibfield  {author} {\bibinfo {author} {\bibfnamefont {H.~P.}\ \bibnamefont
  {B\"{u}chler}}, \bibinfo {author} {\bibfnamefont {M.}~\bibnamefont
  {Hermele}}, \bibinfo {author} {\bibfnamefont {S.~D.}\ \bibnamefont {Huber}},
  \bibinfo {author} {\bibfnamefont {M.~P.~A.}\ \bibnamefont {Fisher}}, \ and\
  \bibinfo {author} {\bibfnamefont {P.}~\bibnamefont {Zoller}},\ }\bibfield
  {title} {\bibinfo {title} {Atomic quantum simulator for lattice gauge
  theories and ring exchange models},\ }\href
  {https://doi.org/10.1103%2Fphysrevlett.95.040402} {\bibfield  {journal}
  {\bibinfo  {journal} {Phys. Rev. Lett.}\ }\textbf {\bibinfo {volume} {95}}
  (\bibinfo {year} {2005})}\BibitemShut {NoStop}%
\bibitem [{\citenamefont {Pichler}\ \emph {et~al.}(2016)\citenamefont
  {Pichler}, \citenamefont {Dalmonte}, \citenamefont {Rico}, \citenamefont
  {Zoller},\ and\ \citenamefont {Montangero}}]{Pichler2016}%
  \BibitemOpen
  \bibfield  {author} {\bibinfo {author} {\bibfnamefont {T.}~\bibnamefont
  {Pichler}}, \bibinfo {author} {\bibfnamefont {M.}~\bibnamefont {Dalmonte}},
  \bibinfo {author} {\bibfnamefont {E.}~\bibnamefont {Rico}}, \bibinfo {author}
  {\bibfnamefont {P.}~\bibnamefont {Zoller}}, \ and\ \bibinfo {author}
  {\bibfnamefont {S.}~\bibnamefont {Montangero}},\ }\bibfield  {title}
  {\bibinfo {title} {Real-time dynamics in {U}(1) lattice gauge theories with
  tensor networks},\ }\href {https://doi.org/10.1103/physrevx.6.011023}
  {\bibfield  {journal} {\bibinfo  {journal} {Phys. Rev. X}\ }\textbf {\bibinfo
  {volume} {6}} (\bibinfo {year} {2016})}\BibitemShut {NoStop}%
\bibitem [{\citenamefont {Kogut}(1983)}]{Kogut83a}%
  \BibitemOpen
  \bibfield  {author} {\bibinfo {author} {\bibfnamefont {J.~B.}\ \bibnamefont
  {Kogut}},\ }\bibfield  {title} {\bibinfo {title} {The lattice gauge theory
  approach to quantum chromodynamics},\ }\href
  {https://link.aps.org/doi/10.1103/RevModPhys.55.775} {\bibfield  {journal}
  {\bibinfo  {journal} {Rev. Mod. Phys.}\ }\textbf {\bibinfo {volume} {55}},\
  \bibinfo {pages} {775--836} (\bibinfo {year} {1983})}\BibitemShut {NoStop}%
\bibitem [{\citenamefont {M\"{u}nster}\ and\ \citenamefont
  {Walzl}(2000)}]{Munster2011}%
  \BibitemOpen
  \bibfield  {author} {\bibinfo {author} {\bibfnamefont {G.}~\bibnamefont
  {M\"{u}nster}}\ and\ \bibinfo {author} {\bibfnamefont {M.}~\bibnamefont
  {Walzl}},\ }\bibfield  {title} {\bibinfo {title} {Lattice gauge theory - a
  short primer},\ }\href {https://arxiv.org/abs/hep-lat/0012005} {\bibfield
  {journal} {\bibinfo  {journal} {arXiv:hep-lat/0012005}\ } (\bibinfo {year}
  {2000})}\BibitemShut {NoStop}%
\bibitem [{\citenamefont {Susskind}(1977)}]{Susskind77a}%
  \BibitemOpen
  \bibfield  {author} {\bibinfo {author} {\bibfnamefont {L.}~\bibnamefont
  {Susskind}},\ }\bibfield  {title} {\bibinfo {title} {Lattice fermions},\
  }\href {https://doi.org/10.1103/physrevd.16.3031} {\bibfield  {journal}
  {\bibinfo  {journal} {Phys. Rev. D}\ }\textbf {\bibinfo {volume} {16}},\
  \bibinfo {pages} {3031--3039} (\bibinfo {year} {1977})}\BibitemShut {NoStop}%
\bibitem [{\citenamefont {Brower}\ \emph {et~al.}(2016)\citenamefont {Brower},
  \citenamefont {Fleming}, \citenamefont {Gasbarro}, \citenamefont {Raben},
  \citenamefont {Tan},\ and\ \citenamefont {Weinberg}}]{Brower2016}%
  \BibitemOpen
  \bibfield  {author} {\bibinfo {author} {\bibfnamefont {R.~C.}\ \bibnamefont
  {Brower}}, \bibinfo {author} {\bibfnamefont {G.}~\bibnamefont {Fleming}},
  \bibinfo {author} {\bibfnamefont {A.}~\bibnamefont {Gasbarro}}, \bibinfo
  {author} {\bibfnamefont {T.}~\bibnamefont {Raben}}, \bibinfo {author}
  {\bibfnamefont {C.-I.}\ \bibnamefont {Tan}}, \ and\ \bibinfo {author}
  {\bibfnamefont {E.}~\bibnamefont {Weinberg}},\ }\bibfield  {title} {\bibinfo
  {title} {Quantum finite elements for lattice field theory},\ }\href
  {https://arxiv.org/abs/1601.01367} {\bibfield  {journal} {\bibinfo  {journal}
  {arXiv:1601.01367}\ } (\bibinfo {year} {2016})}\BibitemShut {NoStop}%
\bibitem [{\citenamefont {Schumacher}\ and\ \citenamefont
  {Werner}(2004)}]{SW04}%
  \BibitemOpen
  \bibfield  {author} {\bibinfo {author} {\bibfnamefont {B.}~\bibnamefont
  {Schumacher}}\ and\ \bibinfo {author} {\bibfnamefont {R.~F.}\ \bibnamefont
  {Werner}},\ }\bibfield  {title} {\bibinfo {title} {Reversible quantum
  cellular automata},\ }\href {https://arxiv.org/abs/quant-ph/0405174}
  {\bibfield  {journal} {\bibinfo  {journal} {arXiv:quant-ph/0405174}\ }
  (\bibinfo {year} {2004})}\BibitemShut {NoStop}%
\bibitem [{\citenamefont {Bisio}\ \emph {et~al.}(2015)\citenamefont {Bisio},
  \citenamefont {D'Ariano},\ and\ \citenamefont {Tosini}}]{Bisio2015}%
  \BibitemOpen
  \bibfield  {author} {\bibinfo {author} {\bibfnamefont {A.}~\bibnamefont
  {Bisio}}, \bibinfo {author} {\bibfnamefont {G.~M.}\ \bibnamefont {D'Ariano}},
  \ and\ \bibinfo {author} {\bibfnamefont {A.}~\bibnamefont {Tosini}},\
  }\bibfield  {title} {\bibinfo {title} {Quantum field as a quantum cellular
  automaton: the {D}irac free evolution in one dimension},\ }\href
  {https://doi.org/10.1016/j.aop.2014.12.016} {\bibfield  {journal} {\bibinfo
  {journal} {Ann. Physics}\ }\textbf {\bibinfo {volume} {354}},\ \bibinfo
  {pages} {244--264} (\bibinfo {year} {2015})}\BibitemShut {NoStop}%
\bibitem [{\citenamefont {Farrelly}\ and\ \citenamefont
  {Short}(2014{\natexlab{a}})}]{Farrelly2014a}%
  \BibitemOpen
  \bibfield  {author} {\bibinfo {author} {\bibfnamefont {T.~C.}\ \bibnamefont
  {Farrelly}}\ and\ \bibinfo {author} {\bibfnamefont {A.~J.}\ \bibnamefont
  {Short}},\ }\bibfield  {title} {\bibinfo {title} {Discrete spacetime and
  relativistic quantum particles},\ }\href
  {https://doi.org/10.1103/physreva.89.062109} {\bibfield  {journal} {\bibinfo
  {journal} {Phys. Rev. A}\ }\textbf {\bibinfo {volume} {89}} (\bibinfo {year}
  {2014}{\natexlab{a}})}\BibitemShut {NoStop}%
\bibitem [{\citenamefont {Farrelly}\ and\ \citenamefont
  {Short}(2014{\natexlab{b}})}]{Farrelly2014b}%
  \BibitemOpen
  \bibfield  {author} {\bibinfo {author} {\bibfnamefont {T.~C.}\ \bibnamefont
  {Farrelly}}\ and\ \bibinfo {author} {\bibfnamefont {A.~J.}\ \bibnamefont
  {Short}},\ }\bibfield  {title} {\bibinfo {title} {Causal fermions in discrete
  space-time},\ }\href {https://doi.org/10.1103/physreva.89.012302} {\bibfield
  {journal} {\bibinfo  {journal} {Phys. Rev. A}\ }\textbf {\bibinfo {volume}
  {89}} (\bibinfo {year} {2014}{\natexlab{b}})}\BibitemShut {NoStop}%
\bibitem [{\citenamefont {Zapp}\ and\ \citenamefont {Or\'us}(2017)}]{ZO17}%
  \BibitemOpen
  \bibfield  {author} {\bibinfo {author} {\bibfnamefont {K.}~\bibnamefont
  {Zapp}}\ and\ \bibinfo {author} {\bibfnamefont {R.}~\bibnamefont {Or\'us}},\
  }\bibfield  {title} {\bibinfo {title} {Tensor network simulation of qed on
  infinite lattices: learning from $(1+1)\mathrm{d}$, and prospects for
  $(2+1)\mathrm{d}$},\ }\href {\doibase 10.1103/PhysRevD.95.114508} {\bibfield
  {journal} {\bibinfo  {journal} {Phys. Rev. D}\ }\textbf {\bibinfo {volume}
  {95}},\ \bibinfo {pages} {114508} (\bibinfo {year} {2017})}\BibitemShut
  {NoStop}%
\bibitem [{\citenamefont {Osterloh}\ \emph {et~al.}(2005)\citenamefont
  {Osterloh}, \citenamefont {Baig}, \citenamefont {Santos}, \citenamefont
  {Zoller},\ and\ \citenamefont {Lewenstein}}]{Osterloh2005}%
  \BibitemOpen
  \bibfield  {author} {\bibinfo {author} {\bibfnamefont {K.}~\bibnamefont
  {Osterloh}}, \bibinfo {author} {\bibfnamefont {M.}~\bibnamefont {Baig}},
  \bibinfo {author} {\bibfnamefont {L.}~\bibnamefont {Santos}}, \bibinfo
  {author} {\bibfnamefont {P.}~\bibnamefont {Zoller}}, \ and\ \bibinfo {author}
  {\bibfnamefont {M.}~\bibnamefont {Lewenstein}},\ }\bibfield  {title}
  {\bibinfo {title} {Cold atoms in non-{A}belian gauge potentials: From the
  {H}ofstadter ``{M}oth" to lattice gauge theory},\ }\href
  {https://doi.org/10.1103/physrevlett.95.010403} {\bibfield  {journal}
  {\bibinfo  {journal} {Phys. Rev. Lett.}\ }\textbf {\bibinfo {volume} {95}}
  (\bibinfo {year} {2005})}\BibitemShut {NoStop}%
\bibitem [{\citenamefont {Ahlbrecht}\ \emph {et~al.}(2011)\citenamefont
  {Ahlbrecht}, \citenamefont {Alberti}, \citenamefont {Meschede}, \citenamefont
  {Scholz}, \citenamefont {Werner},\ and\ \citenamefont {Werner}}]{AADSWW11}%
  \BibitemOpen
  \bibfield  {author} {\bibinfo {author} {\bibfnamefont {A.}~\bibnamefont
  {Ahlbrecht}}, \bibinfo {author} {\bibfnamefont {A.}~\bibnamefont {Alberti}},
  \bibinfo {author} {\bibfnamefont {D.}~\bibnamefont {Meschede}}, \bibinfo
  {author} {\bibfnamefont {V.~B.}\ \bibnamefont {Scholz}}, \bibinfo {author}
  {\bibfnamefont {A.~H.}\ \bibnamefont {Werner}}, \ and\ \bibinfo {author}
  {\bibfnamefont {R.~F.}\ \bibnamefont {Werner}},\ }\bibfield  {title}
  {\bibinfo {title} {Bound molecules in an interacting quantum walk},\ }\href
  {https://arxiv.org/abs/1105.1051} {\bibfield  {journal} {\bibinfo  {journal}
  {arXiv:1105.1051}\ } (\bibinfo {year} {2011})}\BibitemShut {NoStop}%
\bibitem [{\citenamefont {Goerbig}(2009)}]{Goerbig_review}%
  \BibitemOpen
  \bibfield  {author} {\bibinfo {author} {\bibfnamefont {M.~O.}\ \bibnamefont
  {Goerbig}},\ }\bibfield  {title} {\bibinfo {title} {Quantum {H}all effects},\
  }\href {https://arxiv.org/abs/0909.1998} {\bibfield  {journal} {\bibinfo
  {journal} {arXiv:0909.1998v2}\ } (\bibinfo {year} {2009})}\BibitemShut
  {NoStop}%
\bibitem [{\citenamefont {Tong}(2016)}]{Tong2016}%
  \BibitemOpen
  \bibfield  {author} {\bibinfo {author} {\bibfnamefont {D.}~\bibnamefont
  {Tong}},\ }\bibfield  {title} {\bibinfo {title} {Lectures on the quantum
  {H}all effect},\ }\href {https://arxiv.org/pdf/1606.06687.pdf} {\bibfield
  {journal} {\bibinfo  {journal} {arXiv:1105.1051}\ } (\bibinfo {year}
  {2016})}\BibitemShut {NoStop}%
\bibitem [{\citenamefont {Konschuh}\ \emph {et~al.}(2010)\citenamefont
  {Konschuh}, \citenamefont {Gmitra},\ and\ \citenamefont
  {Fabian}}]{Konschuh2010}%
  \BibitemOpen
  \bibfield  {author} {\bibinfo {author} {\bibfnamefont {S.}~\bibnamefont
  {Konschuh}}, \bibinfo {author} {\bibfnamefont {M.}~\bibnamefont {Gmitra}}, \
  and\ \bibinfo {author} {\bibfnamefont {J.}~\bibnamefont {Fabian}},\
  }\bibfield  {title} {\bibinfo {title} {Tight-binding theory of the spin-orbit
  coupling in graphene},\ }\href {https://doi.org/10.1103/physrevb.82.245412}
  {\bibfield  {journal} {\bibinfo  {journal} {Phys. Rev. B}\ }\textbf {\bibinfo
  {volume} {82}} (\bibinfo {year} {2010})}\BibitemShut {NoStop}%
\bibitem [{\citenamefont {Groh}\ \emph {et~al.}(2016)\citenamefont {Groh},
  \citenamefont {Brakhane}, \citenamefont {Alt}, \citenamefont {Meschede},
  \citenamefont {Asb\'oth},\ and\ \citenamefont {Alberti}}]{Groh2016}%
  \BibitemOpen
  \bibfield  {author} {\bibinfo {author} {\bibfnamefont {T.}~\bibnamefont
  {Groh}}, \bibinfo {author} {\bibfnamefont {S.}~\bibnamefont {Brakhane}},
  \bibinfo {author} {\bibfnamefont {W.}~\bibnamefont {Alt}}, \bibinfo {author}
  {\bibfnamefont {D.}~\bibnamefont {Meschede}}, \bibinfo {author}
  {\bibfnamefont {J.~K.}\ \bibnamefont {Asb\'oth}}, \ and\ \bibinfo {author}
  {\bibfnamefont {A.}~\bibnamefont {Alberti}},\ }\bibfield  {title} {\bibinfo
  {title} {Robustness of topologically protected edge states in quantum walk
  experiments with neutral atoms},\ }\href
  {https://link.aps.org/doi/10.1103/PhysRevA.94.013620} {\bibfield  {journal}
  {\bibinfo  {journal} {Phys. Rev. A}\ }\textbf {\bibinfo {volume} {94}},\
  \bibinfo {pages} {013620} (\bibinfo {year} {2016})}\BibitemShut {NoStop}%
\bibitem [{\citenamefont {Shikano}\ \emph {et~al.}(2014)\citenamefont
  {Shikano}, \citenamefont {Wada},\ and\ \citenamefont {Horikawa}}]{SWH14}%
  \BibitemOpen
  \bibfield  {author} {\bibinfo {author} {\bibfnamefont {Y.}~\bibnamefont
  {Shikano}}, \bibinfo {author} {\bibfnamefont {T.}~\bibnamefont {Wada}}, \
  and\ \bibinfo {author} {\bibfnamefont {J.}~\bibnamefont {Horikawa}},\
  }\bibfield  {title} {\bibinfo {title} {Discrete-time quantum walk with
  feed-forward quantum coin},\ }\href
  {https://www.nature.com/articles/srep04427} {\bibfield  {journal} {\bibinfo
  {journal} {Sci. Rep.}\ }\textbf {\bibinfo {volume} {4}},\ \bibinfo {pages}
  {4427} (\bibinfo {year} {2014})}\BibitemShut {NoStop}%
\bibitem [{\citenamefont {Fowler}\ \emph {et~al.}(2012)\citenamefont {Fowler},
  \citenamefont {Mariantoni}, \citenamefont {Martinis},\ and\ \citenamefont
  {Cleland}}]{Fowler2012}%
  \BibitemOpen
  \bibfield  {author} {\bibinfo {author} {\bibfnamefont {A.~G.}\ \bibnamefont
  {Fowler}}, \bibinfo {author} {\bibfnamefont {M.}~\bibnamefont {Mariantoni}},
  \bibinfo {author} {\bibfnamefont {J.~M.}\ \bibnamefont {Martinis}}, \ and\
  \bibinfo {author} {\bibfnamefont {A.~N.}\ \bibnamefont {Cleland}},\
  }\bibfield  {title} {\bibinfo {title} {Surface codes: towards practical
  large-scale quantum computation},\ }\href
  {https://doi.org/10.1103/physreva.86.032324} {\bibfield  {journal} {\bibinfo
  {journal} {Phys. Rev. A}\ }\textbf {\bibinfo {volume} {86}} (\bibinfo {year}
  {2012})}\BibitemShut {NoStop}%
\bibitem [{\citenamefont {Inaba}\ \emph {et~al.}(2014)\citenamefont {Inaba},
  \citenamefont {Tokunaga}, \citenamefont {Tamaki}, \citenamefont {Igeta},\
  and\ \citenamefont {Yamashita}}]{Inaba2014}%
  \BibitemOpen
  \bibfield  {author} {\bibinfo {author} {\bibfnamefont {K.}~\bibnamefont
  {Inaba}}, \bibinfo {author} {\bibfnamefont {Y.}~\bibnamefont {Tokunaga}},
  \bibinfo {author} {\bibfnamefont {K.}~\bibnamefont {Tamaki}}, \bibinfo
  {author} {\bibfnamefont {K.}~\bibnamefont {Igeta}}, \ and\ \bibinfo {author}
  {\bibfnamefont {M.}~\bibnamefont {Yamashita}},\ }\bibfield  {title} {\bibinfo
  {title} {High-fidelity cluster state generation for ultracold atoms in an
  optical lattice},\ }\href
  {https://link.aps.org/doi/10.1103/PhysRevLett.112.110501} {\bibfield
  {journal} {\bibinfo  {journal} {Phys. Rev. Lett.}\ }\textbf {\bibinfo
  {volume} {112}} (\bibinfo {year} {2014})}\BibitemShut {NoStop}%
\bibitem [{\citenamefont {Einstein}\ \emph {et~al.}(1935)\citenamefont
  {Einstein}, \citenamefont {Podolsky},\ and\ \citenamefont
  {Rosen}}]{Einstein1935}%
  \BibitemOpen
  \bibfield  {author} {\bibinfo {author} {\bibfnamefont {A.}~\bibnamefont
  {Einstein}}, \bibinfo {author} {\bibfnamefont {B.}~\bibnamefont {Podolsky}},
  \ and\ \bibinfo {author} {\bibfnamefont {N.}~\bibnamefont {Rosen}},\
  }\bibfield  {title} {\bibinfo {title} {Can quantum-mechanical description of
  physical reality be considered complete?}\ }\href
  {https://doi.org/10.1103/physrev.47.777} {\bibfield  {journal} {\bibinfo
  {journal} {Phys. Rev.}\ }\textbf {\bibinfo {volume} {47}},\ \bibinfo {pages}
  {777--780} (\bibinfo {year} {1935})}\BibitemShut {NoStop}%
\bibitem [{\citenamefont {Barnett}(2009)}]{book_Barnett}%
  \BibitemOpen
  \bibfield  {author} {\bibinfo {author} {\bibfnamefont {S.~M.}\ \bibnamefont
  {Barnett}},\ }\href
  {https://global.oup.com/academic/search?q=quantum+information&cc=fr&lang=en}
  {\emph {\bibinfo {title} {{Q}uantum {I}nformation}}}\ (\bibinfo  {publisher}
  {Oxford University Press},\ \bibinfo {year} {2009})\BibitemShut {NoStop}%
\bibitem [{\citenamefont {Junge}\ and\ \citenamefont
  {Palazuelos}(2011)}]{Junge2011}%
  \BibitemOpen
  \bibfield  {author} {\bibinfo {author} {\bibfnamefont {M.}~\bibnamefont
  {Junge}}\ and\ \bibinfo {author} {\bibfnamefont {C.}~\bibnamefont
  {Palazuelos}},\ }\bibfield  {title} {\bibinfo {title} {Large violation of
  {B}ell inequalities with low entanglement},\ }\href
  {https://doi.org/10.1007/s00220-011-1296-8} {\bibfield  {journal} {\bibinfo
  {journal} {Commu. Math. Phys.}\ }\textbf {\bibinfo {volume} {306}},\ \bibinfo
  {pages} {695--746} (\bibinfo {year} {2011})}\BibitemShut {NoStop}%
\bibitem [{\citenamefont {Vidick}\ and\ \citenamefont
  {Wehner}(2011)}]{Vidick2011}%
  \BibitemOpen
  \bibfield  {author} {\bibinfo {author} {\bibfnamefont {T.}~\bibnamefont
  {Vidick}}\ and\ \bibinfo {author} {\bibfnamefont {S.}~\bibnamefont
  {Wehner}},\ }\bibfield  {title} {\bibinfo {title} {More nonlocality with less
  entanglement},\ }\href {https://doi.org/10.1103/physreva.83.052310}
  {\bibfield  {journal} {\bibinfo  {journal} {Phys. Rev. A}\ }\textbf {\bibinfo
  {volume} {83}} (\bibinfo {year} {2011})}\BibitemShut {NoStop}%
\bibitem [{\citenamefont {Liang}\ \emph {et~al.}(2011)\citenamefont {Liang},
  \citenamefont {V{\'{e}}rtesi},\ and\ \citenamefont {Brunner}}]{Liang2011}%
  \BibitemOpen
  \bibfield  {author} {\bibinfo {author} {\bibfnamefont {Y.-C.}\ \bibnamefont
  {Liang}}, \bibinfo {author} {\bibfnamefont {T.}~\bibnamefont
  {V{\'{e}}rtesi}}, \ and\ \bibinfo {author} {\bibfnamefont {N.}~\bibnamefont
  {Brunner}},\ }\bibfield  {title} {\bibinfo {title} {Semi-device-independent
  bounds on entanglement},\ }\href {https://doi.org/10.1103/physreva.83.022108}
  {\bibfield  {journal} {\bibinfo  {journal} {Phys. Rev. A}\ }\textbf {\bibinfo
  {volume} {83}} (\bibinfo {year} {2011})}\BibitemShut {NoStop}%
\bibitem [{\citenamefont {Heisenberg}(1930)}]{book_Heisenberg}%
  \BibitemOpen
  \bibfield  {author} {\bibinfo {author} {\bibfnamefont {W.}~\bibnamefont
  {Heisenberg}},\ }\href {http://store.doverpublications.com/0486601137.html}
  {\emph {\bibinfo {title} {The Physical Principles of Quantum Theory}}}\
  (\bibinfo  {publisher} {University of Chicago Press},\ \bibinfo {year}
  {1930})\BibitemShut {NoStop}%
\bibitem [{\citenamefont {Bohm}(1952{\natexlab{a}})}]{Bohm1952I}%
  \BibitemOpen
  \bibfield  {author} {\bibinfo {author} {\bibfnamefont {D.}~\bibnamefont
  {Bohm}},\ }\bibfield  {title} {\bibinfo {title} {A suggested interpretation
  of the quantum theory in terms of ``hidden" variables {I}},\ }\href
  {https://doi.org/10.1103/physrev.85.166} {\bibfield  {journal} {\bibinfo
  {journal} {Phys. Rev.}\ }\textbf {\bibinfo {volume} {85}},\ \bibinfo {pages}
  {166--179} (\bibinfo {year} {1952}{\natexlab{a}})}\BibitemShut {NoStop}%
\bibitem [{\citenamefont {Bohm}(1952{\natexlab{b}})}]{Bohm1952II}%
  \BibitemOpen
  \bibfield  {author} {\bibinfo {author} {\bibfnamefont {D.}~\bibnamefont
  {Bohm}},\ }\bibfield  {title} {\bibinfo {title} {A suggested interpretation
  of the quantum theory in terms of ``hidden" variables {II}},\ }\href
  {https://doi.org/10.1103/physrev.85.180} {\bibfield  {journal} {\bibinfo
  {journal} {Phys. Rev.}\ }\textbf {\bibinfo {volume} {85}},\ \bibinfo {pages}
  {180--193} (\bibinfo {year} {1952}{\natexlab{b}})}\BibitemShut {NoStop}%
\bibitem [{\citenamefont {Aspect}\ \emph {et~al.}(1982)\citenamefont {Aspect},
  \citenamefont {Dalibard},\ and\ \citenamefont {Roger}}]{Aspect1982}%
  \BibitemOpen
  \bibfield  {author} {\bibinfo {author} {\bibfnamefont {A.}~\bibnamefont
  {Aspect}}, \bibinfo {author} {\bibfnamefont {J.}~\bibnamefont {Dalibard}}, \
  and\ \bibinfo {author} {\bibfnamefont {G.}~\bibnamefont {Roger}},\ }\bibfield
   {title} {\bibinfo {title} {Experimental test of {B}ell's inequalities using
  time-varying analyzers},\ }\href {\doibase 10.1103/physrevlett.49.1804}
  {\bibfield  {journal} {\bibinfo  {journal} {Phys. Rev. Lett.}\ }\textbf
  {\bibinfo {volume} {49}},\ \bibinfo {pages} {1804--1807} (\bibinfo {year}
  {1982})}\BibitemShut {NoStop}%
\bibitem [{\citenamefont {Vaidman}(2012)}]{Vaidman2012}%
  \BibitemOpen
  \bibfield  {author} {\bibinfo {author} {\bibfnamefont {L.}~\bibnamefont
  {Vaidman}},\ }\bibfield  {title} {\bibinfo {title} {Role of potentials in the
  {A}haronov-{B}ohm effect},\ }\href
  {https://doi.org/10.1103/physreva.86.040101} {\bibfield  {journal} {\bibinfo
  {journal} {Phys. Rev. A}\ }\textbf {\bibinfo {volume} {86}} (\bibinfo {year}
  {2012})}\BibitemShut {NoStop}%
\bibitem [{\citenamefont {Pauli}(1941)}]{Pauli1941}%
  \BibitemOpen
  \bibfield  {author} {\bibinfo {author} {\bibfnamefont {W.}~\bibnamefont
  {Pauli}},\ }\bibfield  {title} {\bibinfo {title} {Relativistic field theories
  of elementary particles},\ }\href {\doibase 10.1103/revmodphys.13.203}
  {\bibfield  {journal} {\bibinfo  {journal} {Rev. Mod. Phys.}\ }\textbf
  {\bibinfo {volume} {13}},\ \bibinfo {pages} {203--232} (\bibinfo {year}
  {1941})}\BibitemShut {NoStop}%
\bibitem [{\citenamefont {Heisenberg}(1932{\natexlab{a}})}]{Heisenberg1932I}%
  \BibitemOpen
  \bibfield  {author} {\bibinfo {author} {\bibfnamefont {W.}~\bibnamefont
  {Heisenberg}},\ }\bibfield  {title} {\bibinfo {title} {\"{U}ber den {B}au der
  {A}tomkerne {I}},\ }\href {\doibase 10.1007/bf01342433} {\bibfield  {journal}
  {\bibinfo  {journal} {Z. Physik}\ }\textbf {\bibinfo {volume} {77}},\
  \bibinfo {pages} {1--11} (\bibinfo {year} {1932}{\natexlab{a}})}\BibitemShut
  {NoStop}%
\bibitem [{\citenamefont {Heisenberg}(1932{\natexlab{b}})}]{Heisenberg1932II}%
  \BibitemOpen
  \bibfield  {author} {\bibinfo {author} {\bibfnamefont {W.}~\bibnamefont
  {Heisenberg}},\ }\bibfield  {title} {\bibinfo {title} {\"{U}ber den {B}au der
  {A}tomkerne {II}},\ }\href {\doibase 10.1007/bf01337585} {\bibfield
  {journal} {\bibinfo  {journal} {Z. Physik}\ }\textbf {\bibinfo {volume}
  {78}},\ \bibinfo {pages} {156--164} (\bibinfo {year}
  {1932}{\natexlab{b}})}\BibitemShut {NoStop}%
\bibitem [{\citenamefont {Heisenberg}(1933)}]{Heisenberg1932III}%
  \BibitemOpen
  \bibfield  {author} {\bibinfo {author} {\bibfnamefont {W.}~\bibnamefont
  {Heisenberg}},\ }\bibfield  {title} {\bibinfo {title} {\"{U}ber den {B}au der
  {A}tomkerne {III}},\ }\href {\doibase 10.1007/bf01335696} {\bibfield
  {journal} {\bibinfo  {journal} {Z. Physik}\ }\textbf {\bibinfo {volume}
  {80}},\ \bibinfo {pages} {587--596} (\bibinfo {year} {1933})}\BibitemShut
  {NoStop}%
\end{thebibliography}

\end{document}